\documentclass[longauth]{aa} 

\usepackage{threeparttable}
\usepackage{longtable}

\usepackage[utf8]{inputenc}
\usepackage[T1]{fontenc}
\usepackage{xspace}
\usepackage{hyperref}
\usepackage{graphicx}
\usepackage{color}
\usepackage{lineno}
\usepackage{graphics}
\usepackage{subcaption}
\usepackage[section]{placeins}
\usepackage{longtable}
\usepackage[export]{adjustbox}
\usepackage{natbib}
\usepackage{txfonts}
\usepackage{tikz}
\usepackage{caption}
\usepackage{pdflscape}
\usepackage{geometry}
\usepackage{amsmath}
\usepackage{bm}
\usepackage{float}
\usepackage{rotating}
\usepackage{pdflscape}
\makeatletter
\newcommand\subsubsubsection{\@startsection{paragraph}{4}{\z@}%
  {-3.25ex \@plus -1ex \@minus -.2ex}%
  {1.5ex \@plus .2ex}%
  {\normalfont\normalsize}}
\makeatother

\newcommand{\exofasttwo}{{\tt EXOFASTv2}}
\newcommand{\bjdtdb}{\ensuremath{\rm {BJD_{TDB}}}}

\newcommand{\msun}{\ensuremath{\,M_\odot}}
\newcommand{\rsun}{\ensuremath{\,R_\odot}}
\newcommand{\lsun}{\ensuremath{\,L_\odot}}

\begin{document}

    \title{Characterization of seven transiting systems including four warm Jupiters from SOPHIE and TESS}
    
   \author{N.~Heidari\inst{\ref{iap},\thanks{CNES postdoctoral fellowship}},
          G.~H\'ebrard\inst{\ref{iap},\ref{OHP}},
          E. ~Martioli\inst{\ref{eder},\ref{iap}},
          J.~D.\ Eastman \inst{\ref{jason}},
          J.~M.~Jackson\inst{\ref{Jackson}},
          X.~Delfosse \inst{\ref{gronobl}},
          A.~ Jord\'an\inst{\ref{Jordan1},\ref{Jordan2}},
          A.~C.~M.~Correia \inst{\ref{correia},\ref{laskar}},
          S. Sousa \inst{\ref{santos1},\ref{santos2}},
          D. Dragomir \inst{\ref{diana}},
          T.~Forveille \inst{\ref{gronobl}},
          I. Boisse\inst{\ref{lam},\ref{OHP}},
          S.~A.~Giacalone\inst{\ref{triceratop},\ref{triceratop2}},
          R.~ F. D\'iaz \inst{\ref{diaz}},
          R. Brahm\inst{\ref{Jordan1},\ref{Jordan2}},
          D.~Almasian \inst{\ref{iran}},
          J.~M.~Almenara\inst{\ref{gronobl},\ref{geneva}},
          A.~Bieryla\inst{\ref{alyson}},
          K.~Barkaoui\inst{\ref{Khalid},\ref{Khalid2},\ref{Khalid3}},
          D.~Baker\inst{\ref{baker}},
          S.~C.~C~.~Barros\inst{\ref{santos1},\ref{santos2}},
          X.~Bonfils\inst{\ref{gronobl}},
          A.~Carmona\inst{\ref{gronobl}},
          K. A. Collins\inst{\ref{alyson}},
          P.~Cort\'es-Zuleta \inst{\ref{pia1},\ref{pia2}},
          M.~Deleuil\inst{\ref{lam}},
          O. D. S. Demangeon \inst{\ref{santos1},\ref{santos2}},
          B.~Edwards\inst{\ref{billy}},
          J.~Eberhardt\inst{\ref{Eberhardt}},
          N.~Espinoza\inst{\ref{noster}},
          N. Eisner\inst{\ref{nora},\ref{nora1}},
          D.~L.~Feliz\inst{\ref{feliz}},
          A.~C.~Frommer\inst{\ref{jason}},
          A.~Fukui\inst{\ref{japan},\ref{enrik}},
          F.~Grau\inst{\ref{grou}},
          A.~F.~Gupta\inst{\ref{arvind}},
          N.~Hara\inst{\ref{lam}},
          M.~J.~Hobson\inst{\ref{geneva}},
          T.~ Henning\inst{\ref{Eberhardt}},
          S.~B.~Howell\inst{\ref{nasa}},
          J.~ M.~ Jenkins\inst{\ref{nasa}},
          F.~Kiefer \inst{\ref{Lesia}},
          D.~M.~LaCourse\inst{\ref{Daryll}},
          J.~Laskar\inst{\ref{laskar}},
          N.~Law\inst{\ref{law}},
          A.~W.~Mann\inst{\ref{law}},
          F.~Murgas\inst{\ref{Murgas1},\ref{Murgas2}},
          C.~Moutou \inst{\ref{Toulouse}},
          N.~Narita \inst{\ref{japan},\ref{japan2},\ref{enrik}},
          E.~Palle\inst{\ref{enrik},\ref{enrik1}},
          H.~M.~Relles\inst{\ref{alyson}},
          K.~G.\ Stassun\inst{\ref{Stassun}},
          J.~Serrano Bell\inst{\ref{diaz}},
          R.~ P.~Schwarz\inst{\ref{alyson}},
          G.~Srdoc\inst{\ref{Croatia}},
          P.~ A. Str{\o}m \inst{\ref{wilson1},\ref{wilson2}},
          B.~Safonov\inst{\ref{stefano}},
          P.~Sarkis\inst{\ref{Eberhardt}},
          M.~Schlecker\inst{\ref{martin}},
          M.~Tala Pinto\inst{\ref{Jordan1},\ref{Jordan2}},
          J.~Pepper\inst{\ref{jashua}},
          F. I. Rojas\inst{\ref{chili},\ref{Jordan2}},
          J.~D.~Twicken\inst{\ref{nasa},\ref{jpseph}},
          T.~Trifonov\inst{\ref{Eberhardt},\ref{Trifonov},\ref{Trifonov1}},
          S.~Villanueva Jr\inst{\ref{Villanueva}},
          C.~ N.~Watkins\inst{\ref{alyson}},
          J.~N.~Winn\inst{\ref{Josh}},
          C.~Ziegler\inst{\ref{Ziegler}}}

   \institute{Institut d'astrophysique de Paris, UMR 7095 CNRS université pierre et marie curie, 98 bis, boulevard Arago,  75014, Paris \label{iap}
   \and
   Observatoire de Haute-Provence, CNRS, Universit\'e d'Aix-Marseille, 04870 Saint-Michel-l'Observatoire, France \label{OHP}
   \and
   Laboratório Nacional de Astrofísica, Rua Estados Unidos 154, 37504-364 Itajubá, MG, Brazil \label{eder}
   \and
    Center for Astrophysics \textbar \ Harvard \& Smithsonian, 60 Garden St, Cambridge, MA 02138, USA \label{jason}
    \and
    Astronomy Department and Van Vleck Observatory, Wesleyan University, 96 Foss Hill Drive, Middletown, CT 06459, USA \label{Jackson}
    \and
    Univ. Grenoble Alpes, CNRS, IPAG, 38000 Grenoble, France\label{gronobl} 
    \and
    Facultad de Ingenier\'{i}a y Ciencias, Universidad Adolfo Ib\'{a}\~{n}ez, Av. Diagonal las Torres 2640, Pe\~{n}alol\'{e}n, Santiago, Chile \label{Jordan1}
    \and
     Millennium Institute for Astrophysics, Chile \label{Jordan2}
    \and
    CFisUC, Departamento de F\'isica, Universidade de Coimbra, 3004-516 Coimbra, Portugal\label{correia}
    \and
    ASD, IMCCE, Observatoire de Paris, PSL Universit\'e, 77 Av. Denfert-Rochereau, 75014 Paris, France\label{laskar}
    \and
     Instituto de Astrof\'isica e Ci\^encias do Espa\c{c}o, Universidade do Porto, CAUP, Rua das Estrelas, 4150-762 Porto, Portugal.\label{santos1}
    \and     
     Departamento de F\'isica e Astronomia, Faculdade de Ci\^encias, Universidade do Porto, Rua do Campo Alegre, 4169-007 Porto, Portugal.\label{santos2}
    \and
    Department of Physics \& Astronomy, University of New Mexico, 1919 Lomas Boulevard NE, Albuquerque, NM 87131, USA \label{diana}
    \and
     Aix Marseille Univ, CNRS, CNES, LAM, Marseille, France.\label{lam}
     \and
     Department of Astronomy, California Institute of Technology, Pasadena, CA 91125, USA\label{triceratop}
     \and
     NSF Astronomy and Astrophysics Postdoctoral Fellow\label{triceratop2}
     \and
    International Center for Advanced Studies (ICAS) and ICIFI (CONICET), ECyT-UNSAM, Campus Miguelete, 25 de Mayo y Francia, (1650) Buenos Aires, Argentina.\label{diaz}
        \and   
         Department of Physics, Shahid Beheshti University, Tehran, Iran \label{iran}
    \and
     Observatoire de Gen\`eve,  Universit\'e de Gen\`eve, Chemin Pegasi, 51, 1290 Sauverny, Switzerland\label{geneva}
     \and
    Center for Astrophysics \textbar \ Harvard \& Smithsonian, 60 Garden
Street, Cambridge, MA 02138, USA \label{alyson}
    \and
    Astrobiology Research Unit, Universit\'e de Li\`ege, 19C All\'ee du 6
     Ao\^ut, 4000 Li\`ege, Belgium \label{Khalid}
     \and
    Department of Earth, Atmospheric and Planetary Science, Massachusetts Institute of Technology, 77 Massachusetts Avenue, Cambridge, MA 02139, USA\label{Khalid2}
    \and
    Instituto de Astrof\'isica de Canarias (IAC), Calle V\'ia L\'actea s/n,
38200, La Laguna, Tenerife, Spain\label{Khalid3}
   \and
   Physics Department, Austin College, Sherman, TX 75090, USA\label{baker}
   \and
   Cerro Tololo Inter-American Observatory, Casilla 603, La Serena, Chile\label{chili2}
  \and
  SUPA, School of Physics and Astronomy, University of St Andrews, North Haugh, St Andrews, KY169SS, UK \label{pia1}
\and
  Centre for Exoplanet Science, University of St Andrews, North Haugh, St
Andrews, KY169SS, UK\label{pia2}
\and
SRON, Netherlands Institute for Space Research, Niels Bohrweg 4, NL-2333 CA, Leiden, The Netherlands\label{billy}
\and
 Max-Planck-Institut für Astronomie, Königstuhl 17, D-69117 Heidelberg, Germany\label{Eberhardt}
 \and
Space Telescope Science Institute, 3700 San Martin Drive, Baltimore,
MD 21218, USA\label{noster}
\and
Center for Computational Astrophysics, Flatiron Institute, 162 Fifth Avenue, New York, NY 10010, USA\label{nora}
\and
Department of Astrophysical Sciences, Princeton University, Princeton, NJ 08544, USA\label{nora1}
\and
American Museum of Natural History, New York, NY, USA\label{feliz}
\and
Observatory de Ca l'Ou, Sant Martí Sesgueioles, GEECAT, Barcelona Spain\label{grou}
\and
U.S. National Science Foundation National Optical-Infrared Astronomy Research Laboratory, 950 N.\ Cherry Ave., Tucson, AZ 85719, USA\label{arvind}
\and
NASA Ames Research Center, Moffett Field, CA 94035, USA\label{nasa}
\and
 LESIA, Observatoire de Paris, Université PSL, CNRS, Sorbonne Université, Université de Paris, 5 place Jules Janssen, 92195, Meudon, France \label{Lesia}
 \and
Department of Physics and Astronomy, The University of North Carolina at Chapel Hill, Chapel Hill, NC 27599-3255, USA\label{law}
\and
7507 52nd Pl NE, Marysville, WA 98270, USA\label{Daryll}
\and
Instituto de Astrof\'isica de Canarias (IAC), E-38205 La Laguna,
Tenerife, Spain\label{Murgas1}
\and
Departamento de Astrof\'isica, Universidad de La Laguna (ULL), E-38206
La Laguna, Tenerife, Spain\label{Murgas2} 
\and
 Univ. de Toulouse, CNRS, IRAP, 14 avenue Belin, 31400 Toulouse, France \label{Toulouse}
\and
Komaba Institute for Science, The University of Tokyo, 3-8-1 Komaba,
Meguro, Tokyo 153-8902, Japan\label{japan}
\and
Astrobiology Center, 2-21-1 Osawa, Mitaka, Tokyo 181-8588, Japan\label{japan2}
\and
Instituto de Astrof\'\i sica de Canarias (IAC), 38205 La Laguna,
Tenerife, Spain\label{enrik}
\and
Departamento de Astrof\'\i sica, Universidad de La Laguna (ULL), 38206,
La Laguna, Tenerife, Spain\label{enrik1}  
\and  
   Vanderbilt University, Department of Physics \& Astronomy, 6301 Stevenson Center Ln., Nashville, TN 37235, USA \label{Stassun}
\and
Kotizarovci Observatory, Sarsoni 90, 51216 Viskovo, Croatia\label{Croatia}
\and
Department of Physics, University of Warwick, Coventry, CV4 7AL, UK\label{wilson1}
\and
 Centre for Exoplanets and Habitability, University of Warwick, Gibbet Hill Road, Coventry, CV4 7AL, UK \label{wilson2}
 \and
 Sternberg Astronomical Institute of Lomonosov Moscow State University, Moscow, 119234 Russia\label{stefano}
 \and
 Steward Observatory and Department of Astronomy, The University of Arizona, Tucson, AZ 85721, USA\label{martin}
\and
 Department of Physics, Lehigh University, 16 Memorial Drive East, Bethlehem, PA 18015, USA\label{jashua}
 \and
Instituto de Astrof\'isica, Facultad de F\'isica, Pontificia
Universidad Cat\'olica de Chile, Chile\label{chili}
\and
SETI Institute, Mountain View, CA 94043 USA\label{jpseph}
\and
Department of Astronomy, Sofia University ``St Kliment Ohridski'', 5 James Bourchier
Blvd, BG-1164 Sofia, Bulgaria\label{Trifonov}
\and
Landessternwarte, Zentrum f\"ur Astronomie der Universit\"at Heidelberg, K\"onigstuhl 12, D-69117 Heidelberg, Germany\label{Trifonov1}
\and
Department of Physics and Kavli Institute for Astrophysics and Space Research, Massachusetts Institute of Technology, Cambridge, MA 02139, USA\label{Villanueva}
\and
Princeton University, Princeton, NJ, USA\label{Josh}
\and
Department of Physics, Engineering and Astronomy, Stephen F. Austin State University, 1936 North Street, Nacogdoches, TX 75962, USA\label{Ziegler}
}
   \date{Received XX, 2024; accepted XX, 2024}

  \abstract
   {While several thousand exoplanets are now confirmed, the number of known transiting warm Jupiters ($10 ~\text{d} < \text{period} < 200 ~ \text{d}$) remains relatively small. These planets are generally believed to have formed outside the snowline and migrated to their current orbits. Because they are sufficiently distant from their host stars, they mitigate proximity effects and so offer valuable insights into planet formation and evolution. Here, we present the study of seven systems, three of which—TOI-2295, TOI-2537, and TOI-5110—are newly discovered planetary systems. Through the analysis of TESS photometry, SOPHIE radial velocities, and high-spatial resolution imaging, we found that TOI-2295b, TOI-2537b, and TOI-5110b are transiting warm Jupiters with orbital periods ranging from 30 to 94 d, masses between 0.9 and 2.9 $M_{\rm{J}}$, and radii ranging from 1.0 to 1.5 $R_{\rm{J}}$. Both TOI-2295 and TOI-2537 each harbor at least one additional, outer planet. Their outer planets —TOI-2295c and TOI-2537c— are characterized by orbital periods of 966.5$^{+4.3}_{-4.2}$ and  1920$^{+230}_{-140}$ d, respectively, and minimum masses of 5.61$^{+0.23}_{-0.24}$ and 7.23$^{+0.52}_{-0.45}$ $M_{\rm{J}}$, respectively. We also investigated and characterized the two recently reported warm Jupiters TOI-1836b and TOI-5076b, which we independently detected in SOPHIE RVs. Our new data allow for further discussion of their nature and refinement of their parameters. Additionally, we study the planetary candidates TOI-4081.01 and TOI-4168.01. For TOI-4081.01, despite our detection in radial velocities, we cannot rule out perturbation by a blended eclipsing binary and thus exercise caution regarding its planetary nature. On the other hand, we identify TOI-4168.01 as a firm false positive; its radial velocity curve exhibits a large amplitude in an anti-phase relation with the transit ephemeris observed by TESS, indicating that the detected event is the eclipse of a secondary star rather than a planetary transit. Finally, we highlight interesting characteristics of these new planetary systems. The transits of TOI-2295b are highly grazing, with an impact parameter of 1.056$^{+0.063}_{-0.043}$. This leaves its radius uncertain but potentially makes it an interesting probe of gravitational dynamics in its two-planet system, as transit shapes for grazing planets are highly sensitive to even small variations in inclination. TOI-2537b, in turn, is a temperate Jupiter with an effective temperature of 307$\pm$15 K and can serve as a valuable low-irradiation control for models of hot Jupiter inflation anomalies. We also detected significant transit timing variations (TTVs) for TOI-2537b, which are likely caused by gravitational interactions with the outer planet TOI-2537c. Further transit observations are needed to refine the analysis of these TTVs and enhance our understanding of the system’s dynamics. Finally, TOI-5110b stands out due to its orbital eccentricity of 0.745$^{+0.030}_{-0.027}$, one of the highest planetary eccentricities discovered thus far. We find no conclusive evidence for an external companion, but an unseen planet with a semi-amplitude smaller than 10 m/s could nonetheless still be exciting its eccentricity.}

\titlerunning{long-period planet detection and Characterization}
\authorrunning{N. Heidari \& G.~H\'ebrard et al.}
\maketitle

\section{Introduction}

The origin of warm Jupiters defined here as giant planets with orbital periods ranging from 10~d to 200~d \citep{2018ARAA..56..175D}, remains an incompletely understood puzzle. They may have formed in situ \citep{huang2016warm} or migrated instead from beyond the ice line \citep{antonini2016dynamical}, or may originate from a combination of these two channels \citep[see sect. 4.3 of ][]{2018ARAA..56..175D}. If they migrated, the potential mechanisms include gravitational interactions between the planet and its parent protoplanetary disc \citep{2014prpl.conf..667B}, as well as high-eccentricity migration resulting from gravitational interactions between two planets or between the planet and its host star \citep{petrovich2015hot}. 

Their unique orbital characteristics make warm Jupiters valuable targets for scientific investigation. Unlike hot Jupiters ($R > 6 ~ R_{\oplus}$, $\text{period} < 10 ~\text{d}$), these planets maintain greater distances from their host stars, thereby reducing the influence of tidal and other proximity effects (e.g., atmospheric evaporation) to erase potential migration footprints. Consequently, warm Jupiters can be key diagnostics between various scenarios and pathways in the formation and evolution of both hot and warm Jupiters. In a similar vein, the hot Jupiter radius inflation phenomenon, that is the unexpected observed enlarged radius of giant planets with equilibrium temperatures above 1000 K, remains an unsolved puzzle \citep{miller2011heavy, demory2011lack,thorngren2024hot}. Numerous proposed solutions suggest that additional anomalous energy sources heat the convective interiors of these planets, leading to their larger-than-anticipated sizes \citep[e.g.][]{batygin2011evolution,arras2010thermal,sarkis2021evidence}. Since warm Jupiters have longer orbital periods and consequently lower equilibrium temperatures than hot Jupiters, they represent a valuable control group for models of the mechanisms behind hot Jupiter radius inflation.

Among the warm Jupiter population, those that transit bright host stars provide crucial opportunities for (1) obtaining precise mass and radius measurements to determine their bulk composition, and (2) conducting detailed atmospheric characterization. These aspects are essential for understanding the formation and evolution of these planets \citep{2016ApJ...832...41M,espinoza2017metal,raymond2022planet}. Despite the availability of many ground- and space-based telescopes, such as the Kepler mission and the Transiting Exoplanet Survey Satellite \citep[TESS,][]{ricker2015transiting}, the discovery and characterization of transiting warm Jupiters—particularly those with longer orbital periods (>20 days)—remain limited, accounting for approximately 2\% of all discovered planets as of August 2024.

The observational strategy of the TESS covers a large portion ($>$93\%\footnote{\protect\url{https://science.nasa.gov/mission/nustar/stories/}}) of the sky, but mostly with only about 27~d of baseline coverage every few years. This makes identifying planets with periods above approximately 20~d challenging, as they often appear in TESS light curves as single-transit events. The extended TESS mission often identifies a second transit in a much later sector, making what is known as a duo-transit, but the long elapsed time between the two transits leaves the planet's true period ambiguous. Determining it then requires additional follow-up studies like ground-based photometry and radial velocity (RV).

Here, we present the discovery and characterization of three new transiting warm Jupiters with long orbital periods, TOI-2295b, TOI-2537b, and TOI-5110b, the two first ones being in multi-planetary systems. Furthermore, we investigated and characterized TOI-5076b and TOI-1836b, two planets independently announced in recent papers (\cite{montalto2024gaps} and \cite{chontos2024tess}, respectively), and warrant caution regarding the nature of TOI-4081.01. Lastly, we classify TOI-4168.01 as a clear false positive. In Section \ref{observation}, we describe the observational datasets, including TESS photometric data, RV follow-up observations, and high-spatial resolution imaging. Sections \ref{stellar} and \ref{ident} respectively present the characterization of the host stars and the photometric and spectroscopic analysis of the systems. Finally, we discuss the planetary systems in a broader context in Section \ref{discuss} and provide a summary in Section \ref{summary}.

\section{Observational datasets}

\label{observation}
Here, we describe the diverse datasets utilized for this study, including TESS photometry, follow-up photometric and spectroscopic observations, and high-spatial resolution imaging.

\subsection{TESS}
\label{TESS-photometry}

All seven stars discussed in this study were observed by the TESS space mission. We included the data that had been gathered until August 13, 2024, encompassing TESS Sectors 1 to 79. Some of the targets are scheduled for additional observations in subsequent TESS sectors. Details of the TESS observations are provided in Table\ref{tess_obs}.

\begin{figure*}[h!]
\centering
\includegraphics[width=1.6\columnwidth]{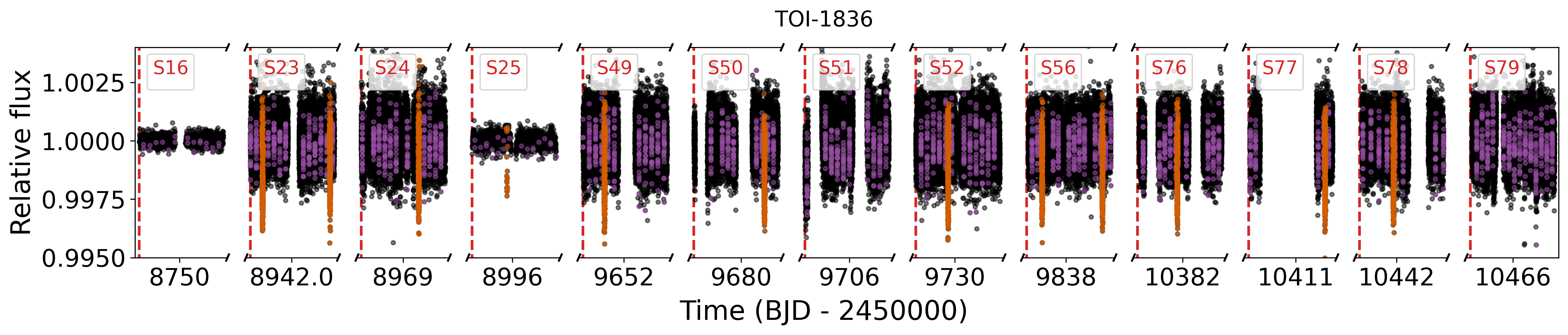}\\
\includegraphics[width=1.6\columnwidth]{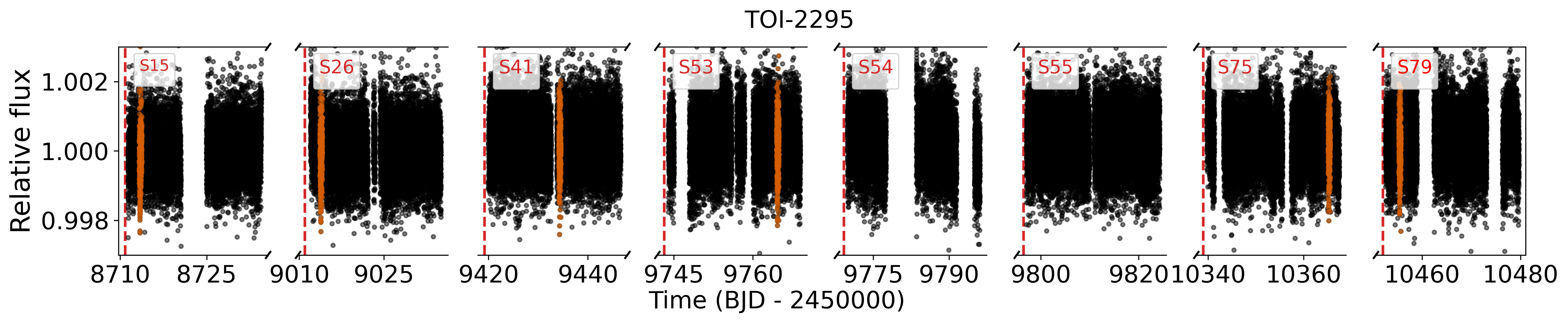}\\
\includegraphics[width=1.6\columnwidth]{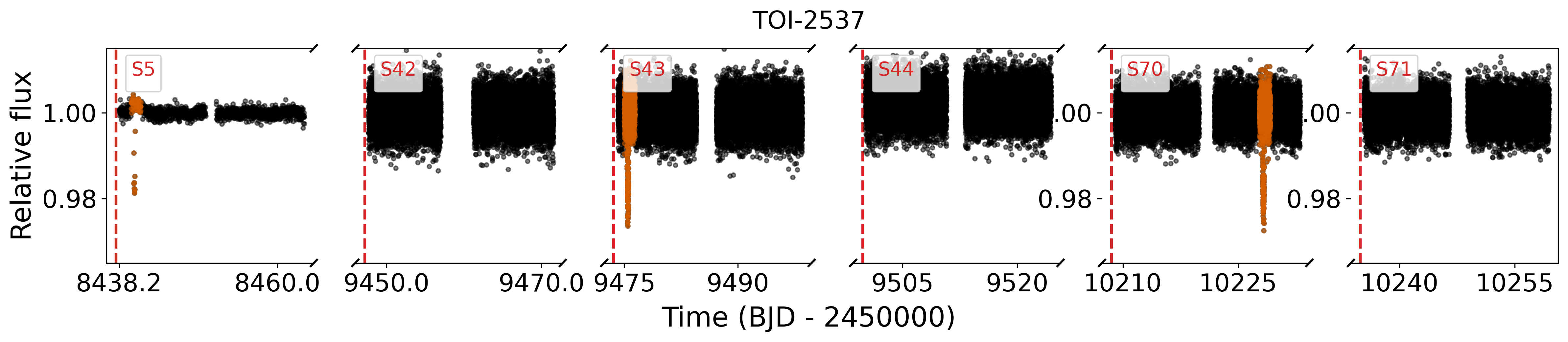}\\
\includegraphics[width=1.6\columnwidth]{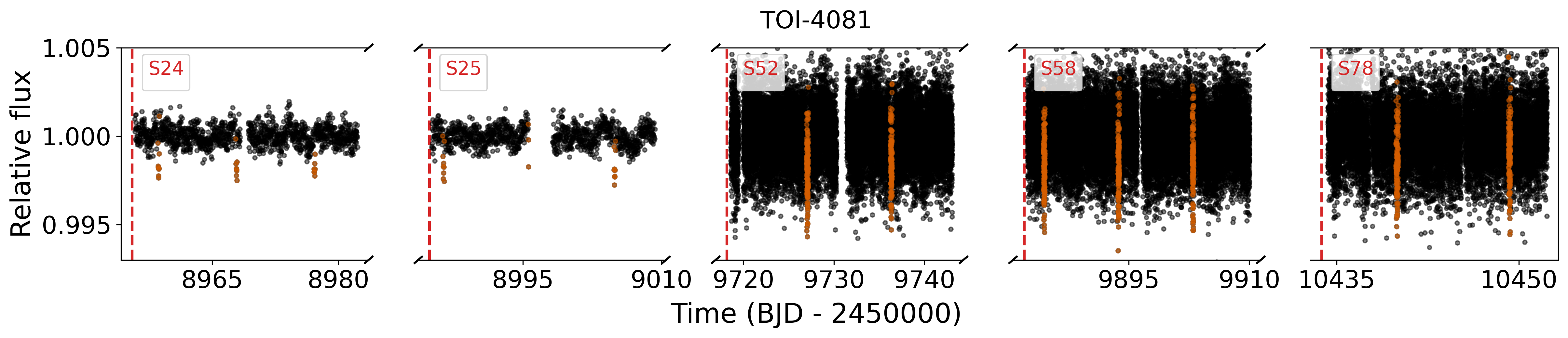}\\
\includegraphics[width=1.6\columnwidth]{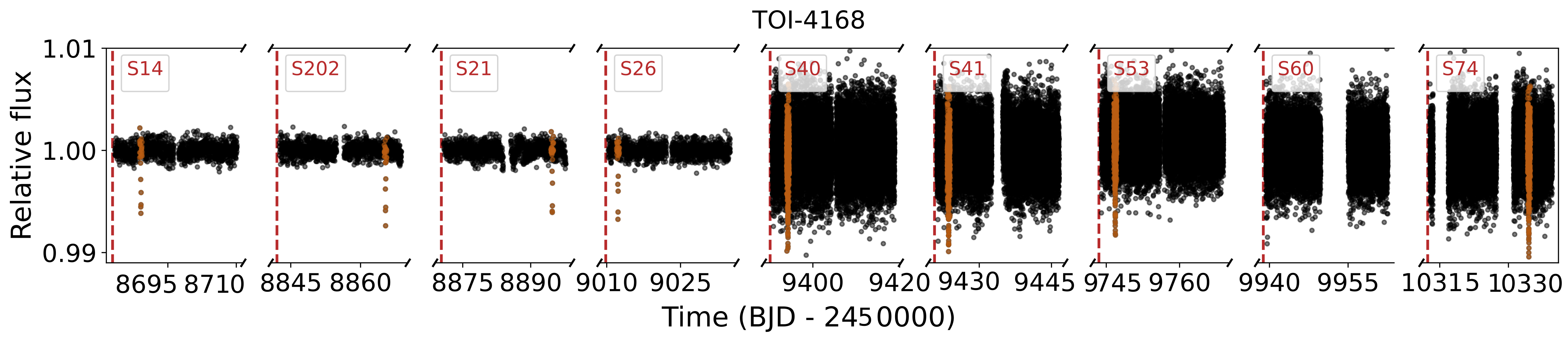}\\
\includegraphics[width=1.6\columnwidth]{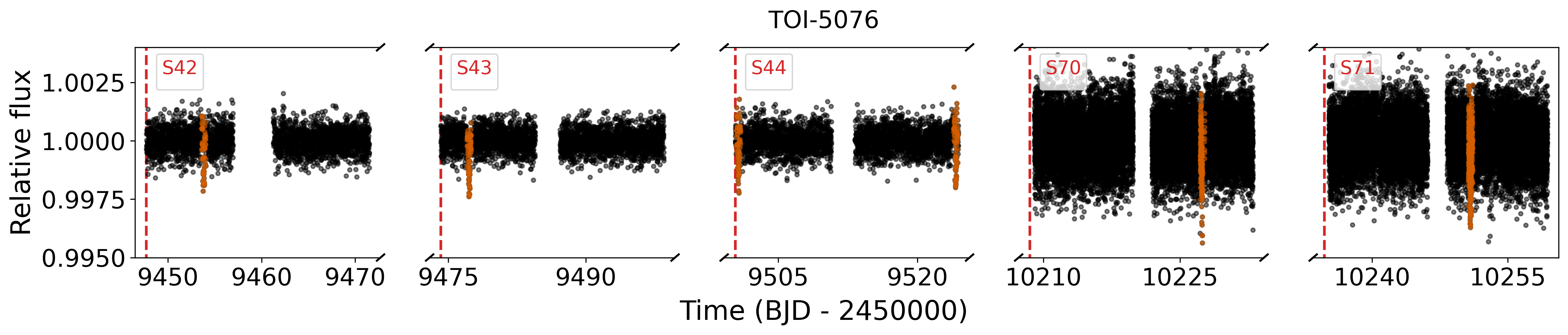}\\
\includegraphics[width=1.6\columnwidth]{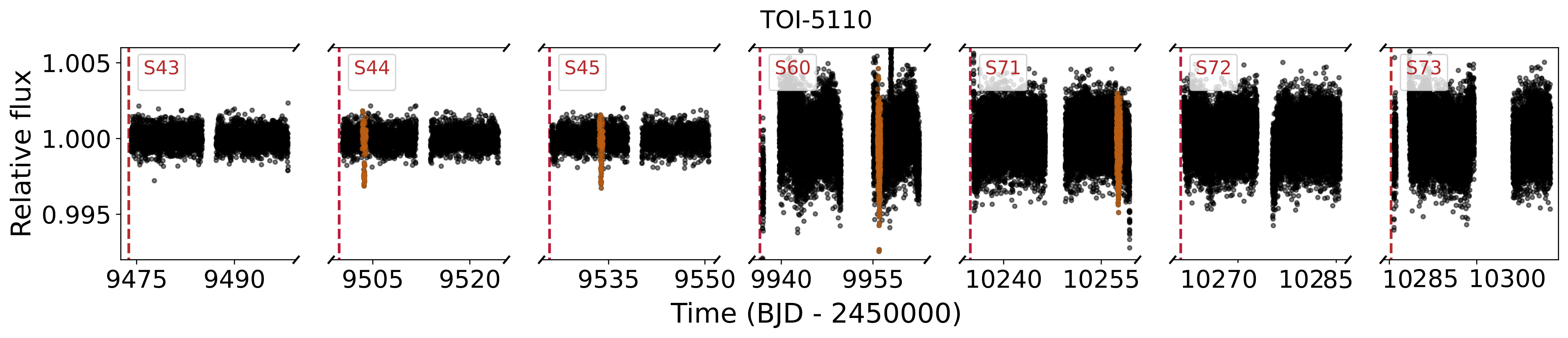}\\
\caption{TESS PDC-SAP light curves for the seven stars analyzed in this paper. Dashed red vertical lines denote the start of individual TESS sectors, labeled by their respective sector number. Data points during transits are highlighted in orange for the first planet candidate and in purple for the second planet candidate (if present).}
\label{all_lc}
\end{figure*}

The data used in this paper, encompassing both 2-minute cadence data and Full Frame Images (FFI), were processed by the Presearch Data Conditioning-Simple Aperture Photometry (PDC-SAP) pipeline \citep{Stumpe2012,stumpe2014multiscale,smith2012kepler,caldwell2020tess}, provided by the Science Processing Operations Center \citep[SPOC;][]{jenkins2016tess} at NASA Ames Research Center. The TESS light curves were retrieved using the \texttt{lightkurve} package \citep{collaboration2018lightkurve} from the Mikulski Archive for Space Telescopes (MAST\footnote{\url{https://mast.stsci.edu/portal/Mashup/Clients/Mast/Portal.html}}). Figure \ref{all_lc} plots the resulting light curves, after masking data points with NaN values or non-zero data quality flags.

\begin{table}[h]
\centering
\begin{tabular}{lll}
\hline
System & Sectors (FFI) & Sectors (2-minutes) \\
\hline
TOI-1836 & 16, 25 & 23, 24, 49, 50, 51, 52, 56,76, 77, \\
&&78, 79\\
TOI-2295 & --- & 15, 26, 41, 53, 54, 55,75, 79 \\
TOI-2537 & 5 & 42, 43, 44, 70, 71 \\
TOI-4081 & 24, 25 & 52, 58, 78 \\
TOI-4168 & 14, 20, 21, 26 & 40, 41, 53, 60, 74 \\
TOI-5076 & 42, 43, 44 & 70, 71 \\
TOI-5110 & 43, 44, 45 & 60, 71, 72, 73\\
\hline
\end{tabular}
\caption[TESS Observations Summary]{TESS observations summary. This table is based on TESS-point Web Tool.\footnote{\url{https://heasarc.gsfc.nasa.gov/wsgi-scripts/TESS/TESS-point_Web_Tool/TESS-point_Web_Tool/wtv_v2.0.py}} The FFIs have exposure times of 1800 seconds for all targets except TOI-5110 and TOI-5076, which have exposure times of 600 seconds.}
\label{tess_obs}
\end{table}

The light curves underwent an automated search for transit-like signals, which are then vetted by the SPOC \citep{2002ApJ...575..493J,2010SPIE.7740E..0DJ} or MIT Quick Look Pipeline \citep[QLP,][]{2020RNAAS...4..204H,2020RNAAS...4..206H} teams. Objects exhibiting planetary candidate signals that successfully passed all tests outlined in the Data Validation report \citep{Twicken:DVdiagnostics2018,2019PASP..131b4506L} including the odd-even transit depth test, the ghost diagnostic test, and the difference-image centroiding test, triggered an alert issued by the TESS Science Office. Subsequently, these targets were assigned TESS Object of Interest (TOI) numbers \citep{guerrero2021tess}. The 7 stars discussed in this paper were identified and alerted as TOIs. No signatures of additional transiting planets were detected in any of the SPOC or QLP runs. 

Two of the targets analyzed in this study initially displayed a single transit in the TESS data. These stars, namely TOI-2295 and TOI-2537, were initially identified and vetted as single-transit candidates by various community groups rather than by the TESS Science Office. TOI-2295 caught the attention of the Planet Hunters TESS (PHT) citizen science project \citep{eisner2021planet}, while TOI-2537 was scrutinized by the TSTPC \citep[Tess Single Transit Planetary Candidate,][]{harris2023separated,burt2021toi} group and by the WINE \citep[Warm GIaNts with tEss,][]{brahm2023three} team. Both stars were subsequently officially recognized as CTOIs (Community TESS Objects of Interest) and later as TOIs.

\begin{figure*}
\centering
\includegraphics[width=0.5\columnwidth]{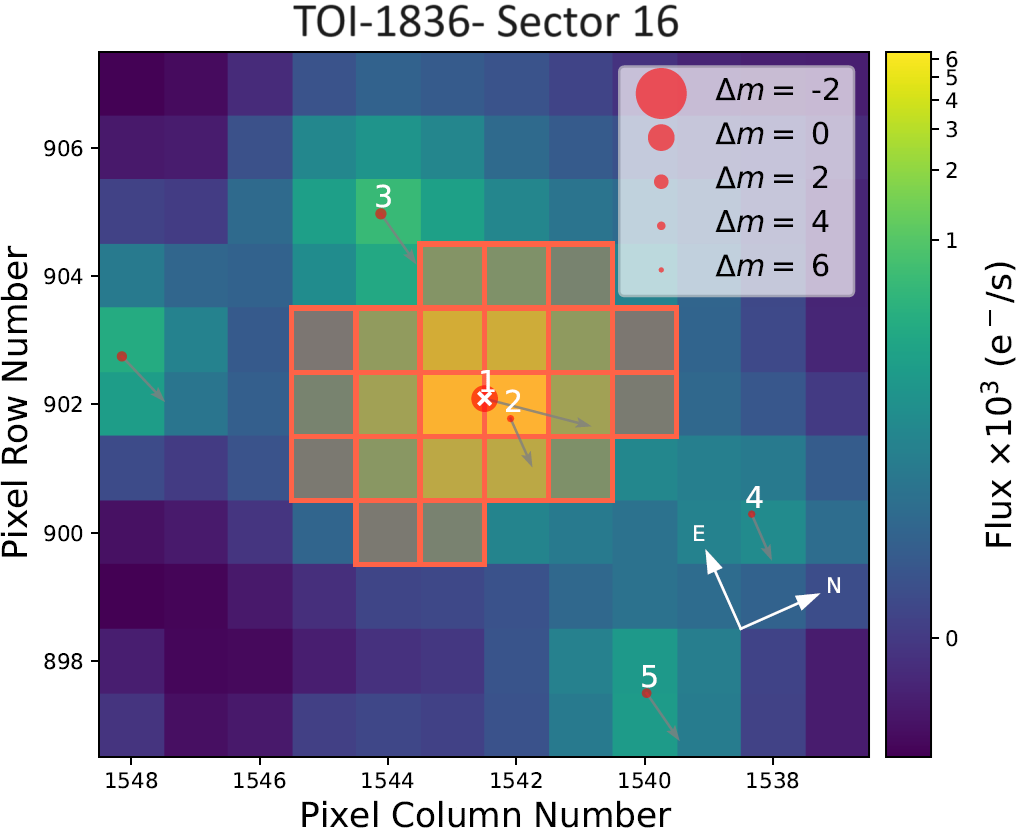}
\includegraphics[width=0.5\columnwidth]{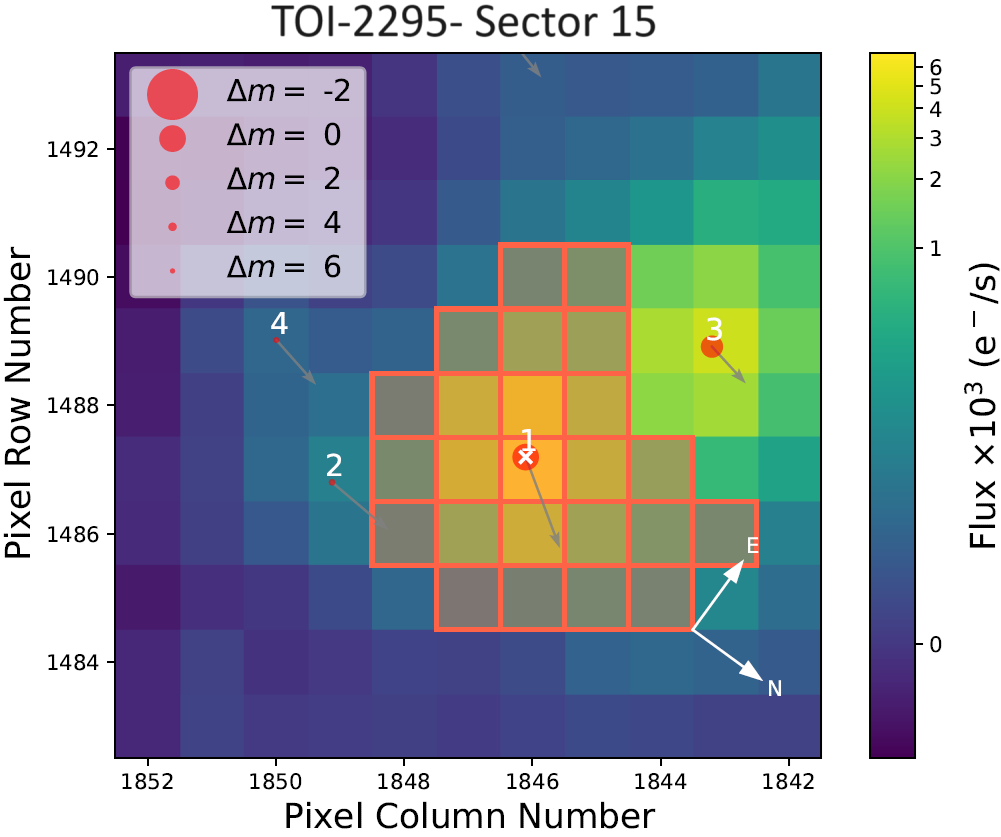}
\includegraphics[width=0.5\columnwidth]{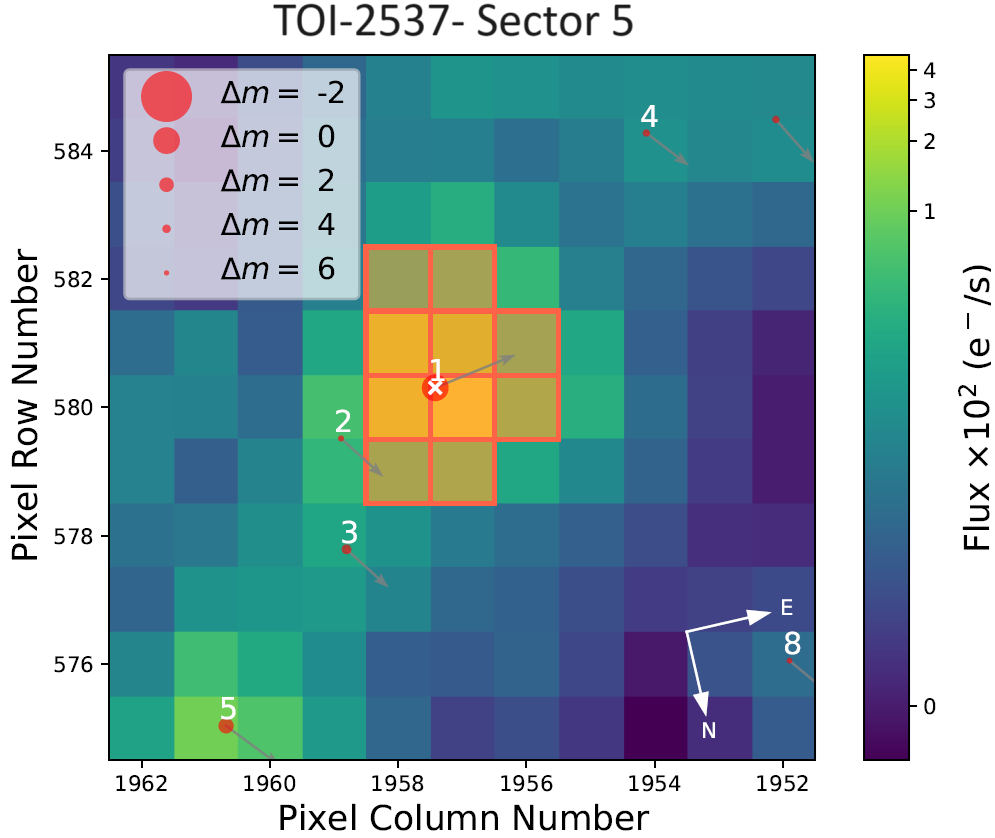}
\includegraphics[width=0.5\columnwidth]{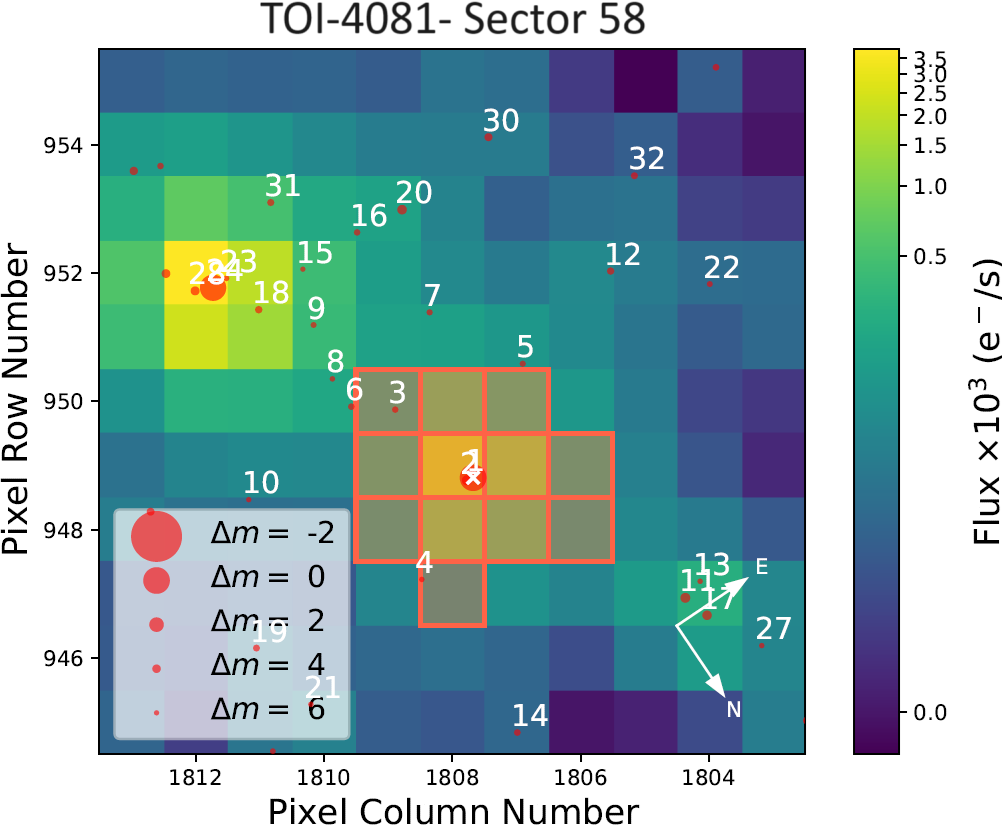}
\includegraphics[width=0.5\columnwidth]{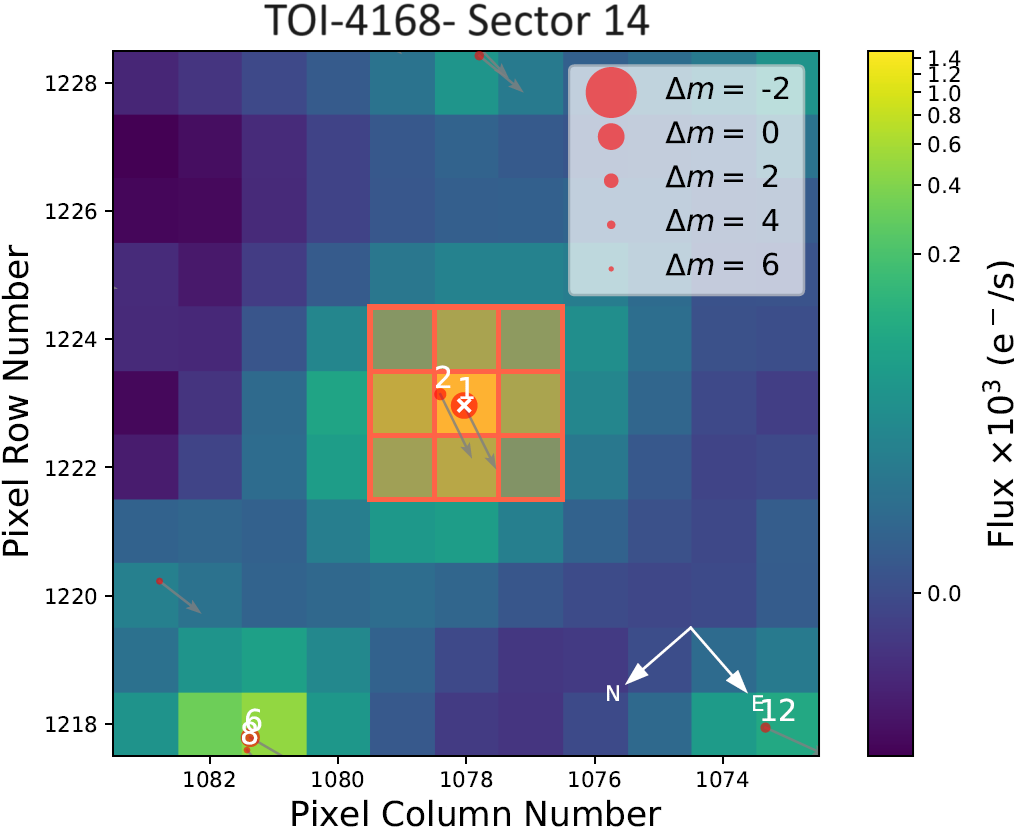}
\includegraphics[width=0.5\columnwidth]{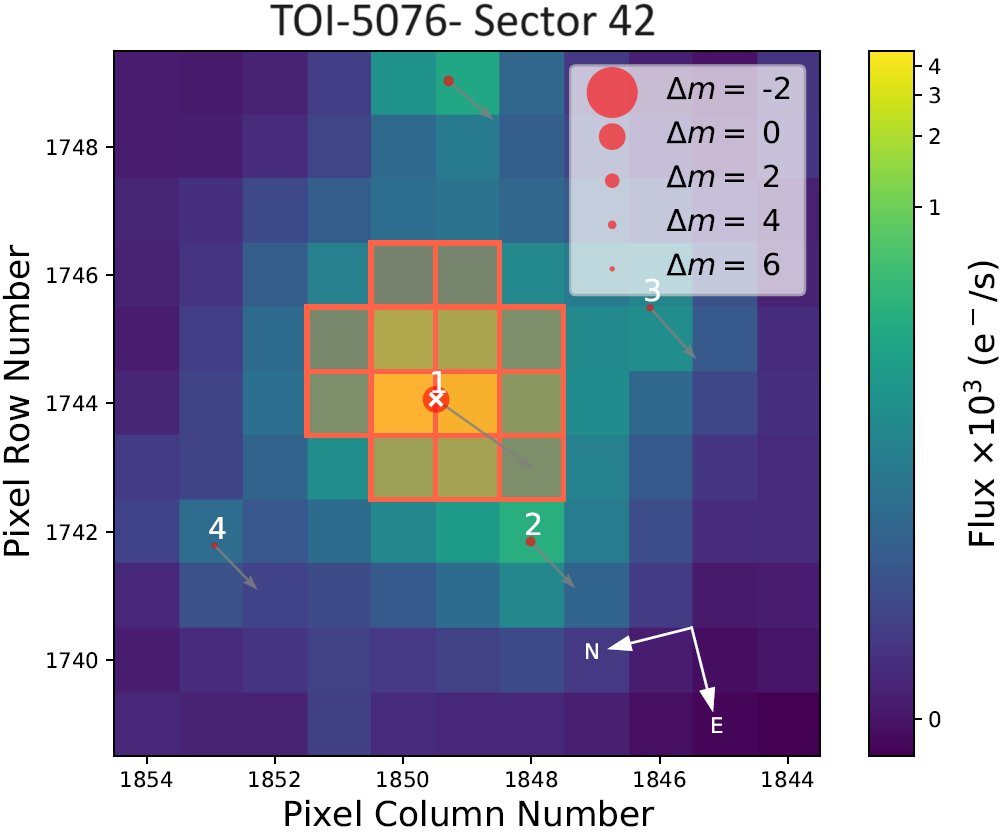}
\includegraphics[width=0.5\columnwidth]{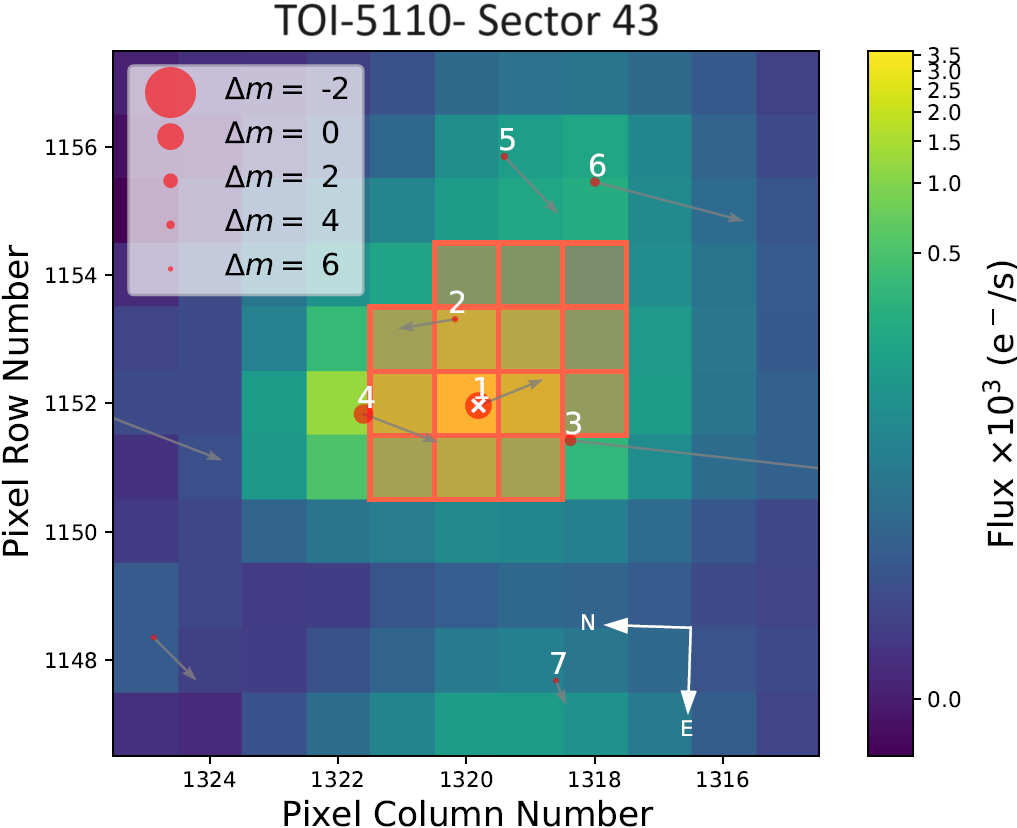}
\caption{Target Pixel Files (TPFs) images from the first observed TESS sector for the seven targets in our analysis, created by \texttt{tpfplotter} \citep{2020AA...635A.128A}. The images depict electron counts, with red-contoured pixels marking the SPOC aperture mask. Additionally, red circles highlight the primary target (numbered as 1) and nearby sources (rest of the numbers) at their Gaia DR3 positions. The area of these circles reflects the relative magnitudes of the sources in comparison to the target star. Arrows represent the proper motion of each star.}
\label{tpfplotter}
\end{figure*}

We employed \texttt{tpfplotter} \citep{2020AA...635A.128A} to plot target pixel files for all targets (Fig. \ref{tpfplotter} for their first TESS observation sector, and appendix \ref{TPf} for the rest of TESS sectors) and subsequently evaluated the light curve contamination from the Gaia Data Release 3 \citep[DR3;][]{Gaia2016,Gaia2021} sources. This investigation extended to a contrast of 6 magnitudes within the aperture employed for extracting the light curves. Above this 6-magnitude threshold, light contamination is negligible, contributing less than 0.4\%. Additionally, following Equation 4 from \cite{2019ApJ...881L..19V}, we note that a nearby star with a contrast greater than 6 magnitudes cannot mimic the planetary depths of the planets presented in this paper. Following \cite{eastman2019}, we estimated the light contamination of Gaia DR3 sources using \(\frac{F_{2}}{F_{1} + F_{2}} \), where \(F_{2}\) is the combined flux of all contaminating sources, and \(F_{1}\) is the flux of the target star. This resulted in the following values: 2.5479$\pm$0.0036\% for TOI-1836, 0\% for TOI-2295, 0.78$\pm$0.011\% for TOI-2537, 9.35$\pm$0.18\% for TOI-4081, 10.359$\pm$ 0.012\% for TOI-4168, 0\% for TOI-5076, and 9.183$\pm$0.011\% for TOI-5110. The contamination levels varied slightly from sector to sector in some cases, and we then report here the maximum contamination. Importantly, the SPOC pipeline already corrects for the contamination attributable to Gaia Data Release 2 \citep[DR2;][]{gaia2018gaia} sources but does not account for sources newly identified in Gaia DR3. This is relevant for the TOI-4081 aperture, which contains a Gaia G= 14.5 mag (Gaia DR3 source 513218084729328640, $\Delta$Gmag= 3.3 with the primary star) at a projected sky distance of $2.17^{\prime\prime}$ from the target. Our joint analysis (Sect. \ref{detection-4081}) accounts for the 6.64$\pm$0.19\% light contamination from this source as a dilution factor. According to Gaia DR3, this star and TOI-4081 exhibit consistent RVs and parallaxes, as well as similar proper motions (Fig. \ref{tpfplotter}). Consequently, the two stars likely form a bound system with a projected sky separation of approximately 980 AU. We note that the secondary star has no derived temperature from the Gaia photometry.

\begin{table*}[ht]
\caption{\label{rv_ob} RV observations summary}
\centering
\resizebox{2\columnwidth}{!}{%
\begin{tabular}{lllllllllll}
\hline
System &Instrument& Mode& Mask&N RVs (used)& Start [UT]/end [UT]& RMS [m/s]& S/N& $\sigma_{RV}$[m/s]& EXP [s]\\
\hline
TOI-1836&SOPHIE&HR&G2&89 (85)&July 2020/October 2023& 13.6& 48& 6.0&1073\\
TOI-2295&SOPHIE&HR&G2&46 (44)&October 2020/October 2023& 81.6& 41&2.9&722\\
TOI-2537&SOPHIE&HE&G2&48 (46)&August 2019/ February 2024& 118.3 & 23 & 19.7& 1643\\
TOI-2537&HARPS&HR&G2&22 (21)&Janurary 2021/ November 2023&94.1 & 20 &14.9&1800 \\
TOI-2537&FEROS&HR&G2&19 (19)&December 2020/ March 2022&116.8& 50&11.6& 1500\\
TOI-4081&SOPHIE&HR&G2&43(38)&August 2021/ October 2023&182.2&28&32.5&1500\\
TOI-4168&SOPHIE&HR&G2&15(15)&December 2021/ February 2024&23029.7&19&8.6&1646\\
TOI-5076&SOPHIE&HR&K5&45(39)&March 2022/February 2024&11.0&33 &4.0&1118 \\
TOI-5110&SOPHIE&HR&G2&23 (23)&February 2022/ October 2023& 133.5& 29&6.5&1318 \\
\hline
\end{tabular}
}
\tablefoot{'HR' and 'HE' represent the high-resolution and high-efficiency observation modes of SOPHIE. The observing start/end column does not indicate continuous observation. 'N RVs' represents the total number of RV observations, with the number of observations used in this study shown in parentheses. $\sigma_{RV}$ represents RV error bars. Additionally, 'RMS' indicates the root mean square of final RVs, calculated after processing mentioned in Sections \ref{sophie_observation}, \ref{harps_observation}, and \ref{feros_observation}. The S/N of SOPHIE and HARPS are measured per pixel at 550 nm, whereas the S/N of FEROS is measured per resolution element. The columns for S/N, exposure time ('EXP'), and $\sigma_{RV}$ represent the average values for each star.}
\end{table*}

\subsection{SOPHIE}
\label{sophie_observation}

The seven targets in this study were observed using the high-resolution, high-precision, fiber-fed SOPHIE spectrograph mounted on the 1.93 m telescope at the Haute Provence Observatory \citep[OHP,][]{perruchot2008sophie, bouchy2013sophie+}. These targets were selected from the TOI catalog, with a focus on long-period or single-transit candidates. We prioritized systems with host stars bright enough to produce strong RV signals detectable by SOPHIE. The observations were conducted as part of a program dedicated to RV follow-up of TESS transiting candidates \citep[e.g.,][]{bell2024toi,martioli2023toi, heidari2022hd,konig2022warm,moutou2021toi}. The SOPHIE aperture is 3~arcsec in diameter. Its high-resolution mode (with a resolving power of $\lambda / \Delta \lambda \approx 75,000$) was employed for all presented targets, except for TOI-2537, which was observed using the High-Efficiency mode (with a resolving power of $\lambda / \Delta \lambda \approx 40,000$). A second fiber aimed at the sky was used to monitor and later remove any contributions from Moon-reflected sunlight. The average exposure times varied between targets, ranging from 722 seconds to 1646 seconds, and resulted in an average signal-to-noise ratio (S/N) per pixel between 19 and 48 at 550~nm. Detailed information about each target observation can be found in Table \ref{rv_ob}.

The SOPHIE RVs are extracted using the SOPHIE Data Reduction System \citep[DRS,][]{bouchy2009sophie}, which cross-correlates the spectra with a binary mask that we select to approximately match the spectral types of each star. These masks assign a value of 1 to regions where narrow absorption lines are typically present in a stellar spectrum, and a value of 0 elsewhere. The specific regions are determined by the star's atmospheric properties. In this study, we used the G2, K5, and M5 masks, which contain approximately 3500, 5000, and 5400 lines, respectively. Subsequently, the DRS fits a Gaussian profile to the cross-correlation function (CCF) as described by \cite{pepe2002coralie} and \cite{baranne1996elodie}. Our procedures follow the optimized methods outlined by \cite{heidari:tel-04043297} and \cite{heidari2023sophie}, which involve: 1) correcting for CCD charge transfer inefficiency \citep{2009EAS....37..247B}; 2) applying template (color) correction for targets observed in HR mode; 3) addressing moonlight contamination by utilizing the simultaneous sky spectrum obtained from the second SOPHIE fiber aperture. Contaminated spectra are first identified based on the significance of the moon's CCF extracted from the second fiber, and the proximity of this CCF to the target CCF \citep[see][]{heidari:tel-04043297}. Subsequently, we applied the correction method outlined in \cite{pollacco2008wasp}; 4) extracting the CCF bisectors using the methodology outlined by \cite{boisse2011disentangling} and calculating their uncertainties, typically twice those of the RV uncertainties \citep{santerne2015pastis}; 5) correcting instrumental nightly variations by employing frequent wavelength calibration observations to interpolate the drift to the precise time of each observation; and 6) applying RV constant master corrections for instrumental long-term drifts for HR and HE mode \citep{courcol2015sophie, heidari:tel-04043297, heidari2023sophie}.

Lastly, we excluded poor measurements for each star based on four criteria: 1) low S/N per pixel at 550 nm ($\sim$ < half of mean S/N), which shows a dependency on the S/N, 2) significant contamination from lunar light ($>$ 60 m/s), 3) large error bars ($\sim$ > three times of mean error bars), and 4) high estimated nightly drift (> 15 m/s). In total, 5.5\% of the gathered data were excluded from this study due to these criteria. The number of data points used for each star is detailed in Table \ref{rv_ob}. 

\begin{figure*}[h!]
\centering
\includegraphics[width=2\columnwidth]{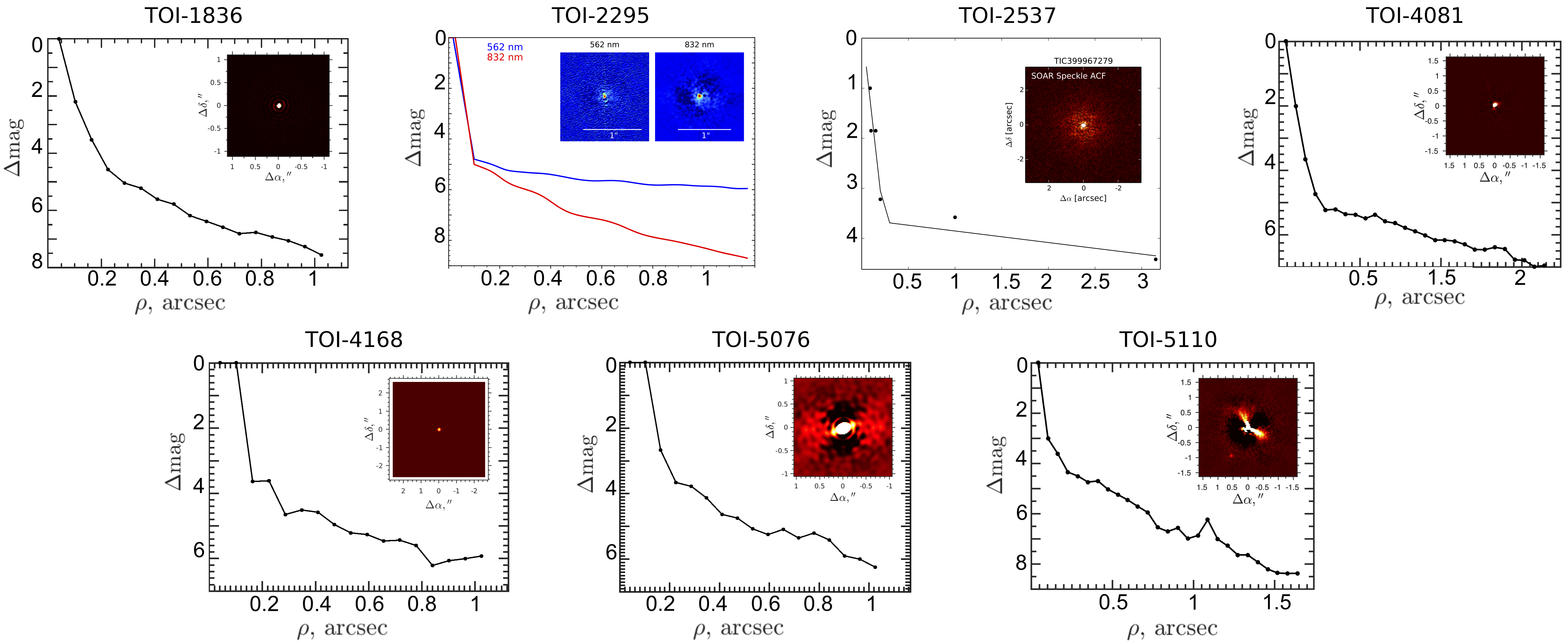}
\caption{Contrast curves for the speckle interferometry observations of TOI-1836, TOI-2295, TOI-2537, TOI-4081, TOI-4168, TOI-5076, and TOI-5110 are shown with black, blue, and red solid lines. The name of each star is indicated on the corresponding plot. TOI-1836, TOI-4081, TOI-4168, TOI-5076, and TOI-5110 were observed by SPP in 625 or 880 nm band. Meanwhile, TOI-2295 was observed using Gemini 'Alopeke, which provides simultaneous speckle imaging in two bands: 562 nm (blue line) and 832 nm (red line). Additionally, TOI-2537 was observed by SOAR in 880~nm band. Each plot includes the final reconstructed image inset in the upper right corner.}
\label{spp}
\end{figure*}

\subsection{HARPS}
\label{harps_observation}
TOI-2537 was also observed with the High Accuracy Radial velocity Planet Searcher \citep[HARPS, ][]{harps} in the context of the WINE collaboration \citep{brahm:2019,jordan:2020,brahm:2020,schlecker:2020,hobson:2021,trifonov:2021,trifonov:2023,bozhilov:2023,brahm:2023,hobson:2023,eberhardt:2023,jones:2024} which focuses on discovering and characterizing transiting warm giant planets. HARPS is a stabilized high-resolution (R=115,000) spectrograph fibre-fed by the 3.6m telescope at the ESO La Silla Observatory, in Chile. We obtained 22 HARPS spectra of TOI-2537 between January 2021 and November 2023 (program IDs 106.21ER.001, 108.22A8.001, and 112.25W1.001). The adopted exposure time was 1800~s and we used the simultaneous calibration mode, where the second fiber is illuminated by light filtered a Fabry-Perot interferometer for tracing subtle instrumental velocity drift variations. The S/N per pixel of these observations ranged between 10 and 25 at 550 nm. The typical RV uncertainty per point was 15 m/s. HARPS data was processed with the \texttt{ceres} pipeline \citep{ceres}, which delivers calibrated spectra, precision RVs extracted by the CCF method, bisector span measurements, and an estimation of the stellar atmospheric parameters. We removed one data point from HARPS data, due to its error bars being twice of the mean error bars.

\subsection{FEROS}
\label{feros_observation}
We also used the Fiber-fed Extended Range Optical Spectrograph \citep[FEROS, ][]{feros} to monitor the RV variations of TOI-2537 in the context of the WINE collaboration. FEROS is a high-resolution (R=48,000) echelle spectrograph fibre-fed to the MPG 2.2m telescope installed at the ESO La Silla Observatory, in Chile. We obtained 19 spectra between December 2020 and March 2022 (program IDs: 0104.A-9007(A), 0107.A-9003(A), and 0108.A-9003(A)) with an exposure time of 1500 s. These spectra achieved a typical S/N per resolution element of 50 and a mean RV uncertainty of 12 m/s. We again adopted the simultaneous calibration technique where the second fiber is here injected with the light of a ThAr lamp. We used the \texttt{ceres} pipeline \citep{ceres} to process the FEROS data and obtain precision RVs with the cross-correlation technique, where a G2-type binary mask was used as a template.

\begin{figure*}
    \centering
    \includegraphics[width=0.5\columnwidth]{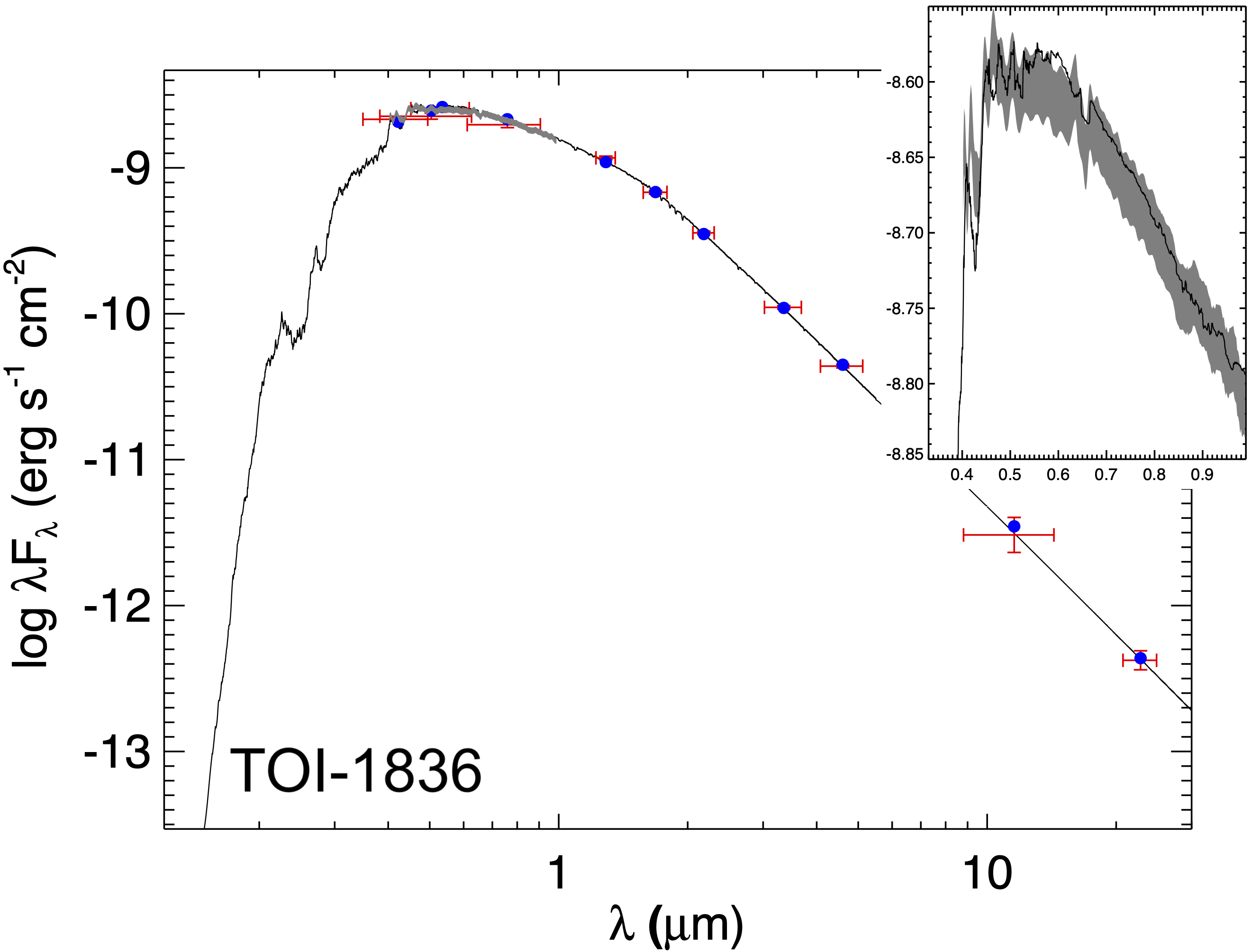}
    \includegraphics[width=0.5\columnwidth]{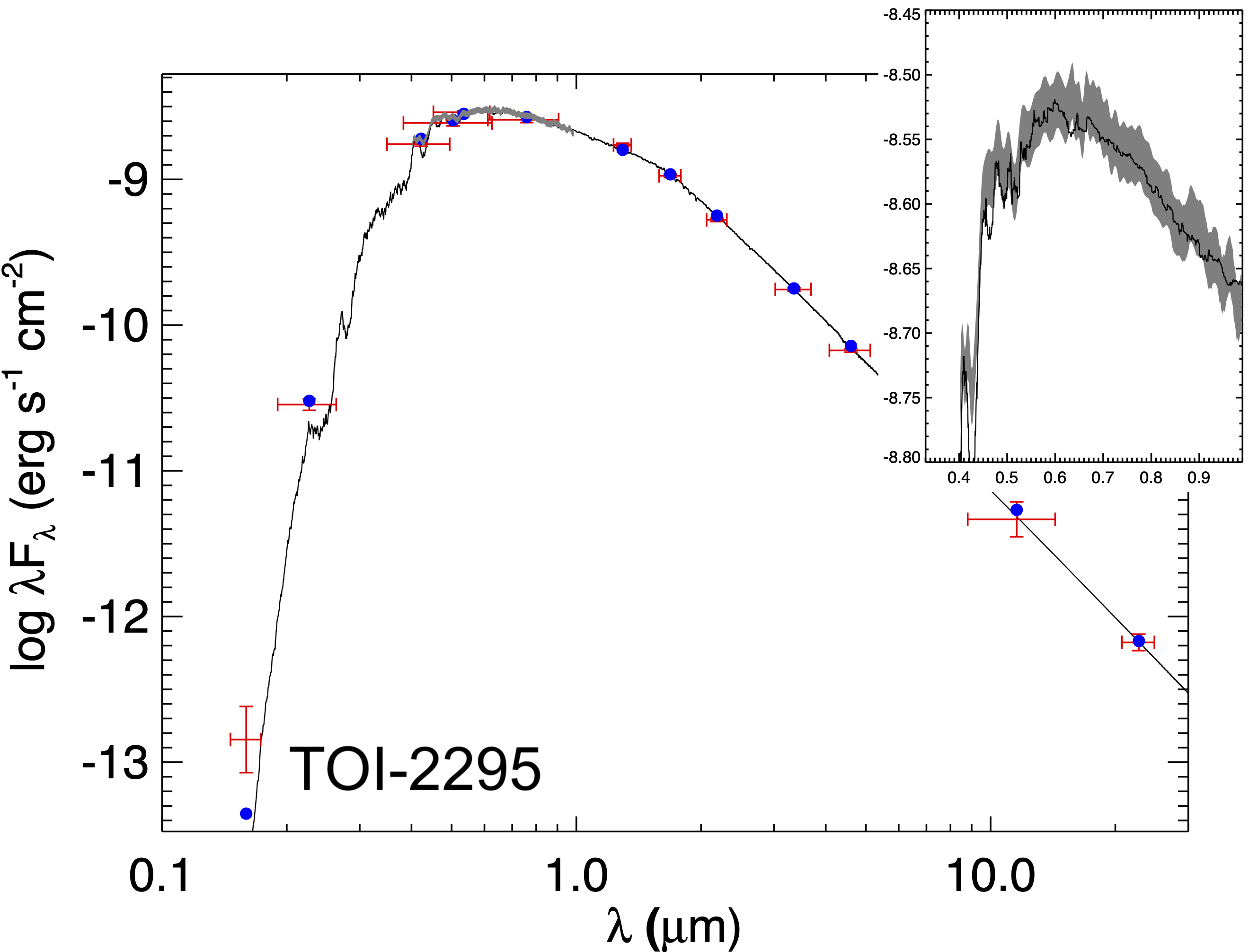}
    \includegraphics[width=0.5\columnwidth]{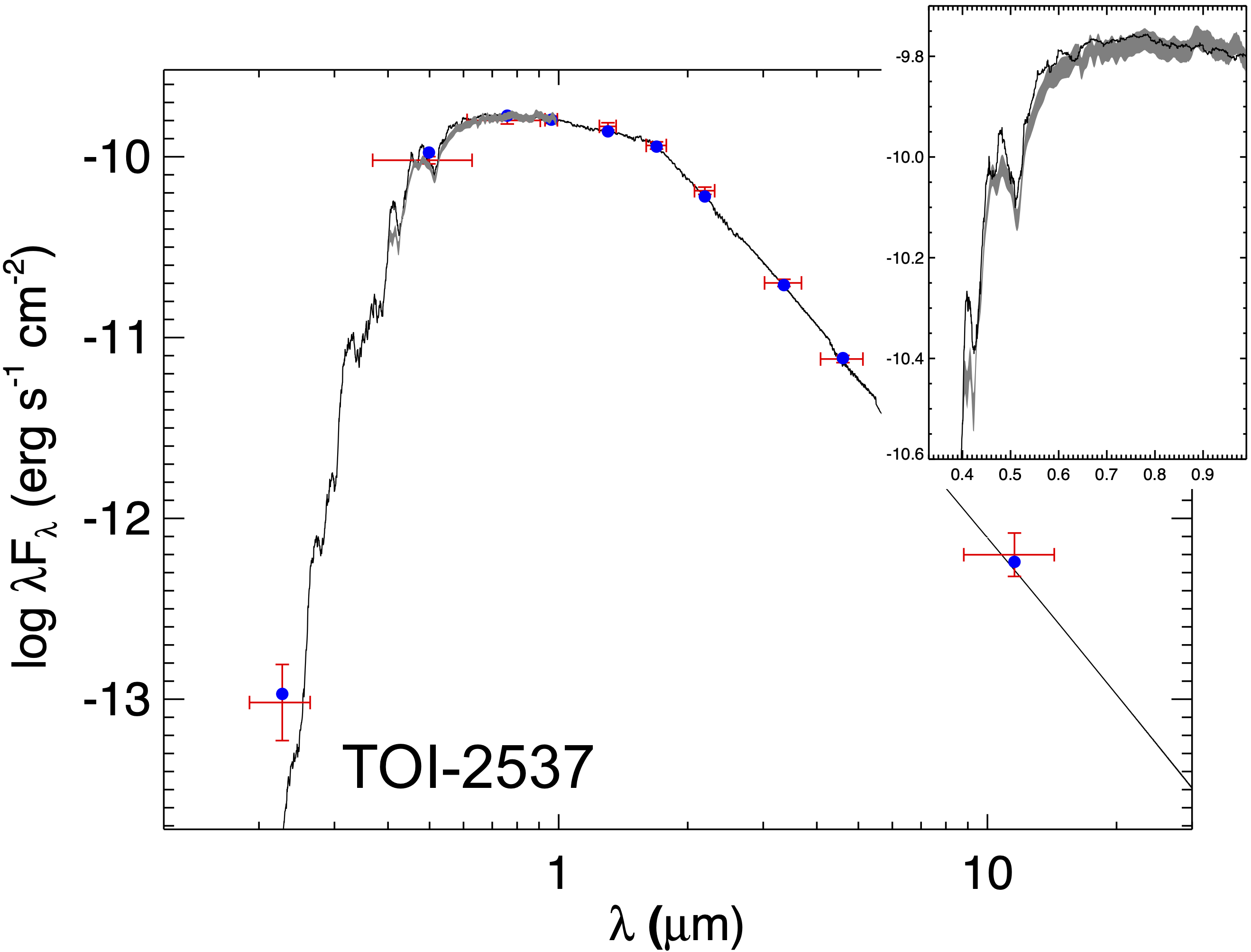}
    \includegraphics[width=0.5\columnwidth]{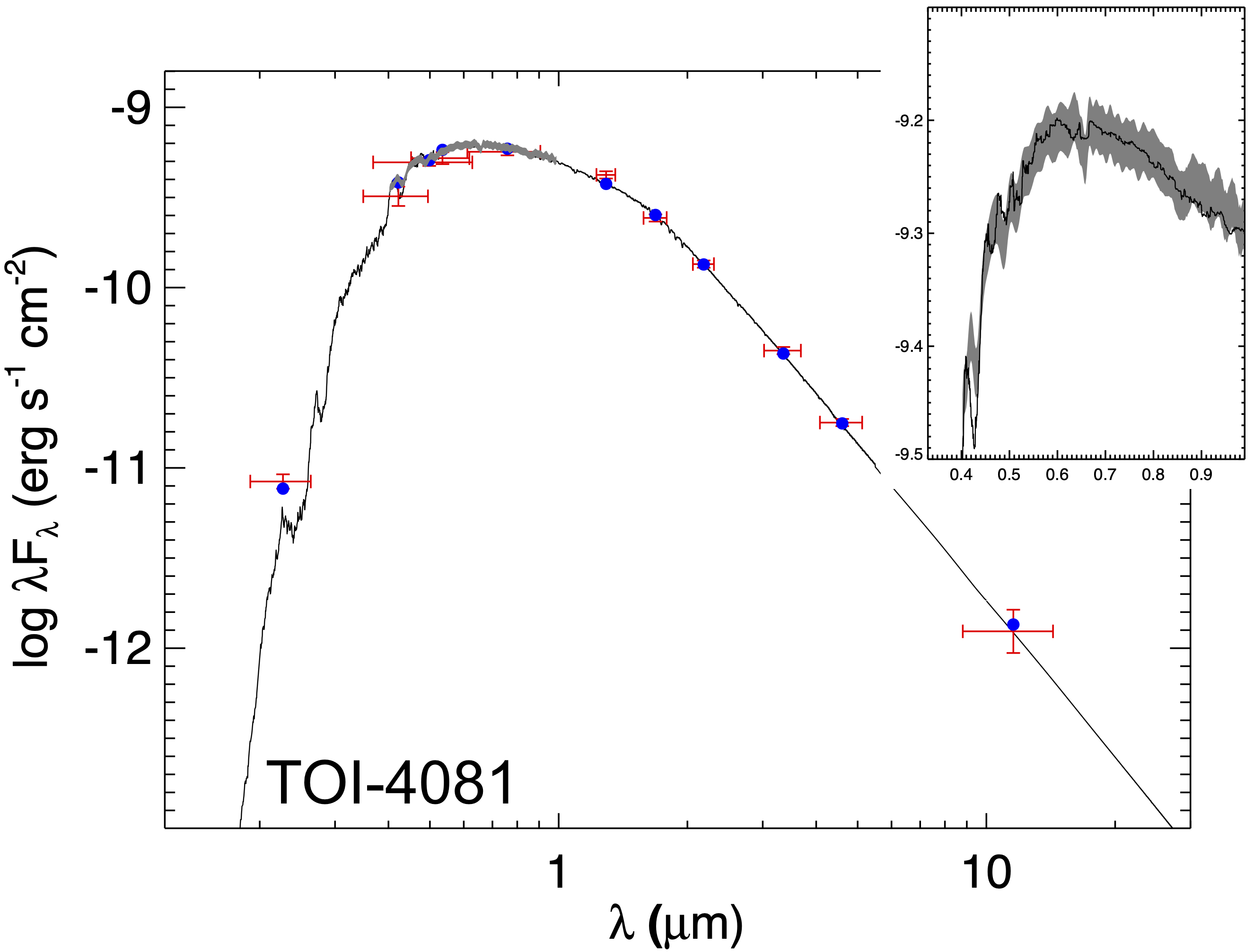}
    \includegraphics[width=0.53\columnwidth]{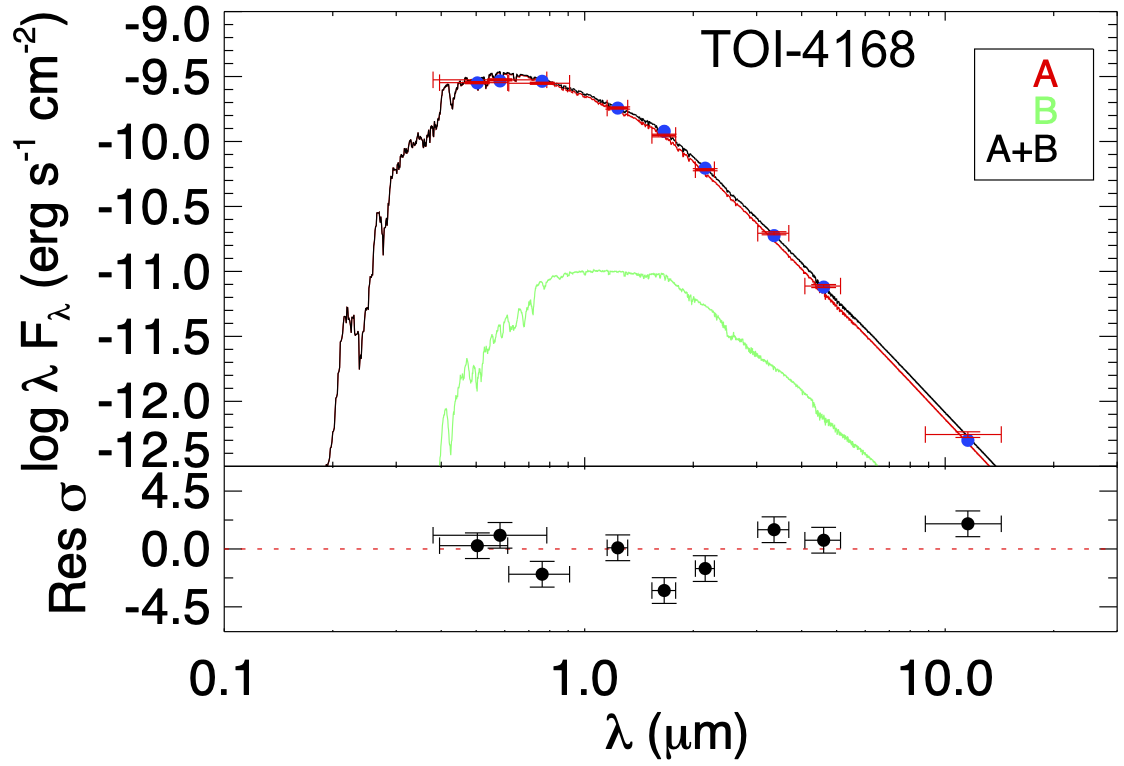}
     \includegraphics[width=0.5\columnwidth]{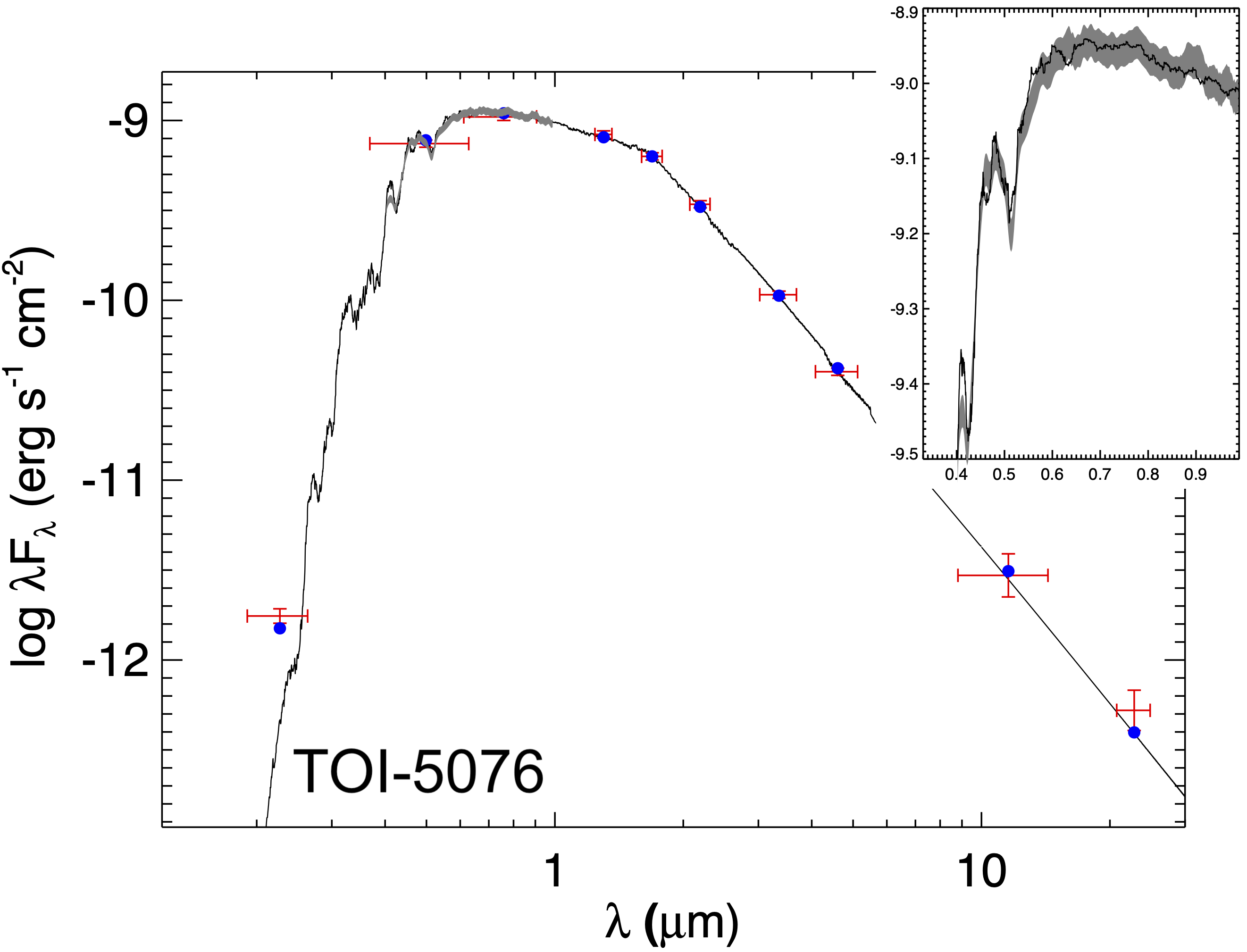}
    \includegraphics[width=0.5\columnwidth]{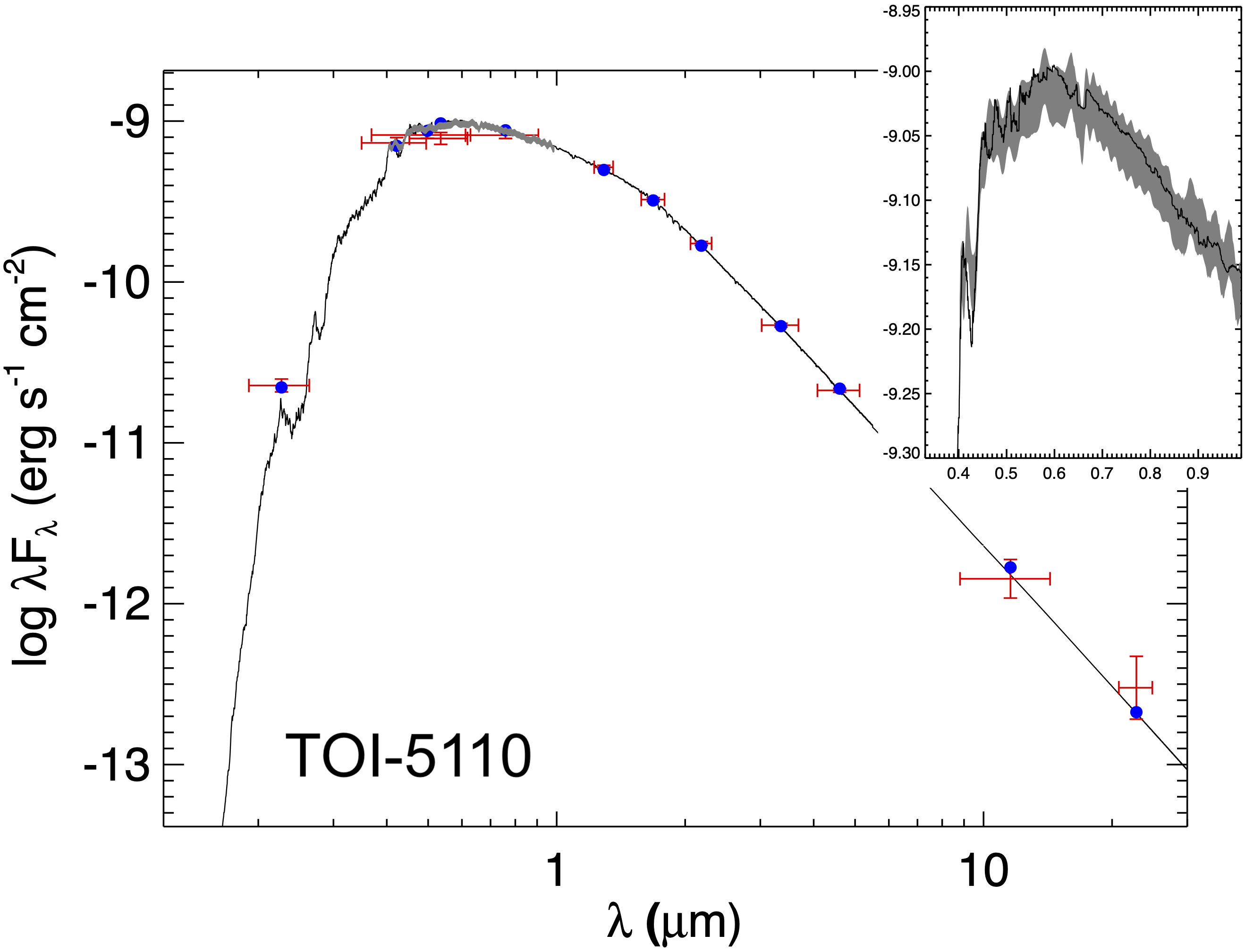}
\caption{Spectral Energy Distributions (SED) for TOI-1836, TOI-2295, TOI-2537, TOI-4081, TOI-4168, TOI-5076 and TOI-5110. The stars' names are indicated on each panel. All SED analyses are performed following the methodology presented in Sect. \ref{stellar}, except for TOI-4168, which is detailed in Sect. \ref{join_4168}. The red symbols represent the observed photometric measurements, and the horizontal bars represent the effective width of the passband. Blue symbols are the model fluxes from the maximum a posteriori} PHOENIX atmosphere model (black). The inset shows the absolute flux-calibrated {\it Gaia} low-resolution spectrum as a grey swathe overlaid on the model (black). \label{fig:sed}
\end{figure*}

\subsection{Ground-based light curve follow-up}

The \textit{TESS} pixel scale is approximately $21\arcsec$ per pixel, and photometric apertures typically extend out to roughly 1 arcminute, generally causing multiple stars to blend in the \textit{TESS} photometric aperture. To attempt to determine the true source of the TESS detections, and to constrain the true period in the case of TOI-2295.01, we acquired ground-based follow-up photometric observations using several facilities for TOI-2295.01 with transit depth = $10.7^{+15.0}_{-6.1}$ parts per thousand (ppt), 
TOI-5110.01 with transit depth = $2.217^{+0.091}_{-0.085}$ ppt, and TOI-1836.01 with transit depth = $2.164^{+0.036}_{-0.037}$ ppt as part of the \textit{TESS} Follow-up Observing Program \citep[TFOP;][]{collins:2019}\footnote{https://tess.mit.edu/followup}. To schedule our transit observations, we used the {\tt TESS Transit Finder}, a customized version of the {\tt Tapir} software package \citep{Jensen:2013}. 

\begin{table*}
\centering
\caption{Summary of ground-based light curve follow-up}
\label{tab:log_of_observations}
\begin{tabular}{ccccc}
\hline
Target & Mid-transit date & Telescope & Band & Coverage \\
\hline
TOI-2295.01 & 2022-06-04 & LCOGT (TEID) & $z$-short & Partial \\
TOI-5110.01 & 2023-03-12 & LCOGT (TEID) & $z$-short & Partial \\
TOI-5110.01 & 2023-12-09 & LCOGT (TEID, CTIO) & $z$-short & Partial \\
TOI-5110.01 & 2023-12-09 & FLWO (KeplerCam) & Sloan-i & Full \\
TOI-1836.01 & 2021-06-16 & LCOGT (McD) & $z$-short & Partial \\
TOI-1836.01 & 2022-05-08 & LCOGT (TEID) & $z$-short & Partial \\
TOI-1836.01 & 2022-08-17 & LCOGT (TEID) & $z$-short & Partial \\
TOI-1836.01 & 2023-07-10 & LCOGT (McD) & $z$-short & Partial \\
TOI-1836.01 & 2021-06-16 & Austin College (Adams Observatory) & $I$ & Partial \\
TOI-5076.01 & 2023-03-12 & CHEOPS & 0.4--1.1 $\mu$m & Full \\
TOI-5076.01 & 2023-12-09 & CHEOPS & 0.4--1.1 $\mu$m & Full \\
\hline
\end{tabular}
\end{table*}

At the start of TOI-2295.01 light curve follow-up, the orbital period was constrained to be 300.344382 d, or an integer divisible harmonics of 300.33113 d (e.g. 150.165565, 100.110377, ..., 37.541391, 33.370126, 30.033113 d). Whereas first SOPHIE data also allowed constraints to be put on the true period (Sect. \ref{detection_toi2295}), we collected follow-up photometry at the 100.110377, 37.541391, and 33.370126 d orbital period harmonics and ruled out the event at those harmonics (see Sect. \ref{MuSCAT2} and \ref{LCOGT} below). We followed up at the orbital period harmonic 30.033113 d and confirmed the event on-target (see Sect. \ref{LCOGT} below). Later, thanks to subsequent TESS observations, the actual period was independently confirmed by the TESS data.

We describe below the four facilities that we used for ground-based photometry. All of the follow-up light curve data are available on the {\tt EXOFOP-TESS} website\footnote{\href{https://exofop.ipac.caltech.edu/tess}{https://exofop.ipac.caltech.edu/tess}}. We note that our joint model relies solely on space-based photometry from TESS and CHEOPS. This decision was made for several reasons: first, to maintain simplicity; second, because TESS provides multiple full-transit detections, while ground-based observations capture mostly partial transits; and third because TESS offers superior precision in its photometry compared to the ground-based data. Nevertheless, for the on-target detections, the median models (see Sect. \ref{ident}) are overplotted on the ground-based light curves (Fig. \ref{lc_more_1836}) and show they are consistent. Incorporating additional models based solely on the ground-based data would not significantly alter our conclusions.

For all lightcurve reductions, except for the MUSCAT2 lightcurves, we conduct parametric detrending of each lightcurve by considering the improvement to the transit model fit after iterating over airmass, time, full with half of the maximum (FWHM), sky background, target x-position, target y-position, and total comparison star ensemble counts as a proxy for common mode systematics. The best zero, one, or two parametric detrend vectors were retained if joint linear fits to them plus a transit model decreased the Bayesian Information Criterion (BIC) of the fit by at least two per detrend parameter. If no detrend vectors are justified by the BIC, we include airmass by default. This process resulted in airmass detrending of all lightcurves, except in one case, as noted below. MUSCAT2 lightcurves are detrended as described in the references below.

\subsubsection{MuSCAT2}
\label{MuSCAT2}
We observed a full transit window of TOI-2295.01, assuming an orbital period harmonic of 100.110377 d, on Coordinated Universal Time (UTC) October 26, 2021 simultaneously in Sloan $g'$, $r'$, $i'$, and Pan-STARRS $z$-short using the MuSCAT2 multi-color imager \citep{Narita:2019} installed at the 1.52~m Telescopio Carlos Sanchez (TCS) in the Teide Observatory, Spain. The photometry was carried out using standard aperture photometry calibration and reduction steps with a dedicated MuSCAT2 photometry pipeline, as described in \citet{Parviainen:2020}. A $\sim$1 ppt event was ruled out at the 100.114794 d harmonics.

\subsubsection{LCOGT}
\label{LCOGT}

We used LCOGT for three candidates studied here TOI-2295.01, TOI-5110.01, and TOI-1836.01. Again for TOI-2295.01, we observed full transit windows at the orbital period harmonics of 33.370126, 30.033113, 37.541391 d on UTC May 15, 2020, June 04, 2022, and July 04, 2022, respectively, using the Pan-STARRS $z$-short band on the Las Cumbres Observatory Global Telescope \citep[LCOGT;][]{Brown:2013} 1\,m network node at Teide Observatory on the island of Tenerife (TEID). The LCOGT 1\,m telescopes are equipped with $4096 \times 4096$ SINISTRO cameras, which have an image scale of $0\farcs389$ per pixel, resulting in a $26\arcmin \times 26\arcmin$ field of view. The images were calibrated by the standard LCOGT {\tt BANZAI} pipeline \citep{McCully:2018}, and differential photometric data were extracted using {\tt AstroImageJ} \citep{Collins:2017}. We used circular photometric apertures with a radius of $6\farcs2$, which excluded all of the flux from the nearest known neighbor in the Gaia DR3 catalog (Gaia DR3 2144184945613590528), located approximately $9\arcsec$ east of TOI-2295. To detrend the light curve, we used the light curves of several comparison stars located within the same field of view. This allowed us to identify and remove common systematic trends that affected the data. The $\sim$1 ppt event was ruled out at the 33.370126 and 37.54139125 d harmonics but was detected on-time and on-target at the 30.033113 d harmonics. 

For TOI-5110.01, we observed a partial transit window on UTC March 12, 2023, using the Pan-STARRS $z$-short band from the LCOGT 1 m network node at TEID. Additionally, two partial transit windows were observed on UTC December 09, 2023, using the Pan-STARRS $z$-short band on the LCOGT network nodes at TEID and Cerro Tololo Inter-American Observatory in Chile (CTIO). In both cases, we used circular photometric apertures with radii of $5\farcs1$--$6\farcs2$ that excluded all of the flux from the nearest known neighbor in the Gaia DR3 catalog (Gaia DR3 3439633217656931712), which is approximately $10\arcsec$ southwest of TOI-5110. An on-time $\sim$2.3 ppt event was detected on-target in all three observations.

For TOI-1836.01, we observed a partial transit window of TOI-1836 on UTC June 16, 2021, using the Pan-STARRS $z$-short band from the LCOGT 1\,m network
node at McDonald Observatory near Fort Davis, Texas, United States (McD). Three partial
transit windows were also observed on UTC May 08 2022 and August 17 2022 in
Pan-STARRS $z$-short band on the LCOGT 1\,m network node at TEID, and on July 10
2023 in Pan-STARRS $z$-short band on the LCOGT 1\,m network node at McD. We used
circular photometric apertures with radius $4\farcs3$--$5\farcs5$ that excluded all
of the flux from the nearest known neighbor in the Gaia DR3 catalog (Gaia DR3
1428639820687958400), which is $\sim10\arcsec$ northwest of TOI-1836. An on-time $\sim$ 2.4 ppt event was detected on-target in all four observations.

\subsubsection{KeplerCam}

We observed one full transit of TOI-5110.01 using the 1.2m telescope with the KeplerCam CCD at the Fred Lawrence Whipple Observatory (FLWO) at Mt. Hopkins, Arizona. Observations were taken in the Sloan-i band on UT 2023 on December 9. KeplerCam has a $4096\times4096$ Fairchild CCD 486 detector with an image scale of $0\farcs672$ per $2\times2$ binned pixel, resulting in a $23\farcm1\times23\farcm1$ field of view, and an image scale of 0.672”/pixel when binned by 2. Transit observations were scheduled using the TESS Transit Finder. Data were reduced using standard IDL routines and aperture photometry was performed using {\tt AstroImageJ}. Similar to LCO observations, we used circular photometric apertures with a radius $5\farcs1$--$6\farcs2$ that excluded all of the flux from the nearest known neighbor in the Gaia DR3 catalog (Gaia DR3 3439633217656931712), which is $\sim10\arcsec$ southwest of TOI-5110. An on-time $\sim$ 2.3 ppt event was detected on-target. 

\subsubsection{Austin College}

For TOI-1836.01, we observed a partial transit window of TOI-1836 on UTC June 16, 2021, using the Johnson/Cousins $I$ band from the Adams Observatory at Austin College in Sherman, TX. The Adams Observatory 0.6\,m telescope is equipped with an FLI Proline PL16803 detector that has an image scale of $0\farcs38$ pixel$^{-1}$, resulting in a $26\arcmin\times26\arcmin$ field of view. We used {\tt AstroImageJ} with circular photometric apertures having radius $4\farcs5$ that excluded all of the flux from the nearest known neighbor in the Gaia
DR3 catalog (Gaia DR3 1428639820687958400). An on-time $\sim$2.4 ppt egress was detected on-target. 

\subsection{CHEOPS}

We used the CHaracterising ExOPlanets Satellite \citep[CHEOPS, ][]{benz2021} to observe two extra transits of TOI-5076.01 (PN: AO3-34; PI: B. Edwards), in addition to the four transits previously observed by TESS. CHEOPS is a small-class ESA mission that launched in December 2019 and is dedicated to the characterization of exoplanets. Operating in a sun-synchronous low-Earth orbit, and with a pixel scale of approximately 1 arcsecond, CHEOPS offers photometric time series at a wavelength range of 0.4 $\mu$m -- 1.1 $\mu$m. Each CHEOPS visit of TOI-5076.01 lasted 9 spacecraft orbits ($\sim$14.7 hours) and the data were acquired with an exposure time of 60 s. We obtained the calibrated lightcurves from the CHEOPS archive\footnote{\url{https://cheops-archive.astro.unige.ch/archive_browser/}}. These have been processed by the CHEOPS Data Reduction Pipeline \citep[][]{hoyer2020} which removes instrumental variability and environmental effects such as background and smearing. The DRP produces files that contain the extracted lightcurves, and ancillary data such as quality flags and spacecraft roll-angle that can be used for further corrections. The signal is extracted for four different aperture radii. In this work, we use the DEFAULT aperture (25 pix) as this provides the light curve with the highest SNR.

\subsection{High angular resolution observations}

In the following subsections, we present ground-based high-spatial resolution imaging with SPeckle Polarimeter (SPP), ‘Alopeke, and HRCam.

\subsubsection{SAI/SPP}
\label{SPP}

To search for close-in companions unresolved in our follow-up
observations, we observed TOI-1836, TOI-4081, TOI-4168, TOI-5076, and TOI-5110 with SPP \citep[][]{2023AstBu..78..234S} on 2021 January 24th, 2021 September 9th, 2022 March 21th, 2022 November 20th, and 2022 December 9th UT, respectively. SPP is a facility instrument on the 2.5 m telescope at the Caucasian Observatory of the Sternberg Astronomical Institute (SAI) of Lomonosov Moscow State University. We used an Electron Multiplying CCD Andor iXon 897 as a detector for the observations of TOI-1836, TOI-4081, and TOI-4168, and a CMOS Hamamatsu ORCA-quest for the observations of TOI-5076 and TOI-5110. The pixel scale is 20.6~mas/pixel, the angular resolution is 89~mas, and the field of view is $5\times5^{\prime\prime}$ centered on the star. For TOI-1836 we used a custom filter with 50~nm FWHM centered on 625~nm, and the rest of the targets were observed with a $I_\mathrm{c}$ filter. The power spectrum was estimated from 4000 frames per star with 30 ms exposure time. The atmospheric dispersion compensator was employed. The resulting sensitivity curves are shown in Fig. \ref{spp}. 

\begin{table}
\centering
\caption{Summary of high-spatial resolution imaging observations}
\label{tab:high_res_imaging_observations}
\resizebox{\columnwidth}{!}{%
\begin{tabular}{ccccc}
\hline
Target & Date (UT) & Instrument & Filter & Detection? \\
\hline
TOI-1836.01 & 2021-01-24 & SPP & 625 nm & No \\
TOI-4081.01 & 2021-09-09 & SPP & $I_C$ & No\\
TOI-4168.01 & 2022-03-21 & SPP & $I_C$ & No \\
TOI-5076.01 & 2022-11-20 & SPP & $I_C$ & No \\
TOI-5110.01 & 2022-12-09 & SPP & $I_C$ & Yes \\
TOI-2295.01 & 2020-06-10 & 'Alopeke & 562 nm, 832 nm & No \\
TOI-2537.01 & 2021-11-20 & HRCam & $I_C$ & No \\
\hline
\end{tabular}%
}
\end{table}

For TOI-1836, we did not detect stellar companions brighter than our detection limits of $\Delta I_C=4.8$ and $7.5$ at $\rho=0\farcs25$ and $1\farcs0$, respectively, where $\rho$ is the separation between the source and a potential companion. We note that \cite{chontos2024tess}, using adaptive-optics imaging with the Palomar Observatory and a narrow-band Br-$\gamma$ filter, reported a nearby star with a separation of $\rho=0.82$\arcsec and a magnitude difference of $\Delta m=5.7$. This source is presumably below our 625~nm contrast limit. Utilizing the similarity between the near-infrared K-band (centered at approximately 2190 nm) and the Br-$\gamma$ filter (at 2165 nm), we estimated the color correction (1.1) and inferred its $\Delta m$ in the Gaia RP band to be 6.8 mag. Assuming the TESS band (600-1000~nm) and Gaia RP band (640–1050~nm) are similar, we estimated a contamination level of about 0.09\%, which is negligible. We investigated false positive scenarios related to this new nearby star in Sect. \ref{toi-1836_for}.

Additionally, no stellar companions were detected for TOI-5076, with detection limits of $\Delta I_C=3.7$ and $6.2$ at separations of $\rho=0\farcs25$ and $1\farcs0$. Similarly, for TOI-4168 we did not detect any companions, with detection limits of $\Delta I_C=4.2$ and $6.0$ at $\rho=0\farcs25$ and $1\farcs0$.

As mentioned in Sect. \ref{TESS-photometry}, TOI-4081 has a stellar companion at a $2.17^{\prime\prime}$ separation which is 3.3 magnitudes fainter in the Gaia G band. We did not detect any closer companions (Fig. \ref{spp}), with detection limits of $\Delta I_C=5.0$ and $6.0$ at $\rho=0\farcs25$ and $1\farcs0$. 

For TOI-5110 we detected a companion (Fig. \ref{spp}), and its parameters were determined by fitting a binary source model to the power spectrum \citep{2017AstL...43..344S}. The separation was found to be $1.084^{\prime\prime}\pm0.009$, the position angle $148.9^{\circ}\pm0.4$, and the brightness difference $6.2~\pm~0.1$ mag. Assuming the TESS bandpass and the $I_\mathrm{c}$ (700-900 nm) filter are similar, the light contamination would be 0.33\%, which is negligible. We examined false positive scenarios associated with this new nearby star in Sect. \ref{5110}.

\begin{table*}[h!]
\caption{Summary of the stellar parameters.}
\centering
\resizebox{2.1\columnwidth}{!}{%
\begin{tabular}{llllllll}
\hline
\hline
Parameter& TOI-1836& TOI-2295& TOI-2537& TOI-4081&TOI-4168& TOI-5076& TOI-5110\\
\hline
Identifiers &  &&&&&& \\
ID (TIC) &207468071& 48018596&399967279&421973253&372758077&303432813&239977528 \\
ID (Gaia DR3)&1428639816392087552 & 2144184945615891712 &12320268307415552&513218084729328512&1130051731468987520&55752218851048320&3439633221952197120 \\
 \hline
\hline
  & &\\
Astrometric properties :& &&&&&\\
Parallax (mas) &5.273$\pm$ 0.013$^{1}$ &7.946$\pm$0.010$^{1}$&5.493$\pm$0.018$^{1}$&2.218$\pm$0.017$^{1}$&3.055$\pm$0.015$^{1}$&12.085$\pm$0.015$^{1}$&2.577$\pm$0.022$^{1}$\\
Distance (pc) $^{3}$ & 189.23$_{+0.51}^{-0.50}$&125.51$_{-0.14}^{+0.16}$ &182.42$_{-0.93}^{+1.05}$&442.81$^{+4.84}_{-7.21}$&319.87$^{+2.48}_{-4.39}$&82.53$_{-0.10}^{+0.09}$&358.92$_{-2.96}^{+1.57}$\\
$ \alpha$  (h m s) & 16:23:37.97$^{1}$ &18:46:18.39$^{1}$&03:37:33.18 $^{6}$&01:26:17.42$^{1}$&11:35:19.93$^{1}$&03:22:02.32$^{1}$&06:17:37.03$^{1}$\\ 
$ \delta$ (d m s)& +54:41:24.28 $^{1}$ & +50:29:32.46$^{1}$&+10:03:27.89$^{6}$&+64:50:24.21$^{1}$&+79:24:46.88&+17:14:23.23.88$^{1}$& +31:36:37.22$^{1}$\\ 
&  &&&&&&\\
Photometric properties: & && &&&&\\
 B-V &0.45$^{2}$&0.76$^{2}$&1.01$^{6}$&0.75$^{2}$&0.75$^{2}$&0.93$^{6}$&0.35$^{2}$\\
 V(mag) & 9.77 $\pm$0.03$^{2}$&9.60$\pm$0.03$^{2}$&13.24 $\pm$ 0.08$^{6}$&11.45$\pm$0.08$^{2}$&12.04$\pm$ 0.15$^{2}$&10.90$\pm$0.03$^{6}$&11.06$\pm$0.09$^{2}$\\
 Gaia(mag) &9.6509 $\pm$0.0028$^{1}$&9.4678$\pm$0.0028$^{1}$&12.7191$\pm$0.0028$^{3}$&11.1647$\pm$0.0028$^{1}$&11.8651$\pm$0.0028$^{1}$&10.5928$\pm$0.0028$^{1}$&10.6840$\pm$0.0028$^{1}$\\
J(mag) & 8.799$\pm$0.021 $^{4}$&8.379$\pm$0.019$^{4}$&11.050$\pm$0.022$^{6}$&9.907$\pm$0.023$^{4}$&10.825$\pm$ 0.021$^{4}$&9.164$\pm$0.025$^{4}$&9.687$\pm$0.022$^{4}$\\
H(mag) & 8.586$\pm$0.021$^{4}$&8.105$\pm$0.020$^{4}$&10.511$\pm$ 0.027$^{6}$&9.701$\pm$0.032$^{4}$&10.551$\pm$0.026$^{4}$&8.666$\pm$0.027$^{4}$&9.386$\pm$0.024$^{4}$\\
K$_{s}$(mag) & 8.531$\pm$0.018&8.111$\pm$0.013$^{4}$&
10.387 $\pm$ 0.021$^{6}$&9.591$\pm$0.023$^{4}$&10.460$\pm$0.020 $^{4}$&8.582$\pm$0.020$^{4}$&9.318$\pm$0.021$^{4}$\\
W$_{1}$ (mag) &8.487$\pm$
0.023$^{7}$&7.984$\pm$0.025$^{7}$ &10.341$\pm$ 0.023$^{7}$&9.470$\pm$0.023$^{7}$&10.371 $\pm$ 0.022$^{7}$&8.519$\pm$0.023$^{7}$&9.268$\pm$
0.024$^{7}$\\
W$_{2}$(mag) &8.514$\pm$
0.020$^{7}$&8.050$\pm$0.020$^{7}$ &10.413$\pm$ 0.020$^{7}$&9.488$\pm$0.022$^{7}$&10.403$\pm$ 0.020$^{7}$&8.609$\pm$0.021$^{7}$&9.301$\pm$
0.021$^{7}$\\
W$_{3}$(mag) &8.483$\pm$
0.020$^{7}$&8.026$\pm$0.018$^{7}$&10.196$\pm$ 0.076$^{7}$&9.460$\pm$0.037$^{7}$&10.426$\pm$0.054$^{7}$&8.517$\pm$0.026$^{7}$&9.306$\pm$0.041$^{7}$\\
W$_{4}$(mag)& 8.48$\pm$0.16$^{7}$&7.98$\pm$0.14$^{7}$&8.83$^{7}$&8.79$^{7}$&9.377$^{7}$&8.24$\pm$0.28$^{7}$&8.84$\pm$0.49$^{7}$\\
&  &&&&&&\\
 Spectroscopic properties:&  &&&&&&\\

$\log~g$ (cm s$^{-2}$)&4.350$^{+0.040}_{-0.108}$ & 4.151 $^{+0.052}_{-0.113}$&4.334$^{+0.374}_{-0.388}$&----&4.367$^{+0.053}_{-0.113}$&4.160$^{+0.259}_{-0.278}$&3.950$^{+0.040}_{-0.108}$\\
 $\log (R'_{HK})$ &-5.0$\pm$0.1$^{11}$&-5.5$\pm$0.1$^{11}$&-4.5$\pm$0.1$^{11}$&-5.0$\pm$0.1$^{11}$&-4.4$\pm$0.1$^{11}$&-5.1$\pm$0.1$^{11}$& <- 5.2\\
 v$\sin i $ (km s$^{-1}$)& 4.0$\pm$1.0$^{9}$&3.9$\pm$1.0$^{9}$&3.0$\pm$1.0$^{9}$&25.2$\pm$2.0$^{9}$&3.8$\pm$1.0$^{9}$&2.3$\pm$1.0$^{9}$&>2.3\\
 $[Fe/H]$$^{10}$ & -0.098$^{+0.024}_{-0.047}$&0.316$^{+0.016}_{-0.043}$&  0.081$^{+0.066}_{-0.077}$& 0.0 $\pm$ 0.3$^{10}$ 
 &0.153$^{+0.026}_{+0.048}$ &0.072$^{+0.046}_{-0.061}$&0.067$^{+0.027}_{-0.049}$\\
T$_{\rm eff}$ (K)$^{10}$ &6369$^{+35}_{-69}$[153]$^{11}$&5733$^{+20}_{-63}$[138]$^{11}$&4843$^{+140}_{-153}$& 
6040 $\pm$ 100$^{10}$[145]$^{11}$ &5959$^{+33}_{-68}$[143]$^{11}$&4832$^{+103}_{-119}$&6154$^{+38}_{-71}$ [148]$^{11}$\\
& &&&&&\\
 Bulk properties:& &&&&&\\
$F_{\rm bol}$ ($10^{-9}$ erg~s$^{-1}$~cm$^{-2}$) & 3.425 $\pm$ 0.040$^{10}$ & 4.141 $\pm$ 0.021$^{10}$& 0.2638 $\pm$ 0.0061$^{10}$ & 1.193 $\pm$ 0.028 $^{10}$&see Table \ref{tab:toi4168} & 1.588  $\pm$  0.037$^{10}$ & 1.528 $\pm$  0.018$^{10}$   \\
$L_{\rm bol}$ (L$_\odot$) & 3.841 $\pm$   0.045$^{10}$ & 2.045 $\pm$ 0.011$^{10}$ & 0.2726 $\pm$     0.0064$^{10}$ & 7.56   $\pm$    0.19$^{10}$&see Table \ref{tab:toi4168}& 0.3389 $\pm$  0.0079 $^{10}$ & 7.17 $\pm$  0.10 $^{10}$  \\
Radius( R$_{\odot} $) &   1.611 $\pm$0.036$^{10}$[0.068]$^{11}$& 1.451 $\pm$ 0.032$^{10}$[0.061]$^{11}$ & 0.774 $\pm$0.050 $^{10}$& 2.511 $\pm$ 0.090$^{10}$[0.105]$^{11}$&see Table \ref{tab:toi4168} & 0.798 $\pm$0.036 $^{10}$& 2.359 $\pm$  0.059$^{10}$[0.099]$^{11}$\\
Mass ( M $_{\odot} $) & 1.29 $\pm$ 0.08 $^{10}$& 1.17 $\pm$ 0.07$^{10}$ & 0.77 $\pm$0.05$^{10}$ & 1.44 $\pm$0.09 $^{10}$&see Table \ref{tab:toi4168}& 0.82  $\pm$0.05$^{10}$ & 1.46$\pm$0.09 $^{10}$  \\
$P_{\rm rot}$(d)  &  20 $\pm$ 3$^{10}$ & 78 $\pm$ 6$^{10}$ & 19 $\pm$ 8$^{10}$&  33 $\pm$ 4$^{10}$& 8 $\pm$ 3$^{10}$& 57 $\pm$ 7$^{10}$ & >14 \\
Age (Gyr)& 6.6 $\pm$ 1.8$^{10}$ & 10.5 $\pm$ 1.4$^{10}$& 1.1 $\pm$ 0.6$^{10}$&  6.6 $\pm$ 1.9$^{10}$&see Table \ref{tab:toi4168}& 8.4$\pm$ 1.9$^{10}$& >4  \\
\hline
\end{tabular}%
}
\tablefoot{$^{1}$ EDR3, $^{2}$Tycho-2 \citep{hog2001tycho}, $^{3}$ Gaia DR3, $^{4}$2MASS, $^{6}$TESS input catalog,$^{7}$WISE, $^{8}$\cite{1993yCat.3135....0C},$^{9}$SOPHIE DRS, $^{10}$This work (see Sect. \ref{stellar}), $^{11}$ Adopted the systematic uncertainty floor as suggested by \cite{tayar2022guide} throughout this study.}
\label{star_info}
\end{table*}

\subsubsection{Gemini-N/‘Alopeke}

We obtained speckle imaging observations of TOI-2295 with Gemini North’s ‘Alopeke
high-resolution imager instrument \cite{scott2021twin}. ‘Alopeke provides
simultaneous speckle imaging in two bands (562 nm and 832 nm) with output data
products including reconstructed images and robust contrast limits on companion
detections \citep[see][]{howell2011speckle}.

The observation, conducted on June~10 2020 at Gemini North, found no detectable companions to TOI-2295 within contrast limits of 5-9 magnitudes (to 5$\sigma$ significance) for separations of 0.02 to 1.2 arcsec (Fig.~\ref{spp}). At the distance to TOI-2295 (d=126.3 pc), these angular limits correspond to 2.5 to 152 AU. We therefore interpret all the flux in the TESS photometric aperture as originating from TOI-2295. This is in agreement with the estimated dilution factors reported above in Sect. \ref{TESS-photometry}.

\subsubsection{SOAR/HRCam}

We searched for stellar companions to TOI-2537 with HRCam \citep{tokovinin2018ten} on the
4.1-m Southern Astrophysical Research (SOAR) telescope on 20 November 2021 UT, observed in the Cousins I-band, which is similar to the TESS bandpass. This observation was sensitive to a 3.7 magnitude fainter star at an angular distance of 1 arcsec from the target. More details of the observations within the SOAR TESS survey are available in \cite{ziegler2019soar}. The 5$\sigma$ detection sensitivity and the speckle auto-correlation functions from the observations are shown in Fig. \ref{spp} The SOAR observation detected no companion within 3\arcsec of TOI-2537.

\section{Stellar properties}
\label{stellar}

We first derived the v$\sin i$ of each star from its average SOPHIE CCF using the calibration of \cite{boisse2010sophie}. For TOI-5110, we only provide an upper limit because its $B-V$ of 0.35~$\pm$0.17 is outside the validity range (0.44 <$B-V$< 1.20) of the \cite{boisse2010sophie} calibration of the non-rotational broadening. To investigate the stellar atmospheric parameters, we averaged the SOPHIE spectra of each star which are unpolluted by moonlight, after correcting for both the RV variation of the star and the barycentric Earth RV. Subsequently, we used the methodology of \cite{santos2013sweet} and \cite{sousa2018sweet} to determine the effective temperature ($T_{\rm eff}$), and metallicity ([Fe/H]). For TOI-4081, we had to derive the $T_{\rm eff}$ and [Fe/H] from its photometry (as described below) because it is a fast rotating star  (v$\sin i$~=~25.2$\pm$2.0 km/s) with strongly broadened spectral lines. The resulting parameters, along with additional stellar information, are detailed in Table \ref{star_info}.

We analyzed the broadband spectral energy distribution (SED) of all stars, except TOI-4168 (see Sect. \ref{join_4168}), along with their  {\it Gaia\/} DR3 parallaxes \citep[with no systematic offset applied; see, e.g.,][]{StassunTorres:2021}. This analysis empirically determines the stellar luminosities, radii, masses, and ages, following the procedures described in \citet{Stassun:2016,Stassun:2017,Stassun:2018}. We pulled the $JHK_S$ magnitudes from {\it 2MASS}, the W1--W4 magnitudes from {\it WISE}, the $G_{\rm BP} G_{\rm RP}$ magnitudes from {\it Gaia}, and when available the FUV and/or NUV magnitudes from {\it GALEX}. The available photometry spans the stellar SED over the wavelength range 0.4--10~$\mu$m in all cases, and over the range 0.2--20~$\mu$m in some cases (Fig.~\ref{fig:sed}).  

\begin{figure*}
\centering
\includegraphics[width=1.2\columnwidth]{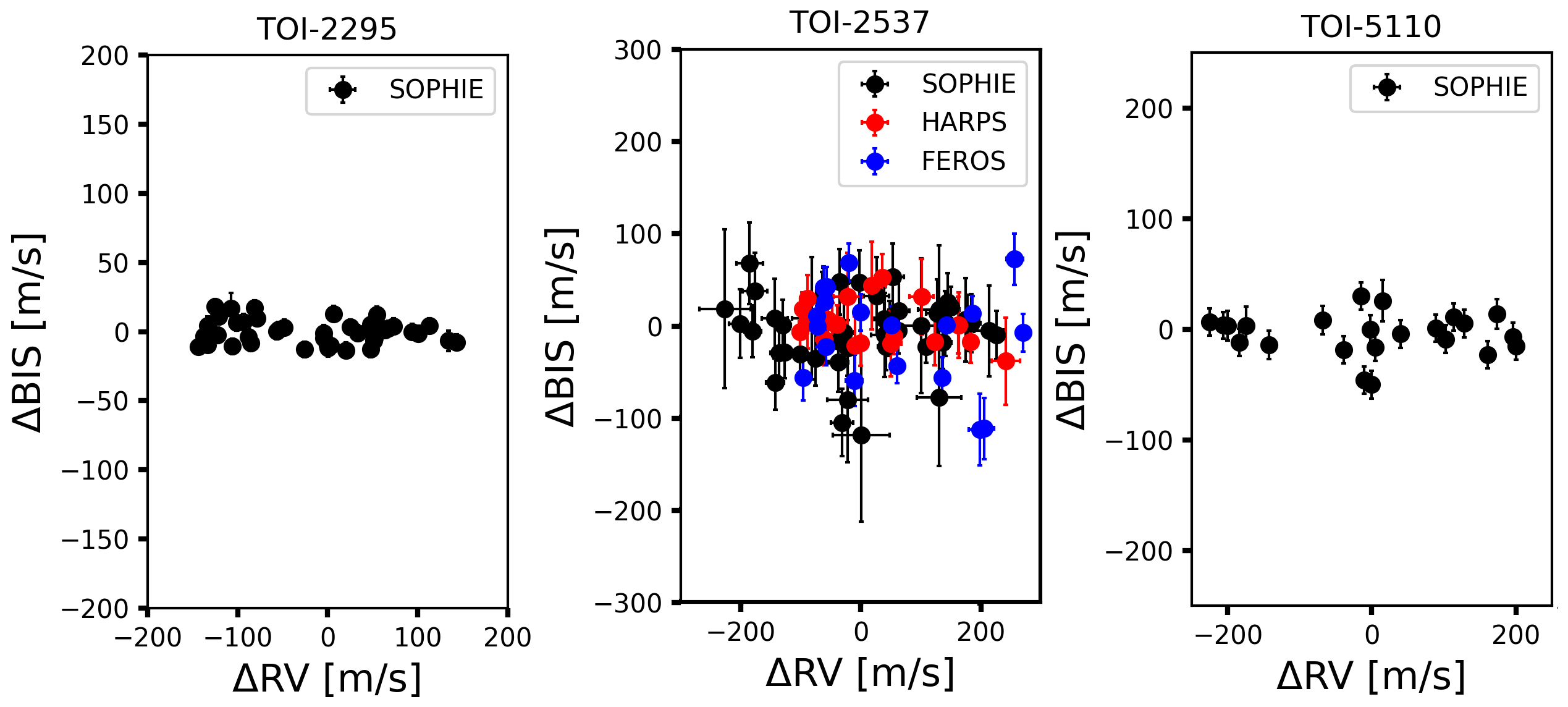}
\includegraphics[width=1.7\columnwidth]{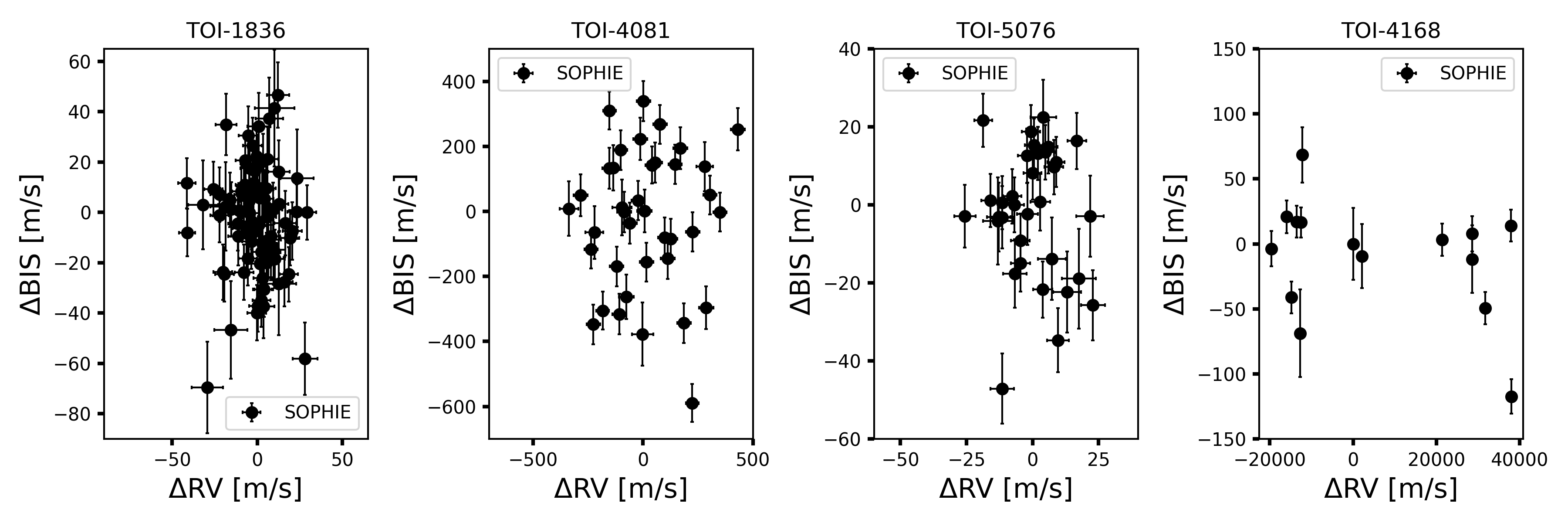}
    \caption{RVs-bisectors diagram of the 7 stars presented in this study. The $x$- and $y$-ranges of each figure are identical, with the exception of TOI-4168 due to its exceptionally large RV range. Notably, the bisectors of the stars in the first row—namely TOI-2295, TOI-2537, and TOI-5076—show no discernible variation. Conversely, the bisector variability of the stars in the second row—specifically TOI-1836, TOI-4081, and TOI-5076— is either larger or comparable to the RV semi-amplitude. Additionally, TOI-4168 displays large RV variability.}
\label{bis-rvcor}
\end{figure*} 

We fit the broadband photometry using PHOENIX stellar atmosphere models of \citet{Husser:2013}, with the extinction $A_V$ as a free parameter limited to at most the extinction to outside the Galaxy along the line of sight to the star in the dust maps of \citet{Schlegel:1998}. We fixed $T_{\rm eff}$ and [Fe/H] to their spectroscopically determined values whenever those were available, and treated them as free parameters for TOI-4081 (where rotational broadening prevents a detailed spectroscopic analysis).

Integrating the (dereddened) model SED gives the bolometric flux at Earth, $F_{\rm bol}$ which, together with the {\it Gaia\/} parallax, directly gives the bolometric luminosity, $L_{\rm bol}$. The stellar radius, $R_\star$, is derived from $L_{\rm bol}$ and $T_{\rm eff}$ via the Stefan-Boltzmann relation. According to the stellar radius ranges defined in \cite{huber2017asteroseismology}, TOI-1836, TOI-4081, and TOI-5110 are subgiants (1.5-3 R$_{\odot}$), while TOI-2295, TOI-2537, TOI-4168, and TOI-5076 are within the main-sequence range (\(\leq\) 1.5 R$_{\odot}$). We also estimate the stellar mass from the empirical eclipsing-binary-based relations of \citet{Torres:2010}. All of these stellar properties are summarized in Table~\ref{star_info}. Additionally, we investigated whether our formal error budget is underestimated. According to \cite{tayar2022guide}, the systematic uncertainty floor is approximately $\approx$ 4.2\% for radius, $\approx$ 5\% for mass, and $\approx$ 2.4\% for temperature. Consequently, we have conservatively adopted these relative uncertainties wherever our formal error bars are smaller, as indicated in brackets in Table~\ref{star_info}, to achieve more realistic stellar parameter errors throughout this study.

To assess the activity levels of each star, we combined the SOPHIE spectra from individual stars and computed the $\log (R'_{HK})$ values, following the methodologies outlined in \cite{noyes1984rotation} and \cite{boisse2009stellar}. For TOI-5110b, due to its B-V range, we provide only an upper limit value. The resulting $\log (R'_{HK})$ values in Table \ref{star_info} indicate that TOI-1836, TOI-2295, TOI-4081, TOI-5076 and TOI-5110 are magnetically quiet, as their $\log (R'{HK})$ values are below the conventional threshold of -4.75 for active stars \citep{henry1996survey}. In contrast, TOI-2537 and TOI-4168 display relatively higher activity levels. 

We estimated the stellar age and the rotational period using the spectroscopic $\log (R'_{HK})$ activity and empirical relations from \citet{Mamajek:2008}. These parameters are also listed in Table~\ref{star_info}. Additionally, we estimated the stellar rotational periods using the method outlined in \citet{mascareno2016magnetic} and compared these values with those estimated from \citet{Mamajek:2008}. The rotational periods were consistent at 2$\sigma$ between the two methods for all stars, except for TOI-2295 and TOI-4081. We estimated periods of 78 $\pm$ 6 d and 33 $\pm$ 4 d for both stars, respectively, using the method from \citet{Mamajek:2008}. In contrast, the recipes from \citet{mascareno2016magnetic} yielded significantly shorter periods of 37 $\pm$ 7 d and 13 $\pm$ 5 d, respectively. Therefore, the rotation periods of TOI-2295 and TOI-4081 remain uncertain.

\section{Detection and characterization of the systems}
\label{ident}

In this section, we first present our general methodology, followed by the results for each system. We categorize these results into three groups: newly characterized planetary systems, incompletely confirmed candidates, and clear false positives. 

\subsection{General method}
\label{general_method}

For each system, we first present evidence for the RV-only detection of the signals. The RV-only analyses were conducted using \texttt{juliet} \citep{espinoza2019juliet}, which employs the \texttt{radvel} package to construct Keplerian models \citep{fulton2018radvel}. \texttt{Juliet} computes the Bayesian log evidence $(\ln Z)$, allowing us to compare different models. A difference in Bayesian log evidence ($\Delta \ln Z$) greater than two indicates a moderate preference for one model over another, while a difference exceeding five suggests strong favorability \citep{trotta2008bayes}. Models with $\Delta \ln Z \leq 2$ are considered statistically indistinguishable. For our analysis, we performed one or two unconstrained Keplerian models per system (depending on the specific case) and a no-planet model. If the unconstrained Keplerian model was statistically indistinguishable from the no-planet model, we proceeded with a more informed model (or models). Table~\ref{table:keplerian_models} summarizes the $\Delta \ln Z$ values for the various models tested across each system.

\begin{table}
\caption{RV-only model comparisons for the targets studied in this paper. For systems where the unconstrained Keplerian model was statistically indistinguishable from the no-planet model ($\Delta \ln Z \leq 2$) or resulted in unsatisfactory convergence, we applied an informed model (or models). In the multi-planet systems TOI-2295 and TOI-2537, an unconstrained one-Keplerian model was applied to the outer planet. The star TOI-4168 is excluded from the table due to bimodality in the posterior distribution of its no-planet model (see Sect.~\ref{TOI4168})}.
\centering
\resizebox{1\columnwidth}{!}{%
\begin{tabular}{lll}
\hline
TOI    & RV-only Model                          & $\Delta \ln Z$ \\ \hline
TOI-2295        & No-planet                              & 0             \\ 
                & Unconstrained one-Keplerian                           & 32.0            \\ 
                & Unconstrained two-Keplerian                           & 62.4            \\ \hline
TOI-2537        & No-planet                              & 0             \\ 
                & Unconstrained one-Keplerian                           & 49.5            \\ 
                & Unconstrained two-Keplerian                           & 981            \\ \hline
TOI-5110        & No-planet                              & 0             \\ 
                & Unconstrained one-Keplerian                           & 16.5            \\ \hline
TOI-5076        & No-planet                              & 0             \\ 
                & Unconstrained one-Keplerian$^{1}$                           & ---            \\ 
                & Informed one-Keplerian (fixed period)        & 1.8             \\
                & Informed one-Keplerian (Gaussian priors)$^{2}$         & 3.4             \\ \hline 
TOI-1836        & No-planet                              & 0             \\ 
                & Unconstrained one-Keplerian                           & 0.7             \\ 
                & Informed one-Keplerian (Gaussian priors)$^{2}$         & 6.2             \\ \hline
TOI-4081        & No-planet                              & 0             \\ 
                & Unconstrained one-Keplerian                           & 2.7             \\ \hline
\end{tabular}%
}
\tablefoot{$^{1}$ The model resulted in unsatisfactory convergence. $^{2}$ Gaussian priors were applied to \(T_c\) and period based on TESS photometry.}
\label{table:keplerian_models}
\end{table}

Additionally, we performed several tests to investigate the nature of the candidate signals. First, we extracted RVs using different numerical masks, including G2, K2, and M5. Variations in the planetary semi-amplitude among RVs extracted with these masks could indicate blending scenarios \citep[e.g.,][]{santerne2011sophie}. The results showed consistent planetary semi-amplitudes within 1$\sigma$ across all tested masks for all candidates, except for TOI-4081.01, where high RV error bars prevented us from a definitive conclusion (see Sect. \ref{detection-4081}). Second, we analyzed the variation and correlation between RVs and bisector spans, as significant bisector variation or a correlation with RVs could also suggest blending. Based on these tests, we either drew conclusions regarding the nature of each signal or performed further analysis.

Next, we performed joint analyses of the RVs and photometry of each system using the \exofasttwo{} package \citep{Eastman2013, Eastman2017,eastman2019}. This modeling software employs a differential evolution Markov Chain coupled with a Metropolis-Hastings Monte Carlo sampler to explore the system parameters. To ensure the convergence of the chains, \exofasttwo{} incorporates a built-in Gelman-Rubin statistic \citep{gelman1992inference, gelman2004bayesian, ford2006improving}; a value below 1.01 indicates that the chains are well-mixed. 

For the joint modeling of the photometry and RVs, we applied Gaussian priors for both \(T_c\) and the period of the planets centered on the values provided by the TESS photometry. We used a wide standard deviation of 0.1~d, which is several orders of magnitude broader than the uncertainties reported by the SPOC and QLP pipelines. We also imposed Gaussian priors on $R_\star$ and $M_\star$ using the results of Sect. \ref{stellar}. The spectroscopic analysis described in Sect. \ref{stellar} informed Gaussian priors on the [Fe/H] and $T_{\rm eff}$ parameters. Throughout this paper, we separated the available FFI and 2-minute light curves and applied a jitter to each. Furthermore, when necessary, we fitted a spline \citep{vanderburg2014technique} to detrend the instrumental systematic on the light curves. It is worth noting that \texttt{EXOFASTv2} inherently integrates priors on the quadratic limb darkening by interpolating the limb darkening models at each step in $\log g$, $T_{\rm eff}$, and [Fe/H] \citep{claret2011gravity}. Table \ref{prior_exofast} summarizes the priors we used and describes the parameters.

\subsection {Planetary systems}

\begin{figure}
\centering
\includegraphics[width=0.48\textwidth]{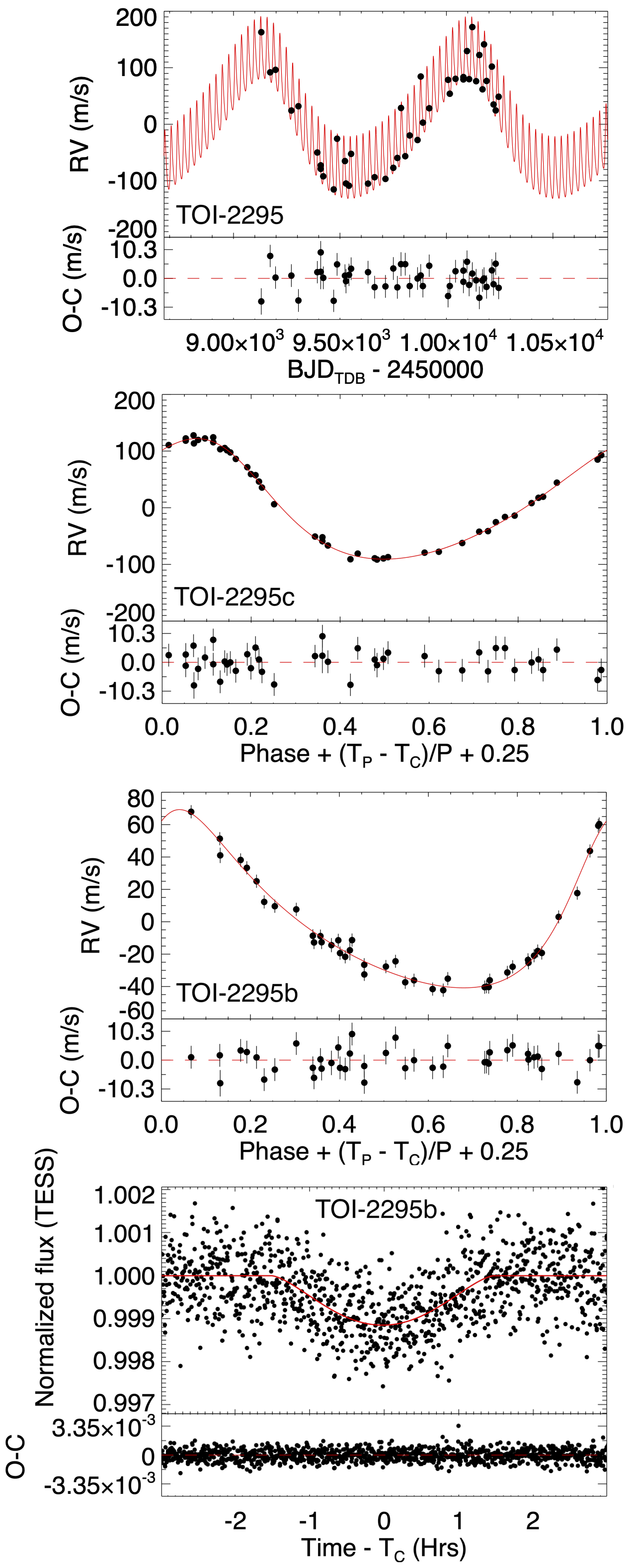}
\caption{SOPHIE RV measurements of TOI-2295 (\emph{top panel}), phase-folded RV measurements for TOI-2295c (\emph{second panel}) and TOI-2295b (\emph{third panel}), along with the phase-folded TESS light curve for TOI-2295b (\emph{bottom panel}). The red lines represent the median of median models using \exofasttwo. Residuals of the data are plotted at the bottom of each panel.}
\label{2295_photo_rv}
\end{figure}

\subsubsection{TOI-2295}
\subsubsubsection{Planetary identification of TOI-2295b and c}
\label{detection_toi2295}
SPOC identified TOI-2295.01 as a planetary candidate in TESS data with a period of 30.03324 $\pm$ 0.00013 d and transit central time of \(T_c\) = 2458713.4518$\pm$ 0.0015 Barycentric Julian Date (BJD). Following the first measurements with SOPHIE, our preliminary RV analysis, fixing \(T_c\), and the period to the derived values from the photometry, revealed RV variations in phase with the 30.03-d signal identified in the TESS data. Furthermore, our RVs also revealed a clear curvature with a possible periodicity. Therefore, we continued to observe this star to better characterize the transiting planet candidate and to investigate the nature of the observed slower drift.

Our final RV datasets comprise 44 measurements collected over 3 years (see Fig. \ref{2295_photo_rv}). The dataset clearly shows a second periodic signal in the RVs. Consequently, we initiated our analysis by applying an unconstrained two-Keplerian RV-only model. For this model, we used wide prior distributions on all parameters (see Table \ref{prior_rv-only} for details of parameter priors). This model detects both signals, revealing a period of 29.998$\pm$ 0.017 d and semi-amplitude of $K$ = 52 $\pm$ 3 m/s in phase with the TESS ephemeris for TOI-2295.01, and a period of 957 $\pm$ 10 d and $K$ = 106 $\pm$ 2 m/s for the second signal. This model is statistically favored over alternative models (see Table~\ref{table:keplerian_models}). The standard deviation of the RV residuals after fitting a two-Keplerian model is approximately 5 m/s, which is comparable to the 3~m/s mean RV error bar.

Additionally, the bisector span displays no significant variations (Fig. \ref{bis-rvcor}), with an 8.5 m/s dispersion which is much smaller than the RV 82 m/s dispersion. Notably, even the 31 m/s RV dispersion after removal of the long-period signal is well above the bisector variance. Furthermore, the CCF bisector spans show no correlation with the RVs (Pearson’s coefficient R = -0.11) or with their residual (R= 0.13) after removing both signals.

These analyses show that the RV variations are attributable to planetary signals, designated as TOI-2295b and TOI-2295c, rather than changes in spectral line profiles due to a blend scenario. We proceed with modeling both the SOPHIE RVs and TESS photometric data for this target using \exofasttwo.

\renewcommand{\arraystretch}{1.2}
\begin{table*}
\caption{Median values with 68\% confidence interval of stellar and planetary parameters derived for TOI-2295, TOI-2537, and TOI-5110 systems, using \exofasttwo.}
\resizebox{\textwidth}{!}{%
\begin{tabular}{lcccccc}
\hline
Parameter & Units & \multicolumn{2}{c}{TOI-2295} & \multicolumn{2}{c}{TOI-2537} & TOI-5110 \\
\hline
\multicolumn{7}{l}{Stellar Parameters:} \\
$M_*$ & Mass ($M_{\odot}$) & \multicolumn{2}{c}{$1.168^{+0.070}_{-0.071}$} & \multicolumn{2}{c}{$0.771\pm0.049$} & $1.469^{+0.089}_{-0.088}$ \\
$R_*$ & Radius ($R_{\odot}$) & \multicolumn{2}{c}{$1.459^{+0.056}_{-0.058}$} & \multicolumn{2}{c}{$0.771^{+0.039}_{-0.040}$} & $2.333^{+0.097}_{-0.096}$ \\
$L_*$ & Luminosity ($L_{\odot}$) & \multicolumn{2}{c}{$2.06^{+0.27}_{-0.25}$} & \multicolumn{2}{c}{$0.301^{+0.052}_{-0.046}$} & $7.03^{+0.97}_{-0.83}$ \\
$\rho_*$ & Density (cgs) & \multicolumn{2}{c}{$0.528^{+0.075}_{-0.060}$} & \multicolumn{2}{c}{$2.37^{+0.41}_{-0.34}$} & $0.163^{+0.024}_{-0.021}$ \\
$\log{g}$ & Surface gravity (cgs) & \multicolumn{2}{c}{$4.176^{+0.042}_{-0.039}$} & \multicolumn{2}{c}{$4.551^{+0.049}_{-0.048}$} & $3.869^{+0.044}_{-0.045}$ \\
$T_{\text{eff}}$ & Effective Temperature (K) & \multicolumn{2}{c}{$5730\pm140$} & \multicolumn{2}{c}{$4870\pm150$} & $6160\pm150$ \\
$[\text{Fe/H}]$ & Metallicity (dex) & \multicolumn{2}{c}{$0.316\pm0.040$} & \multicolumn{2}{c}{$0.080^{+0.079}_{-0.080}$} & $0.065\pm0.050$ \\ 
\\
Planetary Parameters:& & TOI-2295b & TOI-2295c & TOI-2537b & TOI-2537c & TOI-5110b \\
$P$ & Period (days) & $30.033302^{+0.000072}_{-0.000073}$ & $966.5^{+4.3}_{-4.2}$ & $94.1022\pm0.0011^{1}$ & $1920^{+230}_{-140}$ & $30.158577^{+0.000092}_{-0.000095}$ \\
$R_P$ & Radius ($R_{\text{J}}$) & $1.47^{+0.85}_{-0.53}$ & --- & $1.004^{+0.059}_{-0.061}$ & --- & $1.069^{+0.054}_{-0.052}$ \\
$M_P$ & Mass ($M_{\text{J}}$) & $0.875^{+0.042}_{-0.041}$ & --- & $1.307^{+0.091}_{-0.088}$ & --- & $2.90\pm0.13$ \\
$T_C$ & Time of conjunction (BJD$_{TDB}$) & $2458713.4551\pm0.0027$ & $2459302.8\pm3.8$ & $2458440.329\pm0.014^{1}$ & $2460060^{+25}_{-24}$ & $2459503.6790^{+0.0012}_{-0.0013}$ \\
$a$ & Semi-major axis (AU) & $0.1992^{+0.0039}_{-0.0041}$ & $2.018^{+0.040}_{-0.042}$ & $0.3715^{+0.0077}_{-0.0080}$ & $2.78^{+0.22}_{-0.15}$ & $0.2157^{+0.0042}_{-0.0044}$ \\
$i$ & Inclination (Degrees) & $88.16^{+0.14}_{-0.15}$ & undefined & $89.592^{+0.097}_{-0.075}$ & undefined & $85.2^{+2.1}_{-1.2}$ \\
$e$ & Eccentricity & $0.334\pm0.012$ & $0.194\pm0.012$ & $0.364\pm0.039$ & $0.287^{+0.060}_{-0.052}$ & $0.745^{+0.030}_{-0.027}$ \\
$\omega_*$ & Argument of Periastron (Degrees) & $-39.2^{+2.7}_{-2.6}$ & $39.5\pm3.0$ & $75.2^{+6.8}_{-7.1}$ & $-35.6^{+7.8}_{-9.2}$ & $92.2^{+3.1}_{-2.8}$ \\
$T_{\text{eq}}^{2}$ & Equilibrium temperature (K) & $747\pm24$ & $234.9\pm7.6$ & $338\pm14$ & $123.1^{+6.3}_{-6.7}$ & $976^{+33}_{-32}$ \\
$K$ & RV semi-amplitude (m/s) & $54.7\pm1.3$ & $105.8\pm1.1$ & $74.7\pm4.1$ & $145.9^{+5.8}_{-5.1}$ & $218.4^{+13}_{-9.7}$ \\
$M_P\sin i$ & Minimum mass ($M_{\text{J}}$) & $0.875^{+0.042}_{-0.041}$ & $5.61^{+0.23}_{-0.24}$ & $1.307^{+0.091}_{-0.088}$ & $7.23^{+0.52}_{-0.45}$ & $2.89\pm0.13$ \\
$R_P/R_*$ & Radius of planet in stellar radii & $0.104^{+0.057}_{-0.036}$ & --- & $0.1338^{+0.0018}_{-0.0021}$ & --- & $0.04709^{+0.00096}_{-0.00091}$ \\
$\delta$ & Transit depth (fraction) & $0.0107^{+0.015}_{-0.0061}$ & --- & $0.01791^{+0.00049}_{-0.00057}$ & --- & $0.002217^{+0.000091}_{-0.000085}$ \\
$\tau$ & Ingress/egress transit duration (days) & $0.0627\pm0.0024$ & --- & $0.0307^{+0.0033}_{-0.0037}$ & --- & $0.0096^{+0.0018}_{-0.0015}$ \\
$T_{14}$ & Total transit duration (days) & $0.1254\pm0.0048$ & --- & $0.2057^{+0.0031}_{-0.0036}$ & --- & $0.1757^{+0.0029}_{-0.0026}$ \\
$b$ & Transit Impact parameter & $1.056^{+0.063}_{-0.043}$ & --- & $0.480^{+0.064}_{-0.11}$ & --- & $0.43^{+0.12}_{-0.22}$ \\
$\rho_P$ & Density (cgs) & $0.34^{+0.96}_{-0.25}$ & --- & $1.60^{+0.35}_{-0.26}$ & --- & $2.95^{+0.50}_{-0.43}$ \\
$\log{g}_P$ & Surface gravity & $3.00^{+0.39}_{-0.40}$ & $4.038^{+0.069}_{-0.066}$ & $3.507^{+0.060}_{-0.056}$ & $4.159^{+0.074}_{-0.069}$ & $3.799^{+0.047}_{-0.048}$ \\
\hline
\end{tabular}%
}
\tablefoot{$^{1}$ Among all the planets characterized in this paper, only TOI-2537b exhibits TTVs and deviates from purely Keplerian motion. Therefore, its $T_C$ and period do not follow a linear ephemeris and are calculated while accounting for the TTVs. $^{2}$Assumes no albedo and perfect redistribution.}

\label{toi_2295_2537_5110}
\end{table*}

\subsubsubsection{Joint analysis of TOI-2295}
\label{join_2295}

For the joint modeling of photometry and RVs, we followed the general setup described in Sect. \ref{general_method}. Furthermore, we fitted a spline to detrend the instrumental systematic on the light curves. The joint modeling analysis shows that the TOI-2295b light curve exhibits V-shaped and therefore grazing transits (Fig. \ref{2295_photo_rv}). The planet radius and the impact parameter are almost degenerate in such a grazing geometry, with the light curve providing a reliable lower limit on the planetary radius but a much weaker upper one. The most probable solution then often points to a considerably larger radius than would be physically expected \citep[e.g.,][]{grieves2021populating, psaridi2023three,bell2024toi}. Our study encounters that scenario, with an unconstrained radius range of $7.6^{+6.2}_{-4.5}$ R$_{J}$ centered on values that would be stellar rather than planetary. If this were a valid result, we would expect to observe a distinct secondary eclipse, which is not seen. 

To address this issue, we introduced a Gaussian prior on the ratio of planetary radius to the stellar radius ($R_P/R_\star$, see Table \ref{prior_exofast}), informed by the square of the transit depth value supplied by SPOC. We employed a wide standard deviation, assuming that the maximum planetary radius would not exceed three times the radius of Jupiter. We then re-ran the \exofasttwo. The resulting parameters are presented in Tables \ref{toi_2295_2537_5110} and \ref{tab:TOI_parameters}. The probability distribution function (PDF) of $R_P/R_*$ and $R_P$, with and without applying a prior on $R_P/R_*$, is provided in Appendix \ref{corner_2295}. The median model for the RV and photometric data is shown in Fig. \ref{2295_photo_rv}.

The final results interpret TOI-2295b as a warm giant planet with a radius of $1.47^{+0.85}_{-0.53}$ $R_J$, with low precision due to its grazing observing geometry. The planet has a well-constrained mass of $0.875^{+0.042}_{-0.041}$ $M_{\rm{J}}$ and transits its host star every $30.033302^{+0.000072}_{-0.000073}$ d. The empirical mass-radius relation of \cite{2017ApJ...834...17C} for Jovian planets predicts a planetary radius of 1.23$\pm$0.21 $R_J$. Therefore, this prediction and our measured radius are compatible within 1$\sigma$. The outer planet, TOI-2295c, has a minimum mass of 5.61$^{+0.23}_{-0.24}$ $M_J$ and it orbits the host star every $966.5^{+4.3}_{-4.2}$ d.

We used the Gaia DR3 catalog to set an upper limit to the true mass of TOI-2295c. The 5-parameter solution listed for TOI-2295 has a renormalized unit weight error (RUWE) of 0.919. Since it is smaller than 1, Gaia detects no sign of astrometric motion, which sets an upper limit on the mass of the planet TOI-2295\,c. We used the GaiaPMEX tool \citep{kiefer2024searching} to explore the companion masses and orbital periods that remain compatible with the Gaia DR3 RUWE value. Fig.~\ref{fig:PMEXresult_TOI2295} shows the confidence maps on the mass and orbital period of companions around TOI-2295 obtained by GaiaPMEX. Planet b is in the 68.3\% confidence region, while planet c is in the 95.4\% confidence region. As the detection of transits already constrains the inclination of planet b to edge-on, Gaia brings no useful constraints on this planet at the corresponding period of 0.08\,yr. By contrast, the $M\sin i$ of the planet c is already close (in log-space) to the exclusion region located beyond the 99.7\% confidence region at $\sim$25\,$M_{\rm{J}}$. Therefore, while Gaia does not provide further constraints on the planet b's mass beyond what the RV and transit data offer, it establishes an upper limit on the mass of planet c.

\subsubsection{TOI-2537}
\subsubsubsection{Planetary identification of TOI-2537b and c}
\label{Planetary_identification_toi2537}

\begin{figure}
\centering
\includegraphics[width=0.95\columnwidth]{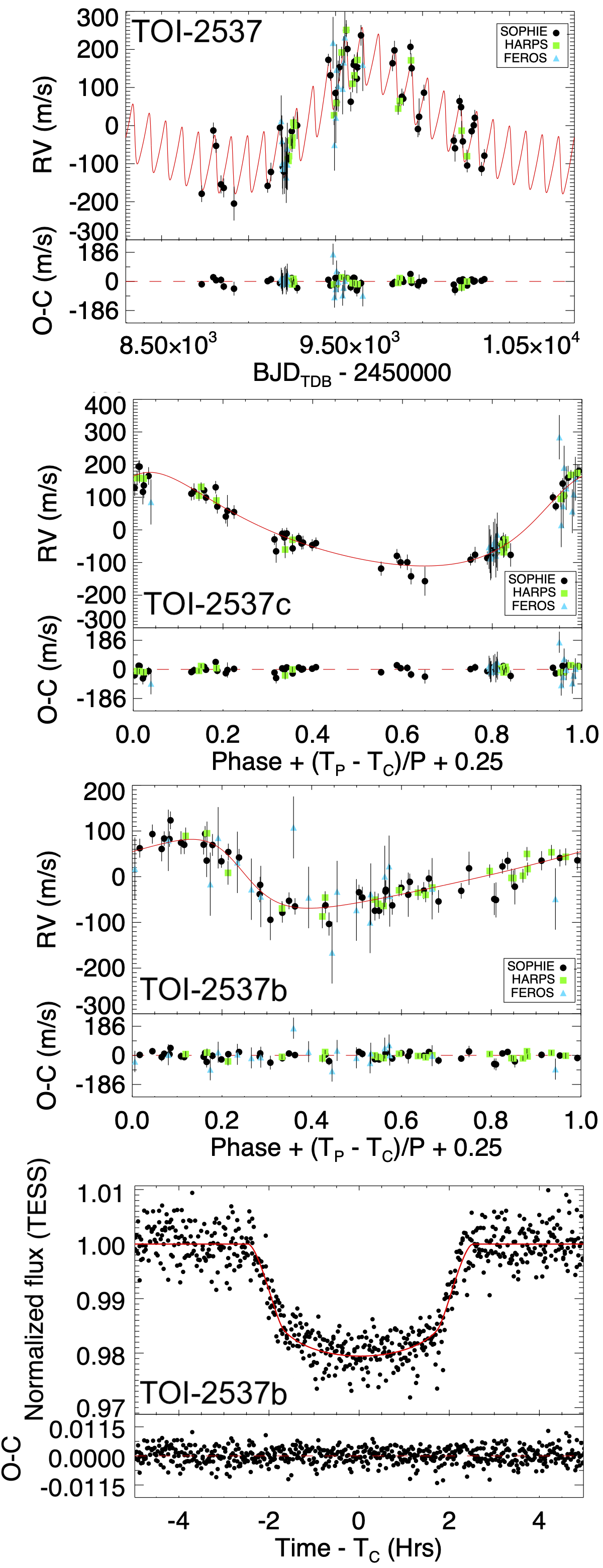}
\caption{RV measurements of TOI-2537 (\emph{top}), phase-folded RVs for TOI-2537c (\emph{second panel}), phase-folded RVs (\emph{third panel}) and light curves for TOI-2537b (\emph{bottom}). The red lines represent the median models, accounting for TTVs as determined by \exofasttwo{}. Residuals are displayed at the bottom of each panel.}
\label{TOI2537_rv_phot}
\end{figure}

The TESS FFI observations of sector 5 and the 2-minute cadence observations of sector 43 each cover one transit of TOI-2537, making it a duo-transit candidate. SPOC lists transit epochs \(T_c\) of the two transits of 2458440.324843 $\pm$0.0000091 BJD (sector 5) and 2459475.46593 $\pm$ 0.00099 BJD (sector 43), with a separation of 1035.14 d. We started our SOPHIE RV follow-up campaign after the identification of the first transit candidate and soon complemented it with HARPS and FEROS. The 94-d periodic variation in those RVs is compatible with the time of the first transit, and also with the time of the second transit when TESS eventually observed it. The RV data also revealed an additional long-term signal. After four years of dedicated observations, we have obtained a total of 83 RV data points, of which 3 data points were excluded due to lower quality (see Table \ref{rv_ob} and Sect.\ref{observation} for details on each instrument's observations and data exclusion). The RV data depicted in Fig. \ref{TOI2537_rv_phot} reveal a clear 94-d periodic signal, accompanied by longer-term variations. Recently, TESS observed a third transit of this system in sector 70, which is consistent in depth, and duration with both previously detected transits.

We initiated our analysis by applying an unconstrained two-Keplerian RV-only model using \texttt{juliet}, employing very wide priors on all parameters (Table. \ref{prior_rv-only}). This model, which was statistically preferred among the tested models (see Table \ref{table:keplerian_models}), converged on two robustly detected periodic signals: the first with period 94.08~$\pm$~0.12 d and amplitude K~=~75~$\pm$~4~m/s, and the second with period 1932$^{+173}_{-120}$ d and amplitude K~=~145~$~\pm$~5~m/s. The \(T_c\) = 2458439 $\pm$ 2 BJD for the inner planet candidate is fully consistent with the SPOC \(T_c\), and the separation between the three TESS transits is a multiple of the inner 94.1-d RV period. Together, the consistency between the RV and SPOC \(T_c\) and between the periodicity of the transit events and the RV period make clear that the 94.1~d RV planet candidate is the TESS transiting candidate. After removing the two RV signals, the dispersion of the residuals is 19 m/s for SOPHIE, to be compared with a 20~m/s mean uncertainty; for FEROS, it is 52 m/s, for a mean uncertainty of 12 m/s, and for HARPS, it is 15 m/s, for a mean uncertainty of 16 m/s. While the dispersions of the SOPHIE and HARPS residuals are compatible with their respective RV uncertainties, the dispersion of the FEROS residuals is approximately four times its mean RV uncertainty. This suggests either that the FEROS RV error bars are underestimated, or/and that its RV measurements are more affected by the stellar activity jitter (see the discussion below about its correlation with bisectors).

Additionally, we observed no significant variations in the bisector (see Fig. \ref{bis-rvcor}), with dispersions of 38 m/s for SOPHIE, 45 m/s for FEROS, and 20 m/s for HARPS, notably lower than the corresponding RV dataset dispersions: 118 m/s for SOPHIE, 117 m/s for FEROS, and 94 m/s for HARPS. Even after removing the outer planet candidate from the RVs, the dispersion of the bisectors remained lower than that of the RV residuals, with values of 57 m/s for SOPHIE, 60 m/s for FEROS, and 43 m/s for HARPS.

Further examination shows no significant correlation between RVs and bisector values, with correlation coefficients of R= -0.005 for SOPHIE, R= 0.2 for FEROS, and R= -0.3 for HARPS. An investigation of the correlation between RV residual after removing the outer planet signal and bisector yielded a correlation coefficient of R=0.2 for SOPHIE, R= -0.5 for HARPS, and R= 0.4 for FEROS. While SOPHIE RVs exhibited low correlation, HARPS and FEROS showed moderate correlation. Subsequent investigation into correlation coefficients between RV residuals after removing both signals and bisector exhibited a correlation coefficient of R=0.2 for SOPHIE, R=0.8 for FEROS, and R=-0.4 for HARPS (see Fig. \ref{rv_bis_res_2537}).

 As this correlation persists in our final RV residual, and also with the fact that TOI-2537 is one of the most active stars in our study sample with $\log (R'_{HK})$= -4.5 $\pm$ 0.1, we attributed this correlation to the activity induced “jitter” of the star and not a blended eclipsing binary scenario. Additionally, we note that the semi-amplitude of the outer signal notably exceeds what could be attributed to a long-term activity cycle of a main sequence star, as outlined by \cite{lovis2011harps}. Therefore, we confidently conclude that TOI-2537b and c have planetary natures.

Before proceeding with the joint analysis, we investigate the effect of stellar activity on our planet parameters using the empirical relationship between $\log (R'_{HK})$ and stellar jitter following \cite{hojjatpanah2020correlation}. This assessment yielded a stellar jitter of approximately 11 m/s, which is smaller than the mean RV errors of all datasets and is significantly smaller than the semi-amplitude of both planets (K = 75 m/s and K = 145 m/s for TOI-2537b and c, respectively). Additionally, our RV data cover multiple stellar rotations, which, to some extent, helps mitigate this challenge as we average over epochs with varying activity levels. Finally, the periods of both planets significantly exceed the estimated stellar rotational period (19 $\pm$8 d), which contributes to minimizing the impact of stellar activity. Considering these factors, we conclude that the influence of stellar activity on planetary parameters will be negligible here. Consequently, we proceeded with joint analysis without employing a Gaussian process.

\subsubsubsection{Joint analysis of TOI-2537}
\label{joint_2537}
For the joint modeling of TESS photometric data along with SOPHIE, HARPS, and FEROS RVs, we applied Gaussian priors for the same parameters as those outlined in the general setup (see Table \ref{prior_exofast} and Sect. \ref{general_method}). The only difference was that since SPOC did not provide a precise period for this planet, we used the period derived from RVs as the center of the Gaussian prior, with a standard deviation of 0.5 d. For TOI-2537, we did not detrend the 2-minute light curves, as they are fairly flat; however, we did detrend the FFI light curve.

Initially, in our joint modeling, we only included the first two transits, as the third one was detected more recently. Based on this initial model, we estimated the mid-time of the transit and a corresponding model for the newly detected, third transit. However, as shown in Fig.~\ref{lightcurves_ttv_2537}, there is a significant difference ($\sim$ 22 minutes) between the observed mid-transit time and the predicted mid-transit time based on the model of the first two transits, indicating a clear transit timing variation (TTV). To account for this, we incorporated TTV modeling using \exofasttwo{} as part of our joint analysis and re-ran the model with all available data. Each transit event was treated as an independent light curve, with no additional priors for the TTV fit—only allowing the transit timings to shift relative to what is expected from a purely Keplerian fit. The BIC statistic for this joint model, which included the TTV model, demonstrated a stronger statistical preference (\(\Delta \text{BIC} = 197.93\)) and provided an improved fit to the light curves. Consequently, we adopted this model as our final solution. The posterior distributions of all parameters are provided in Table~\ref{toi_2295_2537_5110}. The median model for both RV and photometric data is shown in Fig.~\ref{TOI2537_rv_phot}

\begin{figure}
\centering
\includegraphics[width=0.9\columnwidth]{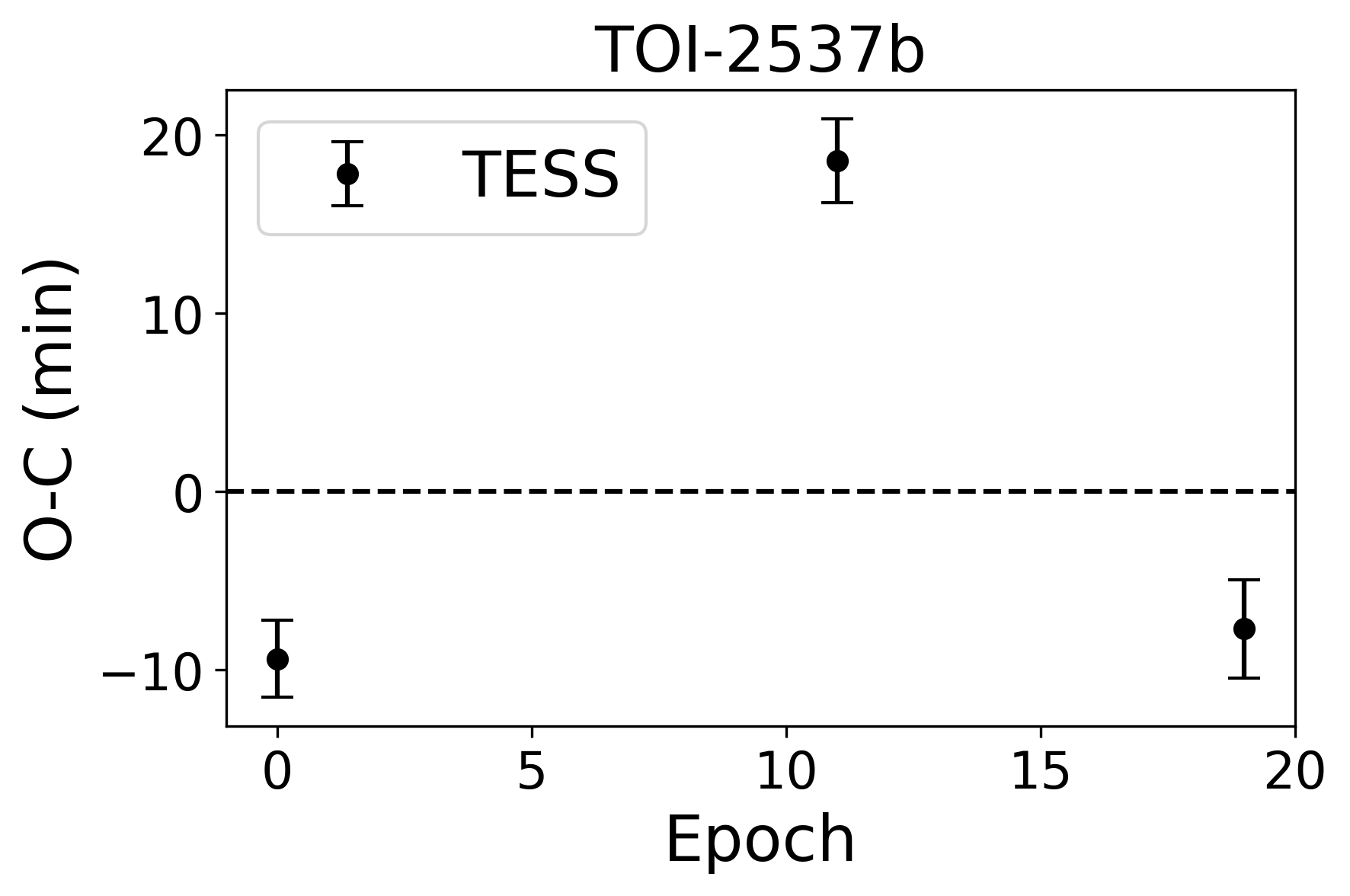}
\caption{The difference between the predicted and observed mid-transit times for TOI-2537b.}
\label{2537_TTV}
\end{figure}

\noindent Incorporating TTV into the joint analysis allowed us to account for deviations of transit mid-times from the linear ephemeris, \(T_C (E) = T_0 + E \cdot P\), where \(T_0\) represents the conjunction time at a reference epoch (\(E\)). By accounting for these deviations, the corresponding mid-transit times for each TESS light curve were measured and are presented in Table~\ref{tab:transitpars}. \exofasttwo{}  also generated an observed-minus-calculated (O-C) plot of the transit mid-times versus epoch for TOI-2537b (see Fig.~\ref{2537_TTV} and Table~\ref{tab:transitpars} for specific values). In this plot, the observed mid-transit times are subtracted from those predicted by the linear ephemeris. The TTVs listed in Table~\ref{tab:transitpars} have an average amplitude of \( 12.08 \pm 1.40 \) minutes. Each TTV deviates from the predicted times based on the linear ephemeris by \( 4\sigma \), \( 8\sigma \), and \( 3\sigma \) for the first, second, and third detected transits, respectively.

Based on our final parameters, the transiting planet TOI-2537b is classified as a giant planet, boasting a radius of $1.004^{+0.059}_{-0.061}$ R$_{\rm{J}}$ and a mass of $1.307^{+0.091}_{-0.088}$ $M_{\rm{J}}$. It has an orbital period of $94.1022\pm0.0011$ d, with an eccentricity of 0.364$\pm$0.039. Additionally, the outer companion exhibits a period of 1920$^{+230}_{-140}$ d, a minimum mass of $7.23^{+0.52}_{-0.45}$ $M_{\rm{J}}$, with a eccentricity of 0.278$^{+0.060}_{-0.052}$. Both planets have significantly eccentric orbits. The dynamical analyses of the system is discussed in Sect.~\ref{Dynamical_analyses}.

 We also highlight that in our final model for this system, the resulting fitted jitter values are 64$^{+14}_{-11}$ m/s, 12.2$^{+4.9}_{-4.5}$ m/s, and 14.2$^{+4.1}_{-3.8}$ m/s for FEROS, HARPS, and SOPHIE, respectively. The fitted jitters in SOPHIE and HARPS data are likely to be mainly due to stellar activity, whose jitter is expected to be of the order of 11 m/s (see Sect. \ref{Planetary_identification_toi2537}). The FEROS jitter significantly exceeds the expected stellar jitter, likely indicating additional instrumental or activity-induced jitter here. 

\begin{figure}
\centering
\includegraphics[width=\columnwidth]{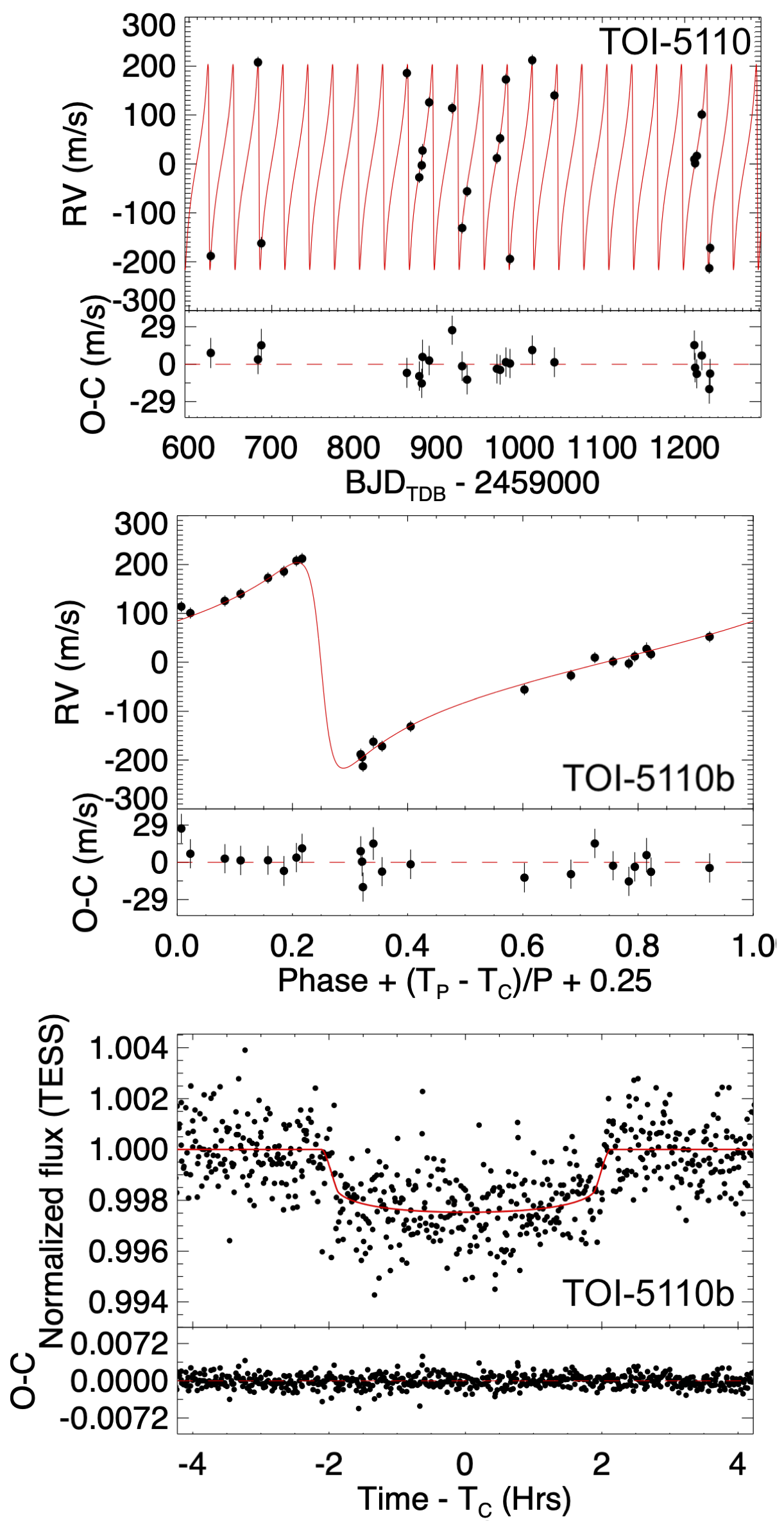}
\caption{SOPHIE RV measurements for TOI-5110, overplotted by the median Keplerlian model (\emph{top}). RVs (\emph{middle}) and TESS photometric data (\emph{bottom}) phase-folded to the orbital period of the planet candidate. Residuals for each dataset are plotted at the bottom of their respective panels.}
\label{5110_rv_phot}
\end{figure}

TOI-2537 has a 5-parameter solution in the Gaia DR3 catalog, with a RUWE of 1.06, which is close to 1, suggesting no significant astrometric motion. However, using GaiaPMEX, we establish an upper limit on the mass of the TOI-2537c as we did for TOI-2295. We show the confidence maps on the mass and orbital period of companions around TOI-2537 obtained by GaiaPMEX in Fig.~\ref{fig:PMEXresult_TOI2295}. Both planets b and c are in the 68.3\% confidence region. Since the inclination of planet b is determined to be edge-on due to the detection of the transit, Gaia does not provide any additional constraints for this planet at its corresponding period of 0.25\,yr. Moreover, at an orbital period of $\sim$5\,yr, the $M\sin i$ of planet c is below the exclusion region located beyond the 99.7\% confidence region at $\sim$150\,$M_{\rm{J}}$. A mass of much larger than 7\,$M_{\rm{J}}$ is thus possible for TOI-2537c.

\subsubsection{TOI-5110}
\subsubsubsection{Planetary identification of TOI-5110b}
\label{5110}

QLP identified a planetary candidate in the TESS lightcurves, namely TOI-5110.01, with a period of 30.159$\pm$0.003 d and \(T_c\)= 2459503.677$\pm$0.003 BJD around the star TOI-5110. Shortly thereafter, we initiated follow-up observations by SOPHIE. We finally collected 23 data points over around 20 months, showing significant RV variations.

We initiated our RV analysis by applying an unconstrained one-Keplerian RV-only model to the data. Uniform priors were defined for both the period and \(T_c\), ranging from 1 d to 50 d and 2459479 d to 2459529 d, respectively. Additionally, we adopted fairly broad prior ranges for the remaining parameters (refer to Table \ref{prior_rv-only}). The model converged to a highly eccentric orbit with a period of 30.136$\pm$ 0.030 d and \(T_c\)=2459504.49$\pm$0.46 BJD, aligning well with the period and \(T_c\) of the transiting candidate reported by QLP. Notably, this model was strongly favored over the no-planet scenario (see Table \ref{table:keplerian_models}). Subsequently, we refined the model by fixing \(T_c\) and period to the values from SPOC, resulting in a semi-amplitude of K= 214$^{+11}_{-7}$ m/s, indicating a significant detection of this signal in our RVs and a planetary mass. After removing this signal, the dispersion of the RV residual is 11 m/s, slightly larger than the mean RV error bars of 7 m/s, which still indicates no significant variation in RV residuals.

Furthermore, the CCF bisector spans show neither variations (see Fig. \ref{bis-rvcor}) nor correlation with RVs (R= -0.06). According to our high-spatial resolution imaging presented in Sect. \ref{SPP}, there is a nearby star with a separation of $1.084^{\prime\prime} \pm 0.009$ from TOI-5110, with a brightness difference of $6.2 \pm 0.1$ mag. This star is within the SOPHIE and TESS apertures. However, based on our bisector and mask effect analysis, we did not find any significant signs of a blend scenario. We note that while this nearby star might be gravitationally bound to TOI-5110, we do not detect any drift in our current data. We acknowledge that if the CCFs from the nearby star and the primary target were fully aligned and blended, it could have hidden any CCF line profile variations or potential RV drift. Without additional information, such as parallax measurements or the systemic velocity of the nearby star, we cannot conclusively rule out this scenario. We note that the star could simply be a projected neighbor unrelated to TOI-5110. Nevertheless, according to Equation 4 of \cite{2019ApJ...881L..19V}, the transit depth of TOI-5110.01 could only be mimicked by a blended eclipsing star with $\Delta\text{m}~\le 0.4$ mag (or $\Delta\text{m}~\le 1.6$ mag at 3$\sigma$ confidence). Thus, this nearby star is not sufficiently bright to contribute to the observed transit signal. Furthermore, given the faintness of this nearby star, any light contamination in both the SOPHIE and TESS data would be negligible.

In summary, we detected TOI-5110.01 in the SOPHIE RVs. No indications of a blend scenario were found in our mask and bisector analyses, and the nearby star is too faint to mimic the transit signal. Accordingly, we rule out background false-positive scenarios. We conclude that the signals observed in both the RV measurements and photometry are attributable to a planet, designated TOI-5110b.

\subsubsubsection{Joint analysis of TOI-5110}
\label{5110_global}

For joint modeling of SOPHIE RVs and TESS photometric data, we applied Gaussian priors on the same parameters as discussed in Sect. \ref{general_method} and listed in Table \ref{prior_exofast}. Additionally, we included a detrending on the 2-minute light curves by fitting a spline simultaneously with our joint modeling. For the FFI data, we did not detrend the light curve as it is fairly flat in time of transit. The posterior distribution of parameters can be found in Tables \ref{toi_2295_2537_5110} and \ref{tab:TOI_parameters}. Furthermore, the median model to the photometric and RV data is shown in Fig. \ref{5110_rv_phot}

Our results indicate that the planet has a period of 30.158577$^{+0.000092}_{-0.000095}$ d, a radius of 1.069$^{0.054}_{0.052}$ \(R_{\text{J}}\), and a mass of 2.90$\pm$0.13 $M_{\rm{J}}$. Additionally, it exhibits a high eccentricity of 0.745$^{+0.030}_{-0.027}$, resulting in highly varying effective temperatures throughout its orbit, ranging from about 1940 K at perihelion to about 740 K at aphelion. 

Although high orbital eccentricity can cause asymmetries in ingress and egress times, such effects are expected to be minimal \citep{hebrard2010observation}. According to equation 17 from \cite{winn2010transits}, the expected ingress/egress asymmetry is of the order of 10$^{-1}$ seconds, whereas our measurement in case of equal ingress and egress durations is 13.8$^{+2.6}_{-2.2}$ minutes. Therefore, no detectable difference in ingress/egress times is possible. Finally, we calculated the eclipse impact parameter to be 2.97$^{+0.54}_{-1.20}$, indicating no eclipse detection. This result is consistent with the absence of an eclipse signal in the TESS light curve.

\begin{figure}
\centering
\includegraphics[width=\columnwidth]{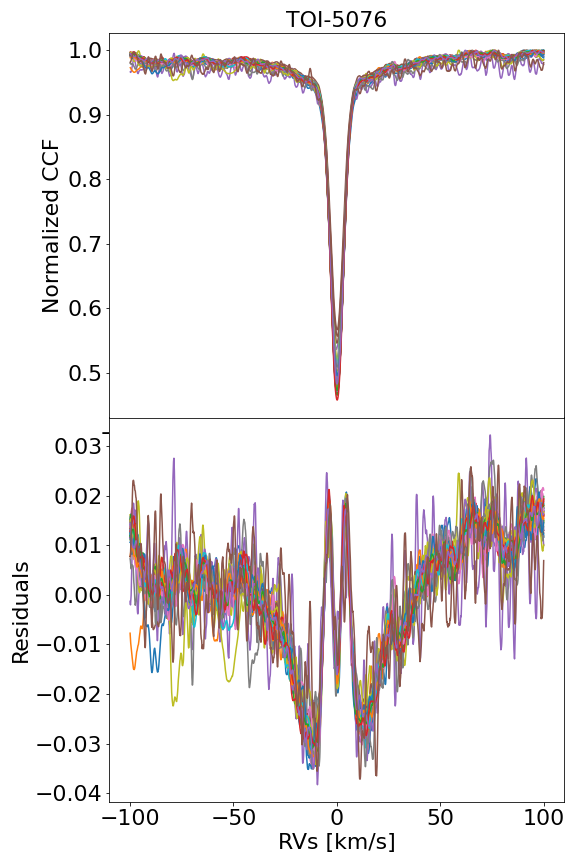}
\caption{TOI-5076's CCFs and its residual. \emph{Top}: The CCFs for each SOPHIE spectrum were computed using a G2V template. All CCFs were shifted from the velocity frame of the solar system barycenter where they are computed, to the frame of
TOI-5076. \emph{Bottom}: A Gaussian fit was subtracted to leave the residual noise. Both panels clearly display second broad CCFs, indicating a contaminated star. The feature in the center of the CCFs is caused by the primary's imperfect Gaussian fit as a result of contamination.}
\label{CCF_5076}
\end{figure}

\begin{figure}
\centering
\includegraphics[width=\columnwidth]{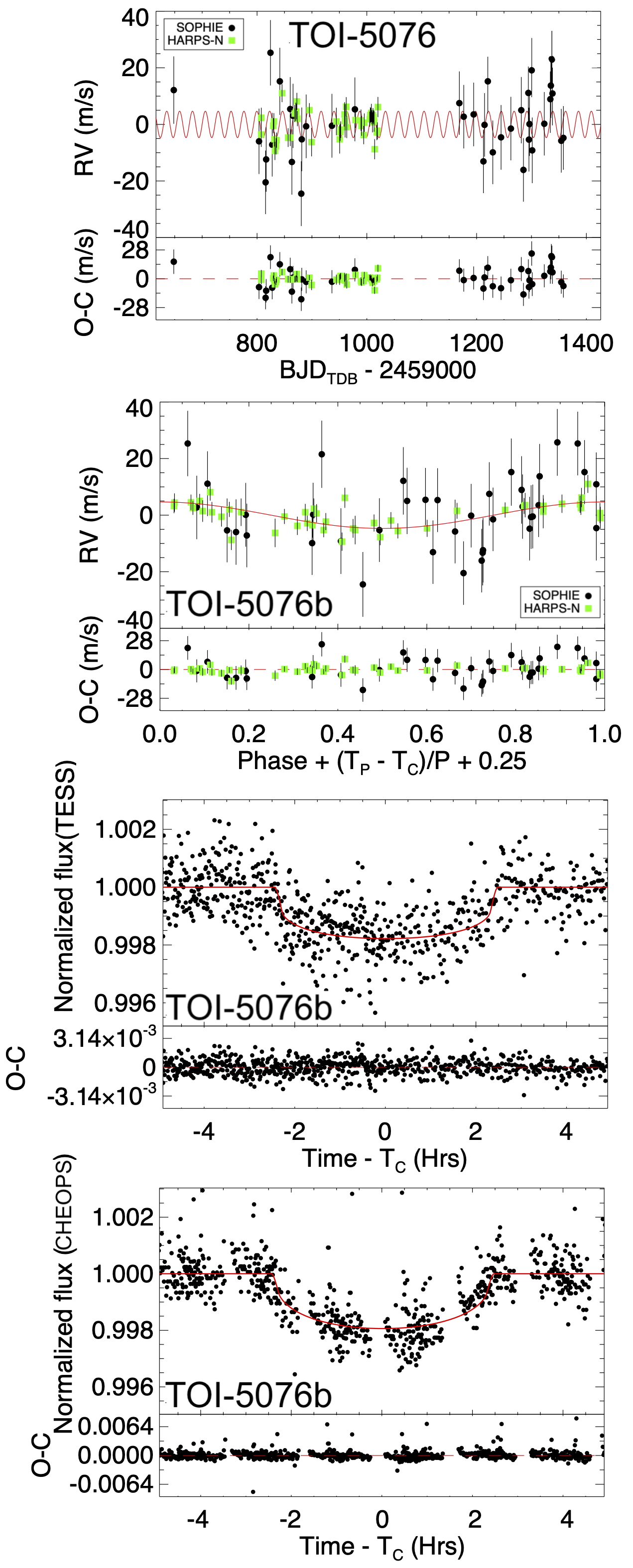}
\caption{SOPHIE and HARPS-N RV measurements for TOI-5076, overplotted by the median Keplerian model (\emph{top}). RVs (\emph{second panel}), along with TESS (\emph{third panel}), and CHEOPS (\emph{fourth panel}) photometric data phase-folded to the orbital period of the TOI-5076b. Residuals for each dataset are plotted at the bottom of their respective panels.}
\label{5076_photo_rv}
\end{figure}

\subsubsection{TOI-5076}
\subsubsubsection{Planetary identification of TOI-5076b}

QLP identified TOI-5076.01 as a planetary transit candidate from the TESS light curves with a period of 23.4435 $\pm$ 0.0004 d and a \(T_c\) of 2460204.01 $\pm$ 0.01 BJD. We collected 39 high-quality SOPHIE spectra (see Table. \ref{rv_ob}) of TOI-5076 over approximately 24 months. Additionally, in the final stages of preparing this study, \cite{montalto2024gaps} characterized this candidate using TESS photometric data and HARPS-N RVs. In this subsection, we provide an independent investigation of the candidate only using SOPHIE RVs. In the following subsection, we consider all available data, already published HARPS-N and TESS data, as well as our new SOPHIE RVs and CHEOPS photometry, to characterize the system.

We began our analysis by applying an unconstrained one-Keplerian model to the SOPHIE RV data. Initially, using wide priors for all Keplerian parameters resulted in unsatisfactory convergence. To address this, we fixed the period to the QLP value and re-ran the model (see Table \ref{prior_rv-only} for the listed model prior). The results show a \(T_c\) of 2460203.9 $\pm$ 1.3 BJD, which aligns well with the \(T_c\) of the TESS planet candidate, confirming its detection in our RV data. Because this model was statistically indistinguishable from the no-planet model (see Table \ref{table:keplerian_models}), we conducted further analysis. We used Gaussian priors centered on the QLP values for both \(T_c\) and the period, with a width of 0.1$\sigma$. A comparison between this model and the no-planet model shows \(\Delta \ln Z = 3.4\), which moderately favors this model. To examine the strength of the signal, we fixed both the \(T_c\) and period to the QLP values, the Keplerian model yields a semi-amplitude of 8.2 $\pm$ 2.6 m/s, indicating a 3$\sigma$ detection of the signal. We acknowledge that the detection falls slightly below the commonly accepted threshold for statistical significance. However, the consistency between the \(T_c\) derived from the RV model and the \(T_c\) from the photometry, along with the moderate preference for the one-Keplerian model with Gaussian priors over the no-planet model, supports the detection of the planet in the SOPHIE RV data.

Upon further investigation, we noticed a particular shape of TOI-5076's CCF, which we did not see in any of the other stars studied here. In Fig. \ref{CCF_5076}, in the top panel, we overlaid all CCFs. These CCFs were shifted from the velocity frame of the solar system barycenter, where they were computed, to the frame of TOI-5076. We then subtracted the fitted Gaussian of each CCF from its corresponding CCF, as shown in the bottom panel of Fig. \ref{CCF_5076}. In both plots, a distinct potential broad secondary CCF is observable, exhibiting a CCF contrast of approximately 4\%, an FWHM of around 68 km/s, and no discernible RV variations.

As seen in the CCF residuals, the Gaussian fit fails to accurately capture the CCF profile of the primary star, resulting in a feature at the center of the residuals. This discrepancy may impact the resulting RV of the primary. To explore this effect, we narrowed the half-window for CCF computation from 100 to 10 km/s, focusing exclusively on the CCF of the primary star. The resulting CCF from this adjustment exhibited a profile that was well-described by the Gaussian fit. Additionally, the planetary semi-amplitude remained consistent (\( K = 8.8^{+2.9}_{-2.4} \) m/s), while the bisector dispersion slightly decreased from 16 m/s to 14 m/s. Given the improved Gaussian fit, we selected this dataset for the rest of our work. The RV dispersion and RV residual dispersion after removing the model are 11 m/s and 10 m/s, respectively, with typical error bars of 4 m/s. The high RV residual could be attributed to remaining contamination from the secondary CCF or another yet undetected planet in the system.

Given the non-Gaussian nature of the CCFs identified in our analysis, we additionally employed the \texttt{sophie-toolkit}\footnote{\url{https://github.com/edermartioli/sophie}} \citep{Martioli2022,martioli2023toi}, as an alternative method for computing CCFs and RVs. This approach effectively calibrated the CCFs of TOI-5076, resulting in residual structures being completely eliminated after subtracting the template CCF. By fixing the period and \(T_c\) to the values reported by QLP, the RV semi-amplitude of this signal was measured at $10.0\pm2.2$ m/s, which is consistent within 1$\sigma$ of the value obtained from our analysis of the standard CCFs reported above and extracted using the DRS (Sect. \ref{sophie_observation}). The RMS of RV residuals is 10~m/s, also consistent with the pipeline results. This consistency between the RVs results from two different methods suggests that despite the non-Gaussian nature of CCFs, the CCFs remain sufficiently stable to yield reliable RVs for planet detection.

\renewcommand{\arraystretch}{1.2}
\begin{table*}
\caption{Median values and the 68\% confidence interval of the stellar and planetary parameters derived for TOI-5076b, TOI-1836b, and TOI-4081.01 using \exofasttwo.}
\resizebox{\textwidth}{!}{%
\begin{tabular}{lcccccc}
\hline
Parameter & Units & TOI-5076 & TOI-1836 & TOI-4081 \\
\hline
\multicolumn{5}{l}{Stellar Parameters:} \\
$M_*$ & Mass ($M_{\odot}$) & $0.789^{+0.046}_{-0.047}$ & $1.307^{+0.077}_{-0.076}$ & $1.448\pm0.090$\\
$R_*$ & Radius ($R_{\odot}$) & $0.844^{+0.020}_{-0.019}$ & $1.577\pm0.060$ & $2.48\pm0.11$ \\
$L_*$ & Luminosity ($L_{\odot}$) & $0.336^{+0.038}_{-0.034}$ & $3.70^{+0.48}_{-0.42}$ & $7.35^{+1.0}_{-0.91}$\\
$\rho_*$ & Density (cgs) & $1.857^{+0.076}_{-0.092}$ & $0.470^{+0.061}_{-0.054}$ & $0.134^{+0.021}_{-0.018}$\\
$\log{g}$ & Surface gravity (cgs) & $4.483^{+0.016}_{-0.019}$ & $4.159\pm0.039$ & $3.809\pm0.047$ \\
$T_{\text{eff}}$ & Effective Temperature (K) & $4780\pm120$ & $6380\pm150$ & $6030^{+140}_{-150}$\\
$[\text{Fe/H}]$ & Metallicity (dex) & $0.082\pm0.079$ & $-0.101\pm0.079$ & $0.01\pm0.30$ \\
$\dot{\gamma}$ & RV slope (m/s/d) & --- &--- & $0.372^{+0.064}_{-0.065}$ \\
\\
Planetary Parameters:&& TOI-5076b & TOI-1836b & TOI-4081b \\
$P$ & Period (days) & $23.443162^{+0.000062}_{-0.000063}$ & $20.380799\pm0.000016$ & $9.258388^{+0.000020}_{-0.000021}$ \\
$R_P$ & Radius ($R_{\text{J}}$) & $0.3113^{+0.0089}_{-0.0084}$ & $0.714\pm0.031$ & $1.193^{+0.066}_{-0.067}$ \\
$M_P$ & Mass ($M_{\text{J}}$) & $0.0508\pm0.0078$ & $0.121\pm0.029$ & $1.89^{+0.35}_{-0.34}$\\
$T_C$ & Time of conjunction (BJD$_{\text{TDB}}$) & $2460204.0018^{+0.0014}_{-0.0015}$ & $2459646.49351^{+0.00042}_{-0.00041}$ & $2458958.6272^{+0.0024}_{-0.0022}$\\ 
$a$ & Semi-major axis (AU) & $0.1481^{+0.0028}_{-0.0030}$ & $0.1597^{+0.0031}_{-0.0032}$ & $0.0977^{+0.0020}_{-0.0021}$\\ 
$i$ & Inclination (Degrees) & $89.84^{+0.11}_{-0.17}$ & $88.74^{+0.24}_{-0.21}$ & $82.6^{+1.1}_{-1.5}$\\
$e$ & Eccentricity & 0 (\text{fixed}) & 0 (\text{fixed}) & $0.25^{+0.22}_{-0.14}$\\
$\omega_*$ & Argument of Periastron (Degrees) & 90 (\text{fixed}) & 90 (\text{fixed}) & $71^{+30}_{-28}$\\
$T_{\text{eq}}^{1}$ & Equilibrium temperature (K) & $550\pm14$ & $966^{+31}_{-30}$ & $1466^{+51}_{-50}$\\
$K$ & RV semi-amplitude (m/s) & $4.23^{+0.62}_{-0.63}$ & $7.5\pm1.8$ & $148^{+31}{-27}$\\ 
$R_P/R_*$ & Radius of planet in stellar radii & $0.03793\pm0.00055$ & $0.04651^{+0.00039}_{-0.00040}$ & $0.0497^{+0.0016}_{-0.0023}$\\
$\delta$ & Transit depth (fraction) & $0.001439\pm0.000042$ & $0.002164^{+0.000036}_{-0.000037}$ & $0.00247^{+0.00016}_{-0.00022}$\\
$\tau$ & Ingress/egress transit duration (days) & $0.00754^{+0.00029}_{-0.00018}$ & $0.0158^{+0.0015}_{-0.0013}$ & $0.0263^{+0.0087}_{-0.011}$\\
$T_{14}$ & Total transit duration (days) & $0.2034^{+0.0020}_{-0.0023}$ & $0.2772^{+0.0017}_{-0.0016}$ & $0.1764^{+0.0069}_{-0.0082}$\\
$b$ & Transit Impact parameter & $0.107^{+0.11}_{-0.075}$ & $0.480^{+0.059}_{-0.074}$ & $0.842^{+0.038}_{-0.11}$\\
$\rho_P$ & Density (cgs) & $2.08^{+0.35}_{-0.34}$ & $0.41^{+0.12}_{-0.11}$ & $1.37^{+0.39}_{-0.31}$\\
$\log{g}_P$ & Surface gravity & $3.113^{+0.063}_{-0.073}$ & $2.77^{+0.10}_{-0.12}$ & $3.516^{+0.092}_{-0.100}$\\
\hline
\end{tabular}%
}
\tablefoot{$^{1}$Assumes no albedo and perfect redistribution}
\label{toi1836_4081_5076}
\end{table*}

While the bisector did not exhibit any correlation with RVs (R=-0.04), it displayed a dispersion of 14 m/s. This dispersion might be due to data accuracy or the presence of an unresolved nearby star, which introduces uncertainty in interpretations of the candidate's nature.

Taking into account that uncertainty, we investigated further the nature of TOI-5076.01 with statistical validation. To do so, we employed \texttt{TRICERATOPS} \citep{giacalone2020vetting}, a statistical tool that calculates the probability of a signal being produced by a transiting planet, an eclipsing binary, a nearby eclipsing binary, or an unresolved companion \citep[see the full list of false positive scenarios in Table 1 of][]{giacalone2020vetting}. This analysis provides two key metrics: the false positive probability (FPP), representing the overall likelihood that the observed signal is attributable to something other than a planet transiting the target star, and the nearby false-positive probability (NFPP), indicating the probability that the signal is caused by a known nearby star blended with the target in the TESS data. To meet validation criteria, a planet must demonstrate FPP < 0.015 and NFPP < 0.001 \citep{giacalone2020vetting}. We ran \texttt{TRICERATOPS} 20 times for TOI-5076.01 and calculated the mean and standard deviation of the resulting FPP and NFPP values. We found FPP = 0.001 $\pm$ 0.002 and NFPP = (1.0$\pm$ 0.3 ) $\times$ 10$^{-5}$. Based on these results, the candidate has a planetary nature.

In summary, we successfully detected the TESS planetary candidate TOI-5076.01 in our SOPHIE RV dataset. Further investigation suggests the presence of a potential secondary shallow and broad CCF within the primary CCF. However, our CCF tests showed that this does not compromise the RV detection of the candidate on the primary CCF. Additionally, we did not observe any RV mask effect, and the bisector analysis indicates no correlation with RV. However, the bisectors show a variation comparable to that of the RVs. Consequently, for further investigation, we used TRICERATOPS, which statistically validated the candidate. Based on all these analyses, we conclude the planetary nature of the candidate and designate it as TOI-5076b.

\subsubsubsection{Joint analysis of TOI-5076}
\label{5076_global}

To perform the joint analysis of the TOI-5076 system, we used our new SOPHIE and CHEOPS data, together with HARPS-N and TESS data. We employed the same setup as used for the previous systems studied in this work to conduct the joint analysis (see Sect. \ref{general_method} and Table \ref{prior_exofast}). Additionally, we detrended the TESS and CHEOPS light curves with a spline simultaneously with our joint analysis. Firstly, we tested a non-zero eccentricity model. From this model, we found that the value for eccentricity (e = $0.094^{+0.073}_{-0.060}$) is not significantly different from 0 as constrained by our data. Therefore, we also performed one circular orbit model. The resulting parameters following this model are consistent with the nonzero eccentricity model. A comparison between the BIC statistics of the two models suggests that the circular model fit is strongly preferred ($\Delta$BIC = 18). Therefore, we adopt the results from the circular orbit model and present them in Table \ref{toi1836_4081_5076} and \ref{tab:TOI_parameters}. Additionally, the median model for both RV and photometric data is illustrated in Fig. \ref{5076_photo_rv}.

The results show TOI-5076b is a planet with a radius of $3.486^{+0.10}_{-0.094}$ R$_{\oplus}$, a mass of $16.1\pm2.4$\ M$_{\oplus}$, and an orbital period of $23.443162^{+0.000062}_{-0.000063}$ d. These results are consistent with the characterized mass of $16\pm2$\ M$_{\oplus}$, period of $23.445\pm0.001$ d, and radius of $3.2\pm0.1$\ R$_{\oplus}$ following \cite{montalto2024gaps}. Based on our analysis, the precision of our derived period has been significantly enhanced by a factor of 16, primarily due to the inclusion of CHEOPS and new TESS data. Additionally, our determination of the planet's radius remains consistent within a 3$\sigma$ confidence interval. The slightly higher radius is attributed to both our fitted parameter for the planet-to-star radius ratio and the stellar radius. Specifically, the planet-to-star radius ratio is $0.03793\pm0.00054$ in our study, compared to $0.037 \pm 0.001$ reported by \cite{montalto2024gaps}. As for the stellar radius, we derive $R_* = 0.844 \pm 0.020$ R$_{\oplus}$, whereas \cite{montalto2024gaps} report 0.78 $\pm$ 0.01 R$_{\oplus}$, consistant in 2.7$\sigma$.

\subsubsection{TOI-1836}

\begin{figure}
\centering
\includegraphics[width=0.48\textwidth]{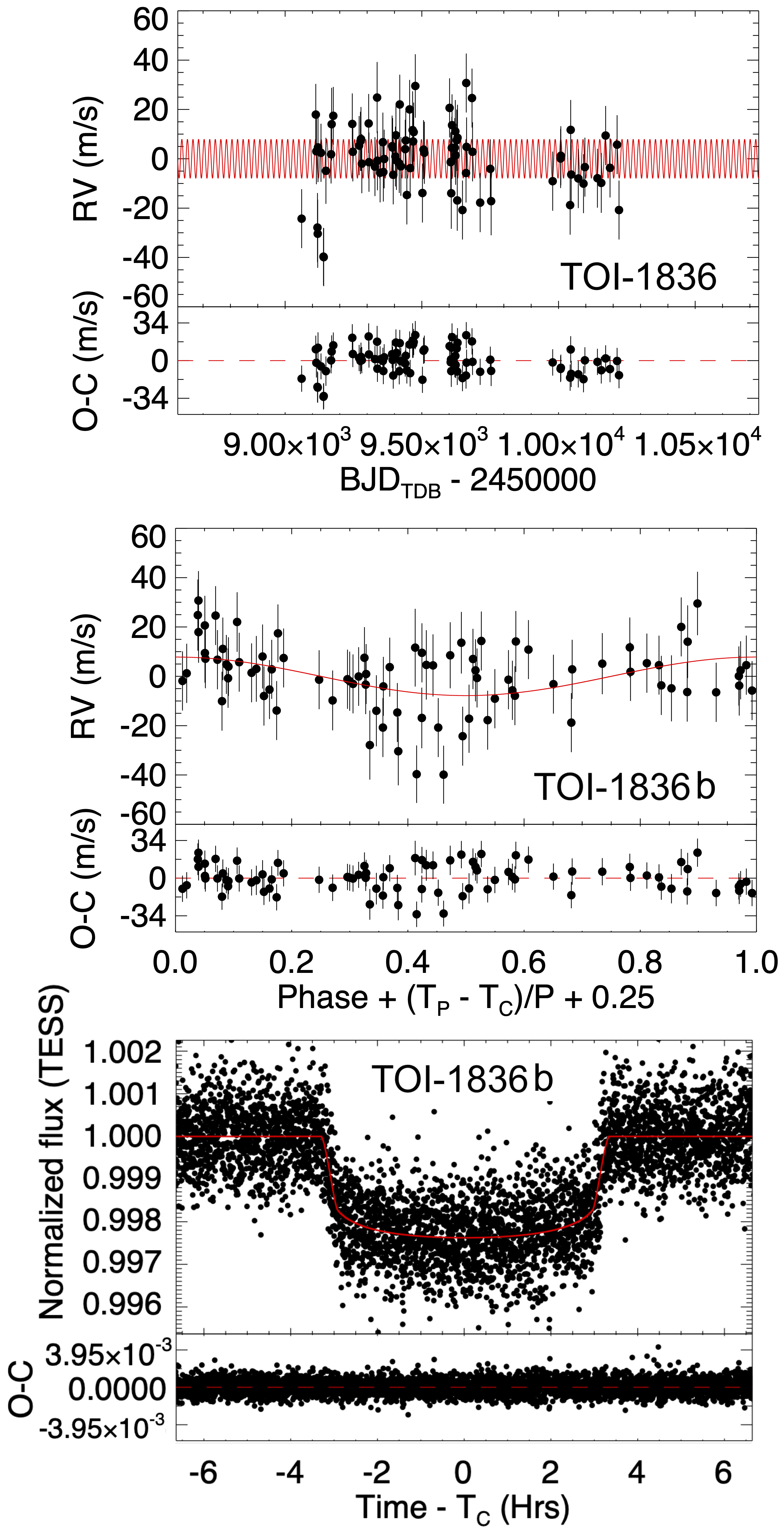}
\caption{The SOPHIE RV time series (\emph{top}), phase-folded RV measurements (\emph{middle}), and light curves (\emph{bottom}) of TOI-1836b. The red lines depict the median models obtained using \exofasttwo. Residuals of the data are presented at the \emph{bottom panels} of their respective panels.}
\label{1836_photo_rv}
\end{figure}

\subsubsubsection{Planetary investigation of TOI-1836 b}
\label{toi-1836_for}

Two planetary candidates are identified by SPOC in the TESS lightcurves around the host star TOI-1836: a sub-Neptune with a period of 1.772750$\pm$0.000008 d and \(T_c\)= 2459739.670$\pm$0.003 BJD namely TOI-1836.02, and a warm Jupiter with a period of 20.38087$\pm$0.00003 d and \(T_c\)= 2459646.492 $\pm$0.002 BJD namely TOI-1836.01. We carried out a 3-year-long campaign of RV follow-up observations by SOPHIE, gathering 89 RV data points (see Table \ref{rv_ob} and Sect.\ref{observation} for data exclusion). At the time of writing this article, \cite{chontos2024tess} has also characterized the parameters of TOI-1836.01 (or HD148193b), utilizing TESS photometric data as well as HIRES and HARPS-N RVs. Since this data is not yet public, we provide an independent conclusion and results based on only TESS and SOPHIE RVs. In the present subsection, our focus is on the planetary identification of the outer planet candidate. Here we do not take into consideration the inner sub-Neptune planet because its expected RV semi-amplitude is substantially smaller (K $\sim$ 3 m/s; TOI-1836.02 is discussed below in Sect. \ref{statistical toi1836b}) than that of the outer candidate, and it is not detected in our RV data.

 We initiated our analysis by performing an unconstrained one-Keplerian RV-only model. For this model, we considered uniform priors for the period between 5 d and 40 d and \(T_c\) ranging from 2459630 d to 2459664 d. Additionally, we adopted wide priors for the rest of the model parameters (see the list of priors in Table. \ref{prior_rv-only}). For simplicity and due to the low amplitude of the RV variation, here we first consider a circular orbit. This model successfully converged at \(T_c = 2459646 \pm 2\) BJD and a period of 20.4$^{+3.0}_{-0.1}$ d, well consistent with the reported planetary period and \(T_c\) values provided by the SPOC mentioned above. This result shows the presence of the outer transiting planet candidate in our RVs. However, this model and the no-planet model are statistically indistinguishable (\(\Delta \ln Z = 0.7\); see Table \ref{table:keplerian_models}), possibly due to perturbations from the inner planet.

As an alternative model, we applied an informed one-Keplerian model using Gaussian priors for \(T_c\) and the period, based on the SPOC values with a 0.1$\sigma$ width. As shown in Table \ref{table:keplerian_models}, this model was statistically strongly preferred over the other tested models. Finally, we refined the model by fixing \(T_c\) and the period to the reported SPOC values and then re-ran our model. The results showed an RV semi-amplitude of \(K = 8 \pm 2\) m/s, indicating a 4$\sigma$ detection of the outer planetary candidate TOI-1836.01 in SOPHIE RVs.

In conclusion, the strong agreement between the detected period and \(T_c\) from our one-Keplerian model with wide priors and the SPOC-reported values, as well as the significant statistical preference for the informed model with Gaussian priors against the no-planet model, supports the detection of the TOI-1836.01 planetary candidate in the SOPHIE RVs. While continued RV monitoring would further refine this detection, the current data already provide an acceptable indication of the presence of this planetary candidate in SOPHIE RVs.

The dispersions of RVs and RV residuals after the Keplerian fit are 13.6 m/s and 12.3 m/s, respectively, while the mean uncertainty of RVs is 6.0 m/s. The bisector dispersion is 22 m/s. This dispersion might be explained by the accuracy of the bisector measurements but warrants caution when interpreting the nature of the signal. Finally, there is no significant correlation between bisectors and RVs (R = -0.03) and their residuals (R= 0.09). We note that \cite{chontos2024tess} did not discuss bisectors of the spectral lines of TOI-1836, nor any other parameter linked to their possible profile variation.  

Following Equation 4 of \cite{2019ApJ...881L..19V}, TOI-1836.01 transit event, could be caused by a blended eclipsing star with $\Delta\text{m}~\le 0.6$ mag (or $\Delta\text{m}~\le 1.2$ mag at 3$\sigma$ confidence). Therefore the event cannot be attributed to the newly detected nearby star within the SOPHIE and TESS apertures, located at a separation of $\rho = 0.82$ arcsec with a magnitude difference of $\Delta m = 5.7$, as reported in \cite{chontos2024tess}. We note that TOI-1836.01 has a radius (details in the following subsection) equal to the upper limit (8 R$_{\oplus}$) for objects that can be statistically validated \citep[e.g.,][]{giacalone2020vetting,mayo2018275}. This limit arises because giant planets, brown dwarfs, and low-mass stars are indistinguishable based on radius alone. Nevertheless, due to its proximity to this limit, and for the sake of completeness, we still conducted a statistical validation test using \texttt{TRICERATOPS}, considering all nearby stars including the newly detected one.

Since this nearby star was detected using the Br-$\gamma$ band, we modeled it as an M dwarf with properties: $M_* = 0.25~ M_{\odot}$, $R_* = 0.25~ R_{\odot}$, and $T_{\mathrm{eff}} = 3300~ K$. Due to its proximity to the primary star, we assumed they were physically bound. For this analysis, we adopted a TESS magnitude of 15, as the actual magnitude remains unknown. These assumptions are conservative, and we note that altering them does not affect the results. After conducting 20 simulation runs, we determined FPP = 0.0003 $\pm$ 0.0004 and NFPP = (8.4 $\pm$ 4.4)$\times$ 10$^{-6}$. These findings indicate that TOI-1836.01 meets the validation criteria for classification as a planet and suggest that the transit event is unlikely to be associated with the newly detected star.

To summarize, we detected TOI-1836.01 in our RVs with 4$\sigma$ confidence. Additionally, the lack of correlation between the bisector and RV, along with the absence of the mask effect, supports the hypothesis of a planetary companion for TOI-1836.01. Using \texttt{TRICERATOPS}, we show that it is unlikely the newly detected nearby star is causing the TOI-1836.01 transit event. While the bisector variation is comparable to that of the RVs, which may raise caution for the possibility of a blend scenario, high-spatial resolution imaging reveals no additional star within the fiber aperture of SOPHIE and TESS. Based on all the analyses above, we conclude that the detected signal in RV and TESS data has a planetary nature, and henceforth, we will refer to it as TOI-1836b.

\subsubsubsection{Planetary investigation of TOI-1836.02}
\label{statistical toi1836b}

The dispersion of RV residuals after removing the outer planetary candidate is 12.3 m/s, consistent with the presence of TOI-1836.02 in our RV data. However, by fixing the period and \(T_c\) to the reported SPOC values for TOI-1836.02, we did not detect the inner candidate in our RVs. Similarly, \cite{chontos2024tess} did not detect TOI-1836.02 in their RV datasets either. We explored the required RV precision for its detection, employing a predicted mass inferred from its radius. According to the SPOC analysis, the inner planet candidate has a radius of 2.7 $\pm$ 1.0 R$_{\oplus}$. Following the methodology of \cite{2017ApJ...834...17C}, we estimated the mass of TOI-1836.02 to be approximately 8 M$_{\oplus}$. Combining this mass with a planetary period of 1.77 d, assuming a circular orbit yields a predicted semi-amplitude of about 3 m/s. Considering the precision of the acquired RVs, with a mean uncertainty of 6 m/s, the absence of a clear detection of TOI-1836.02 in the RVs is not surprising.

Additionally, according to the SPOC, this signal exhibits a relatively low S/N of 11.5 in the TESS photometric data. Consequently, validating this signal using statistical methods may yield unreliable results \citep[e.g., the TRICERATOPS and PASTIS package requires a transit S/N > 15 and S/N> 50 for statistical validation of a transiting planet candidate;][]{2014MNRAS.441..983D}. Therefore, we could not statistically validate the inner planetary signal using photometry. Furthermore, since we also did not detect the signal in RVs, we excluded this candidate from our joint analysis.

\subsubsubsection{Joint analysis of TOI-1836}
\label{1836_global}

While we stayed cautious about the nature of the outer planetary candidate, we performed the joint analysis of the photometric and RV data with the \exofasttwo. Table \ref{prior_exofast} presents a list of informed parameters with Gaussian priors, following a similar approach as presented in Sect. \ref{general_method}. Additionally, we incorporated a spline fitting to detrend both the 2-minute and FFI light curves simultaneously within our joint analysis.

Initially, we performed the joint modeling considering an eccentric orbit. Using this model, we derived the mass of 37$\pm$9 M$_{\oplus}$, eccentricity of $0.20^{+0.12}_{-0.11}$, and $\omega$=$158^{+25}_{-52}$ degrees. As we did not detect eccentricity significantly, we also conducted a circular orbit solution. Following this model, the derived planetary candidate parameters (see Table \ref{toi1836_4081_5076}) are consistent with our previous model. A comparison of the BIC statistic values between these two models indicates that the model with a circular orbit has a higher statistical preference ($\Delta$BIC = 16). Therefore, we selected this particular model as our final choice and listed the results in Table \ref{toi1836_4081_5076} and \ref{tab:TOI_parameters}. Additionally, the median model on RV and photometric data is presented in Fig. \ref{1836_photo_rv}.

Our results show that the outer planet has a mass of $38.47\pm9.22$ M$_{\oplus}$, a radius of $8.00\pm0.35$ R$_{\oplus}$, and a period of 20.380799 $\pm$ 0.000016 d. These values are consistent with the values provided by \cite{chontos2024tess} (mass= 28.4 $\pm$ 4.3 M$_{\oplus}$, radius= 8.38 $\pm$ 0.19 R$_{\oplus}$, period= 20.380850 $\pm$ 0.000025 d). Fig. \ref{1836_photo_rv} illustrates the median transit and RV model. We note that including the inner candidate in our joint analysis, assuming a circular orbit, yielded a consistent mass for the outer planet and a 3$\sigma$ upper limit of about 5 M$_{\oplus}$ for the inner candidate.

\subsection{Planetary candidate TOI-4081}
\subsubsection{Planetary investigation of TOI-4081.01}
\label{detection-4081}

\begin{figure}
\centering
\includegraphics[width=\columnwidth]{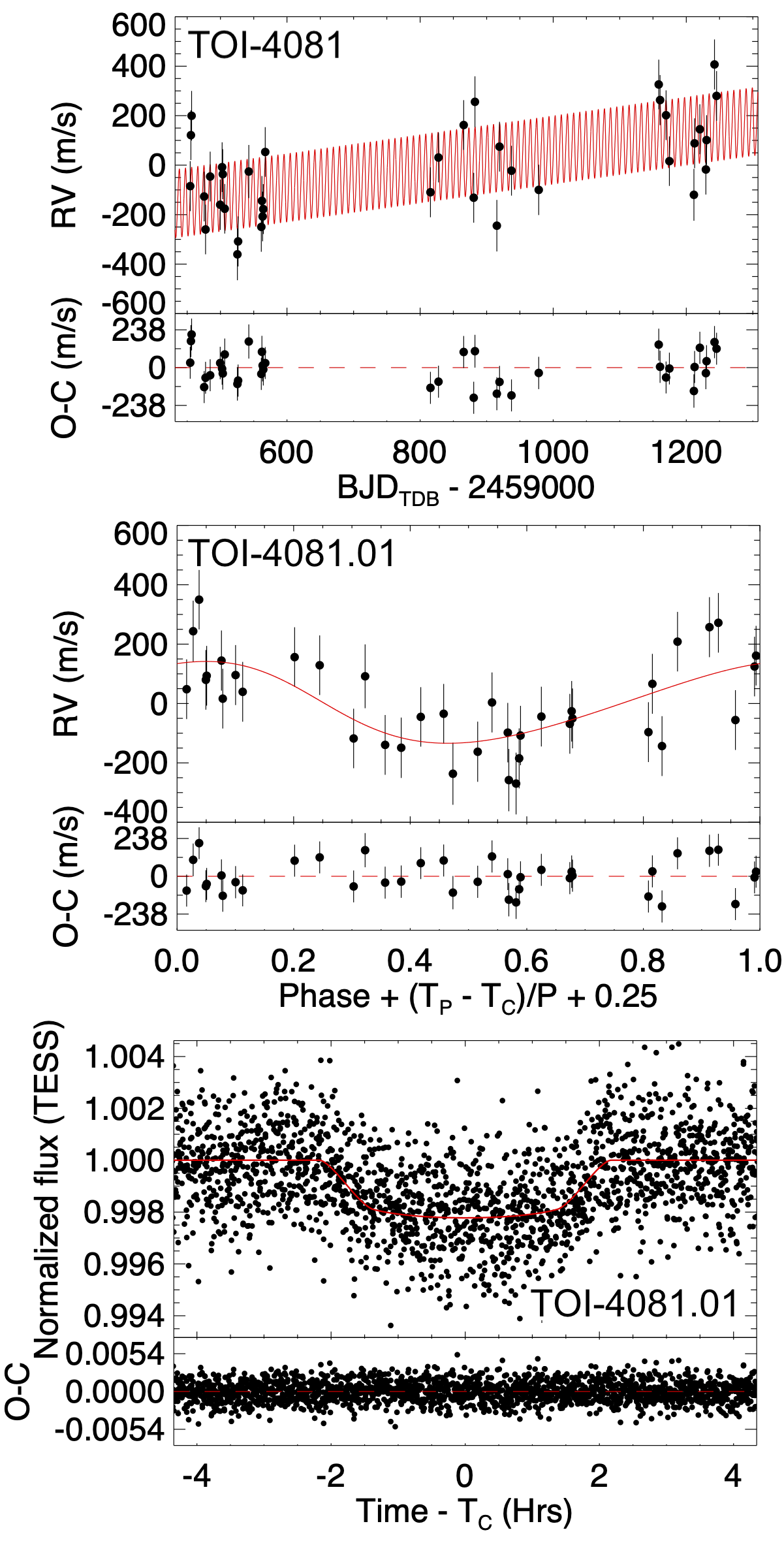}
\caption{RVs for TOI-4081, overplotted by the median Keplerlian model (\emph{top}). SOPHIE RV measurements (\emph{middle}) and TESS photometric data (\emph{bottom}) phase-folded to the orbital period of the planet candidate.}
\label{4081_photo_rv}
\end{figure}

TESS detected a planetary candidate, TOI-4081.01, with a period of 9.2584 $\pm$ 0.0002 d and a \(T_c\)= 2459736.331 $\pm$0.006 BJD around the TOI-4081 star. We conducted follow-up observations with SOPHIE over roughly 50 months, and gathered 38 data points, revealing significant RV variations and a clear linear drift (see Fig. \ref{4081_photo_rv} top). 

We applied an unconstrained one-Keplerian RV-only model, employing broad priors (refer to Table.\ref{prior_rv-only}) and incorporating a linear drift. This model converged to a period= 9.263$\pm$ 0.011 and Tc= 242459736.8$\pm$ 4.6 d, in agreement with the period and phase obtained from the TESS transits. Notably, this model is moderately favored over the no-planet model (see Table \ref{table:keplerian_models}). To assess the strength of the signal, we refined the model by fixing \(T_c\) and period to the values from SPOC, leading to a semi-amplitude of K = 141$^{+22}_{-24}$ m/s. This indicates a significant 6$\sigma$ detection of the TESS signal in our RV data.

The dispersion of RV residuals after removing the refined model is 101 m/s. We attribute this large dispersion to the large RV uncertainty of the data from this star (35 m/s), due to the star’s fast rotation, characterized by a v$\sin i$ of 25 km/s, and a FWHM of 30 km/s. While a convolution between a rotational broadening profile and a Gaussian profile would offer a more precise description of the CCF for fast-rotating stars, here, only a Gaussian profile was employed to describe the RV CCF. We acknowledge that this choice may reduce RV accuracy; however, its application is deemed acceptable, especially given that we are able to detect the signal within our RV dataset with high confidence (6$\sigma$). 

Because of uncertainties caused by the broad, rotational profile of the CCF, we analyzed further the TOI-4081 SOPHIE spectra using an alternative CCF methodology implemented in the \texttt{sophie-toolkit}. This method employs a CCF template matching approach, avoiding reliance on Gaussian fitting for RV determination. Hence, it is less affected by the non-Gaussian profile of a rapidly rotating star. We utilized a G2 mask provided within the package. RVs obtained from this method displayed a linear trend with a slope of $\sim$ 0.51~m/s per day. The RV semi-amplitude was measured at $163\pm26$~m/s with RV residuals having an RMS of 149~m/s, both aligning with pipeline results. This consistency between methods provides further evidence supporting a robust detection of the planet signal in our SOPHIE data, indicating that the rapid rotation of TOI-4081 does not introduce significant systematics.

Similar to other stars discussed in this study, we computed RVs using various stellar masks (G2, M5, or K5). However, for this particular star, the average RV uncertainty was significantly high (36 m/s, 58 m/s, and 231 m/s for G2, K5, and M5 masks, respectively). As a result, although the RVs obtained with G2 and K5 masks show consistent semi-amplitudes within 1$\sigma$ for the detected signal, the substantial error bars associated with the M5 mask make it challenging to confirm this consistency. Additionally, the bisector dispersion is 215 m/s (see Fig. \ref{bis-rvcor}), consistent with the RV dispersion of 182 m/s within the error bars, but surpassing significantly the residual RV dispersion of 147 m/s after the removal of linear drift only (without any Keplerian), where the bisector error bars are typically twice that of the RV. Notably, there is no correlation between the bisector and the RVs (R= 0.04), as well as between the bisector and the RV residuals (R=0.0008) after removing the linear drift.

  According to equation 4 from \cite{2019ApJ...881L..19V}, the transit event of TOI-4081.01 could be caused by a star that is brighter by approximately $\Delta \text{mag} \sim 3.7$. Therefore, the variation in the bisector could be due to a blend scenario, possibly involving Gaia DR3 sources located 2.17$^{\prime\prime}$ from the star with $\Delta G$ = 3.3 mag (see Sect. \ref{TESS-photometry}), or another nearby unresolved star. Nonetheless, rapidly rotating stars have broadened lines, potentially resulting in lines blending, which are isolated in the spectra of slower rotating stars. As a result, these stars might exhibit noisy CCFs, a characteristic particularly noticeable in the CCFs of TOI-4081. While bisector measurements as defined by \cite{queloz2001no} appear unreliable in low S/N regimes, the method employed here, following \cite{boisse2011disentangling}, provides more robust bisector measurements \citep[see discussion in Sect. 2.3 of][]{boisse2011disentangling}. Nevertheless, we still underscore that the rapid rotation of the star and its noisy CCF could indeed have impacted the precision of our bisectors.

In summary, we have detected the signal in our RVs with high confidence (6$\sigma$). However, considering the variability in the bisector and the presence of another star within the TESS and SOPHIE apertures, alongside the challenges posed by the rapid rotation of the star in testing the masking effect, we cannot exclude blend scenarios and definitively determine the nature of this candidate. We also note that performing the statistical validation analysis for this candidate was not possible, as the radius of TOI-4081.01 (see next section) exceeds the upper limit of 8 R$_{\oplus}$ required for such validation. Therefore, while conducting a joint analysis for this system, we exercise caution and account for the uncertain nature of this candidate.

\subsubsection{Joint analysis of TOI-4081}

 Table \ref{prior_exofast} gives a detailed list of priors and Sect. \ref{general_method} discusses these choices for our joint analysis. Additionally, we employed splines to detrend both the 2-minute and FFI light curves simultaneously with the joint analysis. Finally and as discussed in Sect. \ref{TESS-photometry}, the SPOC pipeline does not correct for the light contribution of the source identified near this star by Gaia DR3. To determine the true transit depth of the planet candidate, the model therefore incorporates a dilution factor for the TESS photometry, which is informed by a Normal prior based on the calculated dilution factor (Sect. \ref{TESS-photometry}). Tables\ref{toi1836_4081_5076} and \ref{tab:TOI_parameters} list the posterior parameters. Fig. \ref{4081_photo_rv} illustrates the median model on the RV and photometric data.

 The results show that TOI-4081.01 would have a period of $9.258388^{+0.000020}_{0.000021}$ d, a mass of $1.89^{+0.35}_{-0.34}$ M$_{J}$, and a radius of $1.19^{+0.66}_{-0.67}$ R$_{J}$. Additionally, RVs show a linear trend of $0.3727^{+0.064}_{-0.065}$ m/s per day. This drift could be induced by a second companion, for which the period has not yet been fully observed. 
 

\begin{figure}
\centering
\includegraphics[width=\columnwidth]{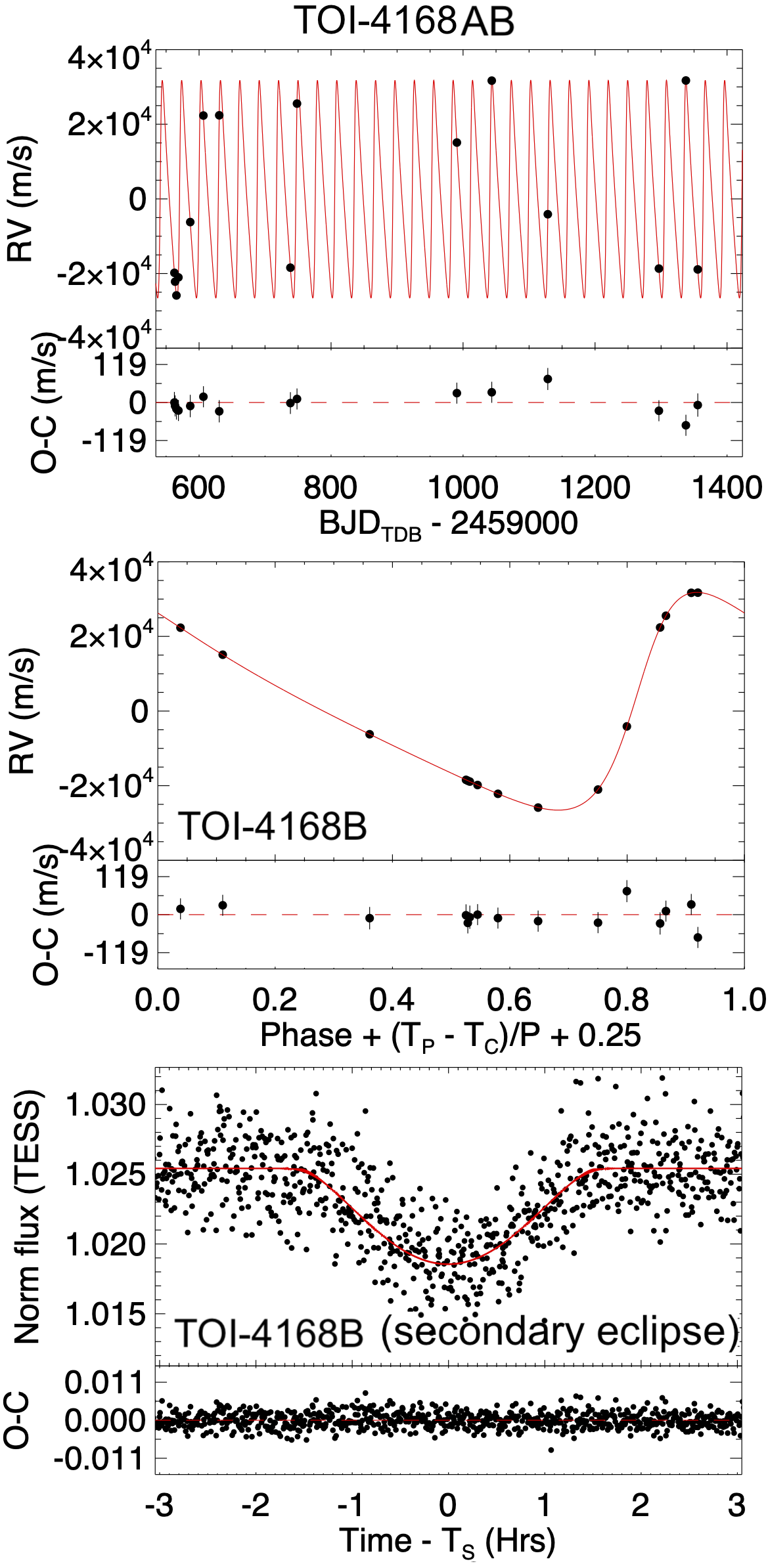}
\caption{SOPHIE RVs for TOI-4168AB, overplotted by the median Keplerlian model (\emph{top}). RVs (\emph{middle}) and the secondary eclipse observed in TESS data (\emph{bottom}) are phase-folded to the orbital period of TOI-4168B. Residuals for each dataset are plotted at the bottom of their respective panels.}
\label{4168_photo_rv}
\end{figure}

\subsection{TOI-4168: false positive identification}

\subsubsection{Planetary investigation of TOI-4168}
\label{TOI4168}

TESS identified TOI-4168.01 as a planetary candidate with a period of 29.3821$\pm$ 0.0002 d and a \(T_c\)= 2459394.257$\pm$ 0.002 BJD. We conducted observations with SOPHIE, collecting 15 RV measurements that exhibit significant variation.

Applying an unconstrained one-Keplerian model to the RV data, employing wide priors for all parameters (see Table. \ref{rv_ob}), revealed a periodic signal with a period of 29.386212$\pm$ 0.000094 d, \(T_c\)= 2459407.40164 $\pm$ 0.00065 BJD, K=29180$^{+6}_{-9}$ m/s, and e=0.423694$\pm$0.000065. Notably, while the period aligns well with that of the TESS planetary candidate TOI-4168.01, the RV curve obtained from SOPHIE exhibits an anti-phase relation with the transit ephemeris observed by TESS, consistent instead with a secondary eclipse. Additionally, the large semi-amplitude and eccentricity, suggest a scenario consistent with an eccentric binary star where the primary transit remains undetected \citep[see discussion in][]{santerne2013contribution}. This scenario is akin to those of KOI-419.01 and KOI-698.01, as discussed in \cite{santerne2012sophie}.

Consequently, the RV variations originate from an eclipsing binary star, hereafter designated as TOI-4168A and TOI-4168B. We note that according to Gaia DR3, the system is likely part of a three-star system with an additional star (Gaia DR3 1130051735765065984) located at 8.65 arcsecs (or 2831 AU) far from TOI-4168AB, with comparable proper motion (see Fig. \ref{tpfplotter}), consistent distance, and consistent RVs. We note that, for this particular system, a model comparison between the unconstrained one-Keplerian model and the no-planet model was not possible due to bimodality in the latter. Nevertheless, the detection of the periodic signal remains unequivocal.

\subsubsection{Joint analysis of TOI-4168.01}
\label{join_4168}

We used the method pioneered by Frommer et. al. (in prep) to model this binary stellar system. Using the linkpars branch of \exofasttwo, both stars were modeled with Mesa isochrones and stellar tracks evolutionary model \citep[MIST,][]{dotter2016mesa, choi2016mesa} and a blended SED model. The two stars were forced to have the same age, initial metalicity, extinction, and distance. The modeled relative flux in the TESS band of the two stars was used to constrain the dilution in the TESS light curve. Fig. \ref{fig:sed} shows the SED model of the binary pair, with TOI-4168AB plotted on top.

In addition to the two stars, we modeled one ``planet,'' but linked its mass and radius to TOI-4168B's mass and radius. This allows the RVs, which normally constrain only the planet's mass, to measure the mass of TOI-4168B. In addition, the dilution from the SEDs coupled with the observed grazing secondary eclipse, helps us to constrain the radii of TOI-4168A and TOI-4168B through the eclipse depth.

The system parameters for TOI-4168AB are listed in Table \ref{tab:toi4168}. Many of the parameters output by \exofasttwo \ assume the companion is dark. For clarity, we have extensively modified the default output table from \exofasttwo \ to remove parameters biased by planetary assumptions and reformat it to remove references to the fictitious planet. Due to the eccentricity and inclination of the system, we can definitively say that there is no observable primary transit, both from the primary transit duration (0) and the impact parameter. Fig. \ref{4168_photo_rv} illustrates the median model for both RV and photometric data.

\section{Discussion}
\label{discuss}

In this section, we discuss the five characterized planetary systems—TOI-2295, TOI-2537, TOI-5110, TOI-1836, and TOI-5076—in context and highlight some of their interesting characteristics.
\newcommand{\toifouronesixeightmstarA}{\ensuremath{1.024^{+0.066}_{-0.058}}}
\newcommand{\toifouronesixeightrstarA}{\ensuremath{1.162^{+0.048}_{-0.044}}}
\newcommand{\toifouronesixeightrstarsedA}{\ensuremath{1.132^{+0.024}_{-0.022}}}
\newcommand{\toifouronesixeightlstarA}{\ensuremath{1.464^{+0.11}_{-0.092}}}
\newcommand{\toifouronesixeightfbolA}{\ensuremath{4.49^{+0.34}_{-0.28}\times 10^{-10}}}
\newcommand{\toifouronesixeightrhostarA}{\ensuremath{0.92^{+0.14}_{-0.12}}}
\newcommand{\toifouronesixeightloggA}{\ensuremath{4.318^{+0.046}_{-0.047}}}
\newcommand{\toifouronesixeightteffA}{\ensuremath{5890\pm140}}
\newcommand{\toifouronesixeightteffsedA}{\ensuremath{5970^{+130}_{-110}}}
\newcommand{\toifouronesixeightfehA}{\ensuremath{0.122^{+0.080}_{-0.081}}}
\newcommand{\toifouronesixeightinitfehA}{\ensuremath{0.153^{+0.073}_{-0.075}}}
\newcommand{\toifouronesixeightageA}{\ensuremath{7.3^{+3.4}_{-3.1}}}
\newcommand{\toifouronesixeighteepA}{\ensuremath{407^{+20}_{-35}}}
\newcommand{\toifouronesixeightlogmstarA}{\ensuremath{0.007^{+0.025}_{-0.024}}}
\newcommand{\toifouronesixeightAvA}{\ensuremath{0.103^{+0.094}_{-0.073}}}
\newcommand{\toifouronesixeighterrscaleA}{\ensuremath{2.12^{+0.92}_{-0.53}}}
\newcommand{\toifouronesixeightparallaxA}{\ensuremath{3.096\pm0.018}}
\newcommand{\toifouronesixeightdistanceA}{\ensuremath{323.0^{+1.9}_{-1.8}}}
\newcommand{\toifouronesixeightmstarB}{\ensuremath{0.506^{+0.018}_{-0.017}}}
\newcommand{\toifouronesixeightrstarB}{\ensuremath{0.481^{+0.024}_{-0.023}}}
\newcommand{\toifouronesixeightrstarsedB}{\ensuremath{0.483^{+0.032}_{-0.031}}}
\newcommand{\toifouronesixeightlstarB}{\ensuremath{0.0408^{+0.0090}_{-0.0076}}}
\newcommand{\toifouronesixeightfbolB}{\ensuremath{1.25^{+0.28}_{-0.23}\times 10^{-11}}}
\newcommand{\toifouronesixeightrhostarB}{\ensuremath{6.41^{+0.92}_{-0.79}}}
\newcommand{\toifouronesixeightloggB}{\ensuremath{4.778^{+0.038}_{-0.036}}}
\newcommand{\toifouronesixeightteffB}{\ensuremath{3740\pm160}}
\newcommand{\toifouronesixeightteffsedB}{\ensuremath{3740^{+180}_{-170}}}
\newcommand{\toifouronesixeightfehB}{\ensuremath{0.17^{+0.10}_{-0.11}}}
\newcommand{\toifouronesixeightinitfehB}{\ensuremath{0.153^{+0.073}_{-0.075}}}
\newcommand{\toifouronesixeightageB}{\ensuremath{7.3^{+3.4}_{-3.1}}}
\newcommand{\toifouronesixeighteepB}{\ensuremath{296^{+13}_{-14}}}
\newcommand{\toifouronesixeightlogmstarB}{\ensuremath{-0.297\pm0.014}}
\newcommand{\toifouronesixeightAvB}{\ensuremath{0.103^{+0.094}_{-0.073}}}
\newcommand{\toifouronesixeightparallaxB}{\ensuremath{3.096\pm0.018}}
\newcommand{\toifouronesixeightdistanceB}{\ensuremath{323.0^{+1.9}_{-1.8}}}
\newcommand{\toifouronesixeightPeriod}{\ensuremath{29.381670^{+0.000049}_{-0.000048}}}
\newcommand{\toifouronesixeightrp}{\ensuremath{4.68^{+0.24}_{-0.23}}}
\newcommand{\toifouronesixeightmp}{\ensuremath{530^{+19}_{-17}}}
\newcommand{\toifouronesixeightmpsun}{\ensuremath{0.506^{+0.018}_{-0.017}}}
\newcommand{\toifouronesixeighttco}{\ensuremath{2459407.090^{+0.012}_{-0.013}}}
\newcommand{\toifouronesixeighttc}{\ensuremath{2459407.089^{+0.012}_{-0.013}}}
\newcommand{\toifouronesixeighttt}{\ensuremath{2460406.061\pm0.012}}
\newcommand{\toifouronesixeighttzero}{\ensuremath{2460406.062\pm0.012}}
\newcommand{\toifouronesixeighta}{\ensuremath{0.2147^{+0.0039}_{-0.0036}}}
\newcommand{\toifouronesixeightideg}{\ensuremath{87.14^{+0.15}_{-0.16}}}
\newcommand{\toifouronesixeighte}{\ensuremath{0.43517^{+0.00045}_{-0.00046}}}
\newcommand{\toifouronesixeightomegadeg}{\ensuremath{-78.056^{+0.088}_{-0.081}}}
\newcommand{\toifouronesixeightomegagr}{\ensuremath{0.0779^{+0.0035}_{-0.0032}}}
\newcommand{\toifouronesixeightteq}{\ensuremath{660^{+12}_{-11}}}
\newcommand{\toifouronesixeighttcirc}{\ensuremath{478^{+120}_{-94}}}
\newcommand{\toifouronesixeightk}{\ensuremath{29176\pm17}}
\newcommand{\toifouronesixeightp}{\ensuremath{0.414^{+0.028}_{-0.027}}}
\newcommand{\toifouronesixeightar}{\ensuremath{39.7\pm1.8}}
\newcommand{\toifouronesixeightdelta}{\ensuremath{0.171^{+0.024}_{-0.022}}}
\newcommand{\toifouronesixeightdepth}{\ensuremath{-}}
\newcommand{\toifouronesixeighttau}{\ensuremath{-}}
\newcommand{\toifouronesixeighttonefour}{\ensuremath{-}}
\newcommand{\toifouronesixeighttfwhm}{\ensuremath{-}}
\newcommand{\toifouronesixeightb}{\ensuremath{2.792^{+0.068}_{-0.070}}}
\newcommand{\toifouronesixeightcosi}{\ensuremath{0.0499^{+0.0027}_{-0.0025}}}
\newcommand{\toifouronesixeightbs}{\ensuremath{1.125^{+0.027}_{-0.028}}}
\newcommand{\toifouronesixeighttaus}{\ensuremath{0.0637\pm0.0016}}
\newcommand{\toifouronesixeighttonefours}{\ensuremath{0.1274^{+0.0033}_{-0.0032}}}
\newcommand{\toifouronesixeighttfwhms}{\ensuremath{0.0637\pm0.0016}}
\newcommand{\toifouronesixeightlogp}{\ensuremath{1.464^{+0.11}_{-0.092}}}
\newcommand{\toifouronesixeightloggp}{\ensuremath{4.318^{+0.046}_{-0.047}}}
\newcommand{\toifouronesixeightsafronov}{\ensuremath{47.4^{+2.7}_{-2.5}}}
\newcommand{\toifouronesixeightfave}{\ensuremath{0.0361^{+0.0026}_{-0.0023}}}
\newcommand{\toifouronesixeighttso}{\ensuremath{2459394.25597\pm0.00082}}
\newcommand{\toifouronesixeightts}{\ensuremath{2459394.25644\pm0.00082}}
\newcommand{\toifouronesixeightte}{\ensuremath{2460422.6151\pm0.0020}}
\newcommand{\toifouronesixeighttezero}{\ensuremath{2460422.6147\pm0.0020}}
\newcommand{\toifouronesixeighttp}{\ensuremath{2459394.6034^{+0.0031}_{-0.0028}}}
\newcommand{\toifouronesixeightta}{\ensuremath{2459397.3588\pm0.0039}}
\newcommand{\toifouronesixeighttd}{\ensuremath{2459419.7950^{+0.0054}_{-0.0057}}}
\newcommand{\toifouronesixeightvcve}{\ensuremath{1.56786^{+0.00093}_{-0.00098}}}
\newcommand{\toifouronesixeightecosw}{\ensuremath{0.09006^{+0.00065}_{-0.00060}}}
\newcommand{\toifouronesixeightesinw}{\ensuremath{-0.42575^{+0.00050}_{-0.00047}}}
\newcommand{\toifouronesixeightsecosw}{\ensuremath{0.13646^{+0.00098}_{-0.00092}}}
\newcommand{\toifouronesixeightsesinw}{\ensuremath{-0.64542^{+0.00042}_{-0.00039}}}
\newcommand{\toifouronesixeightmsini}{\ensuremath{529^{+19}_{-17}}}
\newcommand{\toifouronesixeightq}{\ensuremath{0.494\pm0.013}}
\newcommand{\toifouronesixeightdr}{\ensuremath{56.1^{+2.6}_{-2.5}}}
\newcommand{\toifouronesixeightpt}{\ensuremath{0.01045^{+0.00095}_{-0.00088}}}
\newcommand{\toifouronesixeightptg}{\ensuremath{0.02521^{+0.00086}_{-0.00080}}}
\newcommand{\toifouronesixeightps}{\ensuremath{0.0259^{+0.0023}_{-0.0022}}}
\newcommand{\toifouronesixeightpsg}{\ensuremath{0.0626^{+0.0021}_{-0.0020}}}
\newcommand{\toifouronesixeightuone}{\ensuremath{0.289^{+0.027}_{-0.026}}}
\newcommand{\toifouronesixeightutwo}{\ensuremath{0.285^{+0.019}_{-0.020}}}
\newcommand{\toifouronesixeightthermal}{\ensuremath{25400^{+3800}_{-3300}}}
\newcommand{\toifouronesixeighteclipsedepth}{\ensuremath{25400^{+3800}_{-3300}}}
\newcommand{\toifouronesixeightgamma}{\ensuremath{-29142\pm14}}
\newcommand{\toifouronesixeightjitter}{\ensuremath{43^{+16}_{-10}}}
\newcommand{\toifouronesixeightjittervar}{\ensuremath{1910^{+1600}_{-800}}}
\newcommand{\toifouronesixeightdilute}{\ensuremath{0.0235^{+0.0065}_{-0.0053}}}

\begin{table*}[!htbp]
\centering
\caption{Median values and 68\% confidence interval for TOI4168AB, created using EXOFASTv2.}
\resizebox{0.7\textwidth}{!}{%
\begin{tabular}{lccc}
\hline
Parameters & Description & \multicolumn{2}{c}{Values} \\
\hline
\smallskip\\\multicolumn{2}{l}{Stellar Parameters:}&A&B\smallskip\\
~~~~$M_*$              & Mass (\msun)                   & \toifouronesixeightmstarA              & \toifouronesixeightmstarB    \\
~~~~$R_*$             & Radius (\rsun)                 & \toifouronesixeightrstarA              & \toifouronesixeightrstarB    \\
~~~~$R_{*,SED}$        & Radius$^{1}$ (\rsun)            & \toifouronesixeightrstarsedA           & \toifouronesixeightrstarsedB \\
~~~~$L_*$             & Luminosity (\lsun)             & \toifouronesixeightlstarA              & \toifouronesixeightlstarB    \\
~~~~$F_{Bol}$          & Bolometric Flux (cgs)           & \toifouronesixeightfbolA               & \toifouronesixeightfbolB     \\
~~~~$\rho_*$           & Density (cgs)                   & \toifouronesixeightrhostarA            & \toifouronesixeightrhostarB  \\
~~~~$\log{g}$          & Surface gravity (cgs)          & \toifouronesixeightloggA               & \toifouronesixeightloggB     \\
~~~~$T_{\rm eff}$      & Effective temperature (K)       & \toifouronesixeightteffA               & \toifouronesixeightteffB     \\
~~~~$T_{\rm eff,SED}$  & Effective temperature$^{1}$ (K) & \toifouronesixeightteffsedA            & \toifouronesixeightteffsedB  \\
~~~~$[{\rm Fe/H}]$     & Metallicity (dex)               & \toifouronesixeightfehA                & \toifouronesixeightfehB      \\
~~~~$[{\rm Fe/H}]_{0}$ & Initial Metallicity$^{2}$      & \toifouronesixeightinitfehA            & \toifouronesixeightinitfehB  \\
~~~~$Age$              & Age (Gyr)                       & \toifouronesixeightageA                & \toifouronesixeightageB      \\
~~~~$\sigma_{SED}$     & SED photometry error scaling    & \toifouronesixeighterrscaleA           & --                           \\
~~~~$d$               & Distance (pc)                   & \toifouronesixeightdistanceA           & \toifouronesixeightdistanceB \\
\smallskip\\\multicolumn{2}{l}{Orbital Parameters:}\smallskip\\
~~~~$P$               & Period (days) & \toifouronesixeightPeriod \\
~~~~$T_C$             & Observed Time of conjunction (\bjdtdb)      &\toifouronesixeighttco\\
~~~~$a$               & Semi-major axis (AU)                               &\toifouronesixeighta\\
~~~~$i$               & Inclination (Degrees)                             &\toifouronesixeightideg\\
~~~~$e$               & Eccentricity                                     &\toifouronesixeighte\\
~~~~$\omega_A$        & Arg of periastron (Degrees)                        &\toifouronesixeightomegadeg\\
~~~~$K$               & RV semi-amplitude (m/s)                            &\toifouronesixeightk\\
\smallskip\\\multicolumn{2}{l}{Primary Transit Parameters:}\smallskip\\
~~~~$\delta$          & $\left(R_B/R_A\right)^2$                           & \toifouronesixeightdelta\\
~~~~$T_{14}$          & Total transit duration (days)                     & \toifouronesixeighttonefour\\
~~~~$b$              & Transit impact parameter                           & \toifouronesixeightb \\ 
\smallskip\\\multicolumn{2}{l}{Secondary Eclipse Parameters:}\smallskip\\
~~~~$b_S$             & Eclipse impact parameter                           & \toifouronesixeightbs \\
~~~~$\tau_S$          & In/egress eclipse duration (days)                  & \toifouronesixeighttaus \\
~~~~$T_{S,14}$        & Total eclipse duration (days)                   & \toifouronesixeighttonefours \\
~~~~$\delta_{S}$      & Measured eclipse depth (ppm)                       & \toifouronesixeighteclipsedepth \\
\hline
\end{tabular}%
}
\begin{tablenotes}
   \item[*] $^{1}$ This value ignores the systematic error and is for reference only, $^{2}$The metallicity of the star at birth.
\end{tablenotes}
\label{tab:toi4168}
\end{table*}

\subsection{Comparison to known planets}

Figure \ref{eccenticity_period} (\emph{top} panel) presents the mass–period diagram of transiting exoplanets with known mass and radius from the NASA Exoplanet Data Archive\footnote{\url{https://exoplanetarchive.ipac.caltech.edu/}} as of February 26, 2024. Notably, the newly characterized planets in this study are located in the region of the diagram corresponding to long orbital periods (> 20 d). This region has a lower density of detected planets with determined masses compared to those with shorter orbital periods.

\subsection{Eccentricity}
\label{eccenticity}

\begin{figure}
\centering
\includegraphics[width=0.48\textwidth]{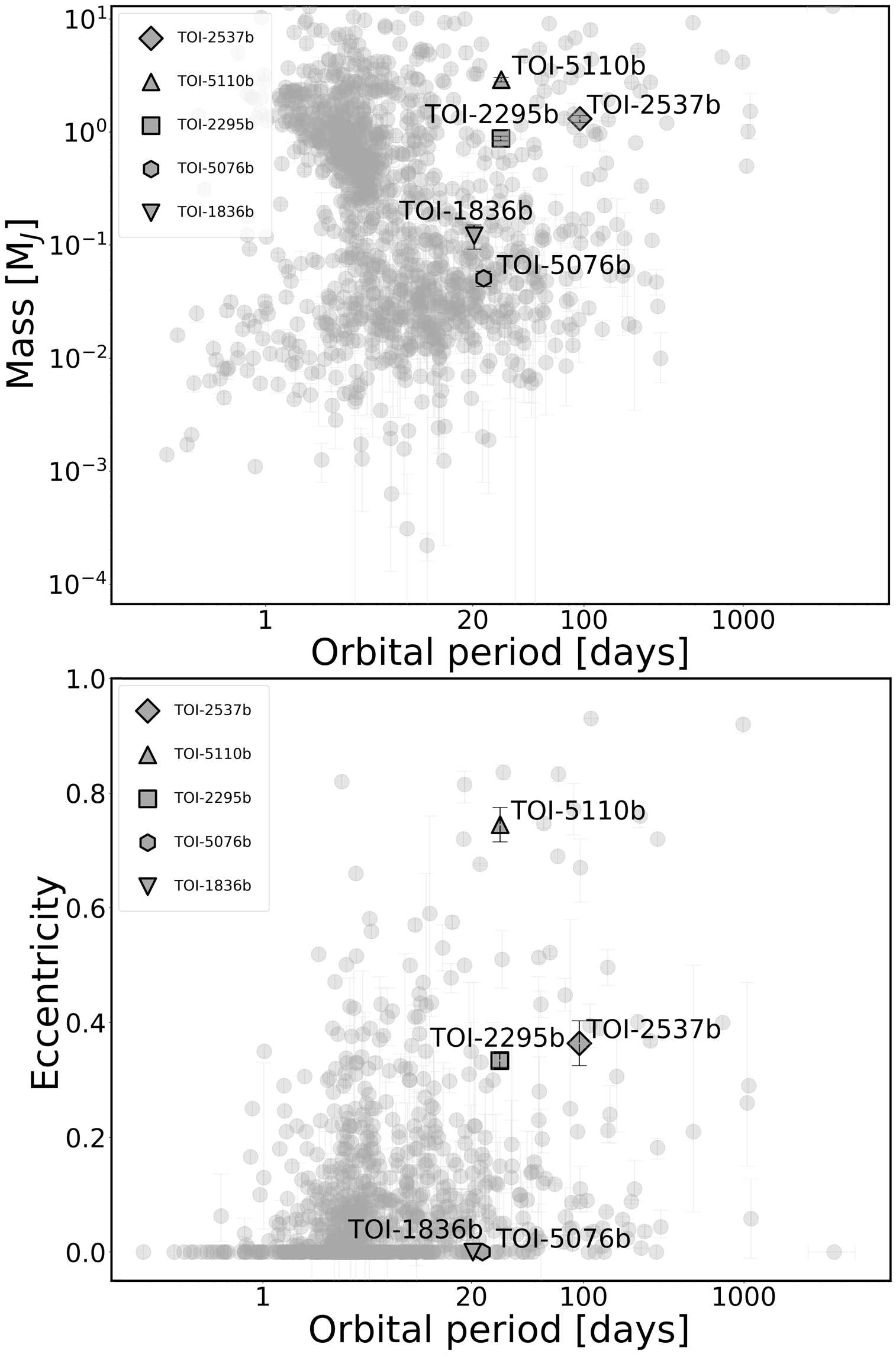}
\caption{Mass-period (\emph{top}) and eccentricity-period (\emph{bottom}) diagrams of characterized transiting planet in this study together with other exoplanets with known mass and radius from the NASA Exoplanet Data Archive (as of February 26, 2024). The \emph{bottom} panel highlights the unique position of TOI-5110b as one of the most eccentric planets known to date. }
\label{eccenticity_period}
\end{figure}

Orbital eccentricity is one of the key features in the study of exoplanet formation and evolution \citep{ribas2007eccentricity,takeda2005high}. In Fig. \ref{eccenticity_period} (\emph{bottom} panel), we compare the orbital eccentricities and periods of the transiting planets characterized in this study to other planets with known mass and radius. Notably, TOI-2295b and TOI-2537b have relatively large eccentricities, whereas TOI-5110b has one of the most eccentric orbits among the transiting planets discovered so far. In the following sections, we discuss in detail the eccentricities of each characterized planetary system and compare different formation scenarios. We note that the planets TOI-1836b and TOI-5076b are not included in this discussion, as their eccentricities are fixed to zero in our analysis.

\subsubsection{TOI-2295 and TOI-2537}

TOI-2295 and TOI-2537 both host eccentric warm Jupiters with massive, long-period companions. These systems may be candidates for perturber-coupled high-eccentricity migration (HEM), in which angular momentum exchange with additional planets in the systems periodically boosts their eccentricities enough for efficient tidal dissipation to occur \citep{nagasawa2008formation}. In this case, the warm Jupiters may have formed further out in their systems and migrated inward through tidal circularization during periods of extreme eccentricity, but would spend most of their lifetimes at more moderate eccentricities. For certain orientations of the two planets in each system, these extreme eccentricity phases may be achievable.

However, because the observed eccentricities of the warm Jupiters in these systems are not extreme, a more simple explanation is plausible: TOI-2295b and TOI-2537b may have formed in situ or by low-eccentricity disk migration, and reached their moderate eccentricities through planet-planet scattering or secular interactions with their companions \citep{dittkrist2014impacts}. This explanation is consistent with the growing body of evidence that most warm Jupiters make up a population that is distinct from hot Jupiters (e.g., \citealt{Huang2016,Wu2023,Hu2024}).

\subsubsection{TOI-5110}
\label{eccentricity-5110}

With an eccentricity of $0.745^{+0.030}_{-0.027}$, TOI-5110b is among the eight most eccentric planets known to date (see Fig.\ref{eccenticity_period}). Eccentricities may be excited by interactions with the protoplanetary disk \citep{Duffell2015} or by planet-planet scattering after the disk has dispersed \citep{Rasio1996,Juric2008}. However, these excitation methods tend to be limited to intermediate eccentricities \citep{Goldreich2004,Ida2013,Petrovich2014}. Unlike TOI-2295b and TOI-2537b, the eccentricity of TOI-5110b is extreme enough to discount these mechanisms. Instead, TOI-5110b may have formed on a more distant orbit and migrated inwards via HEM. Other eccentric warm Jupiters in this category, HD 80606b \citep{Naef2001,hebrard2010observation} and TOI-3362 \citep{Dong2021,espinoza2023aligned}, are considered "proto-hot Jupiters" because they are thought to be migrating planets caught en route to their final destination as circular hot Jupiters \citep[e.g., see][]{2003ApJ...589..605W}. Assuming angular momentum is conserved during the HEM process, we can predict the final semi-major axis (i.e., when the orbit is fully circularized), $a_{\rm{final}}$, based on the current semi-major axis, $a$, and eccentricity, $e$:

\begin{align}
    a_{\rm{final}}=a(1-e^2)
\end{align}

From the derived semi-major axis and eccentricity of TOI-5110b, $a_{\rm{final}}=0.0965^{+0.012}_{-0.011}$. This narrowly falls within the hot Jupiter semi-major axis range ($a<0.1$), suggesting TOI-5110b may also be a ``proto-hot Jupiter.'' However, though the efficiency of tidal dissipation is highly uncertain, the rate of circularization is strongly dependent on $a_{\rm{final}}$ \citep{Adams2006}, so when compared to HD 80606b, TOI-5110b may not be quite eccentric enough for HEM to explain its origin.

HEM may still be a compelling explanation of the period and eccentricity of this system's orbit if the warm Jupiter was able to exchange angular momentum with another planet in the system. In this case, the eccentricity of TOI-5110b would oscillate over time and may sometimes be large enough for efficient tidal dissipation. This companion would need to be massive and nearby enough for the eccentricity precession it induces to overcome general relativistic precession \citep{Dong2014}. Thus, we can set constraints on the presence of a perturbing companion in the system and assess whether such a companion should have been detected in our observations. \citet{Jackson2021} demonstrate that, for many TESS planets, these companions are likely to be detectable with RV or astrometric observations.

Following the methods described in \cite{Jackson2021} and applied in \citet{Gupta2023}, we generate a population of perturbing companions capable of inducing eccentricity oscillations in TOI-5110b and assess the detectability of these perturbers given the available observational data. We construct our population of perturbers by drawing planetary masses and orbital parameters from distributions set by observations of long-period planets. The precise properties of the population of long-period planets are not well-known, but \cite{Jackson2021} show their results are robust to different choices for the underlying companion distribution. We draw masses between 0.1-20 $M_{\rm{J}}$ from the power law distribution set by \cite{Cumming2008}, orbital periods between 200-100,000 d from the broken power law distribution set by \cite{Fernandes2019}, eccentricities from a beta distribution ($\alpha=0.74$, $\beta=1.61$), inclinations from an isotropic distribution, and all other orbital parameters from uniform distributions. We note that since this simulation is based on the actual SOPHIE RV measurements, which have an uncertainty of 6.5 m/s, we set the lower mass limit at 0.1 $M_{\rm{J}}$, as companions with lower masses would not be detectable. Additionally, we require that the perturbing planets be long-term stable and that the precession induced in TOI-5110b be faster than general relativistic precession \citep[for a more detailed discussion, see Sect. 2.4 of][]{Jackson2021}.

Next, we calculate the RV detectability of the perturbers by drawing masses and orbital parameters for TOI-5110b from the results of the joint transit and RV model presented in Table \ref{toi_2295_2537_5110}, drawing a companion from our population of simulated perturbers, and modeling the two-planet RV signal of the system. We then fit for and subtract the warm Jupiter signal using the \textsc{idl} package \textsc{mpfit} \citep{Markwardt2009} to arrive at the residual signal from the companion. We then determine the detectability of this signal by measuring the ratio of the slope of the RV signal to the estimated error in the slope ($|\rm{slope}|/e_{\rm{slope}}$). We also measure this slope ratio for fractions of the full baseline to account for orbital periods shorter than the full baseline of observations. We consider a companion to be detectable for each trial in which the slope ratio is greater than 3.5 for any of the tested baselines. 

\begin{figure}
\centering
\includegraphics[width=0.48\textwidth]{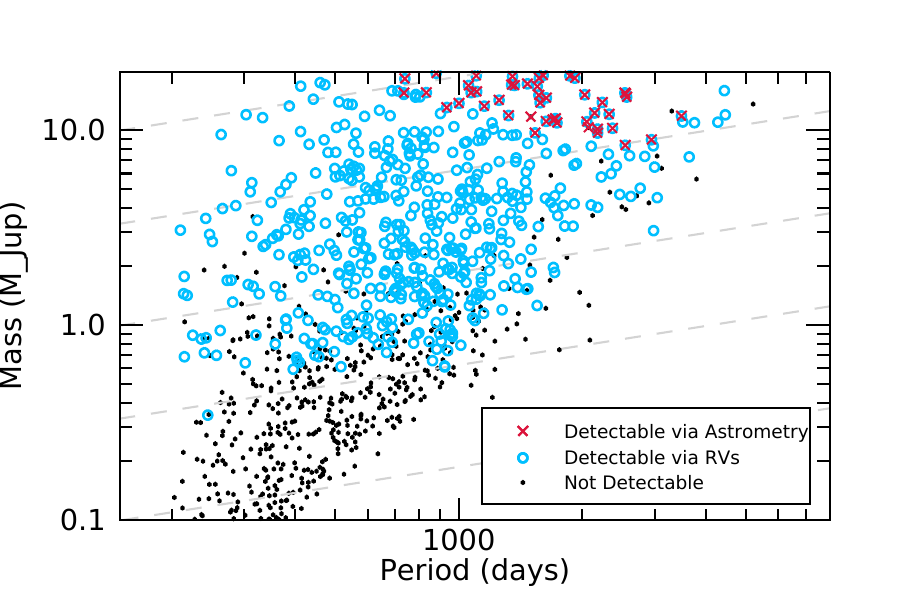}
\caption{Detectability of the simulated population of companions to TOI-5110b capable of exciting the planet's eccentricity leading to high-eccentricity migration. black dots are not detectable with our RV measurements or Gaia astrometry, blue circles are detectable in our RV data, and red crosses are detectable with astrometry. Gray dashed lines represent RV semi-amplitudes of, from bottom to top, 3, 10, 30, 100, and 300 m/s. The shape of the distribution of companions is constrained by stability (in the upper left) and general relativity (in the lower right).}
\label{companions}
\end{figure}

We show the sample of simulated companions in Fig. \ref{companions}. The majority of companions with RV semi-amplitudes greater than $\sim10~\rm{ms}^{-1}$ would have been detected by our observations if they were present in the system, but a substantial fraction of companions also fall below that line and are undetectable with the current RV dataset. Therefore, we cannot rule out angular momentum exchange with a perturber leading to HEM in the case of TOI-5110b - i.e., a short-period, low-mass companion hidden in the data may be capable of inducing significant eccentricity oscillations in the observed warm Jupiter. Additional RV observations with higher precision are necessary to detect such a companion if it exists.

We also calculate the detectability of our simulated companions with Gaia astrometry by modeling the transverse motion of the stellar host due to the gravitational pull of the perturber and subtracting a linear fit to account for proper motion \citep{quirrenbach2010astrometric}. According to the Gaia Data Release 3, TOI-5110 was observed 212 times over its 5-year mission. We sample our model at 212 evenly spaced times and calculate the maximum angular distance, $\Delta\theta$, between model samples. Comparing our results to a conservative detection limit of 100 $\mu\rm{as}$ \citep{Perryman2014}, we find that only the most massive companions would be detectable by this method (see Fig. \ref{companions}).

\subsection{Grazing nature of TOI-2295b}
\label{grazing}

TOI-2295b exhibits a grazing nature (see Sect. \ref{detection_toi2295} and Fig. \ref{2295_photo_rv}), placing it within a subgroup of 76 known grazing planets (b+Rp/R$_{\star}$ $\geq$1), each characterized by an impact parameter precision of at least 30\%. Grazing transits present both advantages and challenges. On the one hand, the radii of the planets are poorly characterized, only ascertainable as lower limits with high confidence. Additionally, they are not ideal candidates for Rossiter-McLaughlin \citep[RM,][]{mclaughlin1924some, rossiter1924detection} measurements and atmospheric characterization due to their limited coverage of the rotating star’s surface compared to fully transiting planets.

On the other hand, grazing planets have transit shapes highly sensitive to small changes in orbital inclination \citep{ribas20085}. This sensitivity offers promising opportunities for studying long-term system dynamics and/or potentially discovering additional undetected planets within the system \citep[e.g.,][]{miralda2002orbital}, prompted by inclination perturbations. Consequently, TOI-2295b, with its grazing nature and bright host star (v= 9.60 $\pm$ 0.03 mag), presents an excellent target for follow-up transit depth or Transit Duration Variations (TDVs) observations.

\subsection{Dynamical analyses of the TOI-2537 system}
\label{Dynamical_analyses}

The orbital analysis presented in Sect.~\ref{ident} for the different planetary systems only considers Keplerian fits, i.e., we neglect any possible mutual gravitational interactions between the planets in multiple systems. This provides good fits of our datasets, except for the timing of the TOI-2537b's transits, as we detected TTVs for them. Such TTVs could not be explained with a Keplerian model, and
the TTVs reported in Sect.~\ref{joint_2537} are only determined as small, observational deviations from the Keplerian model, without any physics constraining them.

In this section, we perform a preliminary dynamical analysis of the TOI-2537 system. The main goal is to evaluate if the outer planet TOI-2537c could be the cause of the TTVs detected in TOI-2537b. Both planets are massive giants on eccentric orbits, making their mutual interactions particularly significant. However, only three transits of TOI-2537b are currently available. While this limited data allows for the detection of TTVs, it is still modest and does not provide a comprehensive view of the system. Therefore, the preliminary analysis presented here should be seen as first
evaluations \citep[see, e.g.,][]{hebrard20}. A more complete dynamical view of that system is expected in forthcoming papers including additional transit light curves \citep[see, e.g.,][]{almenara22}.

\begin{figure}
\centering
\includegraphics[width=9cm, angle=0]{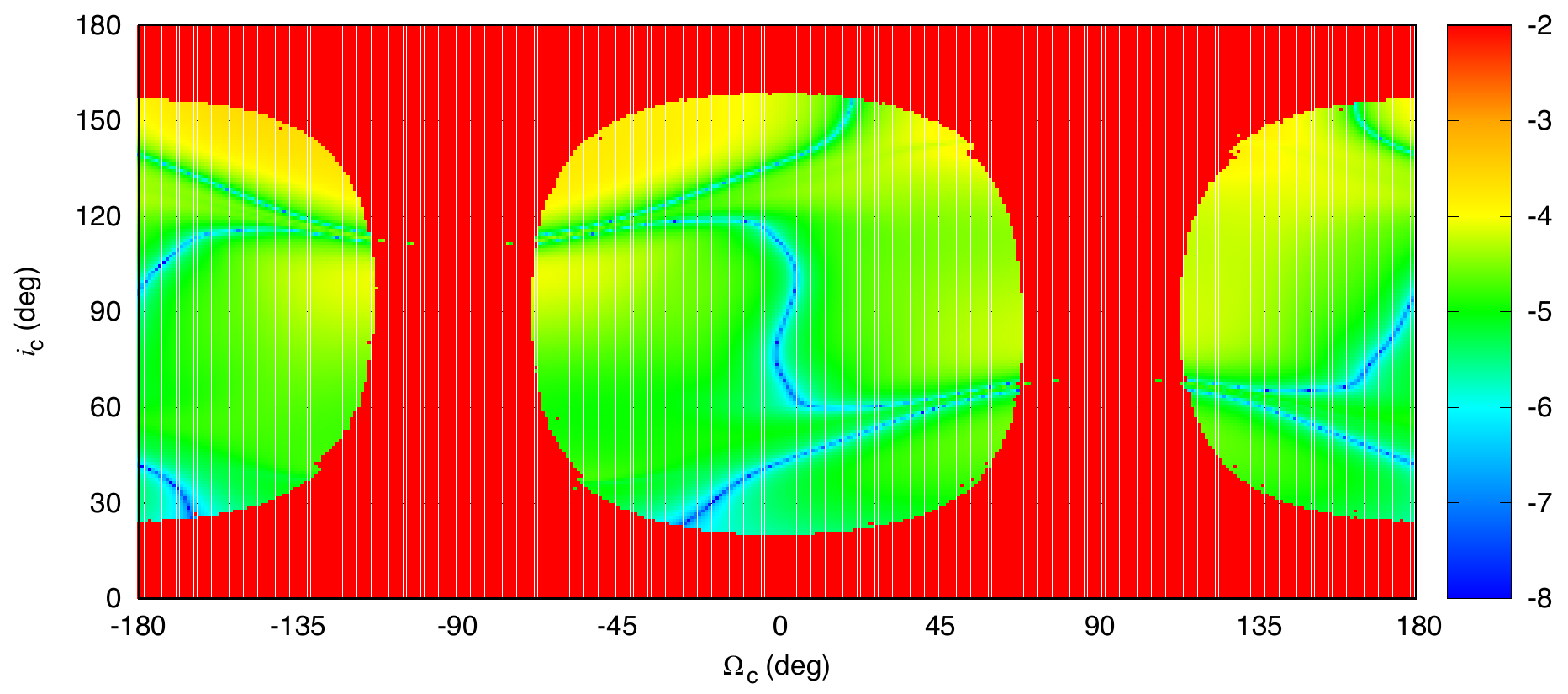}
\caption{Stability analysis of the TOI-2537 planetary system. For fixed initial conditions, the phase space of the system is explored by varying the two unknown parameters of TOI-2537c's orbit, namely the longitude of its ascending node and its inclination, with a step size of $1^\circ$. For each initial condition, the system is integrated over 10~kyr, and a stability criterion is derived with the frequency analysis of the mean longitude of the outer planet. The chaotic diffusion is measured by the variation in the frequencies. The color scale corresponds to values between $-8$ (blue) and $-2$ (red) for the decimal logarithm of the stability index $D$ used in~ \cite{correia10}. The red zone corresponds to highly unstable orbits, while the green-blue regions can be assumed to be stable on a billion-year timescale.}
\label{fig_TOI-2537_stability}
\end{figure}

In Fig.~\ref{fig_TOI-2537_stability} we show the wide vicinity of the TOI-2537 system parameters presented in Table~\ref{toi_2295_2537_5110}, following the approach presented by \cite{correia10}. That plot explores the phase space of the system by varying the two unknown parameters of the orbit of TOI-2537c: the longitude of its ascending node and its inclination.  This allows us to analyze and estimate the stability of the orbital solution. The stability indicator is presented as a color index, with red zones representing strongly chaotic trajectories and dark blue zones indicating extremely stable trajectories. One can see that  $20^{\circ} < i_c < 160^{\circ}$, that is mutual inclinations up to $70^{\circ}$, are stable and thus allowed. Whereas no transits are detected for TOI-2537c, its inclination is unknown, and RVs only constrain its minimum mass $M_P \sin i_c$ value. Our stability analysis thus excludes here inclinations below $i_c = 20^{\circ}$, corresponding to masses larger than $20$\,M$_{\rm Jup}$. That upper limit on the true mass of TOI-2537c is more constraining than the one reported in Sect.~\ref{joint_2537} from Gaia astrometry.

\begin{figure}
\centering
\includegraphics[width=0.5\textwidth]{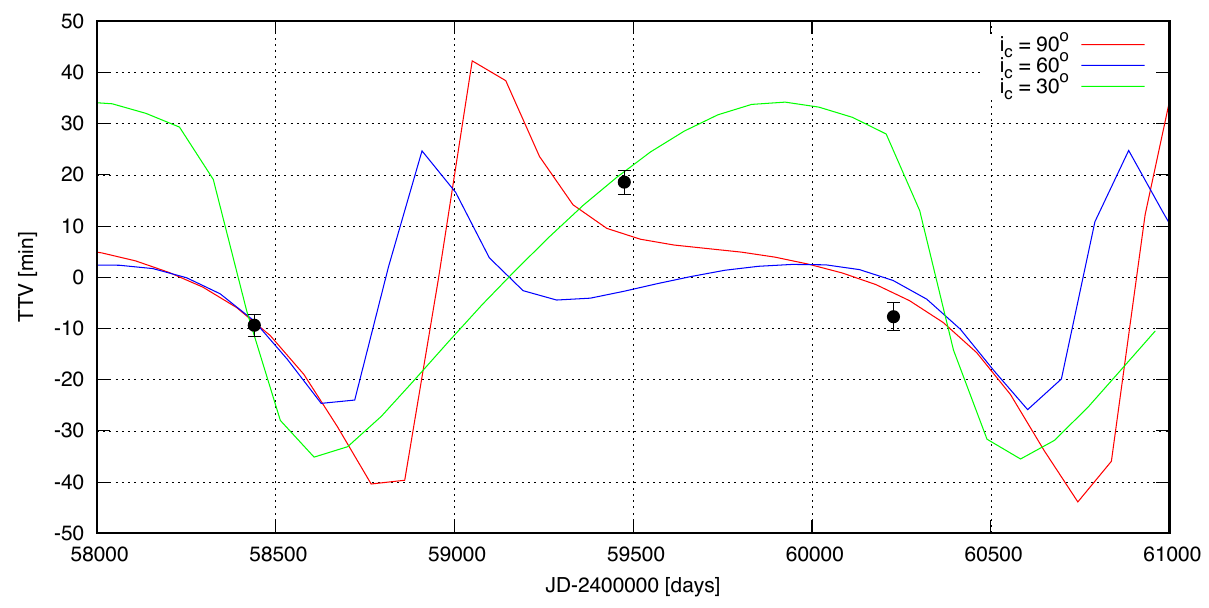}
\caption{Transit-timing variations for the TOI-2537b planet. The three curves correspond to dynamical simulations for the solution shown in Table~\ref{toi_2295_2537_5110}, with orbital inclination of TOI-2537c at $90^{\circ}$ (red, corresponding to an orbit coplanar with that of TOI-2537b), $60^{\circ}$ (blue), and $30^{\circ}$ (green). The measurements correspond to the three measured transits of TOI-2537b (Fig.~\ref{2537_TTV} and Table~\ref{tab:transitpars}). The simulated TTVs have similar orders of magnitude to the measured TTVs.}
\label{fig_TOI-2537_TTV_model}
\end{figure}

To compare the predicted TTVs with the observed ones, we computed the TTVs in three different configurations: $i_c =90^{\circ}$ (corresponding to an orbit coplanar with that of TOI-2537b), $60^{\circ}$, and $30^{\circ}$. The results are displayed in Fig.~\ref{fig_TOI-2537_TTV_model} As expected, the orbital inclination of TOI-2537c, thus its true mass, has an impact on the predicted TOI-2537b TTVs amplitude and shape. In all cases, one can see that the simulated TTVs have similar orders of magnitude to the measured TTVs. We can thus conclude that the TTVs observed in TOI-2537b are likely to be caused by the perturbations of this planet by TOI-2537c. New transit observations and analyses would be required to have a better analysis of those TTVs, and thus of that whole interacting system.

\begin{figure}
\centering
\includegraphics[width=0.5\textwidth]{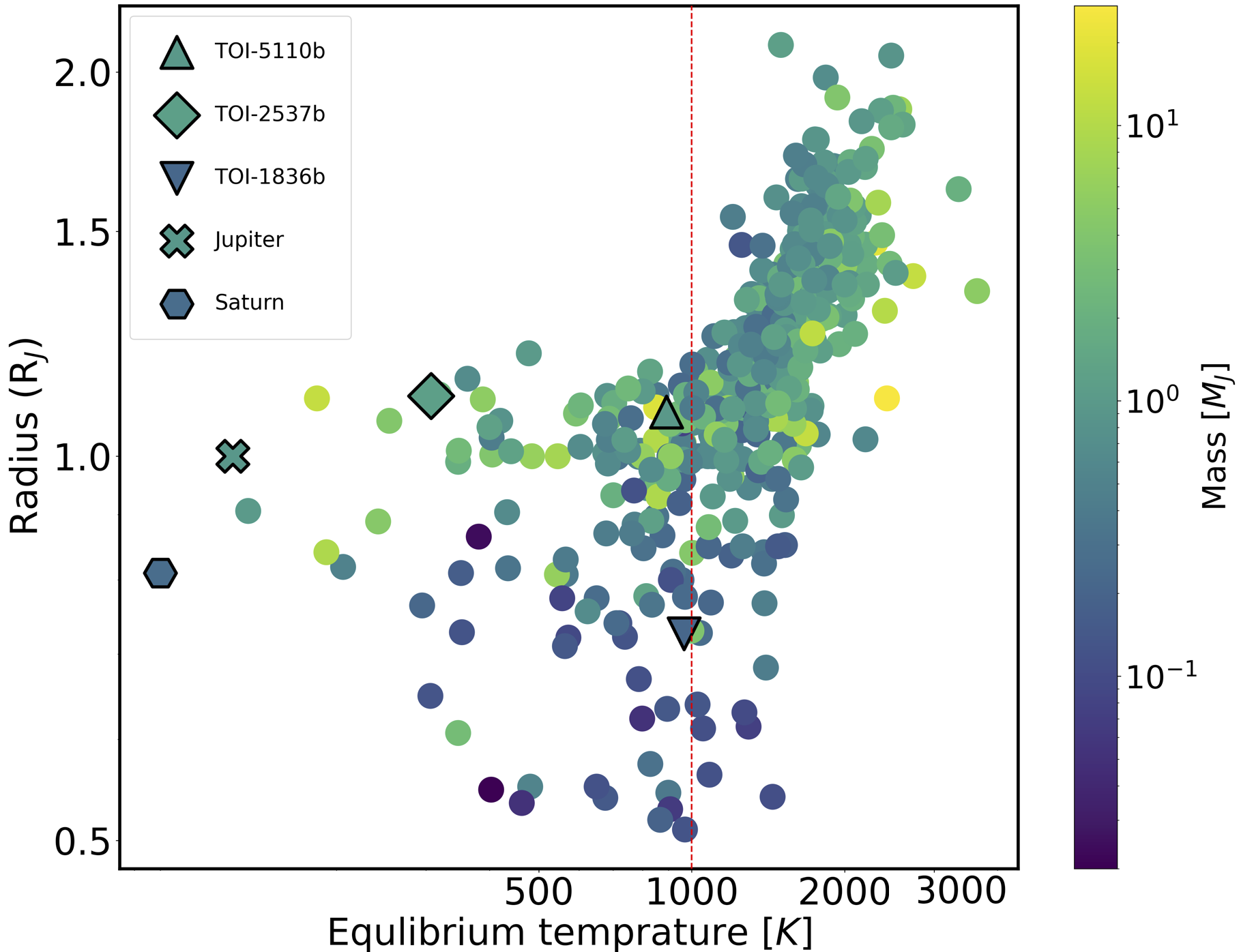}
\caption{The radius-equilibrium temperature of known giant (R > 0.5 R$_{\rm{J}}$) transiting planets from the NASA Exoplanet Data Archive with accurate mass and radii \citep[$\sigma_{M}$/M <= 25\% and $\sigma_{R}$/R <= 8\%,][]{otegi2020revisited}. The red vertical line indicates the empirical inflation boundary \citep{fortney2021hot}, where planet radii are seen to increase with equilibrium temperature. TOI-2537b is well within the low-temperature wing and is a valuable target for controlling the model of hot Jupiter radius anomaly.}
\label{hotjupyter_inflation}
\end{figure}
\subsection{Equilibrium temperature of TOI-2537b}

As TOI-2537b has a fairly eccentric orbit, to determine a more reliable average equilibrium temperature, we used equation 20 in \cite{mendez2017equilibrium}. This results in T$_{eq}$= 307$\pm$15 for TOI-2537b, whereas the other transiting planets presented in this study have equilibrium temperatures above 500 K. Fig. \ref{hotjupyter_inflation} shows the radius-equilibrium temperature distribution of known giant (R > 0.5 R$_{\rm{J}}$) transiting planets from the NASA Exoplanet Data Archive with accurate mass and radii \citep[$\sigma_{M}$/M <= 25\% and $\sigma_{R}$/R <= 8\%,][]{otegi2020revisited}. TOI-2537b is notably situated in the tail of the distribution of observed planets in the plot, within the low-temperature range. Hence, it serves as a valuable target for controlling models aimed at understanding the hot Jupiter radius anomaly.

\begin{figure}
\centering
\includegraphics[width=0.46\textwidth]{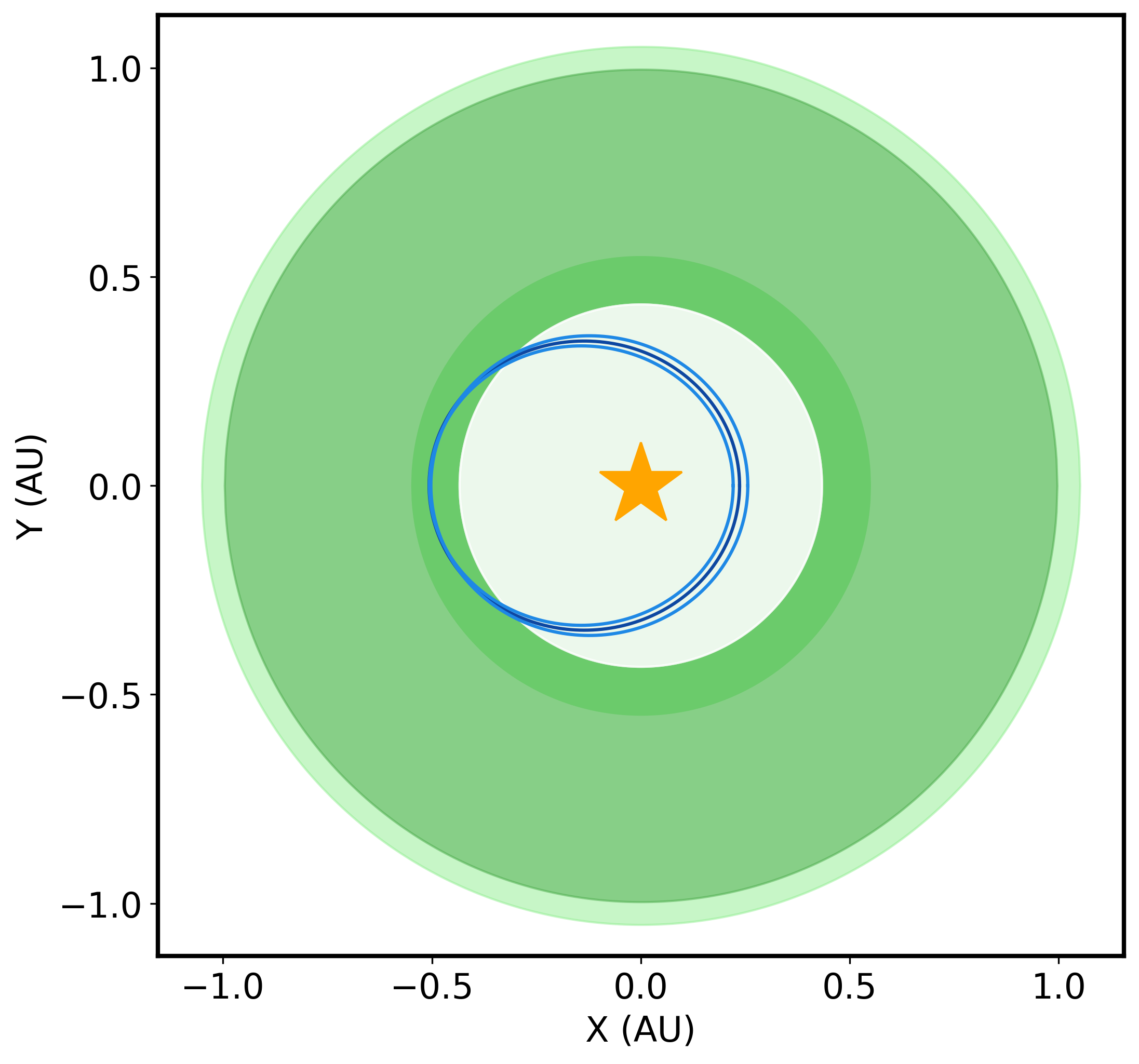}
\caption{The configuration of the TOI-2537b within the habitable zone. The orbit of TOI-2537b is depicted in dark blue, with its uncertainties represented by light blue. The habitable zone boundaries are indicated by green-shaded regions, representing the empirical habitable zone boundaries for recent Venus (inner boundary), runaway greenhouse (middle boundaries), and early Mars (outer boundary). These boundaries are computed following the results from \cite{kopparapu2014habitable}.}
\label{habitable}
\end{figure}

Furthermore, this temperature range might place this planet in the Habitable Zone. However, due to its relatively high eccentricity, the planet experiences changes in its equilibrium temperature throughout its orbit, ranging from about 420 K at perihelion to about 290 K at aphelion. Fig. \ref{habitable}, shows the orbit of TOI-2537b, overplotted by the Habitable Zones based on models outlined by \cite{kopparapu2014habitable}, incorporating both narrow and empirical approaches. The narrow habitable zone is defined by the Runaway Greenhouse limit as its inner boundary and the Maximum Greenhouse limit as its outer boundary. In contrast, the boundaries of the empirical habitable zone are defined by the Recent Venus and Early Mars limits\footnote{The boundary is calculated using \url{https://github.com/Eelt/HabitableZoneCalculator/blob/master/hzcalculator.py}}. As depicted in this figure, TOI-2537b intersects with the habitable zone only during a portion of its orbit.

\subsection{Prospective for RM follow-up}

Warm Jupiters are interesting targets for studying spin-orbit alignment using RM measurements. These planets are likely to have undergone significant inward migration; yet, their distance from their host star prevents tidal or other proximity effects from removing potential migration traces \citep{li2016tidal}. The RM amplitude can be estimated as \citep{gaudi2007prospects}: 

\begin{equation}
    K_{RM} = 52.8  \, \text{m/s}  \, \frac{v \sin(i)}{5  \, \text{km/s}}   \, \left( \frac{R_{\text{pl}}}{R_J}  \right)^{2}  \, \left(\frac{R_{\star}}{R_{\odot}}\right)^{-2},
\end{equation}
\\
where R$_{\text{pl}}$ and R$_{\star}$ represent planet and stellar radius, respectively. Using this formula, we computed the RM amplitudes for the characterized transiting planets in this study. The results are as follows: 8.7$\pm$ 2.4 m/s for TOI-1836b, 41.8$^{+49.6}_{-32.2}$ for TOI-2295b, 3$\pm$2 m/s for TOI-5076b, 53.7$\pm$ 19.8 m/s for TOI-2537b, and $>$ 5 m/s for TOI-5110b as we only have the lower limit of its v$\sin i$ (see Sect. \ref{stellar} and Table \ref{star_info}). The relatively large predicted RM amplitude for some of these targets, combined with their feasible transit durations (< 6.6 h), suggests that they are promising candidates for RM observations, which can be conducted using various ground-based instruments. As discussed in Sect. \ref{grazing}, TOI-2295 has a grazing nature and is not an ideal candidate for RM measurements. 

\subsection{Prospects for atmospheric characterization}

\begin{figure}
\centering
\includegraphics[width=0.48\textwidth]{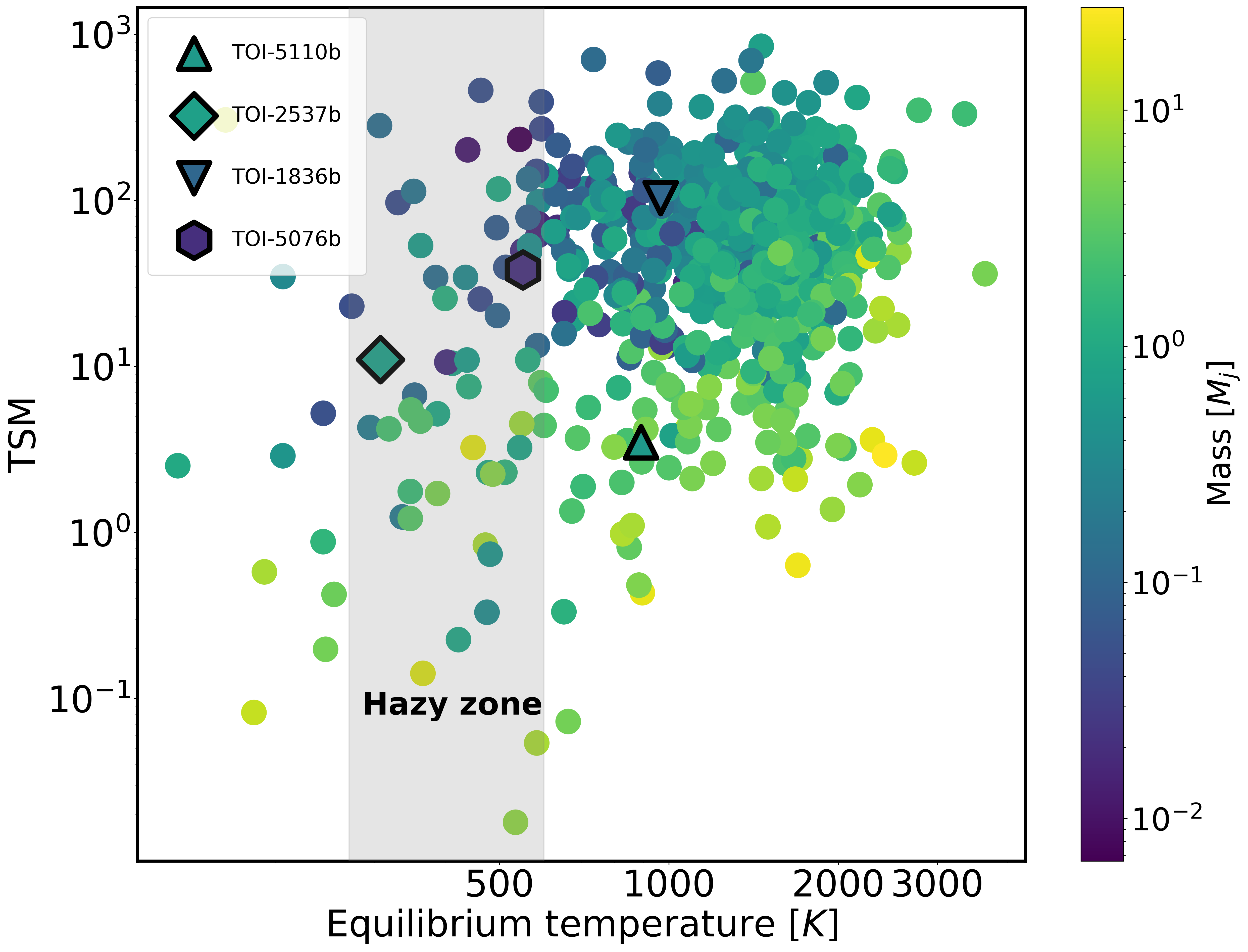}
\caption{TSM- equilibrium temperature diagram for the population of well-characterized \citep[$\sigma_{M}$/M <= 25\% and $\sigma_{R}$/R <= 8\%,][]{otegi2020revisited} giant (R > 0.3 RJ) transiting planets. The gray area highlights the hazy atmosphere zone proposed by \citep{yu2021haze}. Planetary mass is color-coded for each planet.}
\label{atmosphere}
\end{figure}

We investigate the prospects of atmospheric characterization using the Transmission Spectroscopy Metric (TSM), as defined by the formula:

\begin{equation}
TSM = \text{scale factor} \times \left( \frac{T_{eq} \times R_p^3} { M_p \times R_s^2} \right) \times 10^{-m_{\rm{J}}/5},
\end{equation}
\\
following \cite{kempton2018framework}. This metric incorporates several key parameters, including the planet's equilibrium temperature (T$_{eq}$), radius ($R_p$), mass ($M_p$), host star radius ($R_s$), and host star magnitude in the J band (m$_{\rm{J}}$). The scale factor in this metric varies based on the planet's size: it is 1.28 for planets with radii between 2.75 and 4 R$_{\oplus}$, and 1.15 for planets with radii between 4 and 10 R$_{\oplus}$. However, it is not applicable to planets with a radius larger than 10 R$_{\oplus}$, for which we consider a scale factor of 1. Similar to TOI-2537b, both TOI-2295b and TOI-5110b have eccentric orbits. Therefore, we applied Equation 20 from \cite{mendez2017equilibrium} to calculate T${eq}$, resulting in T$_{eq}$= 680$\pm$23 for TOI-2295b, and T$_{eq}$= 892$\pm$32 for TOI-5110b. Finally, we calculate a TSM value of 103.5 $\pm$29.5 for TOI-1836b, 10.9 $\pm$ 2.5 for TOI-2537b, 3.53$\pm$0.64 for TOI-5110b, and 38.2$\pm$ 7.0 for TOI-5076b (see Fig. \ref{atmosphere} and Table \ref{table:TSM_values}). For TOI-2295b, which exhibits grazing transits, not all of its atmosphere may transit the star, leading to a poorly determined TSM value.

\begin{table}[h!]
\centering
\caption{TSM values for the characterized transiting planets in this study. TOI-2295b is excluded due to grazing transits, which result in poorly determined TSM value.}
\label{table:TSM_values}
\begin{tabular}{cc}
\hline
Planet & TSM \\ \hline
TOI-1836b & 103.5 $\pm$ 29.5 \\ 
TOI-2537b & 10.9 $\pm$ 2.5 \\ 
TOI-5110b & 3.53 $\pm$0.64 \\ 
TOI-5076b & 38.2 $\pm$ 7.0 \\ \hline
\end{tabular}
\end{table}

Among all the targets presented in this work, TOI-1836b stands out as a high-quality candidate for atmospheric characterization, exceeding the recommended threshold of TSM= 90 set by \cite{kempton2018framework} for sub-Jovian planets. Remarkably, with this mass, TOI-1836b falls into the sub-Saturn valley, a theoretically predicted but disputed drop in the mass distribution of planets between sub-Neptunes and gas giants~\citep{IdaLin2004,Emsenhuber2021b,Mayor2011,Bennett2021,Schlecker2022}. Therefore, targeting TOI-1836b for atmospheric observation provides a unique opportunity to investigate the nature and origins of the sub-Saturn valley.

Additionally, TOI-2537b and TOI-5076b have effective temperatures of \(307 \pm 15\) K and 550$\pm$ 14 K, respectively, placing them in the hazy atmospheric zone \citep{yu2021haze}. This positioning makes them interesting targets for studying photochemical hazes and the by-products of disequilibrium chemistry \citep{fortney2020beyond}. TOI-2537b, with a TSM value of 11, is amenable for characterization using JWST, comparable to the Cycle 2 target PH-2b, a temperate gas giant with a TSM of 12 (Proposal ID: 3235). Similarly, TOI-5076b, with a TSM of 38.2, is a good target for atmospheric observation.

\section{Summary}
\label{summary}
In this study, we present analyses of seven systems, incorporating various observations such as photometric data, spectroscopic data, and high-spatial resolution imaging. The following is a case-by-case summary of our results, categorized into three groups: newly characterized, candidate, and false positive planetary systems.

\subsection*{Planetary systems:}
\begin{itemize}

\item TOI-2295b and c: The system encompasses two planetary bodies orbiting around the star TOI-2295 ($R_*= 1.459^{+0.056}_{-0.058}~ R_{\odot}$, $V= 9.595 \pm 0.004$ mag, distance= $125.51^{+0.16}_{-0.14}$ pc). The inner planet, TOI-2295b, has a mass of $0.875^{+0.042}_{-0.041}~ M_{\rm{J}}$ and a poorly characterized radius of $1.47^{+0.85}_{-0.53}$~$R_{\rm{J}}$ due to the extremely grazing nature of the transit, with an impact parameter of $1.056^{+0.063}_{-0.043}$. Additionally, its bulk density is $0.34^{+0.96}_{-0.25}~\text{g/cm}^3$. The outer planet, TOI-2295c, has a minimum mass of $5.61^{+0.23}_{+0.24}~ M_{\rm{J}}$ and orbits every $966.5^{+4.3}_{-4.2}$ d around its host star.

\item TOI-2537b and c: TOI-2537b is classified as a giant planet, with a radius of R = 1.004$^{+0.059}{-0.061}$~$R{\rm{J}}$ and a mass of M = 1.307$^{+0.091}{-0.088}$~$M{\rm{J}}$, resulting in a bulk density of 1.60$^{+0.35}{-0.26}$~$\text{g/cm}^3$. It orbits its dwarf star ($R* = 0.771\pm0.049~ R_{\odot}$, $V= 13.236\pm 0.08$ mag, distance = $182.42^{+1.05}{-0.93}$ pc) with a period of $94.1022\pm0.0011$ d. This planet has an equilibrium temperature of T${eq}$= 307$\pm$15 K, placing it partially within the habitable zone, and also within a hazy atmosphere zone. The planet is an interesting target for atmospheric study as well as RM measurements. Additionally, TOI-2537c, the outer planet in the system, has a minimum mass of $7.23^{+0.52}{-0.45}~ M{\rm{J}}$ and an orbital period of $1920^{+230}_{-140}$ d. Preliminary dynamical analyses of the detected TTVs for TOI-2537b suggest that these perturbations are likely due to the gravitational effect of the outer planet. Further transit observations will be necessary to refine the analysis of these TTVs and gain a better understanding of the entire system interaction.

\item TOI-5110b: This planet has a radius of $1.069^{+0.054}_{0.052}~ R_{\rm{J}}$, and a mass of $2.90\pm0.13~ M_{\rm{J}}$, leading to a bulk density of $2.95^{+0.50}_{-0.43}~\text{g/cm}^3$. It orbits every $30.158577^{+0.000092}_{-0.000095}$ d around its sub-giant host star ($R_*= 2.333^{+0.097}_{-0.096} ~R_{\odot}$, $V=11.1 \pm 0.1$ mag, distance= $358.9^{+1.6}_{-3.0}$ pc). It stands out as one of the most eccentric transiting planets detected so far, with an eccentricity of $0.745^{+0.030}_{-0.027}$. This makes it a compelling target for dynamical planetary studies. We investigate the presence of a perturbing companion in the system as a potential eccentricity excitation mechanism, although the current dataset does not provide evidence of any additional body. Further RV observations would be necessary to shed light on this matter.

\item TOI-5076b: We updated and refined the parameters of this known planet \citep{montalto2024gaps}, resulting in a radius of $3.486^{+0.100}_{-0.094}$ R$_{\oplus}$, a mass of $16.1\pm2.4$\ M$_{\oplus}$, and a bulk density of $2.08^{+0.35}_{-0.34} ~\text{g/cm}^3$. It orbits its dwarf host, TOI-5076 $R_*= 0.844^{+0.020}_{-0.019} ~ R{\odot}$, $V=10.9 \pm 0.03$ mag, distance= $82.53^{+0.09}_{-0.10}$ pc), every $23.443162^{+0.000062}_{-0.000063}$ d.

\item TOI-1836b: Similarly, through our refinement of the parameters of this known planet \citep{chontos2024tess}, we determined its radius to be $0.714\pm0.031~R_{\rm{J}}$ and its mass to be $0.121 \pm 0.029~ M_{\rm{J}}$, resulting in a bulk density of $0.41^{+0.12}_{-0.11}~\text{g/cm}^3$. It orbits every $20.380799 \pm0.000016$ d around its sub-giant host, TOI-1836 ($R_*= 1.577\pm0.060~ R_{\odot}$, $V=9.77 \pm 0.03$ mag, distance= $189.23^{+0.50}_{-0.51}$ pc). Remarkably, TOI-1836b is a high-quality target for atmospheric characterization by the James Webb Space Telescope space telescope.
\end{itemize}

\subsection*{Planetary candidate system:}

\begin{itemize}

\item TOI-4081.01: This planet candidate would have a radius corresponding to $1.193^{+0.066}_{-0.067} ~R_{\rm{J}}$ and a mass of $1.89 \pm 0.35 ~ M_{\rm{J}}$, with a bulk density of $1.37^{+0.39}_{-0.31}~\text{g/cm}^3$. It completes an orbit every $9.258388^{+0.000020}_{-0.000021}$ d around its sub-giant host, TOI-4081 ($R_*= 2.48 \pm 0.11 ~ R_{\odot}$, $V=11.45 \pm 0.08$ mag, distance= $442.81^{+4.84}_{-7.21}$ pc). We exercise caution in interpreting the nature of this candidate due to variability observed in its bisectors which exceeds that of RVs, and the presence of another star within the TESS and SOPHIE apertures, alongside the challenges posed by the rapid
rotation of the star in testing the mask effect. 
\end{itemize}

\subsection*{False positive planetary systems:} 
\begin{itemize}

\item TOI-4168B: We show that the transit identified by TESS is actually a secondary stellar eclipse, with the primary transit being undetectable. TOI-4168A has $R_*= 1.162^{+0.048}_{-0.044} ~ R_{\odot}$, $M_*= 1.024^{+0.066}_{-0.058} ~M_{\odot}$, and is at a distance of $323.0^{+1.9}_{-1.8}$ pc. Additionally, TOI-4168B has a $R_*= 0.481^{+0.024}_{-0.023} ~ R_{\odot}$, and $M_*= 0.506^{+0.018}_{-0.017} ~M_{\odot}$.

\end{itemize}

\begin{acknowledgements}

We warmly thank the OHP staff for their support on the observations. We received funding from the French Programme National de Physique Stellaire (PNPS) and the Programme National de Planétologie (PNP) of CNRS (INSU). N.H. acknowledges the Centre National d'\'{E}tudes Spatiales (CNES) postdoctoral funding fellowship. This work was supported by CNES, focused on the PLATO mission. This paper made use of data collected by the TESS mission which is publicly available from the Mikulski Archive for Space Telescopes (MAST) operated by the Space Telescope Science Institute (STScI). Funding for the TESS mission is provided by NASA’s Science Mission Directorate. We acknowledge the use of public TESS data from pipelines at the TESS Science Office and at the TESS Science Processing Operations Center. Resources supporting this work were provided by the NASA High-End Computing (HEC) Program through the NASA Advanced Supercomputing (NAS) Division at Ames Research Center for the production of the SPOC data products.
\\
A.C.M.C. acknowledges support from the FCT, Portugal, through the CFisUC projects UIDB/04564/2020 and UIDP/04564/2020, with DOI identifiers 10.54499/UIDB/04564/2020 and 10.54499/UIDP/04564/2020, respectively. B.S.S. acknowledges the support of M.V. Lomonosov Moscow State University Program
 of Development. The results reported herein benefitted from collaborations and/or information exchange within NASA’s Nexus for Exoplanet System Science (NExSS) research coordination network sponsored by NASA’s Science Mission Directorate under Agreement No. 80NSSC21K0593 for the program ``Alien Earths".  E.M. acknowledges funding from FAPEMIG under project number APQ-02493-22 and a research productivity grant number 309829/2022-4 awarded by the CNPq, Brazil. XD and A.C. acknowledge funding from the French National Research 
Agency in the framework of the Investissements d’Avenir program 
(ANR-15-IDEX-02), through the funding of the “Origin of Life" project of 
the Grenoble-Alpes University. 
A.C. acknowledges funding from the French ANR under contract number
ANR\-18\-CE31\-0019 (SPlaSH). Some of the observations in this paper made use of the High-Resolution Imaging instrument ‘Alopeke and were obtained under Gemini LLP Proposal Number:
GN/S-2021A-LP-105. ‘Alopeke was funded by the NASA Exoplanet Exploration Program and
built at the NASA Ames Research Center by Steve B. Howell, Nic Scott, Elliott P.
Horch, and Emmett Quigley. 'Alopeke was mounted on the Gemini North/South telescopes
of the International Gemini Observatory, a program of NSF’s OIR Lab, which is
managed by the Association of Universities for Research in Astronomy (AURA) under a
cooperative agreement with the National Science Foundation. on behalf of the Gemini
partnership: the National Science Foundation (United States), National Research
Council (Canada), Agencia Nacional de Investigación y Desarrollo (Chile), Ministerio
de Ciencia, Tecnología e Innovación (Argentina), Ministério da Ciência, Tecnologia,
Inovações e Comunicações (Brazil), and Korea Astronomy and Space Science Institute
(Republic of Korea).\\
This work is partly supported by JSPS KAKENHI Grant Numbers
JP24H00017, JP24K00689, and JSPS Bilateral Program Number
JPJSBP120249910.nThis paper is based on observations made with the MuSCAT2 instrument,
developed by ABC, at Telescopio Carlos Sánchez operated on the island
of Tenerife by the IAC in the Spanish Observatorio del Teide.
\\
Based in part on observations obtained at the Southern Astrophysical 
Research (SOAR) telescope, which is a joint project of the 
Minist\'{e}rio da Ci\^{e}ncia, Tecnologia e Inova\c{c}\~{o}es (MCTI/LNA) 
do Brasil, the US National Science Foundation’s NOIRLab, the University 
of North Carolina at Chapel Hill (UNC), and Michigan State University (MSU). DD acknowledges support from the TESS Guest Investigator Program grant 80NSSC22K0185, and from the NASA Exoplanet Research Program grant 18-2XRP18\_2-0136. PCZ acknowledges support from STFC consolidated grant number ST/V000861/1, and UKSA grant number ST/X002217/1. The postdoctoral fellowship of KB is funded by F.R.S.-FNRS grant T.0109.20 and by
the Francqui Foundation.
\end{acknowledgements}
\bibliography{ref}

\begin{appendix}
\label{TPf}
\onecolumn 
\section{TPFs}
\begin{figure*}[h]
\includegraphics[width=0.24\columnwidth]{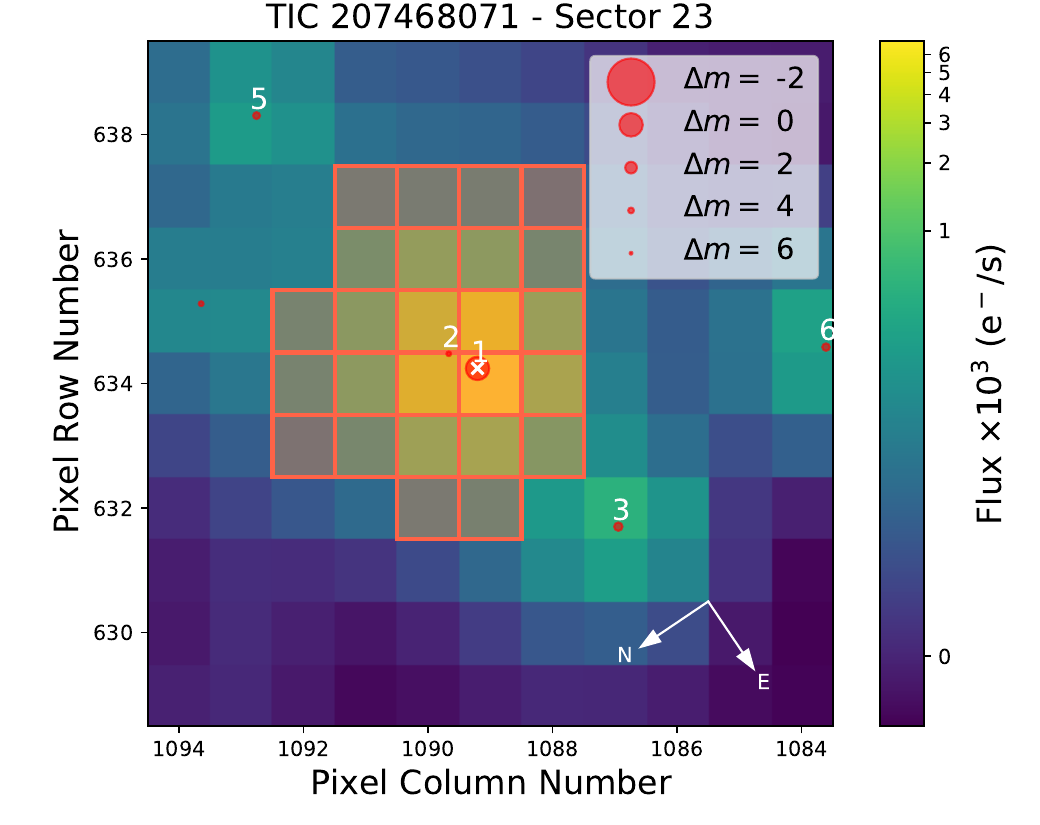}
\includegraphics[width=0.24\columnwidth]{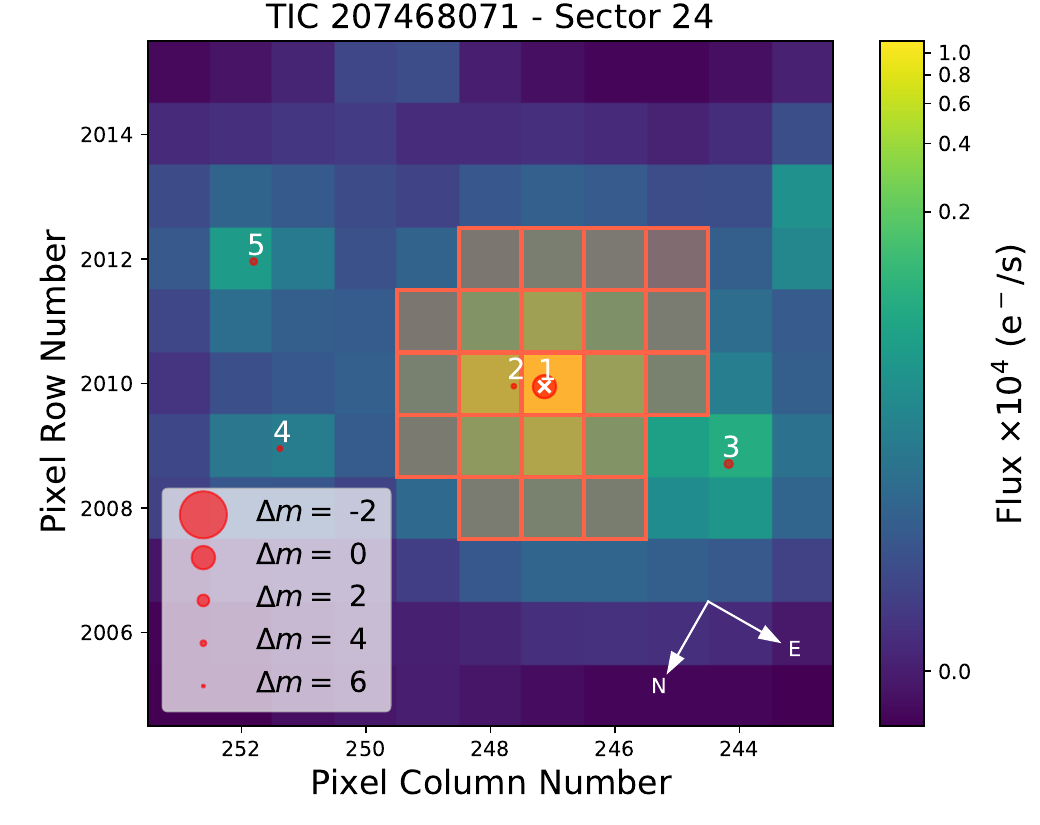}
\includegraphics[width=0.24\columnwidth]{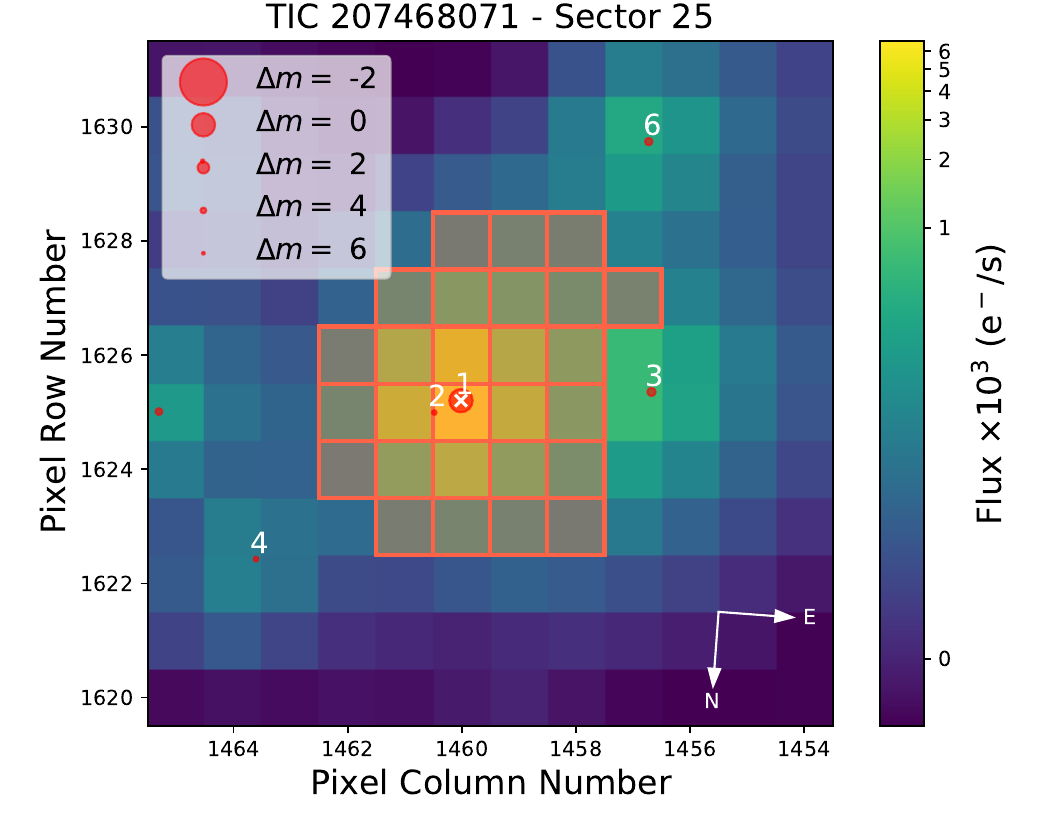}
\includegraphics[width=0.24\columnwidth]{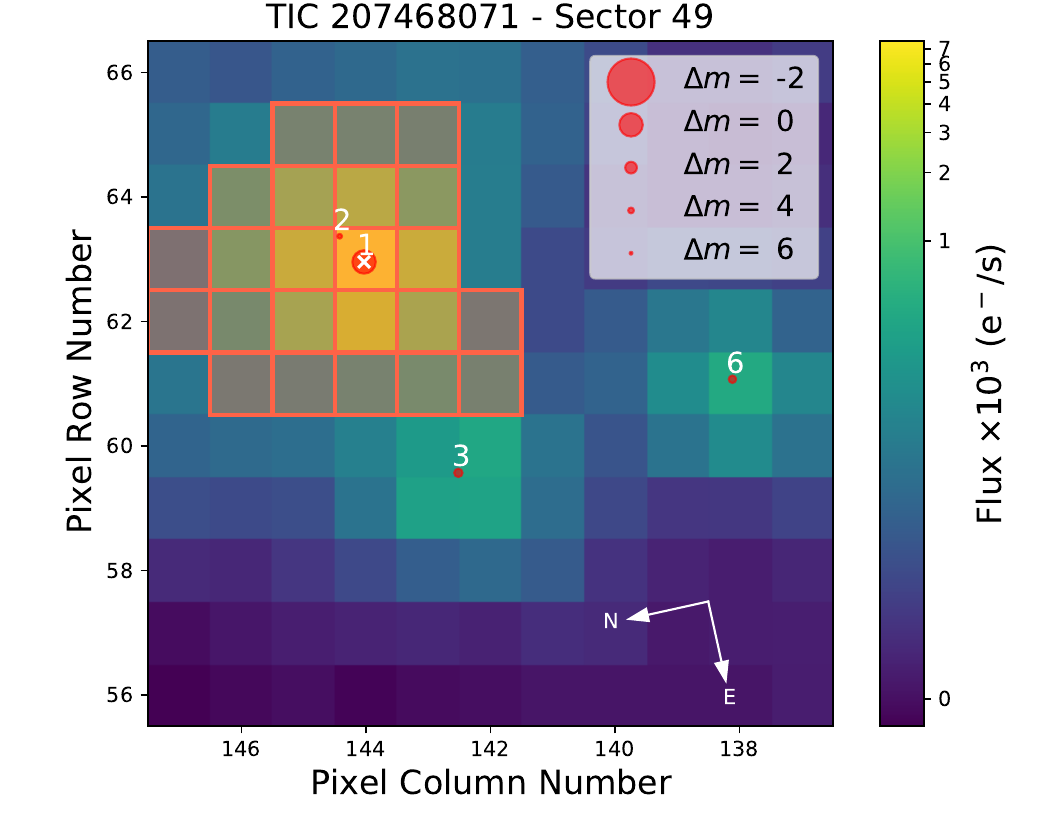}\\
\includegraphics[width=0.24\columnwidth]{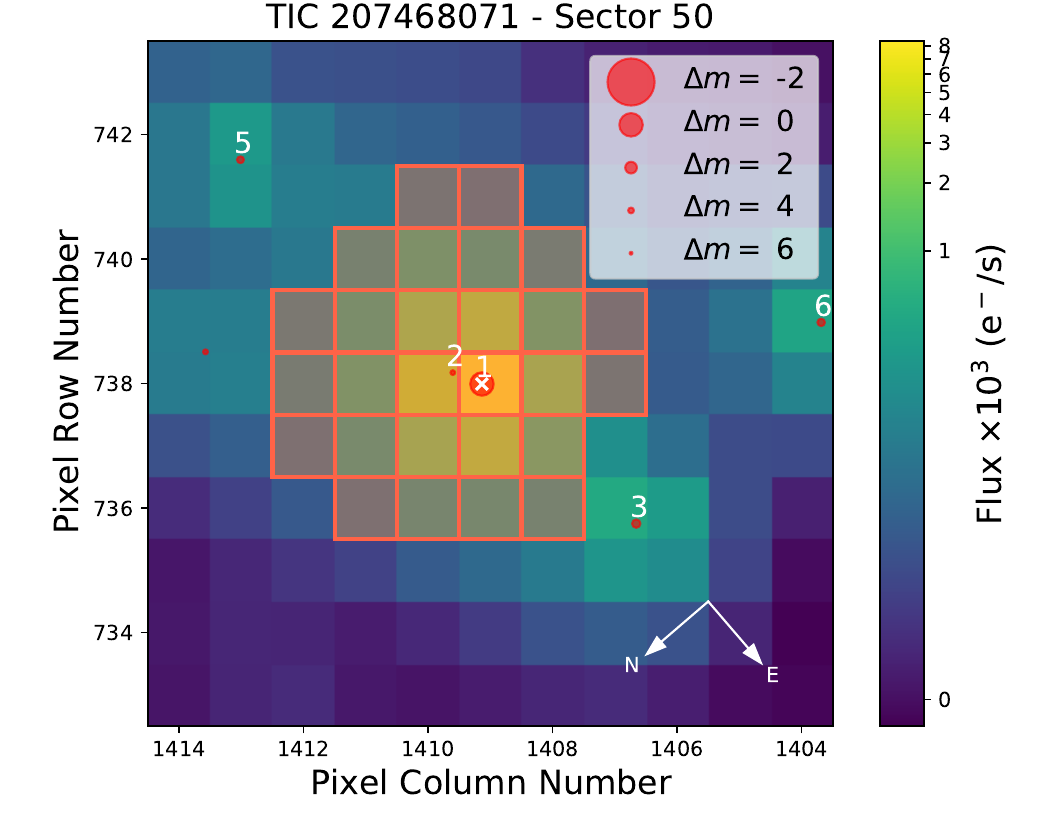}
\includegraphics[width=0.24\columnwidth]{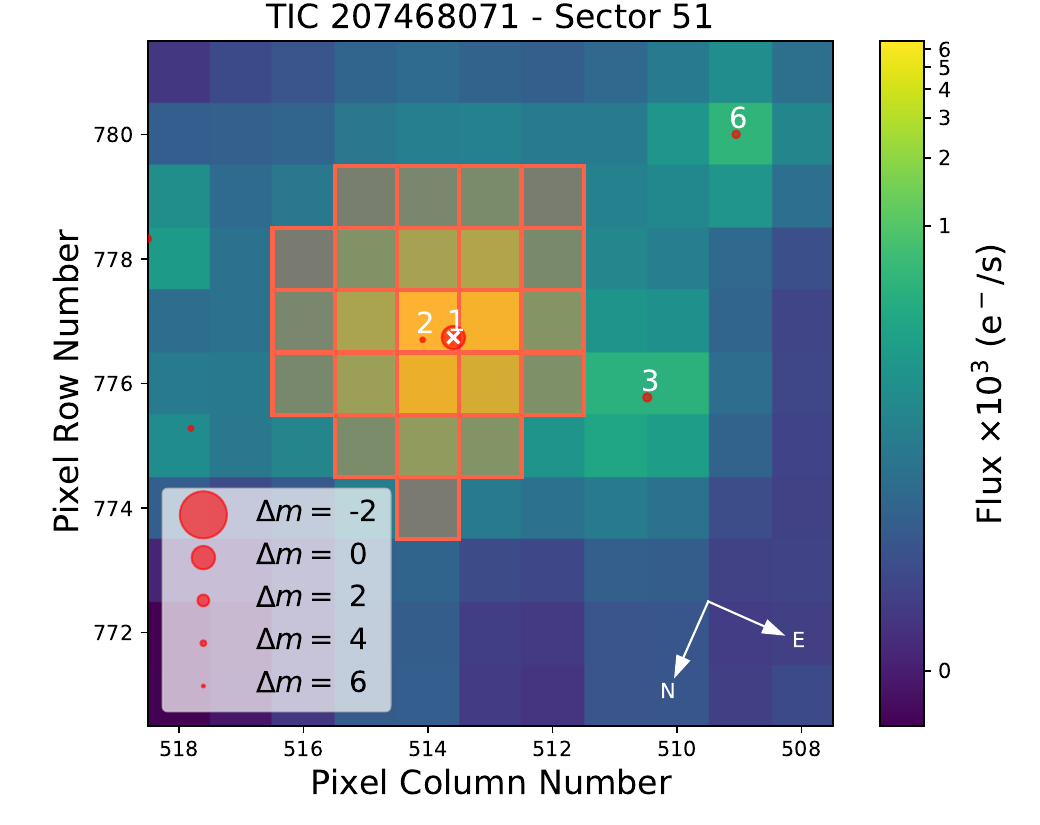}
\includegraphics[width=0.24\columnwidth]{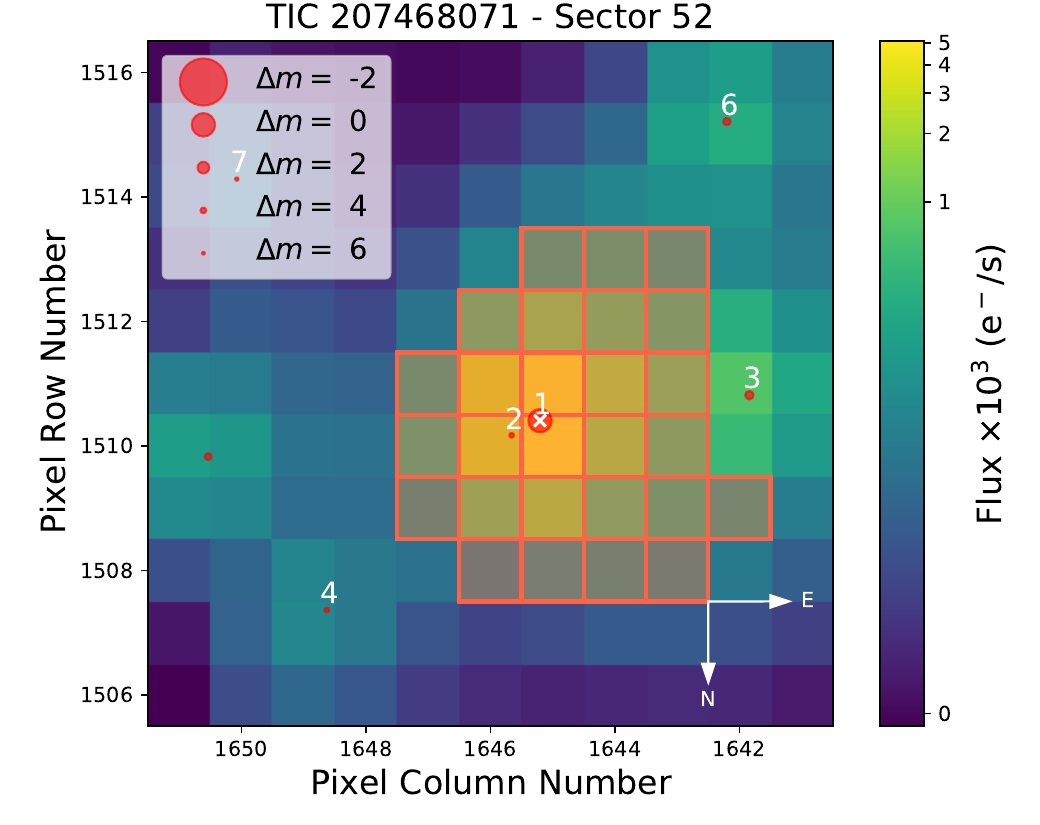}
\includegraphics[width=0.24\columnwidth]{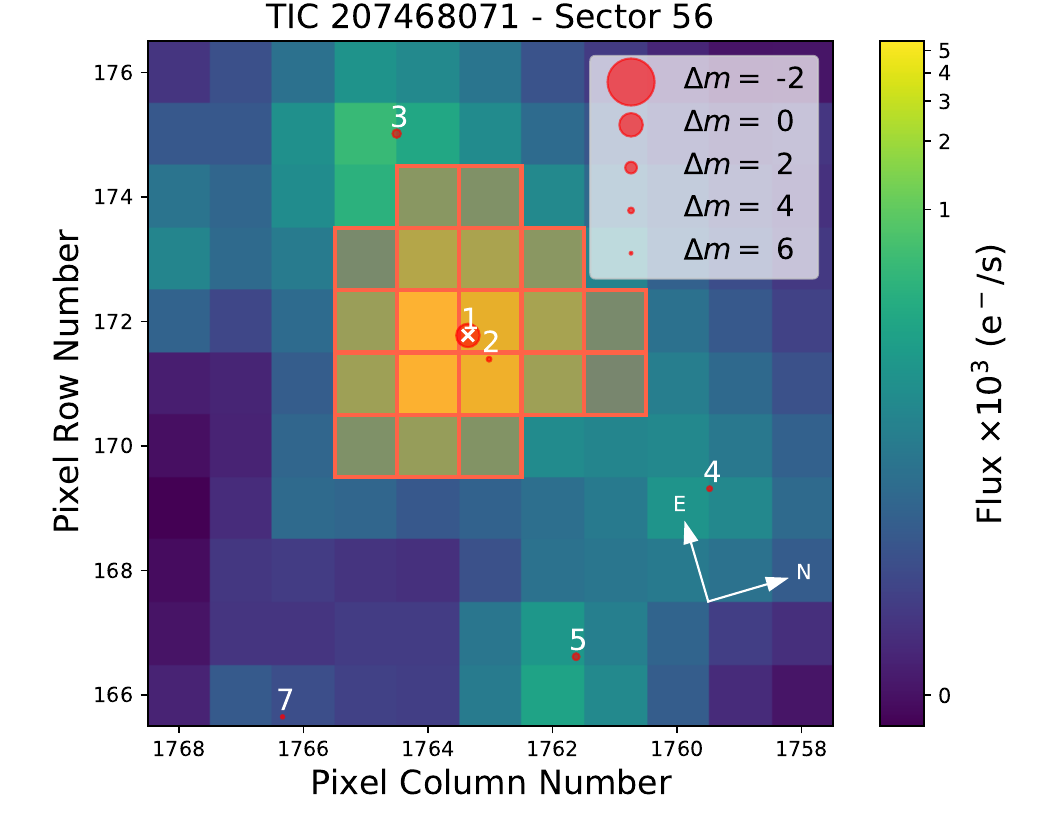}\\
\includegraphics[width=0.24\columnwidth]{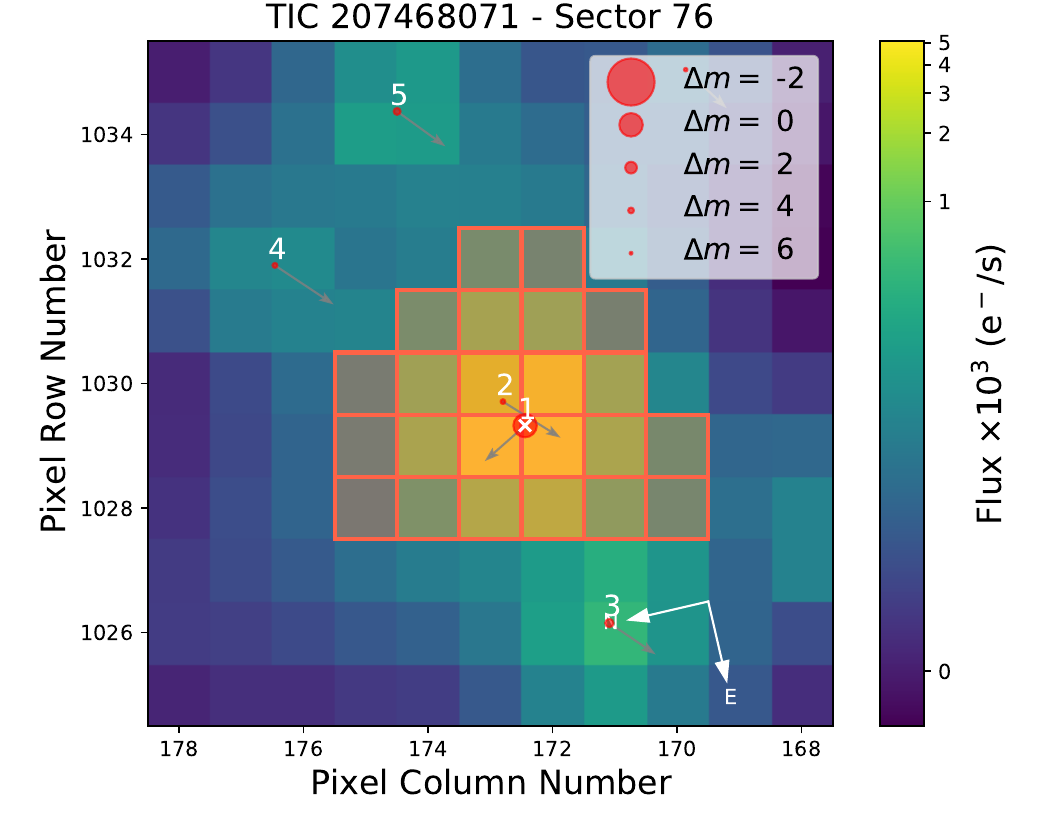}
\includegraphics[width=0.24\columnwidth]{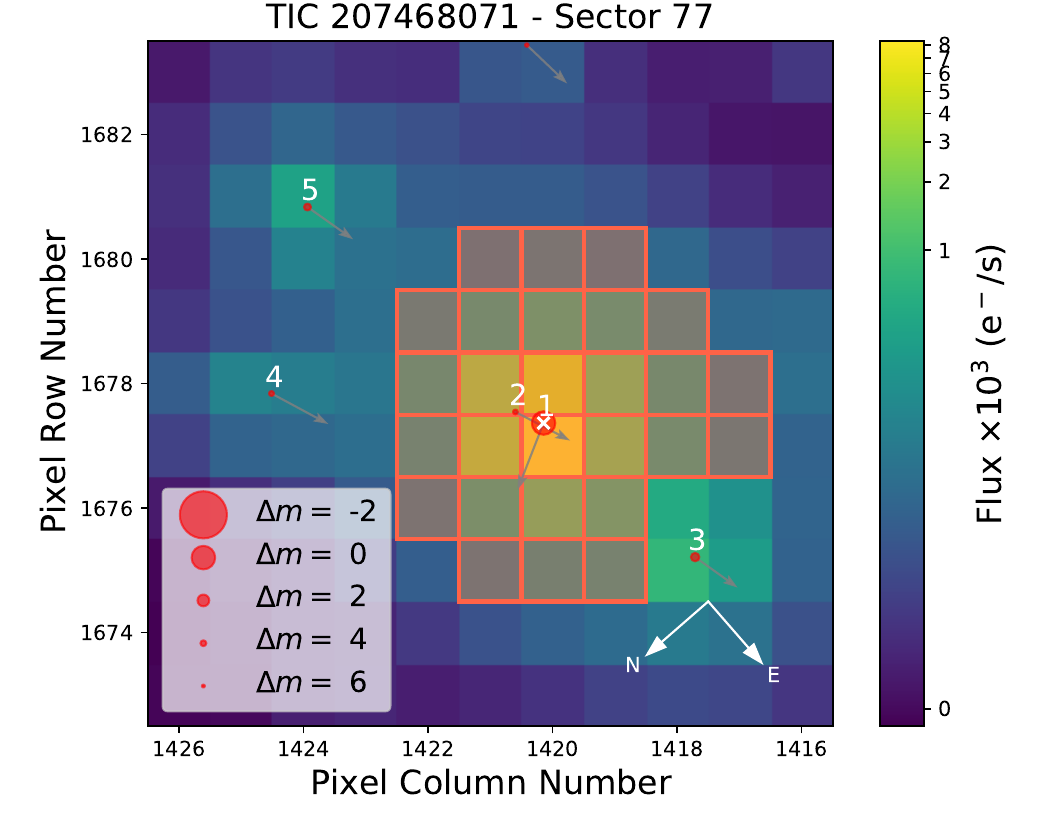}
\includegraphics[width=0.24\columnwidth]{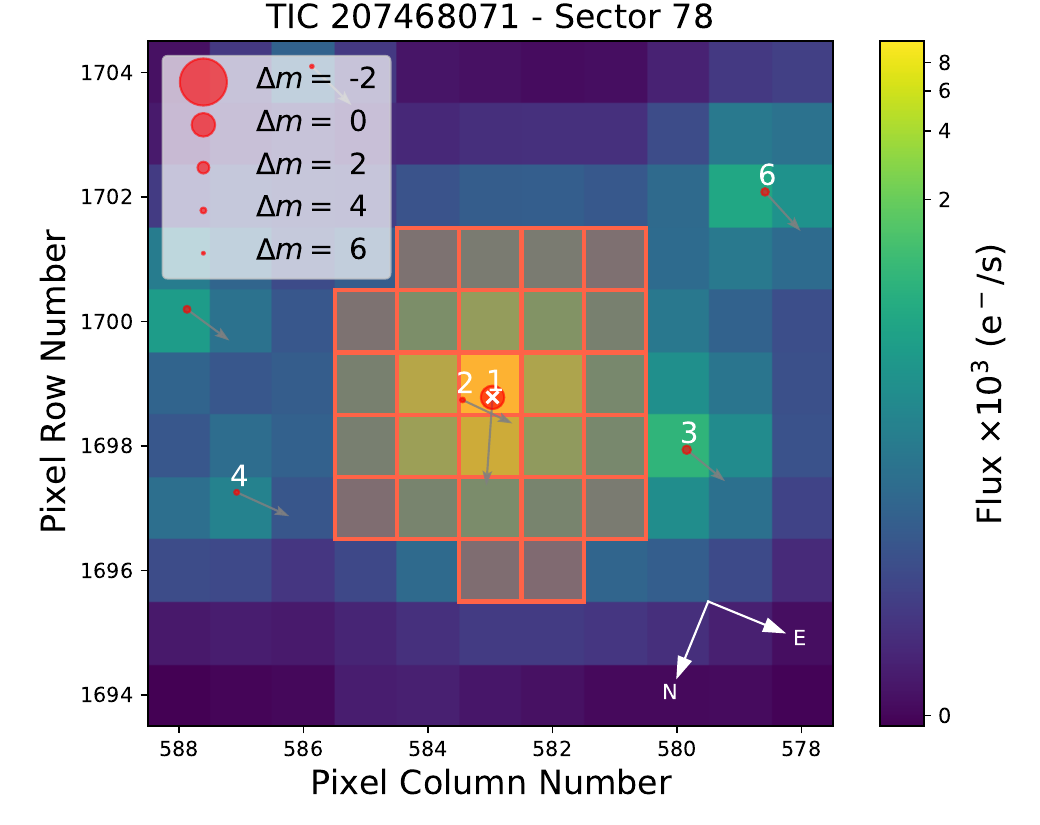}
\caption{The TPFs for TOI-1836, created by \texttt{tpfplotter} \citep{2020AA...635A.128A}. The images display electron counts, with the SPOC aperture mask outlined in red. Red circles are used to indicate the primary target (marked as 1) and nearby sources (rest of the numbers) based on their positions in Gaia DR3. The size of these circles corresponds to the relative magnitudes of the sources compared to the target star. Additionally, arrows illustrate the proper motion of each star.}
\label{tpfplotter_1836}
\end{figure*}

\begin{figure*}[h]
\centering
\includegraphics[width=0.24\columnwidth]{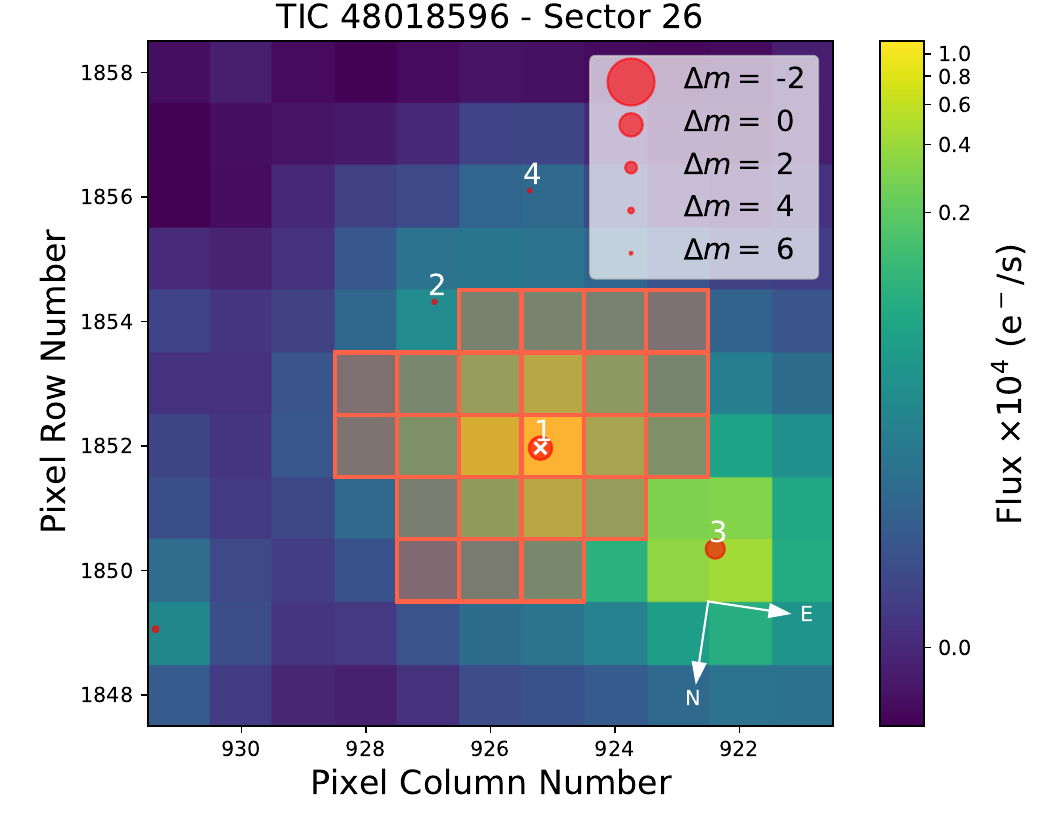}
\includegraphics[width=0.24\columnwidth]{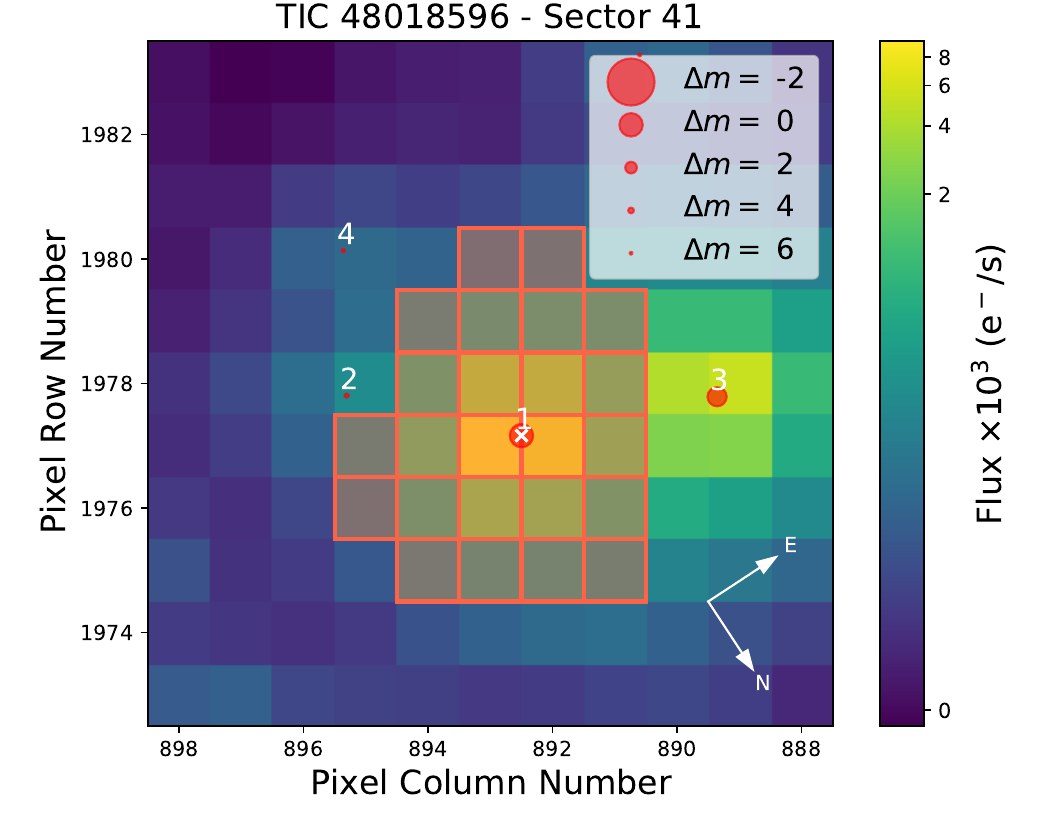}
\includegraphics[width=0.24\columnwidth]{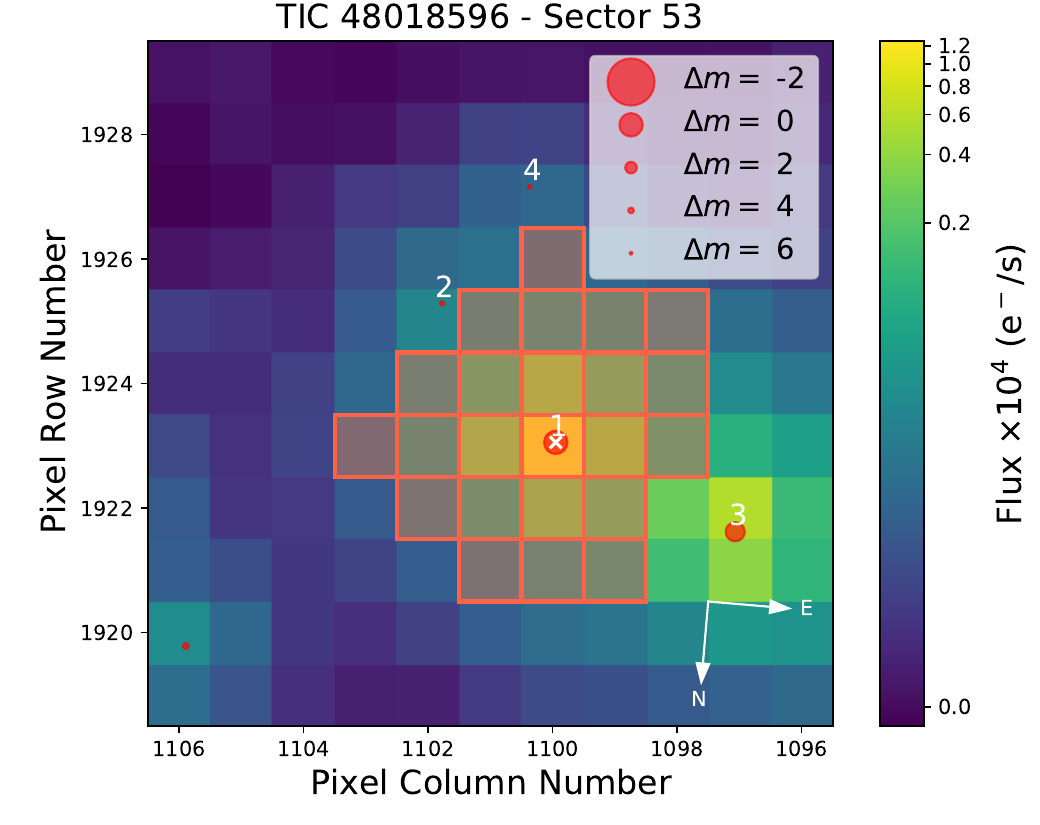}
\includegraphics[width=0.24\columnwidth]{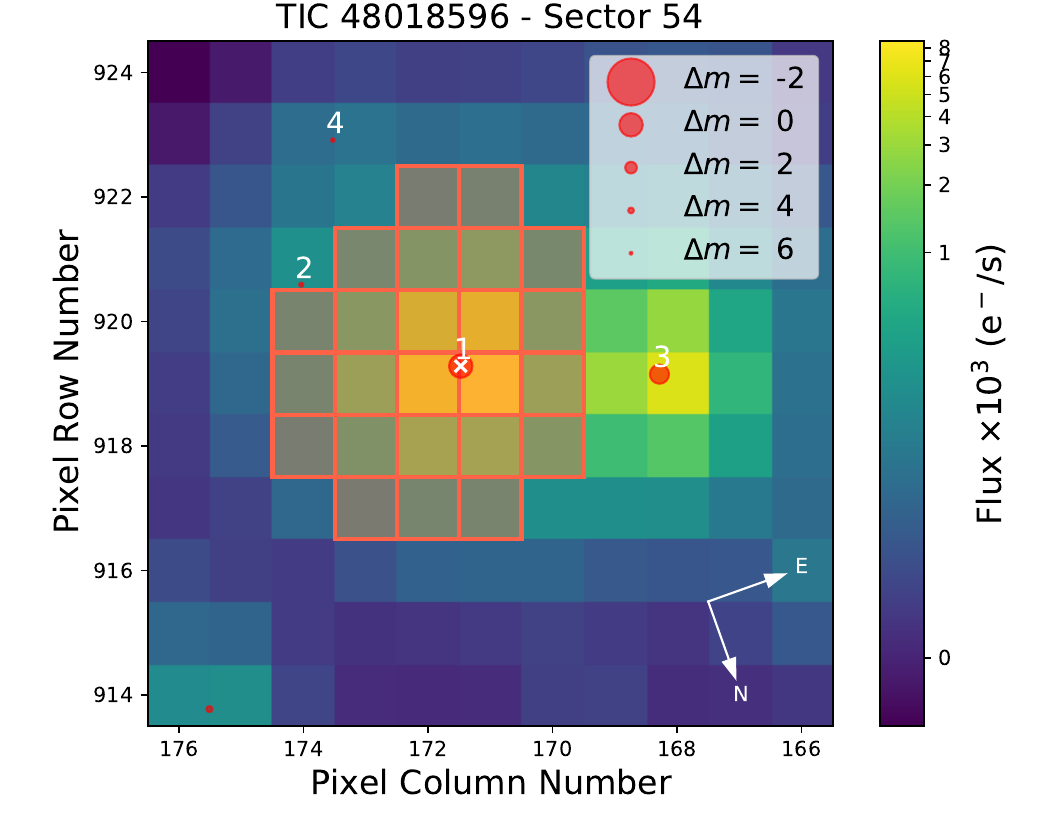}\\
\includegraphics[width=0.24\columnwidth]{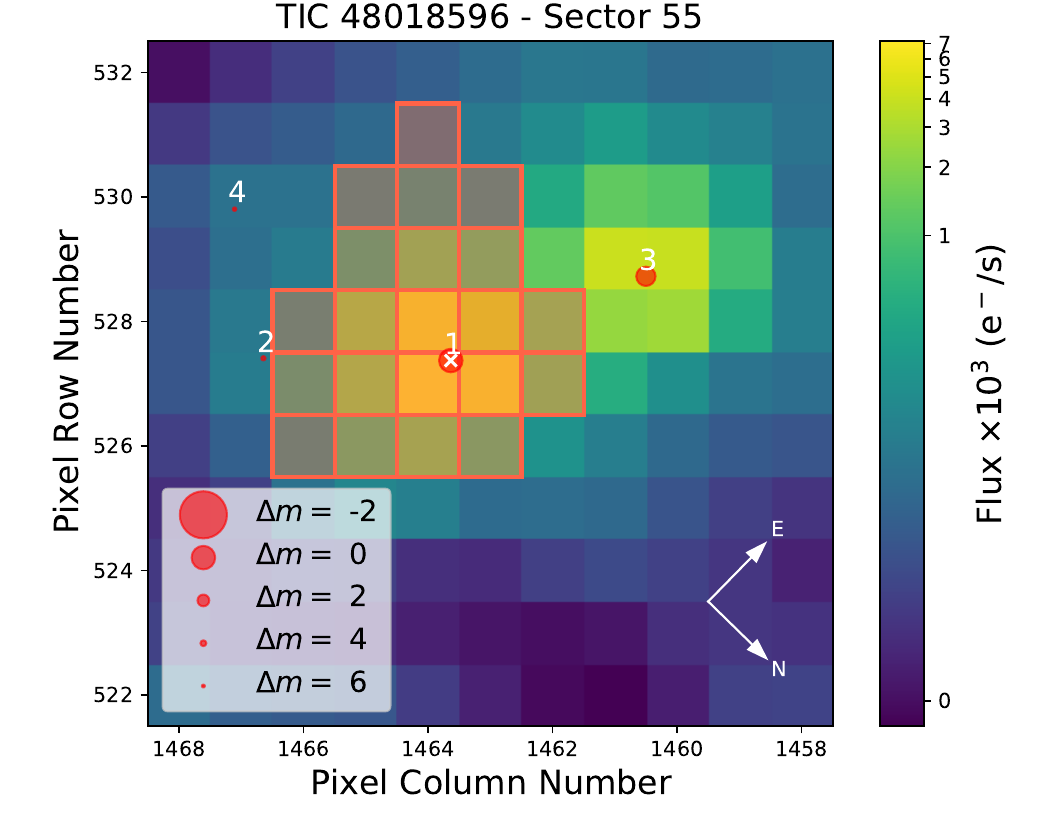}
\includegraphics[width=0.24\columnwidth]{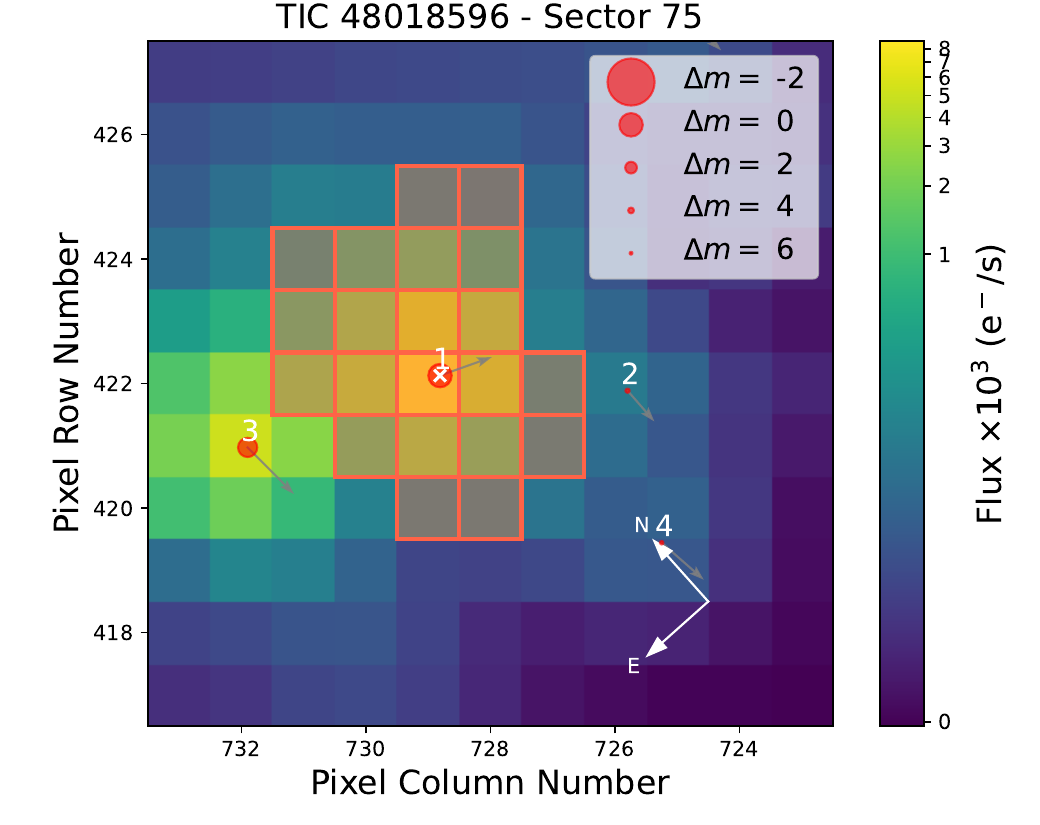}
\includegraphics[width=0.24\columnwidth]{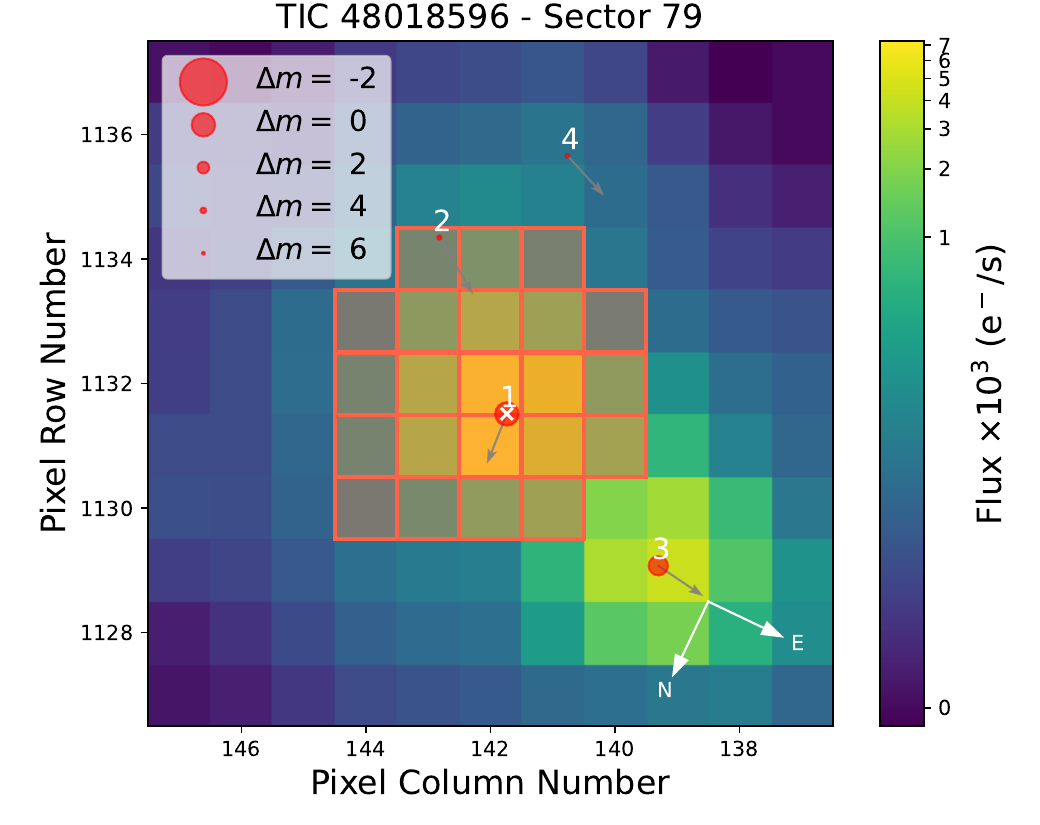}
\caption{The TPFs for TOI-2295. Same as the caption of Fig. \ref{tpfplotter_1836} }
\label{tpfplotter_2295}
\end{figure*}

\begin{figure*}
\centering
\includegraphics[width=0.24\columnwidth]{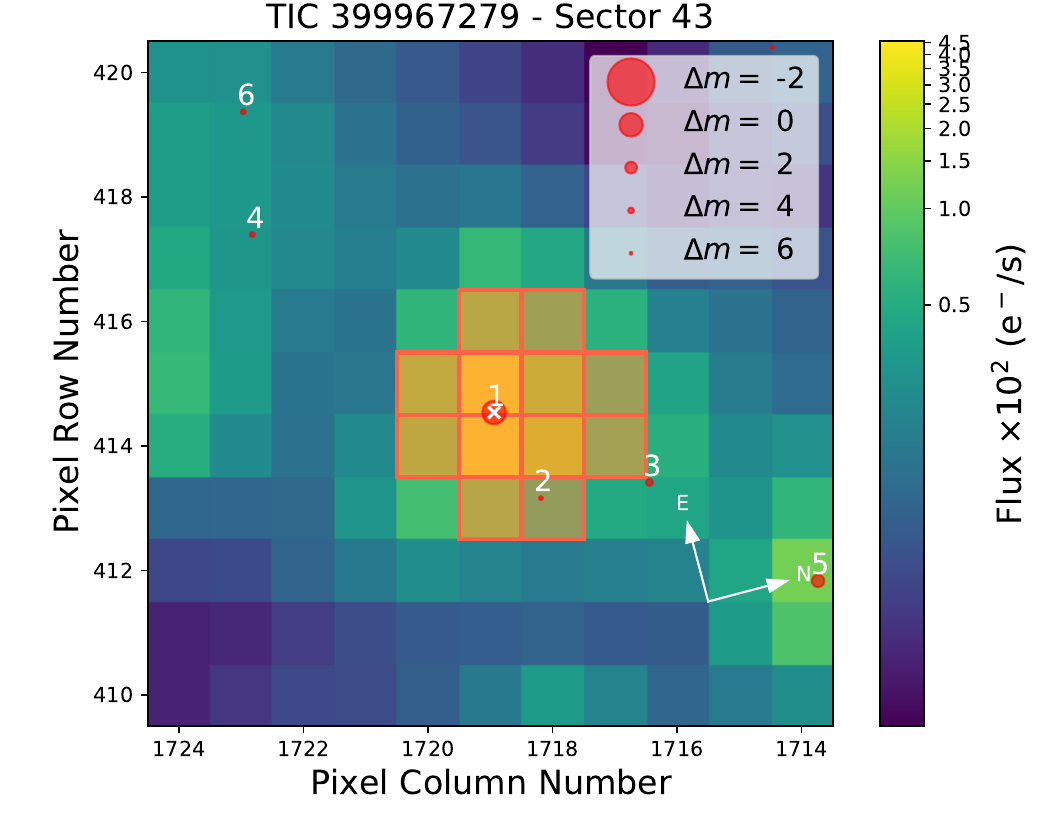}
\includegraphics[width=0.24\columnwidth]{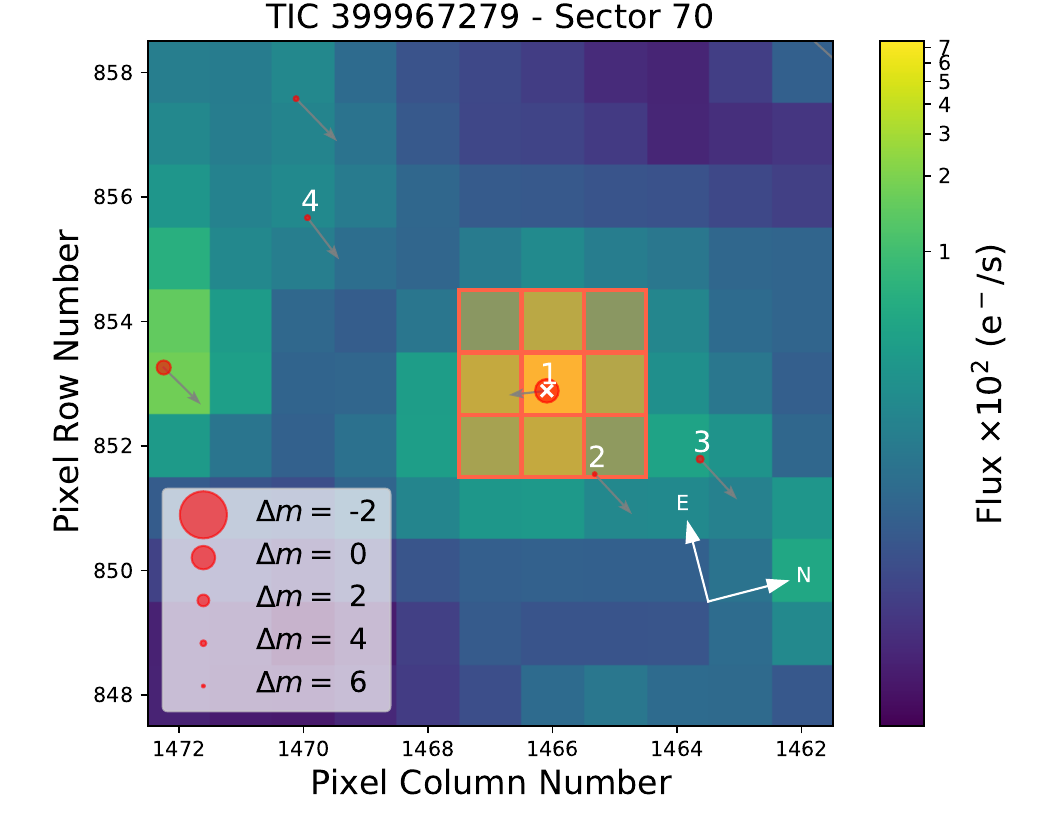}
\caption{The TPFs for TOI-2537. Same as the caption of Fig. \ref{tpfplotter_1836} }
\label{tpfplotter_2537}
\end{figure*}

\begin{figure*}
\centering
\includegraphics[width=0.24\columnwidth]{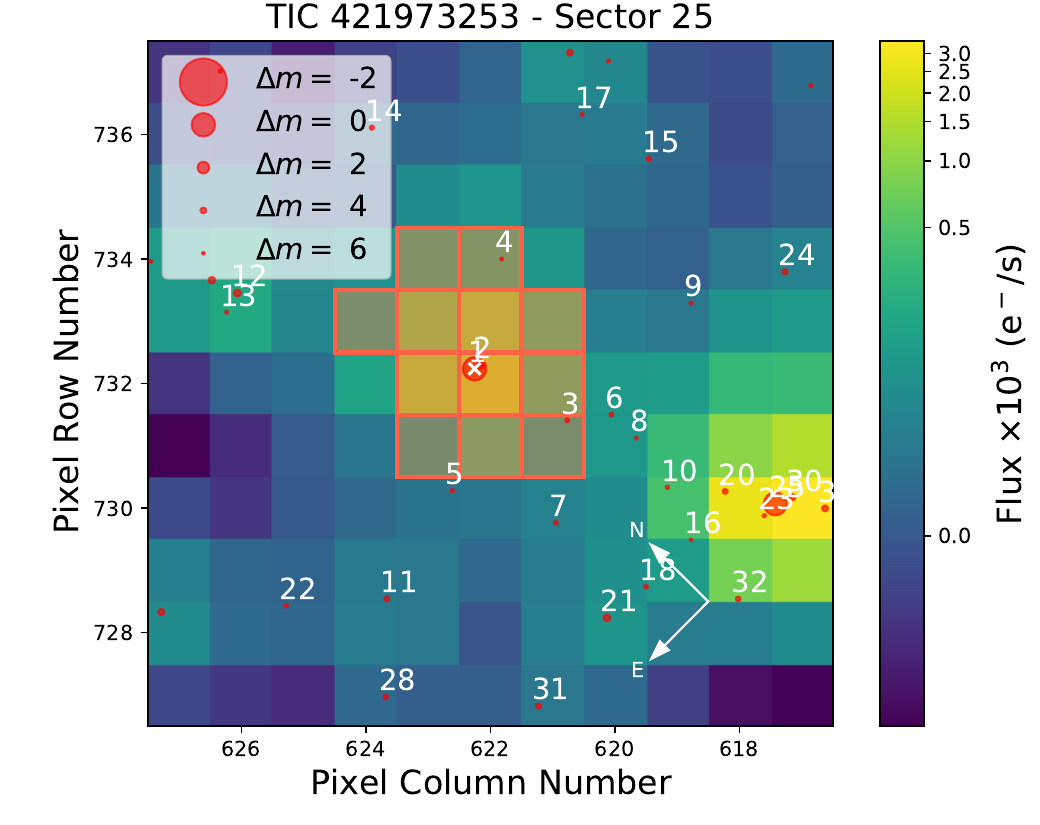}
\includegraphics[width=0.24\columnwidth]{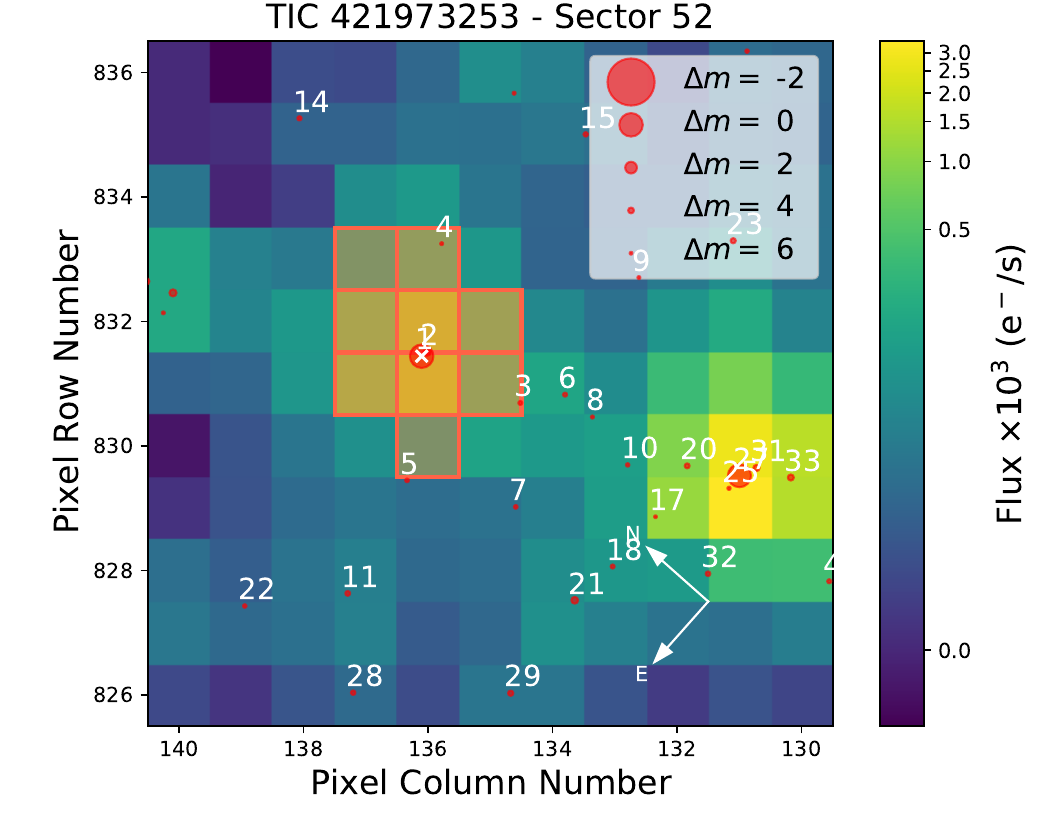}
\includegraphics[width=0.24\columnwidth]{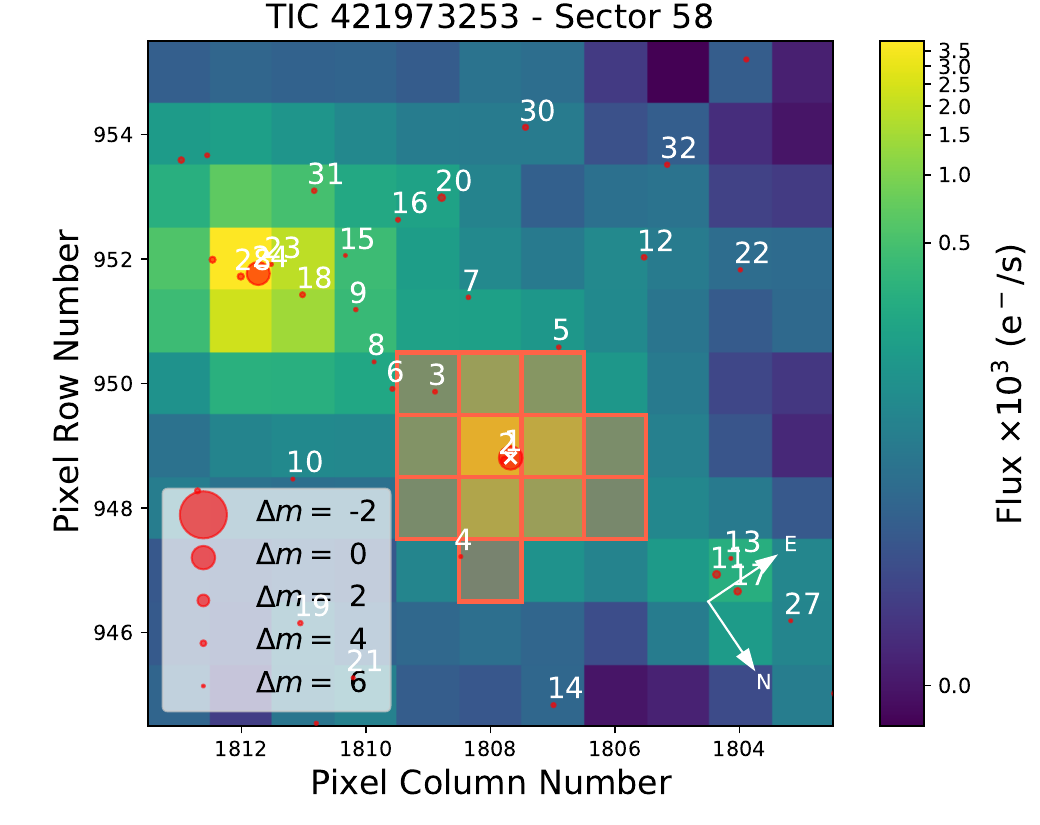}
\includegraphics[width=0.24\columnwidth]{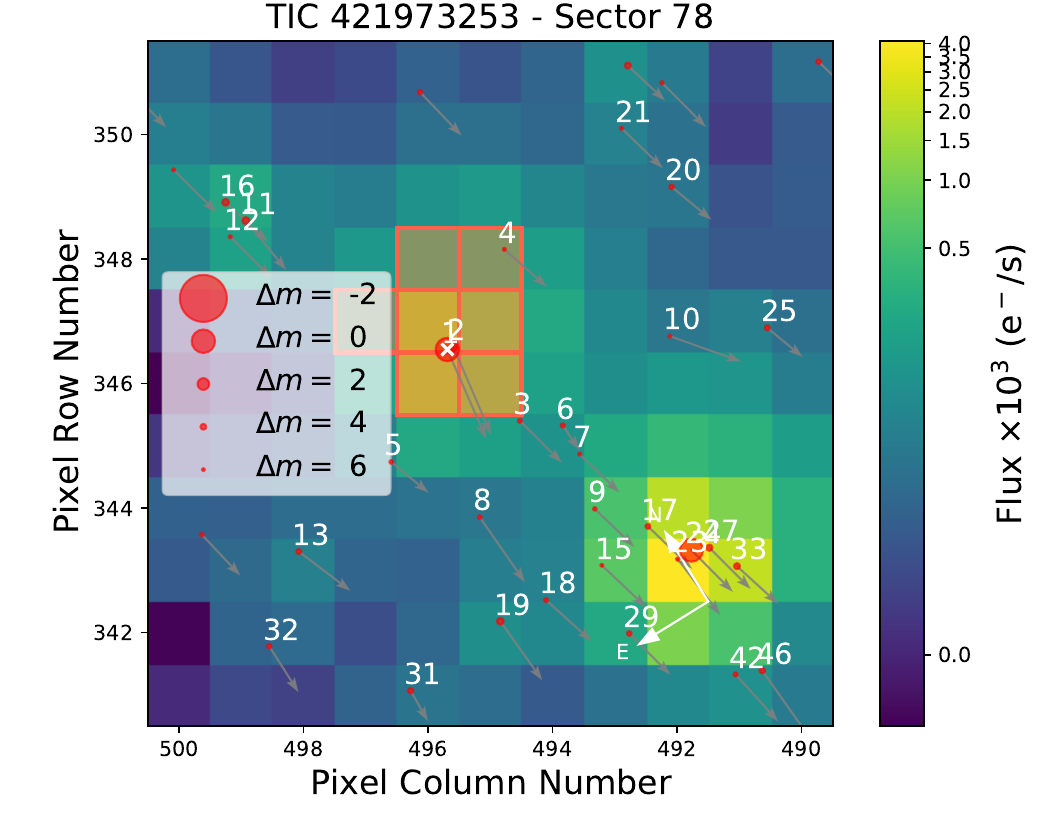}
\caption{The TPFs for TOI-4081. See the caption of Fig. \ref{tpfplotter_1836} for more explanation. }
\label{tpfplotter_4081}
\end{figure*}

\begin{figure*}
\centering
\includegraphics[width=0.24\columnwidth]{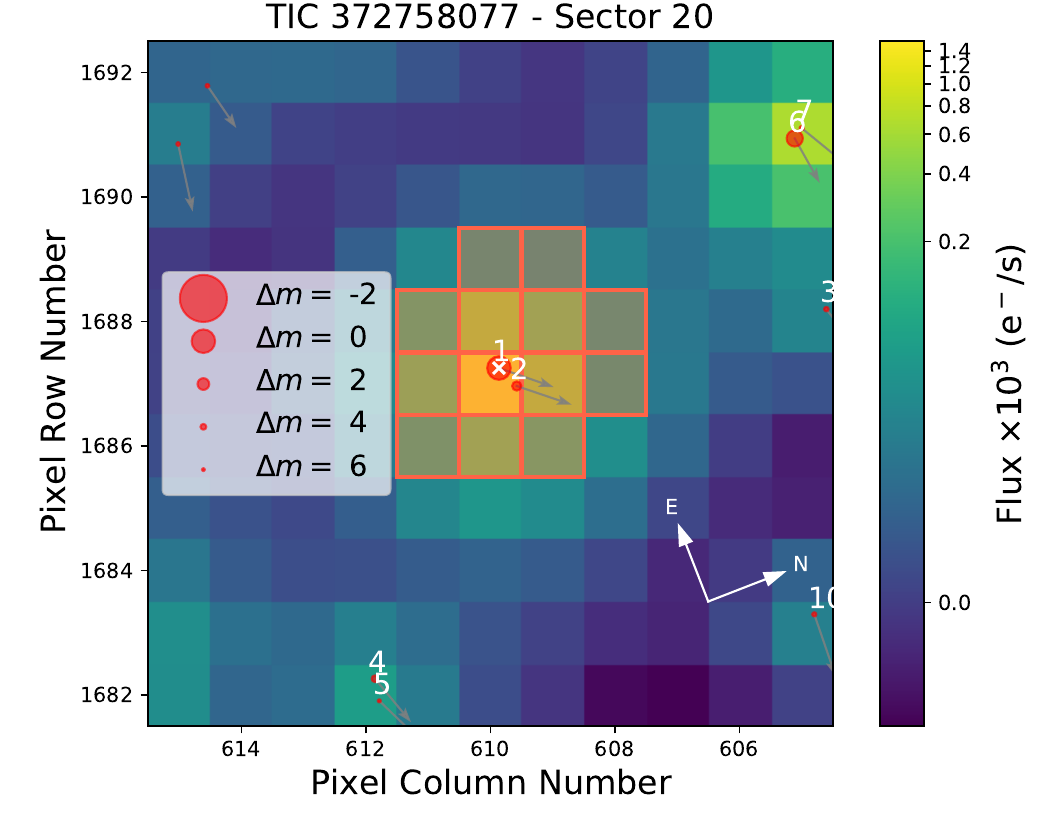}
\includegraphics[width=0.24\columnwidth]{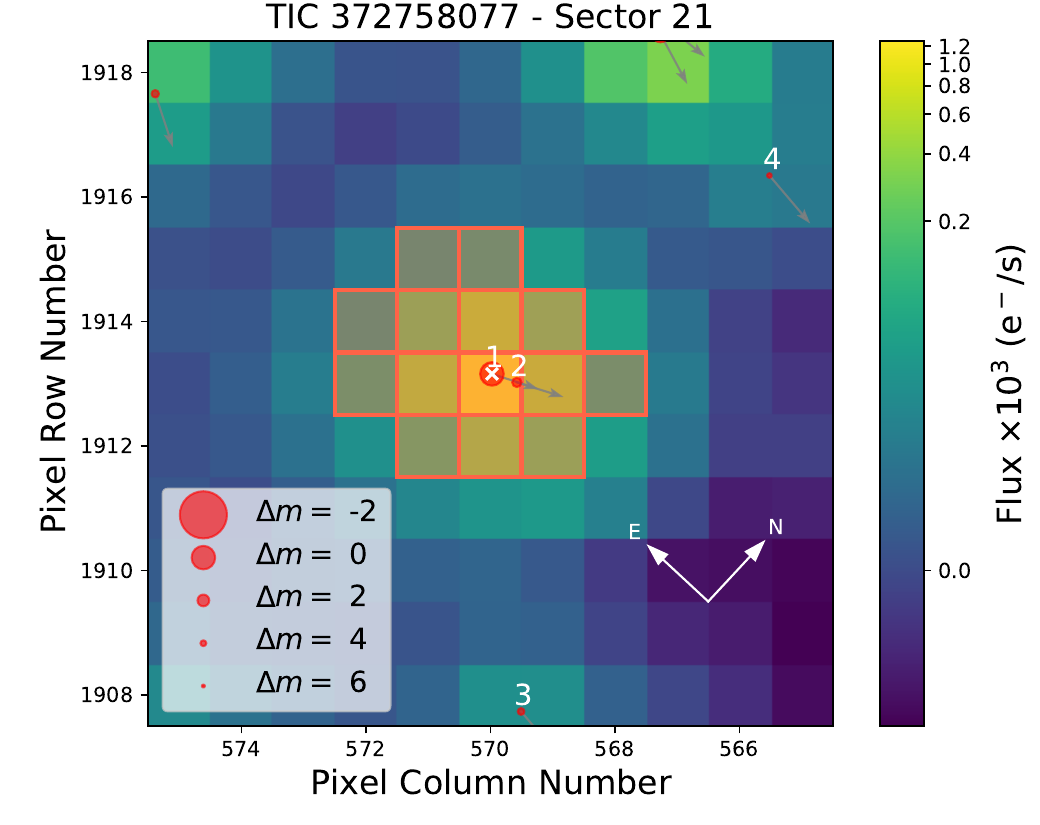}
\includegraphics[width=0.24\columnwidth]{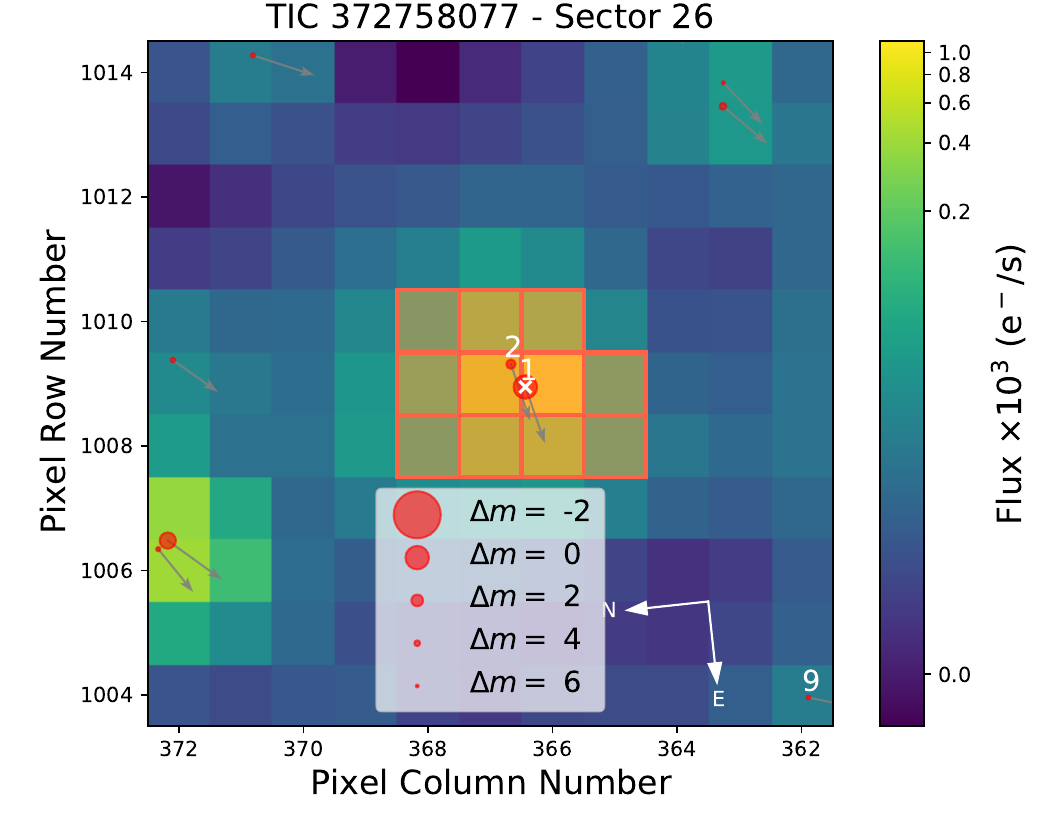}
\includegraphics[width=0.24\columnwidth]{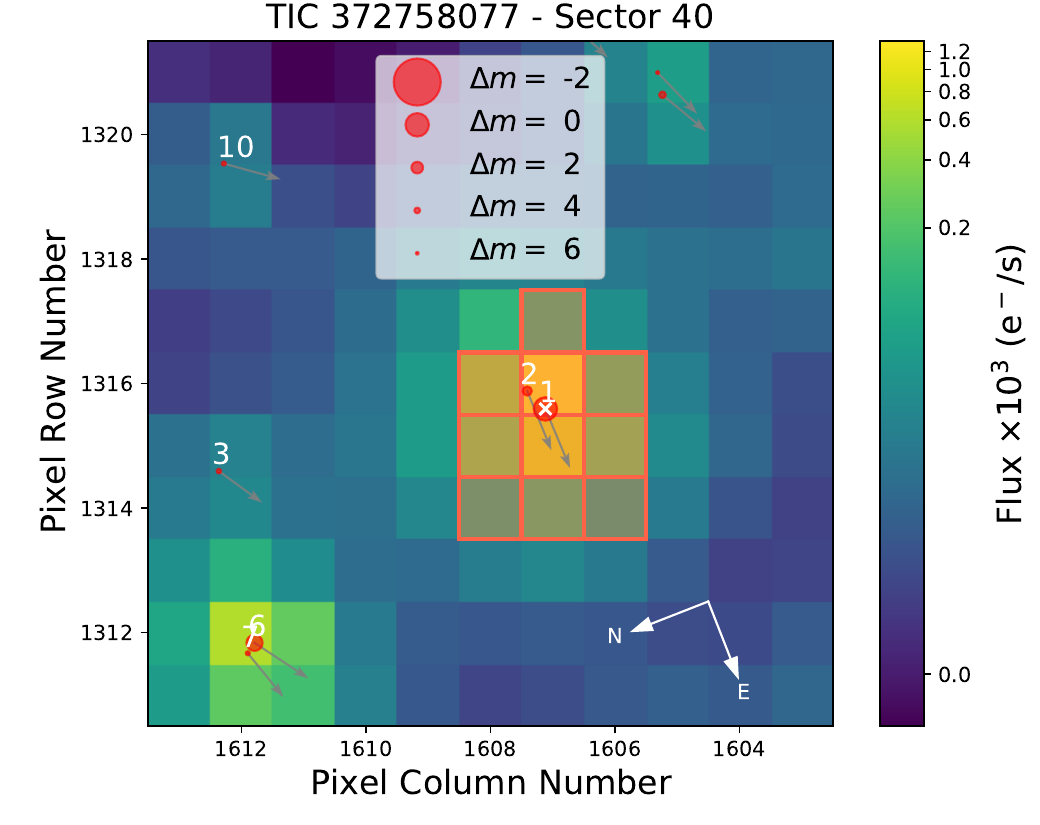}\\
\includegraphics[width=0.24\columnwidth]{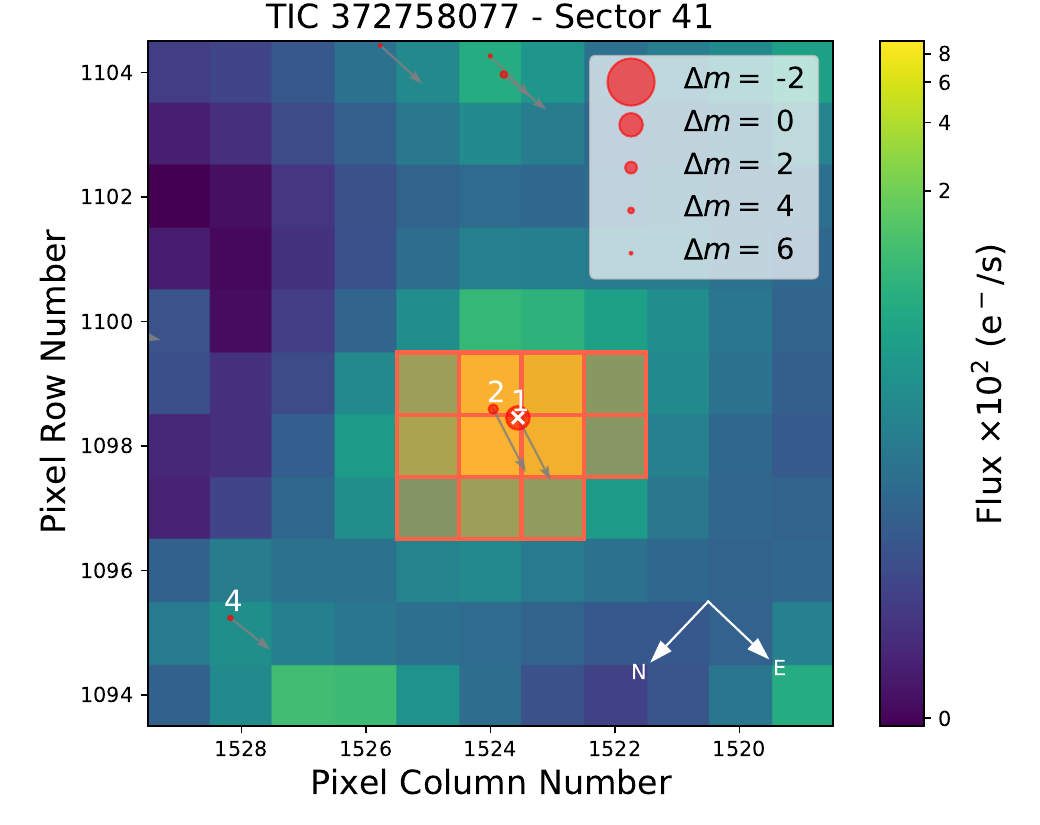}
\includegraphics[width=0.24\columnwidth]{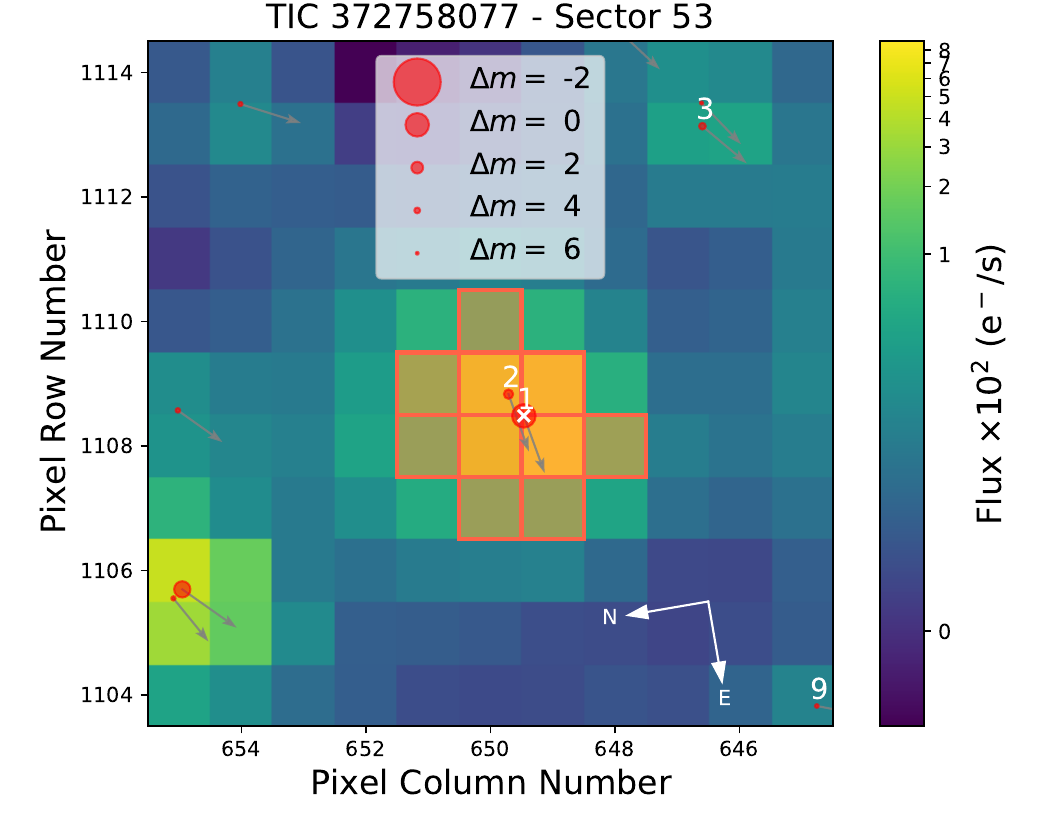}
\includegraphics[width=0.24\columnwidth]{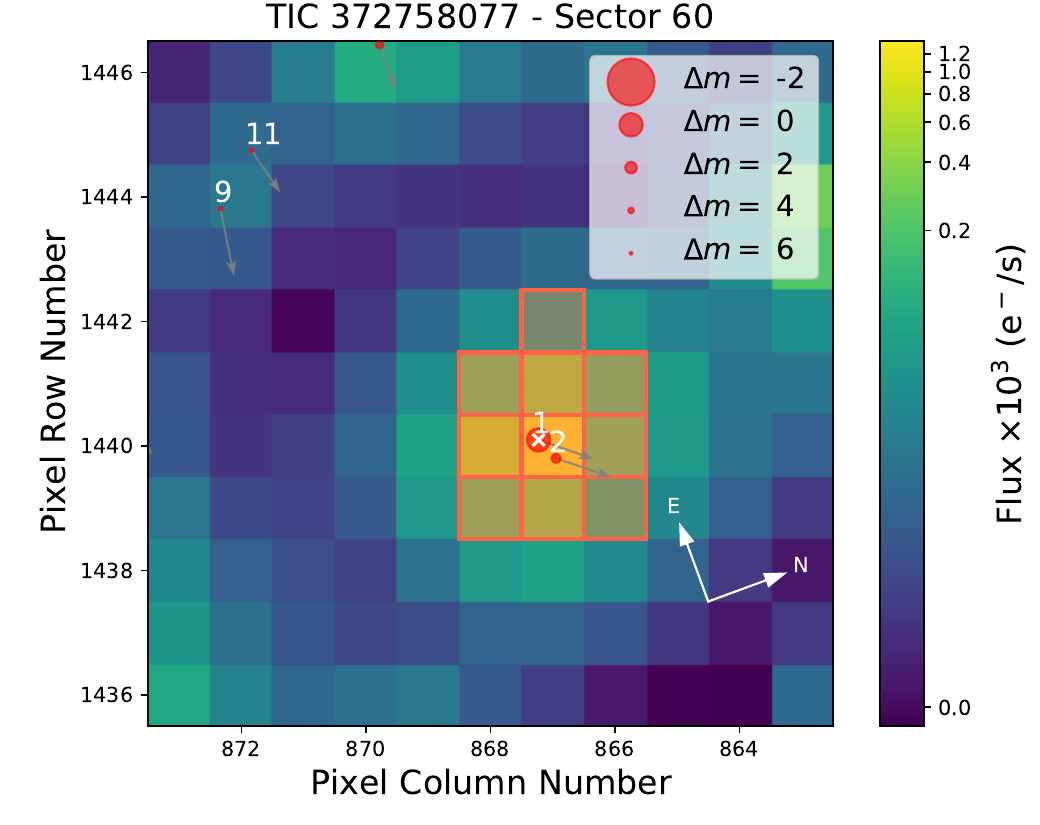}
\includegraphics[width=0.24\columnwidth]{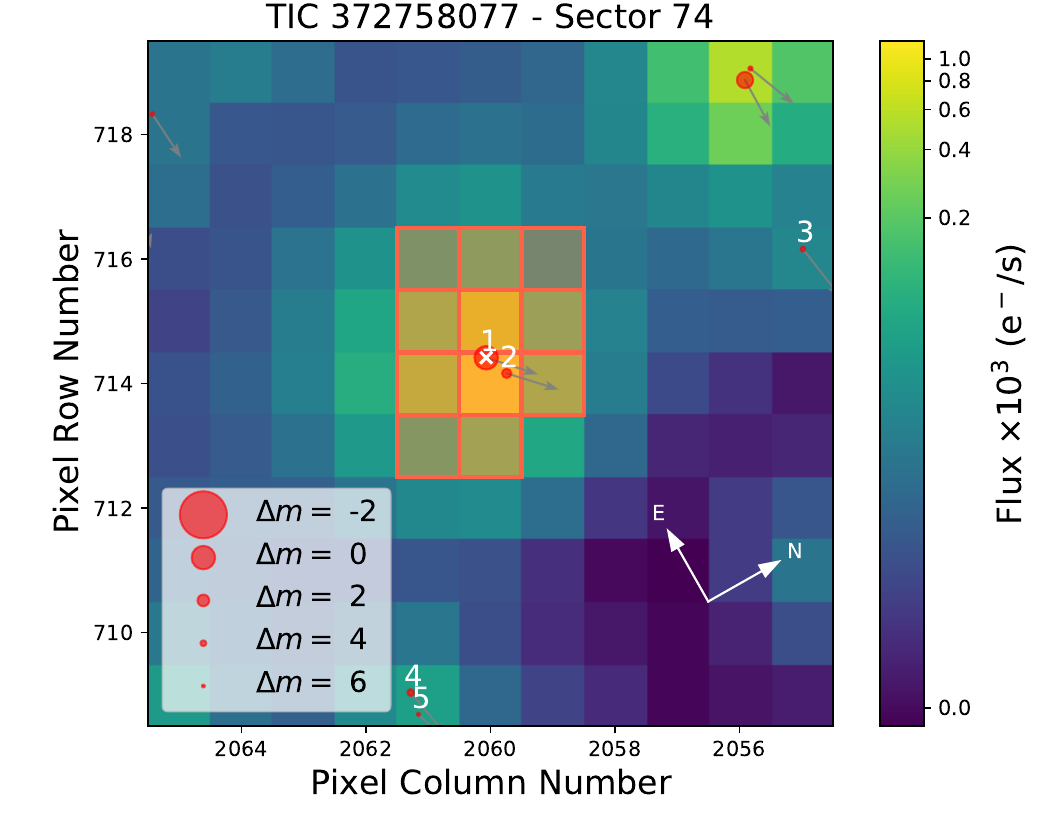}
\caption{The TPFs for TOI-4168. See the caption of Fig. \ref{tpfplotter_1836} for more explanation. }
\label{tpfplotter_4168}
\end{figure*}

\begin{figure*}
\centering
\includegraphics[width=0.24\columnwidth]{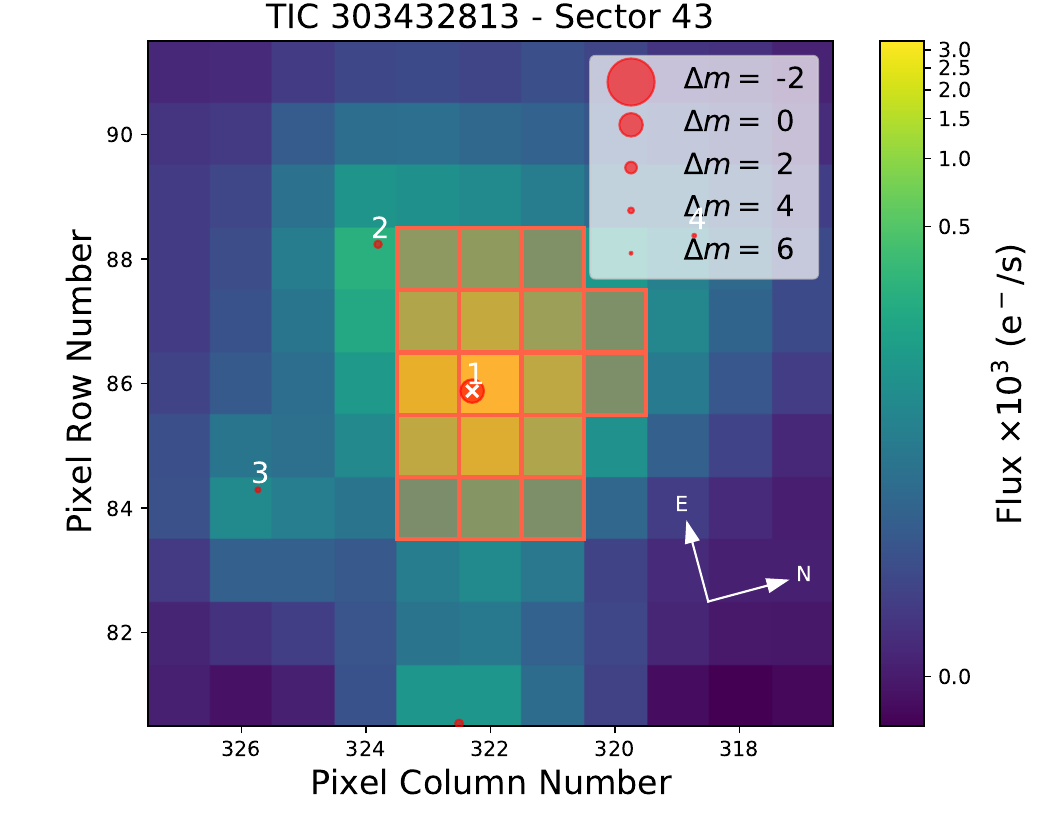}
\includegraphics[width=0.24\columnwidth]{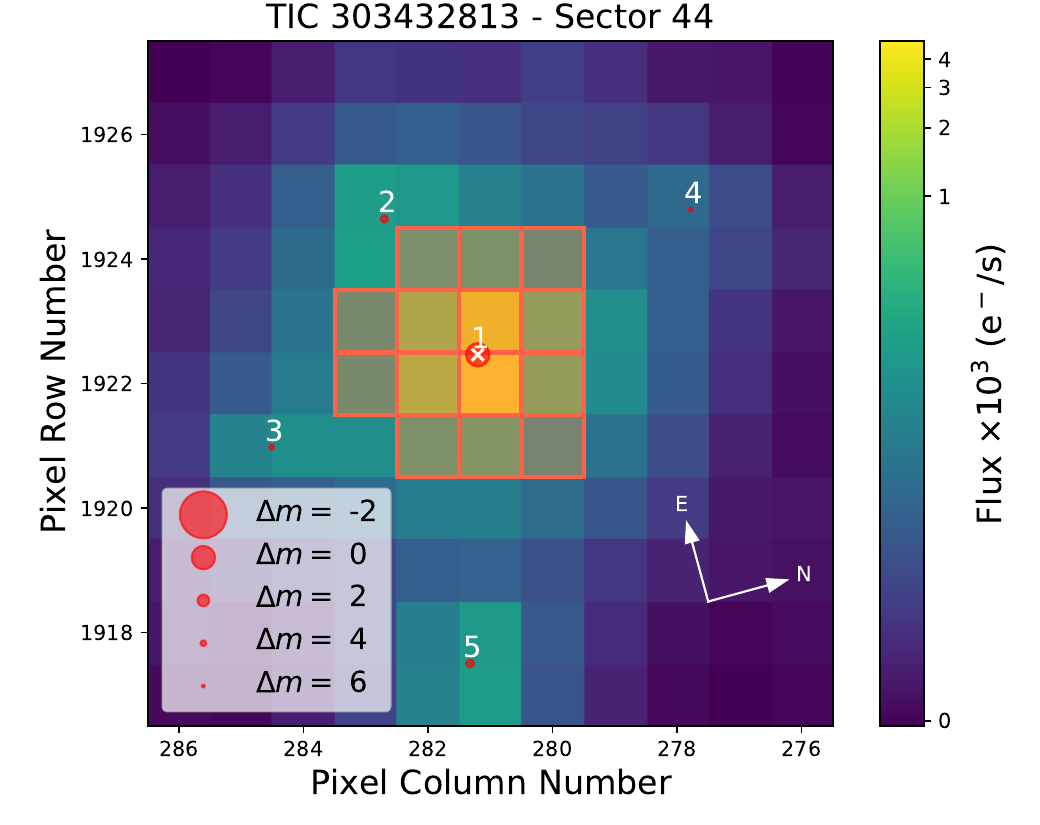}
\includegraphics[width=0.24\columnwidth]{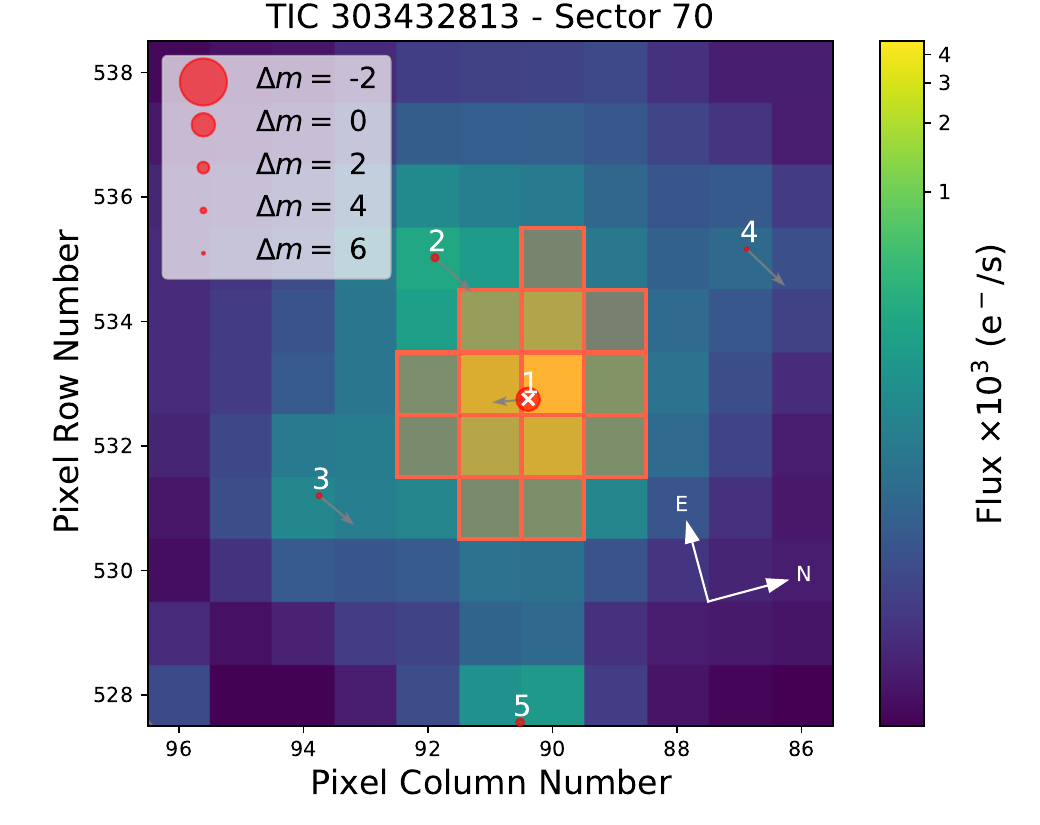}
\includegraphics[width=0.24\columnwidth]{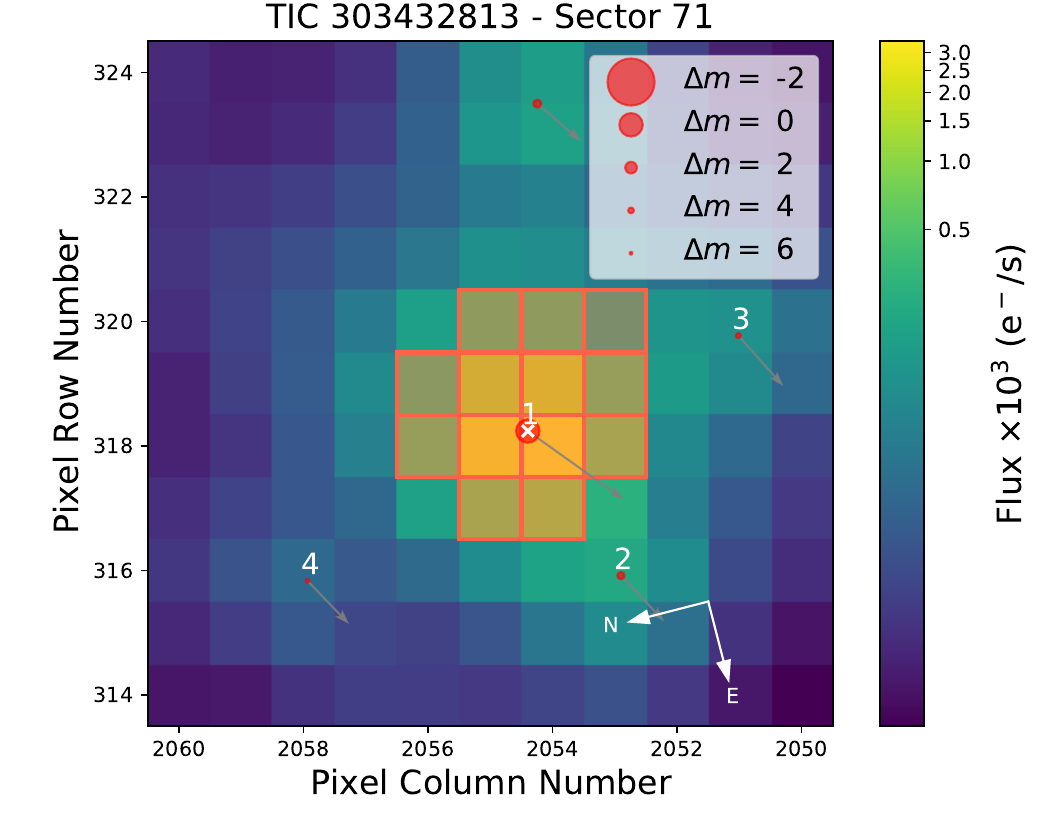}
\caption{The TPFs for TOI-5076. See the caption of Fig. \ref{tpfplotter_1836} for more explanation. }
\label{tpfplotter_5076}
\end{figure*}

\begin{figure*}[H]
\centering
\includegraphics[width=0.24\columnwidth]{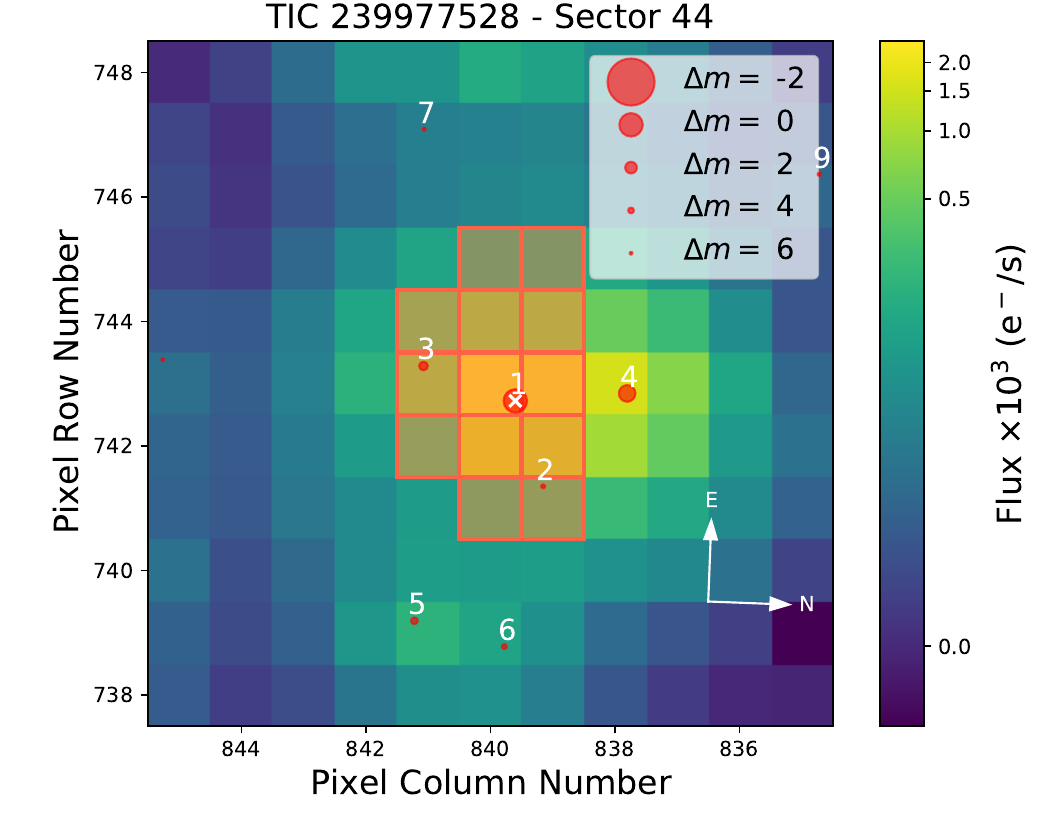}
\includegraphics[width=0.24\columnwidth]{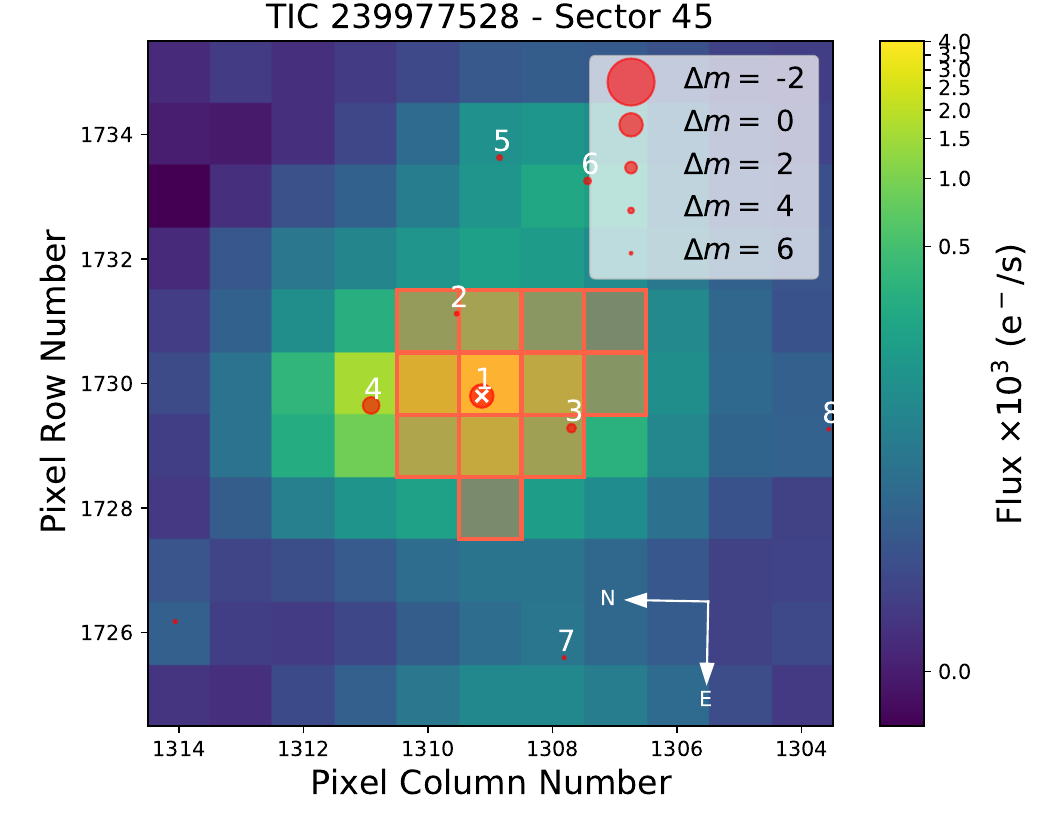}
\includegraphics[width=0.24\columnwidth]{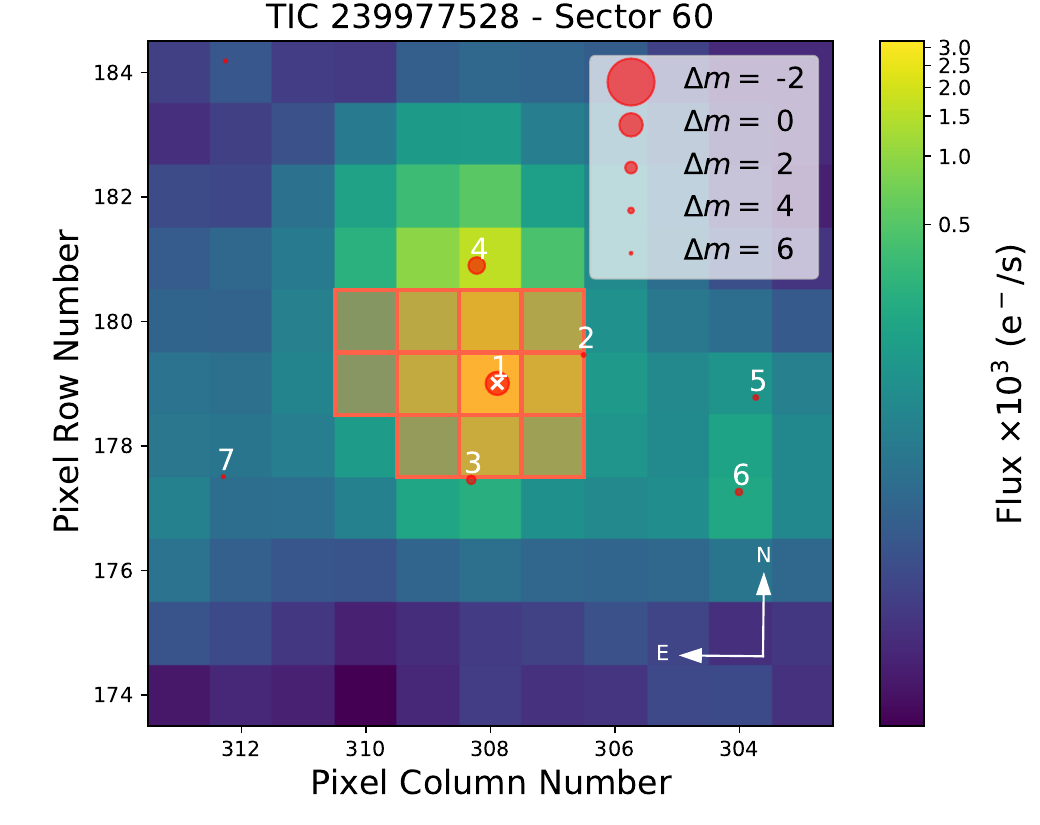}
\includegraphics[width=0.24\columnwidth]{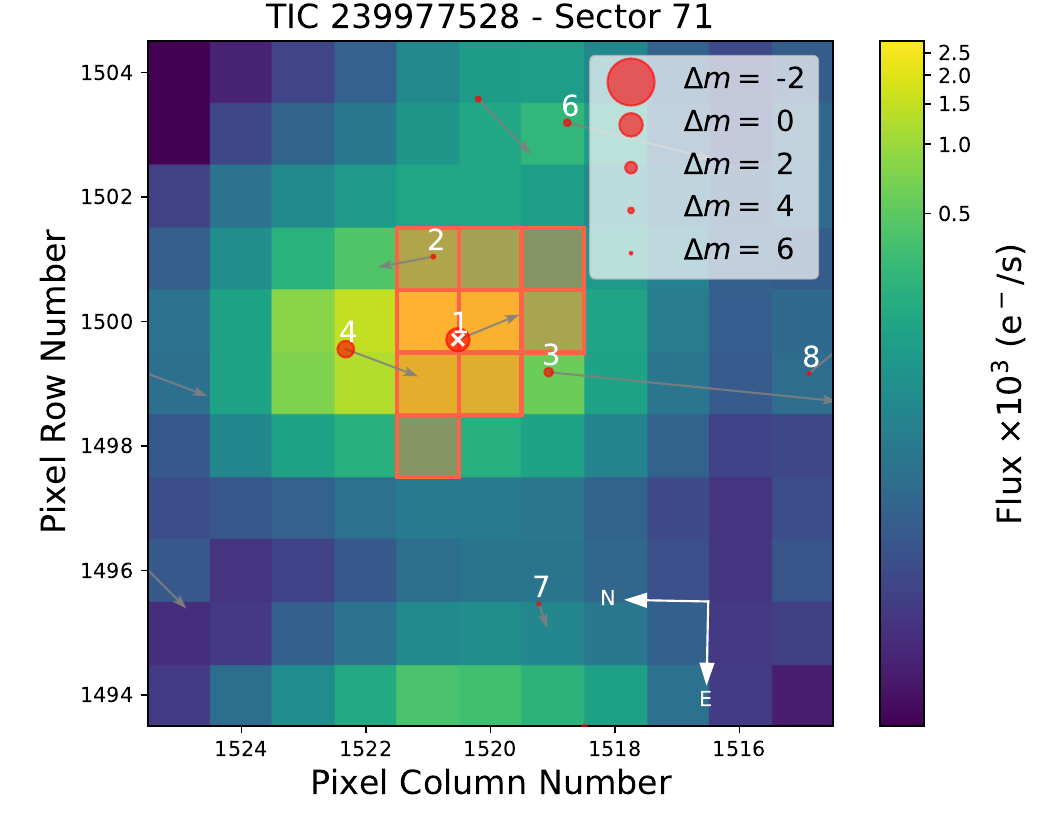}
\caption{The TPFs for TOI-5110. Same as the caption of Fig. \ref{tpfplotter_1836} }
\label{tpfplotter_5110}
\end{figure*}

\twocolumn
\section{RVs}

In this section, we present the RVs used in this study. We note that the bisector error bars are considered to be twice the RV uncertainties.

\begin{table}[h!]
\caption{SOPHIE RVs for TOI-2295.}
\label{tab:rv_2295}
\resizebox{0.45\textwidth}{!}{%
\begin{tabular}{llll}
\hline
BJD (-2400000 d)&       RV (km s$^{-1}$ )  & $\sigma_{RV}$ (m s$^{-1}$ )& BIS (km s$^{-1}$)  \\
\hline
59130.38267 & -38.5705 & 0.0037 & -0.0298 \\
59172.24841 & -38.6415 & 0.0027 & -0.022  \\
59197.21288 & -38.6371 & 0.0027 & -0.0209 \\
59271.70104 & -38.7091 & 0.0028 & -0.0281 \\
59304.65169 & -38.7015 & 0.0028 & -0.0335 \\
59393.49349 & -38.7834 & 0.0027 & -0.0137 \\
59409.4328  & -38.8123 & 0.0055 & -0.0063 \\
59409.5815  & -38.8061 & 0.0028 & -0.0169 \\
59421.4917  & -38.8255 & 0.0028 & -0.0124 \\
59470.47645 & -38.8487 & 0.0027 & -0.0343 \\
59486.33414 & -38.759  & 0.0028 & -0.0219 \\
59523.267   & -38.7985 & 0.0028 & -0.0162 \\
59528.32153 & -38.8382 & 0.0032 & -0.0188 \\
59542.24834 & -38.8421 & 0.0027 & -0.0268 \\
59552.22664 & -38.7858 & 0.0028 & -0.0062 \\
59631.69274 & -38.8384 & 0.0028 & -0.0327 \\
59662.65615 & -38.8271 & 0.0028 & -0.0259 \\
59713.50327 & -38.8301 & 0.0028 & -0.0053 \\
59750.42851 & -38.8105 & 0.0028 & -0.0338 \\
59769.5199  & -38.793  & 0.0028 & -0.0272 \\
59786.60066 & -38.7046 & 0.0031 & -0.0352 \\
59806.47617 & -38.7899 & 0.0028 & -0.0317 \\
59827.39876 & -38.7532 & 0.0028 & -0.0203 \\
59864.2636  & -38.7613 & 0.0028 & -0.0232 \\
59879.25186 & -38.6489 & 0.0028 & -0.0194 \\
59889.30119 & -38.7305 & 0.0028 & -0.0362 \\
59919.21796 & -38.7051 & 0.0027 & -0.0331 \\
60007.68158 & -38.6548 & 0.0028 & -0.0221 \\
60015.68644 & -38.6794 & 0.0028 & -0.02   \\
60042.56447 & -38.653  & 0.0028 & -0.0281 \\
60079.53749 & -38.6541 & 0.0028 & -0.023  \\
60079.60783 & -38.6499 & 0.0028 & -0.0108 \\
60096.58176 & -38.6039 & 0.0028 & -0.0248 \\
60106.50179 & -38.6536 & 0.0028 & -0.0298 \\
60121.45408 & -38.5614 & 0.0028 & -0.0309 \\
60139.4561  & -38.6572 & 0.0028 & -0.0362 \\
60154.48318 & -38.6109 & 0.0028 & -0.0234 \\
60169.36539 & -38.6716 & 0.0028 & -0.0243 \\
60176.47245 & -38.592  & 0.0027 & -0.0189 \\
60188.39096 & -38.6568 & 0.003  & -0.0183 \\
60213.37734 & -38.6314 & 0.0028 & -0.0193 \\
60221.30316 & -38.6984 & 0.0028 & -0.0105 \\
60231.31922 & -38.7089 & 0.0028 & -0.0242 \\
60245.29917 & -38.6847 & 0.0028 & -0.0367\\
\hline
\end{tabular}%
}
\end{table}

\begin{table}
\caption{SOPHIE RVs for TOI-2537.}
\label{tab:rv_2537sophie}
\resizebox{0.46\textwidth}{!}{%
\begin{tabular}{llll}
\hline
BJD (-2400000 d)&       RV (km s$^{-1}$ )  & $\sigma_{RV}$ (m s$^{-1}$ )& BIS (km s$^{-1}$)  \\
\hline
58730.59024 & 61.4184 & 0.0187 & -0.0798  \\
58797.43469 & 61.5845 & 0.0178 & -0.0342  \\
58813.35391 & 61.5442 & 0.0146 & -0.118   \\
58841.47453 & 61.443  & 0.021  & -0.0448  \\
58857.44319 & 61.4338 & 0.0222 & -0.0145  \\
58916.28201&61.3926&0.0431&-0.0637\\
59109.65131 & 61.4391 & 0.0141 & -0.0885  \\
59128.63863 & 61.4757 & 0.0216 & -0.0743  \\
59180.52546 & 61.5918 & 0.0171 & -0.0898  \\
59196.35305 & 61.4889 & 0.0136 & -0.0813  \\
59202.49525 & 61.4766 & 0.0148 & -0.1439  \\
59248.35429 & 61.5825 & 0.0162 & -0.1216  \\
59279.33214 & 61.5978 & 0.0342 & -0.1625  \\
59458.63502 & 61.77   & 0.0112 & -0.062   \\
59470.62766 & 61.7294 & 0.0088 & -0.1049  \\
59499.48431 & 61.6828 & 0.0123 & -0.0875  \\
59506.45463 & 61.659  & 0.0171 & -0.0747  \\
59522.56535 & 61.7506 & 0.0347 & -0.0645  \\
59561.3125  & 61.8472 & 0.013  & -0.0921  \\
59568.42547 & 61.7984 & 0.0152 & -0.0793  \\
59587.30284 & 61.66   & 0.0228 & -0.0921  \\
59604.24429 & 61.7569 & 0.0143 & -0.1003  \\
59608.25057 & 61.7651 & 0.0158 & -0.0565  \\
59622.31303 & 61.7208 & 0.0367 & -0.0821  \\
59626.275   & 61.7504 & 0.0373 & -0.1598  \\
59646.30602 & 61.8343 & 0.0247 & -0.0876  \\
59827.59236 & 61.761  & 0.0176 & -0.0799  \\
59838.56637 & 61.7949 & 0.0226 & -0.0756  \\
59881.52128 & 61.6733 & 0.0181 & -0.029   \\
59889.47907 & 61.6681 & 0.0149 & -0.0851  \\
59930.46325 & 61.8041 & 0.0159 & -0.0795  \\
59937.41218 & 61.7479 & 0.0185 & -0.0686  \\
59973.33085 & 61.5884 & 0.0183 & -0.1874  \\
59980.32242 & 61.6205 & 0.0471 & -0.2007  \\
60008.2872  & 61.6837 & 0.0178 & -0.0659  \\
60179.64656 & 61.5579 & 0.0131 & -0.0451  \\
60186.57343 & 61.5378 & 0.0331 & -0.0741  \\
60212.5682  & 61.6614 & 0.0127 & -0.105   \\
60221.57676 & 61.6459 & 0.0212 & -0.0497  \\
60231.58438 & 61.5558 & 0.0138 & -0.0513 \\
60256.54147& 61.4925& 0.0142& -0.1107\\
60282.38568& 61.5826& 0.0114& -0.0990\\
60294.44127& 61.5981& 0.0151& -0.1061\\
60300.42147& 61.6179& 0.0175& -0.0351\\
60339.37695& 61.4836& 0.0145& -0.1120\\
60355.29273& 61.5183& 0.0134& -0.1133\\
\hline
\end{tabular}%
}
\end{table}

\begin{table}
\caption{HARPS RVs for TOI-2537.}
\label{tab:rv_399}
\resizebox{0.46\textwidth}{!}{%
\begin{tabular}{llll}
\hline
BJD (-2400000 d)&       RV (km s$^{-1}$ )  & $\sigma_{RV}$ (m s$^{-1}$ )& BIS (km s$^{-1}$)  \\
\hline
59225.5762851 & 61559.300 & 8.900  & 18.000  \\
59229.5574822 & 61555.700 & 12.700 & -7.000  \\
59232.5606435 & 61568.100 & 12.700 & 29.000  \\
59244.5885841 & 61609.300 & 10.600 & 3.000   \\
59249.5470031 & 61596.000 & 14.200 & -15.000 \\
59252.5321889 & 61617.200 & 10.500 & 1.000   \\
59257.5406962 & 61656.200 & 12.700 & -19.000 \\
59260.5366339 & 61647.600 & 15.900 & -22.000 \\
59251.5567625 & 61602.100 & 11.500 & 6.000   \\
59491.7923417 & 61675.400 & 23.800 & 43.000  \\
59504.6613020 & 61707.400 & 17.800 & -20.000 \\
59534.7121374 & 61840.300 & 11.500 & -18.000 \\
59561.7125116 & 61899.500 & 23.700 & -39.000 \\
59597.6106865 & 61758.300 & 20.400 & 31.000  \\
59607.5700245 & 61780.200 & 12.700 & -18.000 \\
59625.5622617 & 61820.400 & 17.800 & 0.000   \\
59859.7463690 & 61692.900 & 12.700 & 52.000  \\
59868.7984516 & 61712.800 & 9.600  & -12.000 \\
59933.6462428 & 61819.600 & 15.500 & 0.000   \\
60224.8332078 & 61634.900 & 24.000 & 31.000  \\
60255.6262724 & 61567.900 & 14.200 & 3.000 \\
\hline
\end{tabular}%
}
\end{table}

\begin{table}
\caption{FEROS RVs for TOI-2537.}
\label{tab:rv_399FEROS}
\resizebox{0.46\textwidth}{!}{%
\begin{tabular}{llll}
\hline
BJD (-2400000 d)&       RV (km s$^{-1}$ )  & $\sigma_{RV}$ (m s$^{-1}$ )& BIS (km s$^{-1}$)  \\
\hline
59187.6430111 & 61637.700 & 10.400 & -29.000  \\
59191.6669117 & 61585.300 & 10.000 & -15.000  \\
59194.6038561 & 61527.400 & 10.200 & -52.000  \\
59196.5715686 & 61512.200 & 9.300  & -19.000  \\
59206.6072041 & 61513.600 & 11.100 & -30.000  \\
59212.5850225 & 61528.700 & 10.600 & 13.000   \\
59216.6366852 & 61490.000 & 12.300 & -86.000  \\
59219.6350660 & 61525.800 & 10.600 & -3.000   \\
59222.5836743 & 61565.700 & 10.300 & 39.000   \\
59223.6228979 & 61523.600 & 11.400 & 12.000   \\
59485.7584124 & 61843.100 & 13.900 & 43.000   \\
59493.8380299 & 61575.200 & 13.900 & -89.000  \\
59501.8092966 & 61646.700 & 9.600  & -73.000  \\
59505.8004537 & 61772.500 & 9.500  & -16.000  \\
59514.8221421 & 61729.200 & 9.500  & -29.000  \\
59540.6806318 & 61722.500 & 11.100 & -86.000  \\
59546.5911645 & 61791.800 & 16.600 & -141.000 \\
59553.6105176 & 61857.500 & 10.200 & -37.000  \\
59656.4996819 & 61785.500 & 19.500 & -142.000\\
\hline
\end{tabular}
}
\end{table}

\begin{table}
\caption{SOPHIE RVs for TOI-4168.}
\label{tab:rv_4168}
\resizebox{0.46\textwidth}{!}{%
\begin{tabular}{llll}
\hline
BJD (-2400000 d)&       RV (km s$^{-1}$ )  & $\sigma_{RV}$ (m s$^{-1}$ )& BIS (km s$^{-1}$)  \\
\hline
59562.66417 & -48.9584 & 0.0061 & 0.0114  \\
59563.68571 & -51.3137 & 0.0063 & 0.0153  \\
59565.69649 & -55.0058 & 0.0068 & -0.0093 \\
59568.69687 & -50.1702 & 0.0061 & -0.0467 \\
59586.64028 & -35.3540 & 0.0138 & -0.0056 \\
59606.54786 & -6.8079  & 0.0068 & 0.0022  \\
59630.57207 & -6.7464  & 0.0130 & -0.0174 \\
59738.37593 & -47.5730 & 0.0107 & 0.0628  \\
59748.38983 & -3.6078  & 0.0062 & -0.0549 \\
59990.62535 & -14.0461 & 0.0062 & -0.0023 \\
60043.47585 & 2.5406   & 0.0061 & 0.0084  \\
60128.39420 & -33.2518 & 0.0123 & -0.0150 \\
60296.71684 & -47.8128 & 0.0058 & 0.0109  \\
60337.62165 & 2.5782   & 0.0066 & -0.1229 \\
60355.58164 & -48.0398 & 0.0168 & -0.0743\\
\hline
\end{tabular}%
}
\end{table}

\begin{table}
\caption{SOPHIE RVs for TOI-5110.}
\label{tab:rv_5110}
\resizebox{0.46\textwidth}{!}{%
\begin{tabular}{llll}
\hline
BJD (-2400000 d)&       RV (km s$^{-1}$ )  & $\sigma_{RV}$ (m s$^{-1}$ )& BIS (km s$^{-1}$ )  \\
\hline
59626.37719 & 3.8245 & 0.0067 & 0.0419  \\
59683.33134 & 4.2203 & 0.0063 & 0.0319  \\
59687.35537 & 3.8503 & 0.0085 & 0.0421  \\
59863.6258  & 4.1984 & 0.0067 & 0.0524  \\
59878.66335 & 3.9854 & 0.0062 & 0.0201  \\
59881.68486 & 4.0099 & 0.0062 & 0.0688  \\
59882.61474 & 4.0399 & 0.0094 & 0.0645  \\
59890.69219 & 4.1384 & 0.0062 & 0.0497  \\
59918.57374 & 4.1267 & 0.0062 & 0.0298  \\
59930.57979 & 3.8816 & 0.0064 & 0.0248  \\
59936.53797 & 3.9568 & 0.0064 & 0.0471  \\
59972.47498 & 4.0245 & 0.0063 & -0.0114 \\
59976.39413 & 4.0648 & 0.0063 & 0.0344  \\
59983.42387 & 4.1854 & 0.0061 & 0.0158  \\
59988.35204 & 3.8185 & 0.006  & 0.0424  \\
60015.37496 & 4.2246 & 0.0062 & 0.0236  \\
60042.30927 & 4.1527 & 0.0063 & 0.0441  \\
60211.65267 & 4.0221 & 0.0063 & 0.0388  \\
60212.60277 & 4.014  & 0.0062 & -0.0073 \\
60214.59422 & 4.0291 & 0.006  & 0.0223  \\
60220.63299 & 4.1135 & 0.0062 & 0.0398  \\
60229.66501 & 3.7998 & 0.0062 & 0.0454  \\
60230.66433 & 3.8409 & 0.0061 & 0.0267\\
\hline
\end{tabular}%
}
\end{table}

\begin{table}
\caption{SOPHIE RVs for TOI-5076.}
\label{tab:rv_5076}
\resizebox{0.46\textwidth}{!}{%
\begin{tabular}{llll}
\hline
BJD (-2400000 d)&       RV (km s$^{-1}$ )  & $\sigma_{RV}$ (m s$^{-1}$ )& BIS (km s$^{-1}$)  \\
\hline
59648.33296 & 70.3026 & 0.0052 & -0.0097 \\
59803.61773 & 70.2845 & 0.0043 & -0.0196 \\
59815.60996 & 70.27   & 0.0033 & 0.0107  \\
59816.63756 & 70.2781 & 0.0039 & -0.0033 \\
59824.51153 & 70.3158 & 0.0043 & -0.0242 \\
59827.62803 & 70.2833 & 0.0034 & -0.0081 \\
59841.57065 & 70.3057 & 0.0051 & -0.0114 \\
59860.49069 & 70.2959 & 0.0035 & 0.0023  \\
59863.50478 & 70.2772 & 0.0043 & -0.0041 \\
59865.60809 & 70.2935 & 0.0037 & 0.0     \\
59880.63334 & 70.266  & 0.004  & -0.006  \\
59881.50296 & 70.2852 & 0.0034 & -0.0033 \\
59889.50012 & 70.2898 & 0.0034 & 0.0163  \\
59936.47521 & 70.29   & 0.0035 & 0.0037  \\
59978.34257 & 70.2958 & 0.0035 & -0.0193 \\
60168.61473 & 70.298  & 0.0032 & 0.0034  \\
60176.63613 & 70.2932 & 0.0033 & 0.0093  \\
60194.65866 & 70.294  & 0.0034 & 0.0077  \\
60212.53251 & 70.2774 & 0.0033 & -0.0078 \\
60214.54653 & 70.2903 & 0.0033 & 0.011   \\
60220.52212 & 70.3057 & 0.0036 & 0.012   \\
60229.60612 & 70.2806 & 0.0034 & -0.0025 \\
60244.58895 & 70.2859 & 0.0039 & 0.0026  \\
60262.58747 & 70.289  & 0.0033 & 0.0078  \\
60281.49934 & 70.2955 & 0.0047 & 0.017   \\
60285.43004 & 70.2744 & 0.0033 & -0.0026 \\
60294.41885 & 70.3016 & 0.004  & -0.032  \\
60295.45234 & 70.2851 & 0.0035 & -0.0193 \\
60296.43702 & 70.2906 & 0.0034 & 0.0176  \\
60300.44314 & 70.312  & 0.0051 & 0.0032  \\
60301.45327 & 70.2813 & 0.0044 & -0.0362 \\
60323.43091 & 70.2907 & 0.0033 & 0.0208  \\
60334.42258 & 70.2994 & 0.0052 & -0.0177 \\
60335.37212 & 70.3042 & 0.004  & -0.0205 \\
60336.3065  & 70.3162 & 0.0049 & 0.0096  \\
60337.3786  & 70.3158 & 0.0034 & -0.013  \\
60338.35946 & 70.3014 & 0.0034 & 0.018   \\
60354.35796 & 70.2847 & 0.0034 & -0.0218 \\
60358.28907 & 70.2857 & 0.0034 & 0.0099 \\
\hline
\end{tabular}%
}
\end{table}

\begin{table}
\caption{SOPHIE RVs for TOI-1836.}
\label{tab:rv_18361}
\resizebox{0.46\textwidth}{!}{%
\begin{tabular}{llll}
\hline
BJD (-2400000 d)&       RV (km s$^{-1}$ )  & $\sigma_{RV}$ (m s$^{-1}$ )& BIS (km s$^{-1}$ )  \\
\hline
59060.42881 & -50.4057 & 0.0054 & 0.0386  \\
59112.30495 & -50.3635 & 0.0064 & 0.025   \\
59114.3336  & -50.3784 & 0.008  & 0.0089  \\
59118.31539 & -50.4093 & 0.0091 & -0.0402 \\
59119.31286 & -50.4118 & 0.0088 & 0.0323  \\
59120.29719 & -50.3768 & 0.0092 & 0.0168  \\
59131.32304 & -50.379  & 0.005  & 0.0243  \\
59140.33899 & -50.4211 & 0.0047 & 0.0213  \\
59141.28337 & -50.4213 & 0.005  & 0.0409  \\
59149.27809 & -50.3863 & 0.0081 & 0.023   \\
59168.22047 & -50.3796 & 0.0051 & -0.0079 \\
59170.22701 & -50.3674 & 0.0102 & 0.001   \\
59176.23431 & -50.364  & 0.0049 & 0.0017  \\
59245.72333 & -50.3673 & 0.0063 & 0.0454  \\
59247.69938 & -50.3786 & 0.0055 & 0.0377  \\
59270.69924 & -50.3761 & 0.0055 & -0.0012 \\
59275.59393 & -50.3743 & 0.0054 & 0.0274  \\
59277.59827 & -50.3734 & 0.0074 & 0.0184  \\
59280.65648 & -50.3834 & 0.0055 & 0.0178  \\
59305.65517 & -50.3671 & 0.0054 & 0.0327  \\
59306.60684 & -50.3828 & 0.0055 & 0.0559  \\
59328.56457 & -50.3846 & 0.0054 & 0.0291  \\
59336.47003 & -50.3566 & 0.0097 & 0.0429  \\
59337.53175 & -50.3822 & 0.0055 & 0.0401  \\
59347.50425 & -50.3871 & 0.0055 & 0.0499  \\
59357.53955 & -50.3747 & 0.0055 & 0.0389  \\
59359.36437 & -50.3868 & 0.0054 & 0.0399  \\
59362.49265 & -50.3815 & 0.0053 & 0.0469  \\
59391.42075 & -50.3763 & 0.0063 & -0.008  \\
59393.40956 & -50.3769 & 0.0054 & 0.0032  \\
59395.41575 & -50.3879 & 0.0055 & 0.0363  \\
59403.51646 & -50.3805 & 0.0058 & 0.0516  \\
59405.47363 & -50.3719 & 0.0055 & 0.0134  \\
59407.4076  & -50.3821 & 0.0054 & 0.0409  \\
59417.45531 & -50.3833 & 0.0055 & 0.0257  \\
59419.37087 & -50.3594 & 0.0057 & 0.0219  \\
59423.43756 & -50.3845 & 0.0054 & 0.0468  \\
59439.44842 & -50.3775 & 0.0053 & -0.0056 \\
59441.38508 & -50.374  & 0.0054 & 0.0169  \\
59445.37278 & -50.3961 & 0.0054 & 0.0343  \\
59455.33369 & -50.3614 & 0.0054 & 0.0047  \\
59457.36665 & -50.3852 & 0.0058 & 0.0598  \\
59466.38033 & -50.3698 & 0.0116 & 0.0707  \\
59468.40794 & -50.3744 & 0.006  & 0.0093  \\
59470.35499 & -50.3706 & 0.0054 & 0.0304  \\
59476.28437 & -50.3519 & 0.0072 & -0.0288 \\
59502.28207 & -50.3953 & 0.0054 & 0.0305  \\
59506.24934 & -50.3777 & 0.0054 & 0.0136  \\
59509.25819 & -50.379  & 0.0067 & 0.0634  \\
\hline
\multicolumn{4}{|r|}{{Continued on next page}} \\ \hline
\end{tabular}%
}
\end{table}

\begin{table}[h!]
\caption*{SOPHIE RVs for TOI-1836 (continued from the previous page).}
\label{tab:rv_18362}
\resizebox{0.46\textwidth}{!}{%
\begin{tabular}{llll}
\hline
BJD (-2400000 d)&       RV (km s$^{-1}$ )  & $\sigma_{RV}$ (m s$^{-1}$ )& BIS (km s$^{-1}$ )  \\
\hline
59601.66618 & -50.3608 & 0.0055 & 0.0192  \\
59605.67569 & -50.3828 & 0.0054 & 0.0362  \\
59606.66983 & -50.3826 & 0.0055 & 0.0212  \\
59607.68505 & -50.3954 & 0.0097 & -0.0174 \\
59609.67266 & -50.377  & 0.0067 & 0.0386  \\
59610.66509 & -50.3678 & 0.0065 & 0.076   \\
59620.66327 & -50.3769 & 0.0055 & -0.0013 \\
59622.67958 & -50.3703 & 0.0069 & 0.0144  \\
59623.68697 & -50.38   & 0.0054 & 0.0482  \\
59627.64632 & -50.3739 & 0.0064 & 0.0505  \\
59629.66005 & -50.3983 & 0.0061 & 0.0642  \\
59630.65587 & -50.3729 & 0.0081 & 0.0666  \\
59648.67556 & -50.4022 & 0.0054 & 0.0282  \\
59661.62558 & -50.3872 & 0.0054 & 0.0402  \\
59662.58554 & -50.3507 & 0.0054 & 0.0293  \\
59663.56562 & -50.3766 & 0.0054 & 0.0497  \\
59683.5669  & -50.3568 & 0.0054 & 0.0296  \\
59685.53689 & -50.3786 & 0.0055 & 0.0351  \\
59713.49222 & -50.3992 & 0.0054 & 0.0047  \\
59750.5952  & -50.3855 & 0.0054 & 0.0111  \\
59753.60296 & -50.3986 & 0.0088 & 0.0317  \\
59978.69629 & -50.3905 & 0.0055 & 0.03    \\
60007.63826 & -50.3813 & 0.0055 & 0.0458  \\
60008.6467  & -50.3802 & 0.0055 & -0.0106 \\
60042.52147 & -50.4002 & 0.0055 & 0.0056  \\
60044.57315 & -50.3697 & 0.0058 & 0.0109  \\
60046.58793 & -50.3878 & 0.0055 & 0.0055  \\
60072.49979 & -50.3894 & 0.0055 & 0.0295  \\
60091.40991 & -50.3915 & 0.0054 & 0.0249  \\
60096.45066 & -50.3848 & 0.0054 & 0.0339  \\
60142.44707 & -50.3893 & 0.0054 & 0.0377  \\
60156.4321  & -50.3912 & 0.0058 & 0.0199  \\
60172.33236 & -50.372  & 0.0055 & 0.0199  \\
60188.34481 & -50.3851 & 0.0055 & 0.0205  \\
60214.29835 & -50.3757 & 0.0054 & 0.0347  \\
60221.27995 & -50.4022 & 0.0054 & 0.0364\\
\hline
\end{tabular}%
}
\end{table}

\begin{table}
\caption{SOPHIE RVs for TOI-4081.}
\label{tab:rv_4081}
\resizebox{0.46\textwidth}{!}{%
\begin{tabular}{llll}
\hline
BJD (-2400000 d)&       RV (km s$^{-1}$ )  & $\sigma_{RV}$ (m s$^{-1}$ )& BIS (km s$^{-1}$)  \\
\hline
59454.557   & -15.1026 & 0.0316 & -0.3034 \\
59455.60477 & -14.8966 & 0.03   & -0.1224 \\
59456.6159  & -14.8183 & 0.0292 & -0.8567 \\
59475.50802 & -15.1445 & 0.0308 & -0.079  \\
59477.58705 & -15.2778 & 0.0295 & -0.3842 \\
59484.51116 & -15.0639 & 0.0302 & -0.2344 \\
59499.56673 & -15.1777 & 0.0348 & -0.1336 \\
59502.47183 & -15.026  & 0.03   & -0.4233 \\
59503.48684 & -15.0543 & 0.0323 & -0.0442 \\
59506.42825 & -15.1941 & 0.0292 & 0.0428  \\
59525.45583 & -15.3783 & 0.0421 & -0.2587 \\
59526.5083  & -15.326  & 0.0322 & -0.2175 \\
59542.57829 & -15.0436 & 0.0486 & -0.645  \\
59561.41295 & -15.2674 & 0.0308 & -0.6152 \\
59562.3428  & -15.1622 & 0.0303 & -0.4367 \\
59563.36096 & -15.2253 & 0.029  & -0.5726 \\
59564.34673 & -15.1953 & 0.0315 & -0.1347 \\
59567.30479 & -14.9646 & 0.03   & 0.0001  \\
59815.57211 & -15.1274 & 0.031  & -0.2692 \\
59827.64422 & -14.987  & 0.0311 & -0.1172 \\
59865.50246 & -14.8557 & 0.0304 & -0.611  \\
59880.59643 & -15.1495 & 0.0311 & -0.5835 \\
59882.40813 & -14.762  & 0.038  & -0.1297 \\
59915.31261 & -15.2625 & 0.0405 & -0.332  \\
59919.33761 & -14.9432 & 0.0308 & -0.3473 \\
59937.313   & -15.0402 & 0.0309 & 0.0714  \\
59978.29687 & -15.1178 & 0.0342 & -0.5303 \\
60158.59521 & -14.6917 & 0.0299 & -0.2697 \\
60160.61016 & -14.7547 & 0.0329 & -0.5631 \\
60169.6205  & -14.8161 & 0.0303 & -0.3306 \\
60174.61814 & -15.0017 & 0.0286 & -0.1247 \\
60211.46642 & -15.1377 & 0.0428 & -0.2555 \\
60212.47724 & -14.9296 & 0.0315 & -0.4124 \\
60220.45764 & -14.8728 & 0.0321 & -0.0727 \\
60229.48857 & -15.0354 & 0.0329 & -0.2655 \\
60230.50006 & -14.9166 & 0.0308 & -0.3517 \\
60242.42848 & -14.6109 & 0.0324 & -0.0147 \\
60245.49554 & -14.7379 & 0.0299 & -0.2167\\
\hline
\end{tabular}%
}
\end{table}

\onecolumn

\section{Ground-based light curves}
\label{lc_ground}

\begin{figure*}[h!]
\centering
\includegraphics[width=0.48\columnwidth]{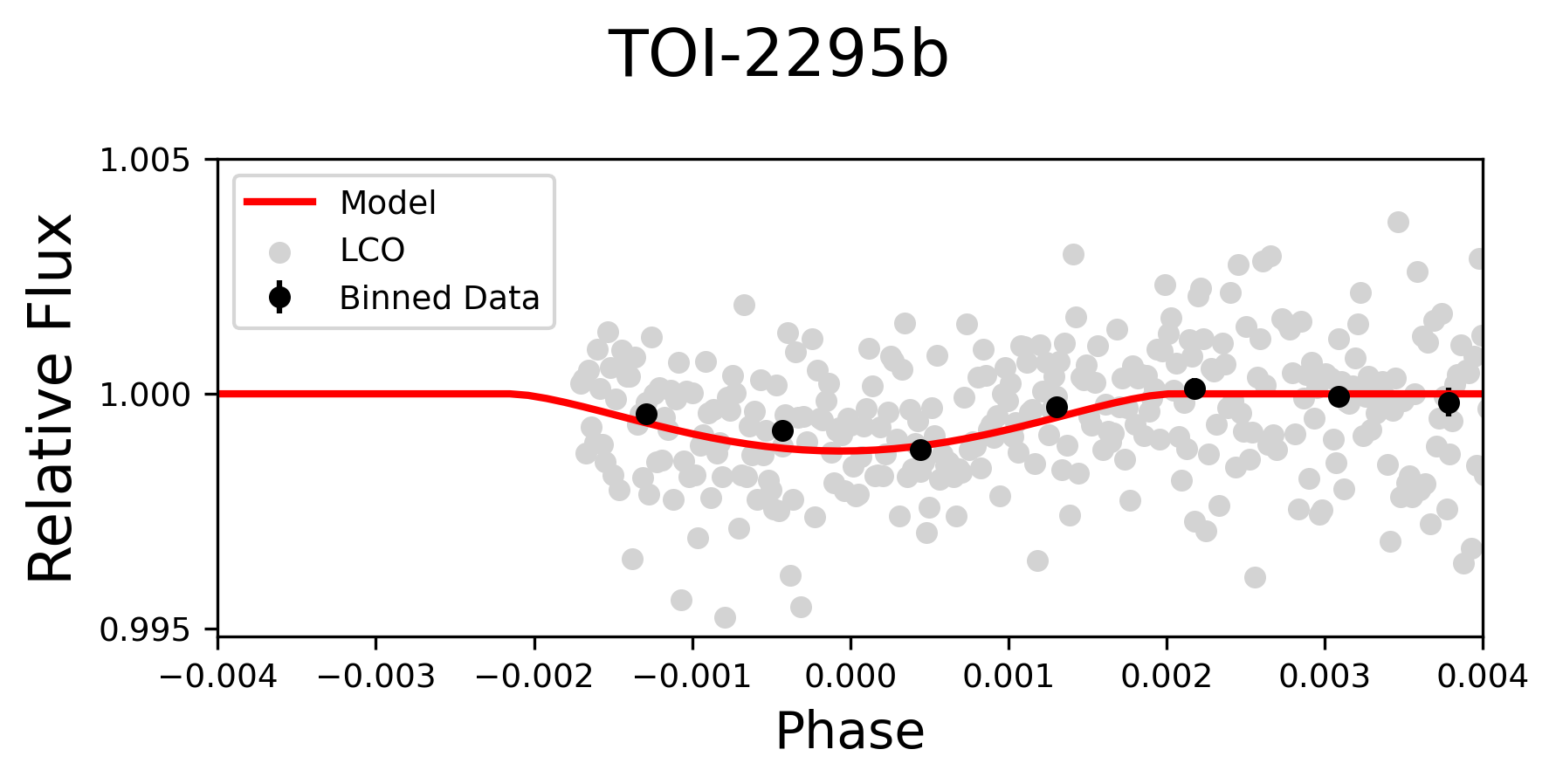}\\
\includegraphics[width=0.48\columnwidth]{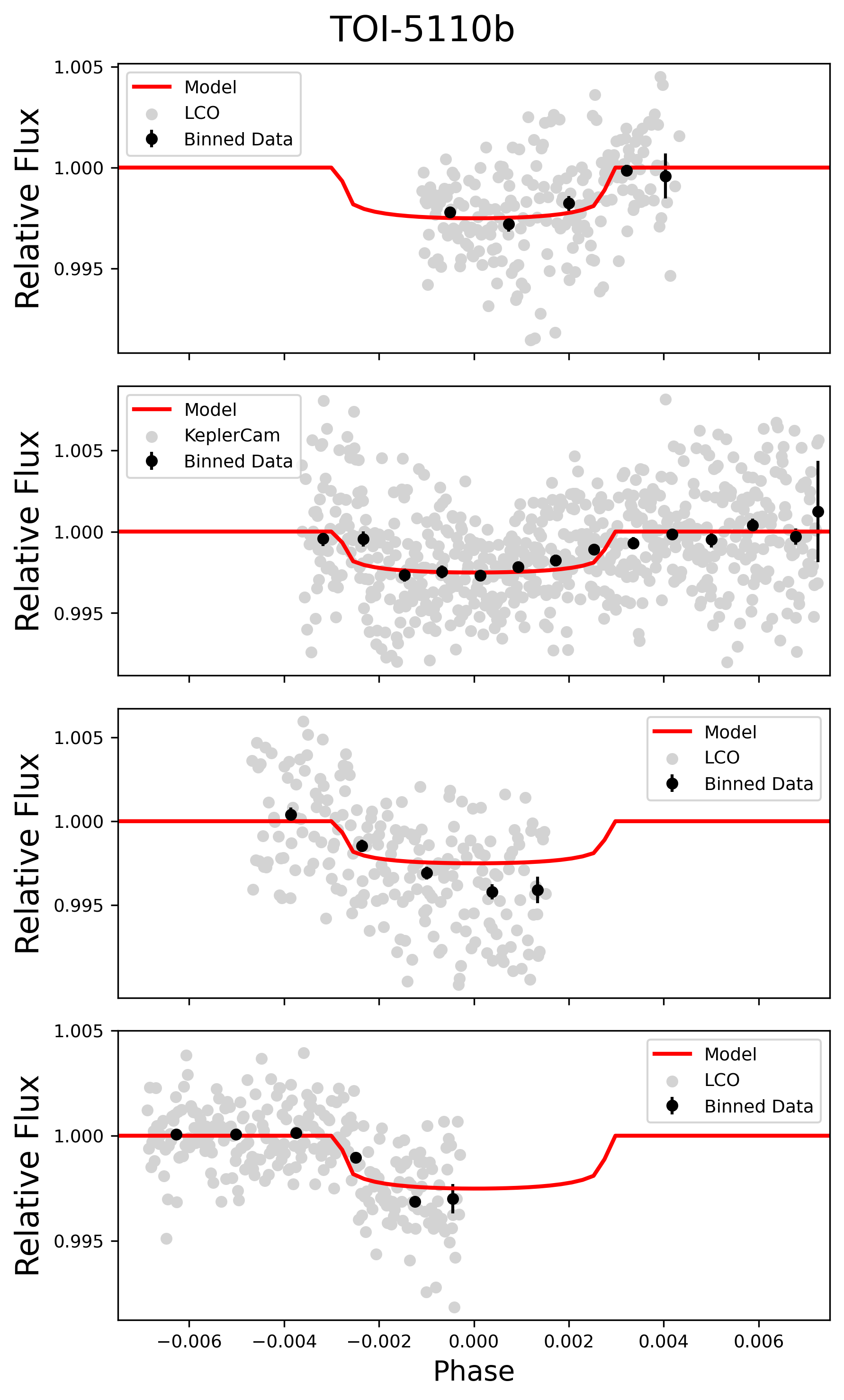}
\includegraphics[width=0.48\columnwidth]{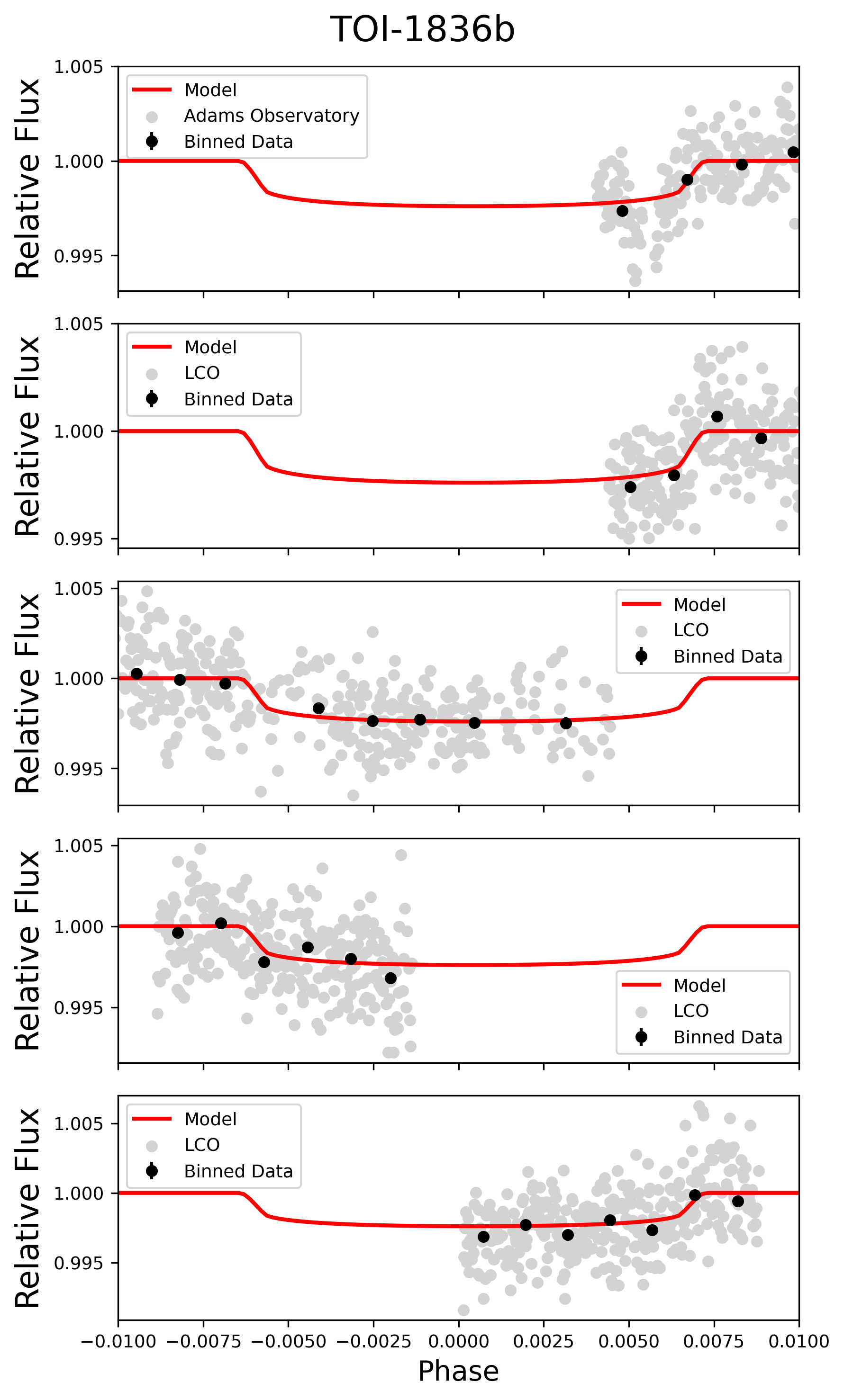}
\caption{Phase-folded ground-based light curves of TOI-2295b, TOI-5110b, and TOI-1836 are shown as gray points. These data are not included in the modeling. The solid red line represents the median model based on the TESS light curve. The data points are binned with a bin size of $\sim$ 0.001 phase.}
\label{lc_more_1836}
\end{figure*}

\section{Priors on RV-only models and joint analysis}

\begin{sidewaystable}
\centering
\caption{Priors and description of parameters used within \texttt{juliet} for RV-only models.}
\resizebox{0.9\textwidth}{!}{%
\begin{tabular}{lcccccccc}
\hline
Parameters & Units & TOI-1836 & TOI-2295 & TOI-2537 & TOI-4081 & TOI-4168 & TOI-5076 & TOI-5110 \\
\hline
Planet parameters: & & .01 & b & b & .01 & B & .01 & b \\
P & Period (d) & $\mathcal{U}(5,40)$ & $\mathcal{U}(1,50)$ & $\mathcal{U}(1,170)$ & $\mathcal{U}(1,15)$ & $\mathcal{U}(1,50)$ & 23.444 (fixed) & $\mathcal{U}(1,50)$ \\
T$_{c}$-2400000 & Center of transit time (d) & $\mathcal{U} (59630,59664)$ & $\mathcal{U} (58688,58738)$ & $\mathcal{U} (58355,58525)$ & $\mathcal{U} (2459729,2459743)$ & $\mathcal{U} (59380,59420)$ & $\mathcal{U} (60193,60215)$ & $\mathcal{U} (59479,59529)$ \\
K & RV semi-amplitude (m/s) & $\mathcal{U}(0,60)$ & $\mathcal{U}(0,70)$ & $\mathcal{U}(0,200)$ & $\mathcal{U}(0,220)$ & $\mathcal{U}(0,30000)$ & $\mathcal{U}(0,30)$ & $\mathcal{U}(0,300)$ \\
e & Eccentricity & 0 (fixed) & $\mathcal{U}(0,1)$ & $\mathcal{U}(0,1)$ & 0 (fixed) & $\mathcal{U}(0,1)$ & 0 (fixed) & $\mathcal{U}(0,1)$ \\
$\omega$ & Argument of periastron (degrees) & 90 (fixed) & $\mathcal{U}(0,360)$ & $\mathcal{U}(0,360)$ & 90 (fixed) & $\mathcal{U}(0,360)$ & 90 (fixed) & $\mathcal{U}(0,360)$ \\
\\
Second planet parameters: & & --- & c & c & --- & --- & --- & --- \\
P & Period (d) & --- & $\mathcal{U}(700,1500)$ & $\mathcal{U}(800,5000)$ & --- & --- & --- & --- \\
T$_{c}$-2400000 & Center of transit time (d) & --- & $\mathcal{U} (58711,59611)$ & $\mathcal{U} (59047,61047)$ & --- & --- & --- & --- \\
K & RV semi-amplitude (m/s) & --- & $\mathcal{U}(0,170)$ & $\mathcal{U}(0,200)$ & --- & --- & --- & --- \\
e & Eccentricity & --- & $\mathcal{U}(0,1)$ & $\mathcal{U}(0,1)$ & --- & --- & --- & --- \\
$\omega$ & Argument of periastron (degrees) & --- & $\mathcal{U}(0,360)$ & $\mathcal{U}(0,360)$ & --- & --- & --- & --- \\
\\
Drift on SOPHIE: & & & && & & &\\
A & Linear RV drift (m/s) & --- & --- & --- & $\mathcal{U}(-100.0,100.0)$ & --- & --- & --- \\
\\
Telescope Parameters:& & & && & & &\\
$\sigma_{SOPHIE}$ & SOPHIE RV jitter (m/s) & $\mathcal{U} (1e-3, 200.)$ & $\mathcal{U} (1e-3, 200.)$ & $\mathcal{U} (1e-3, 200.)$ & $\mathcal{U} (1e-3, 200.)$ & $\mathcal{U} (1e-3, 100.)$ & $\mathcal{U} (1e-3, 100.)$ & $\mathcal{U} (1e-3, 200.)$ \\
mu$_{SOPHIE}$ & SOPHIE instrumental offset (m/s) & $\log \mathcal{U} (-50482,-50282)$ & $\log \mathcal{U} (-38819,-38600)$ & $\log \mathcal{U} (61489,61689)$ & $\log \mathcal{U} (-17000,-14500)$ & $\log \mathcal{U} (-28557,-28357)$ & $\log \mathcal{U} (70000,80000)$ & $\log \mathcal{U} (3925,4125)$ \\
$\sigma_{HARPS}$ & HARPS RV jitter (m/s) & --- & --- & $\mathcal{U} (1e-3, 200.)$ & --- & --- & --- & --- \\
mu$_{HARPS}$ & HARPS instrumental offset (m/s) & --- & --- & $\log \mathcal{U} (61561,61761)$ & --- & --- & --- & --- \\
$\sigma_{FEROS}$ & FEROS RV jitter (m/s) & --- & --- & $\mathcal{U} (1e-3, 200.)$ & --- & --- & --- & --- \\
mu$_{FEROS}$ & FEROS instrumental offset (m/s) & --- & --- & $\log \mathcal{U} (61528,61728)$ & --- & --- & --- & --- \\
\hline
\end{tabular}%
}
\\
\centering
\tablefoot{The prior labels of $\mathcal{U}$ and $\log \mathcal{U}$ indicate uniform, and uniform logarithms of distributions, respectively.}
\label{prior_rv-only}
\end{sidewaystable}

\begin{sidewaystable}
\centering
\caption{Adopted priors in joint modeling using \exofasttwo. $\mathcal{N}[a,b]$ are Gaussian priors, where $a$ and $b$ are the mean and width, respectively.}
\resizebox{0.9\textwidth}{!}{%
\begin{tabular}{lcccccccc}
\multicolumn{4}{c}{} \\
\hline
Parameter & Units & TOI-1836 & TOI-2295& TOI-2237& TOI-4081 & TOI-4168 & TOI-5076 &TOI-5110\\ \hline
\multicolumn{4}{l}{Stellar Parameters:} \\
$M_*$ & Mass (\(M_{\odot}\)) & $\mathcal{N}[1.29,0.08]$ &$\mathcal{N}$[1.17, 0.07]&$\mathcal{N}$[0.770, 0.05]&$\mathcal{N}$[1.44, 0.09] &---&$\mathcal{N}$[0.82, 0.05]&$\mathcal{N}$[1.46, 0.09] \\
$R_*$ & Radius (\(R_{\odot}\)) & $\mathcal{N}[1.611, 0.068]$ &$\mathcal{N}[1.451, 0.061]$&$\mathcal{N}$[0.774, 0.05]&$\mathcal{N}$[2.511, 0.105]&---&$\mathcal{N}$[0.798, 0.036]&$\mathcal{N}$[2.359, 0.099] \\
$T_{\rm eff}$&Effective Temperature (K)&$\mathcal{N}[6369, 153]$ &$\mathcal{N}[5733, 138]$&$\mathcal{N}$[4843, 153]&$\mathcal{N}$[6040, 145]&&$\mathcal{N}$[4832, 119]&$\mathcal{N}$[6154, 148]\\
$[{\rm Fe/H}]$ &Metallicity (dex)&$\mathcal{N}[-0.098, 0.08]$&$\mathcal{N}[0.316, 0.08]$&$\mathcal{N}$[0.081, 0.08]&$\mathcal{N}$[0.0,0.3]&---&$\mathcal{N}$[0.07,0.06]&$\mathcal{N}$[0.067, 0.050]\\
Planetary parameters: & &.01 & b& b& .01& B &.01&b \\
$P$ & Period (days) & $\mathcal{N}[20.4, 0.1]$&$\mathcal{N}[30.0, 0.1]$&$\mathcal{N}$[94.1, 0.5] &$\mathcal{N}$[9.3,0.1]&---&$\mathcal{N}$[23.4,0.1]&$\mathcal{N}$[30.1,0.1]\\
$T_C$ & Time of conjunction$^{1}$ (BJD$_{TDB}$) & $\mathcal{N}[2459646.5, 0.1]$ &$\mathcal{N}[2458713.5, 0.1]$&$\mathcal{N}$[2458440.3, 0.5]&$\mathcal{N}$[2459736.331232,0.1]&---&$\mathcal{N}$[60204.0,0.1]&$\mathcal{N}$[2459503.7, 0.1]\\
$R_P/R_*$&Radius of planet in stellar radii &---&$\mathcal{N}[0.03,0.21]$&---&---&---&---&----\\
\hline
\end{tabular}%
}
\label{prior_exofast}
\end{sidewaystable}
\vspace{-0.1cm}
\section{Wavelength, telescope, and transit parameters derived from \exofasttwo{} fit}

\vspace{-0.1cm}
\begin{sidewaystable}[htbp]
\centering
\caption{Wavelength, telescope, and transit parameters derived from \exofasttwo{} fit.}
\scalebox{0.8}{
\begin{tabular}{lcccccccc}
\hline
Parameter & Units & TOI-1836 & TOI-2295& TOI-2237& TOI-4081 & TOI-4168 & TOI-5076 &TOI-5110\\
\hline
Wavelength Parameters: & & & & & & & & \\
u$_1$-TESS & linear limb-darkening coeff & $0.217^{+0.029}_{-0.030}$ & $0.326\pm0.054$ & $0.447^{+0.035}_{-0.036}$ & $0.263^{+0.041}_{-0.040}$ & $0.289^{+0.027}_{-0.024}$ & $0.687\pm0.051$ & $0.241\pm0.038$ \\
u$_2$-TESS & quadratic limb-darkening coeff & $0.293\pm0.033$ & $0.276\pm0.052$ & $0.182\pm0.033$ & $0.303\pm0.036$ & $0.286^{+0.019}_{-0.020}$ & $0.148^{+0.051}_{-0.052}$ & $0.302\pm0.036$ \\
u$_1$-CHEOPS & & & & & & & $0.468^{+0.038}_{-0.037}$ & \\
u$_2$-CHEOPS & & & & & & & $0.172^{+0.036}_{-0.037}$ & \\
A$_D$ & Dilution from neighboring stars & --- & --- & --- & $0.0660\pm0.0020$ & $0.0231^{+0.0059}_{-0.0048}$ & --- & --- \\
\multicolumn{9}{l}{Telescope Parameters:} \\
$\gamma_{\text{rel}}$-SOPHIE & Relative RV Offset (m/s) & $-50381.5\pm1.4$ & $-38733.10^{+0.84}_{-0.81}$ & $61596.8^{+8.3}_{-12}$ & $-15016^{+20}_{-19}$ & $-29142^{+14}_{-13}$ & $70291.5\pm1.9$ & $4012.2\pm2.6$ \\
$\sigma_J$-SOPHIE & RV Jitter (m/s) & $11.1^{+1.2}_{-1.1}$ & $3.83^{+0.81}_{-0.69}$ & $14.2^{+4.1}_{-3.8}$ & $111^{+17}_{-15}$ & $43^{+15}_{-10}$ &$11.0^{+1.7}_{-1.4}$ & $10.1^{+2.8}_{-2.3}$ \\
$\sigma_J^2$-SOPHIE & RV Jitter Variance (m/s) &$124^{+28}_{-23}$ & $14.7^{+6.8}_{-4.8}$ & $202^{+130}_{-94}$ & $12400^{+4200}_{-3000}$ & $1880^{+1600}_{-790}$ & $121^{+39}_{-28}$ & $102^{+64}_{-41}$\\
$\gamma_{\text{rel}}$-HARPS & Relative RV Offset (m/s) & --- & --- & $61645.3^{+9.3}_{-13}$ & --- & --- & --- & --- \\
$\sigma_J$-HARPS & RV Jitter (m/s) & --- & --- & $12.2^{+4.9}_{-4.5}$ & --- & --- & --- & --- \\
$\sigma_J^2$-HARPS & RV Jitter Variance (m/s) & --- & --- & $148^{+140}_{-89}$ & --- & --- & --- & --- \\
$\gamma_{\text{rel}}$-FEEROS & Relative RV Offset (m/s) & --- & --- & $61621^{+19}_{-20}$& --- & --- & --- & --- \\
$\sigma_J$-FEEROS & RV Jitter (m/s) & --- & --- & $65^{+14}_{-11}$& --- & --- & --- & --- \\
$\sigma_J^2$-FEEROS & RV Jitter Variance (m/s) & --- & --- & $4300^{+2100}_{-1300}$ & --- & --- & --- & --- \\
$\gamma_{\text{rel}}$-HARPS-N & Relative RV Offset (m/s) & --- & --- & --- & --- & --- & $70279.15\pm0.49$ & --- \\
$\sigma_J$-HARPS-N & RV Jitter (m/s) & --- & --- & --- & --- & --- & $2.19^{+0.60}_{-0.58}$ & --- \\
$\sigma_J^2$-HARPS-N & RV Jitter Variance (m/s) & --- & --- & --- & --- & --- & $4.8^{+3.0}_{-2.2}$ & --- \\
\multicolumn{9}{l}{Transit Parameters:} \\
$\sigma^2$-FFI$^{1}$ & Added Variance & $0.0126^{+0.0082}_{-0.0069}$ & --- & see Table \ref{tab:transitpars} & $-0.086^{+0.011}_{-0.010}$ & --- & $0.0174^{+0.0035}_{-0.0034}$&$0.0217^{+0.0063}_{-0.0060}$\\
$F_0$-FFI & Baseline flux & $1.0002\pm0.0036$ & --- & see Table \ref{tab:transitpars} & $1.0002^{+0.0032}_{-0.0033}$ & --- & $1.0000\pm0.0016$ & $1.0000069\pm0.0000074$ \\
$\sigma^{2}$-2 minutes$^{1}$ & Added Variance & $0.0229 \pm 0.0059$ & $0.0506 \pm 0.0045$ & see Table \ref{tab:transitpars} & $-0.128 \pm 0.032$ & --- &$0.00^{+0.0043}_{-0.0040}$ & $0.604^{+0.029}_{-0.028}$ \\
$F_0$-2 minutes & Baseline flux & $1.0002\pm0.0012$ &$1.0000\pm0.0011$ & see Table \ref{tab:transitpars} & $1.0001\pm0.0027$& $1.0000\pm0.0016$ & $1.0000\pm0.0037$ & $1.0000\pm0.0017$ \\
$\sigma^{2}$-CHEOPS$^{1}$ & Added Variance & --- & --- & --- & --- & --- & $0.685^{+0.041}_{-0.038}$ & --- \\
$F_0$-CHEOPS & Baseline flux & --- & --- & --- & --- & --- & $1.0004\pm0.0020$ & --- \\
\hline
\end{tabular}
}
\centering
\\
\tablefoot{$^{1}$ Units are in ppm. The added variance might be a negative value to account for the overestimated photometric errors \citep[][]{eastman2019}}
\label{tab:TOI_parameters}
\end{sidewaystable}

\section{Probability distribution function of TOI-2295b}

\begin{figure*}[h!]
\centering
\includegraphics[width=0.8\columnwidth]{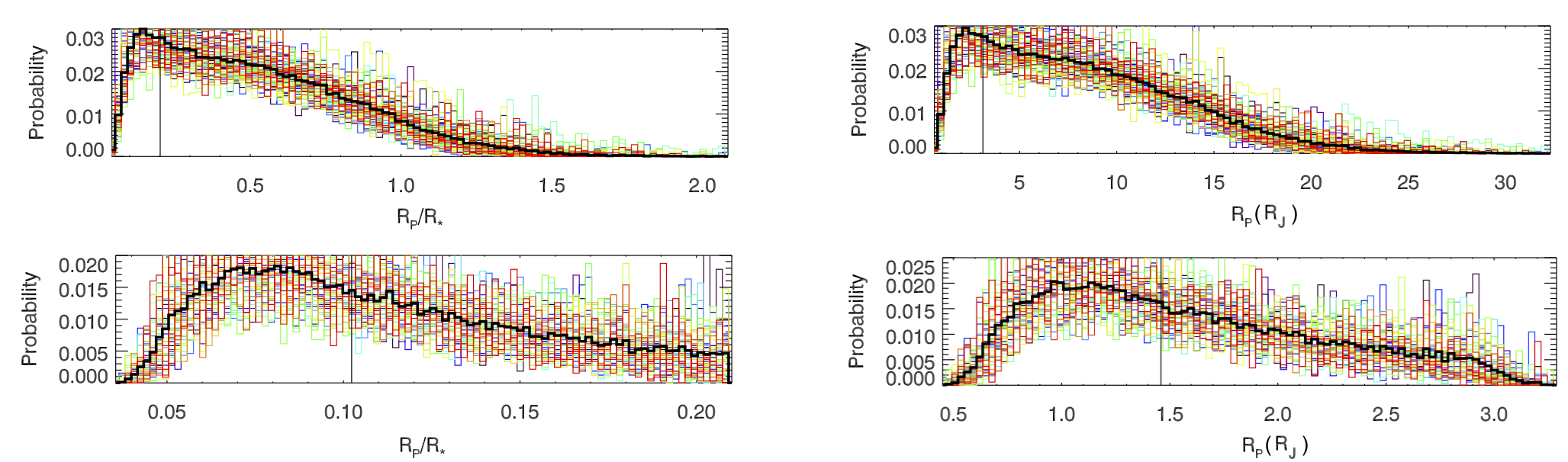}
\caption{Probability distribution function for $R_P/R_*$ and $R_P$ for TOI-2295b, without applying prior on $R_P/R_*$ (\emph{top}) and with applying prior (\emph{bottom}; see Sect. \ref{join_2295}). Each color corresponds to an individual chain, with the thick black probability representing the average of all chains. The thin black vertical line indicates the median value.}
\label{corner_2295}
\end{figure*}

\section{Gaia mass constrain of TOI-2295c and TOI-2537c}

\begin{figure}[h!]
    \centering
    \includegraphics[width=89.3mm,clip=true]{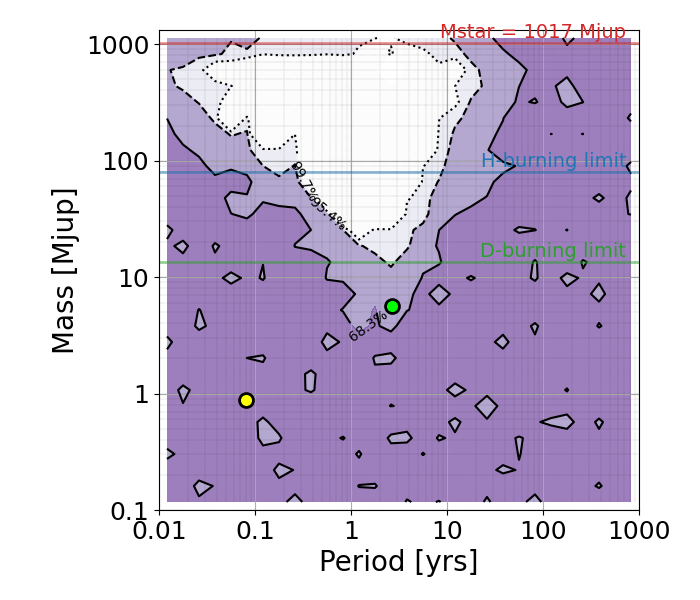}
    \includegraphics[width=89.3mm,clip=true]{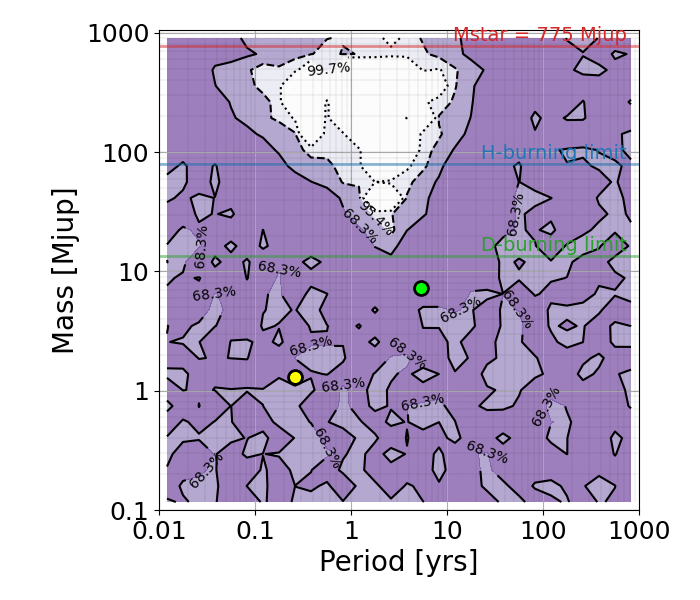}
    \caption{GaiaPMEX confidence map for the mass and orbital period of planets around TOI-2295 (\emph{left}) and TOI-2537 (\emph{right}), constrained by RUWE=0.919 and RUWE=1.06, respectively. The yellow and green circles indicate the mass and orbital period of planets b and c, respectively, as determined from the joint analysis.}
    \label{fig:PMEXresult_TOI2295}
\end{figure}

\vspace{-0.5cm}
\section{RV residuals-bisector}
\vspace{-0.5cm}
\begin{figure*}
\centering
\includegraphics[width=0.8\columnwidth]{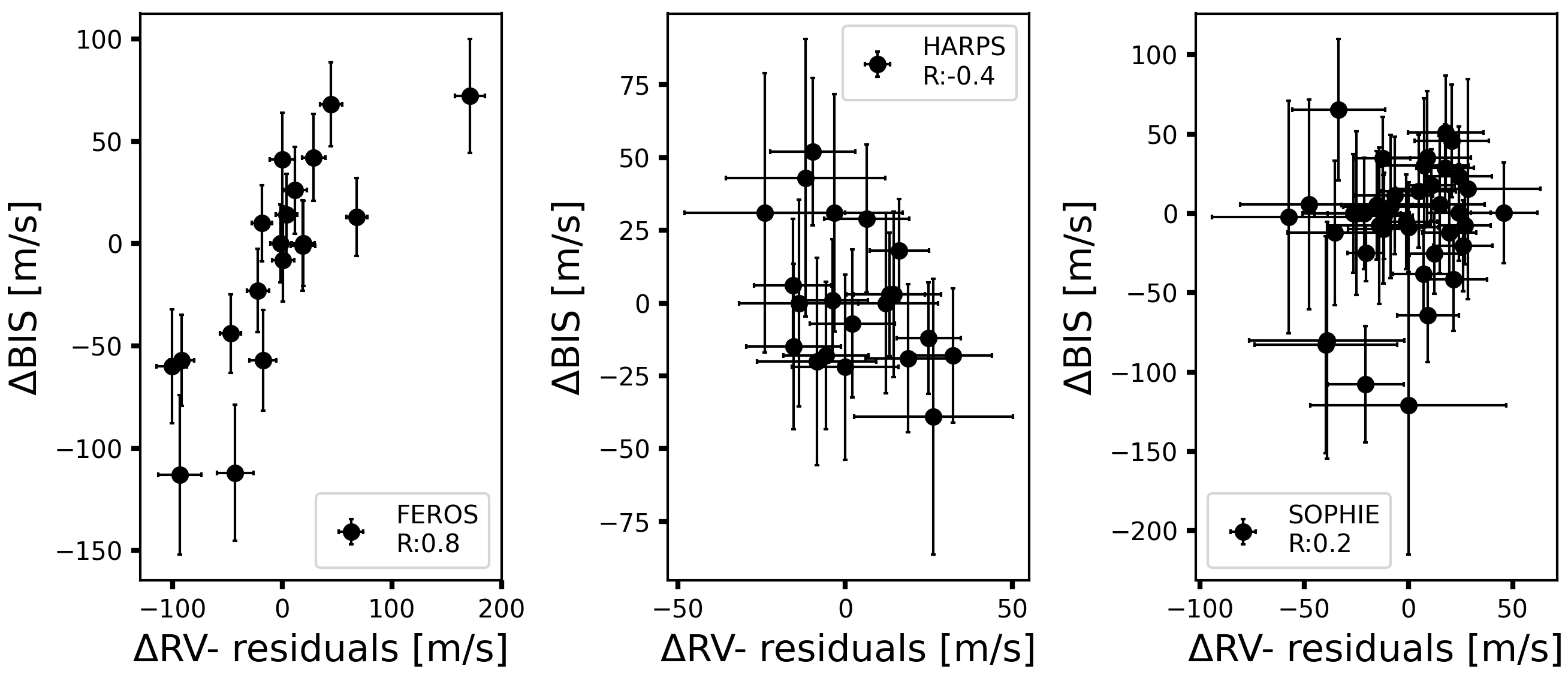}
\caption{RV residuals after removing both planet signals versus bisectors for TOI-2537 data from FEROS, HARPS, and SOPHIE. The instrument names are indicated on the labels.}
\label{rv_bis_res_2537}
\end{figure*}

\section{TTVs of TOI-2537b}

\begin{figure*}[htbp]
\centering
\includegraphics[width=0.4\columnwidth]{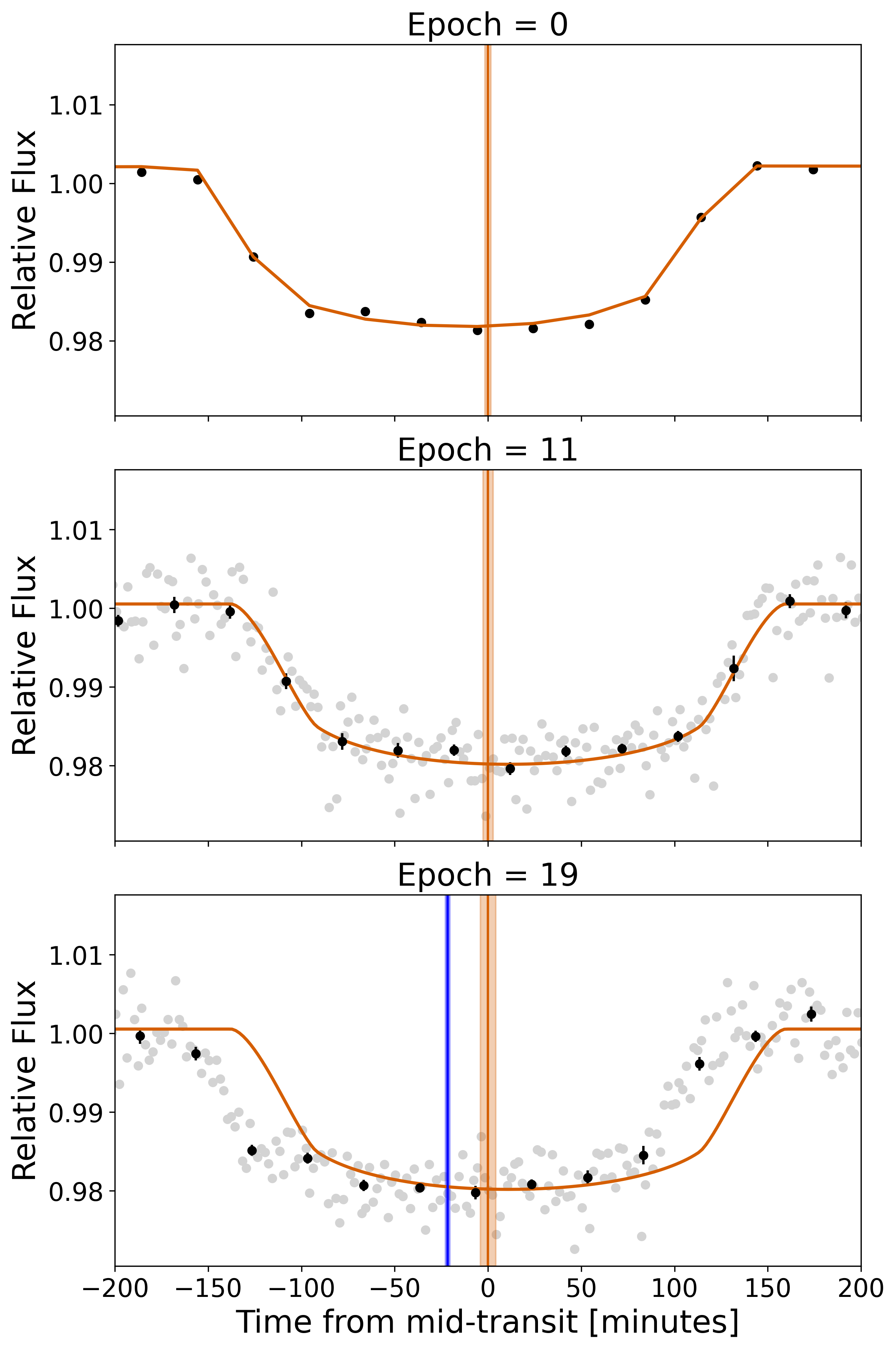}
\caption{Individual TESS transit events of TOI-2537b. The model plotted here is fitted to the first two transits (shown in the \emph{top} and \emph{middle} panels) and is then used to estimate the model for the third detected transit (displayed in the \emph{bottom} panel). Vertical orange lines indicate the predicted mid-transit times, along with their 1$\sigma$ uncertainties, calculated using a linear ephemeris: $T_c = T_0 + E \times P$, where \(T_0\) = 2458440.3331 $\pm$ 0.0011 BJD is the time of conjunction at an arbitrary reference epoch, \(E\) is the epoch, and \(P\) = 94.10245 $\pm$ 0.00014 d is the period. These \(T_0\) and \(P\) values are derived from the fit to the first two transits. The vertical blue line represents the mid-transit time and its 1$\sigma$ uncertainties measured by \texttt{EXOFASTv2} for the \emph{bottom} panel. The transit epochs are labeled at the top of each panel. This plot clearly illustrates the presence of TTVs. Unlike the current fit shown here, the model plotted in Fig.~\ref{TOI2537_rv_phot} and presented in Table.~\ref{toi_2295_2537_5110} accounts for TTVs.
}
\label{lightcurves_ttv_2537}
\end{figure*}
\vspace{-0.5cm}

\begin{table*}
\centering
\caption{Median values and 68\% confidence intervals of the parameters for individual transits of TOI-2537 from the \exofasttwo{} fit. The TTV column shows the difference between the observed (third column) and predicted mid-transit times. The predicted mid-transit times are estimated using a linear ephemeris, with \(T_0 = 2458440.3302 \pm 0.0010\) BJD and \(P = 94.102091 \pm 0.000073\) d, which were obtained by minimizing the covariance between these two parameters. Further details of these calculations are provided in Sect. 18 of \cite{eastman2019}.}
\resizebox{0.9\textwidth}{!}{%
\begin{tabular}{cccccc}
\hline
Transit & Epoch & Mid-transit (BJD) & Added variance$^{1}$ ($\sigma^{2}$) & Baseline flux ($F_0$) & TTV (min) \\
\hline
TESS UT 2018-11-17 (TESS) & 0 & $2458440.3237 \pm 0.0011$ &$-0.094^{+0.071}_{-0.063}$
 & $1.0006^{+0.0054}_{-0.0055}$& $-9.38 \pm 2.16$\\
TESS UT 2021-09-17 (TESS) & 11 & $2459475.4661 \pm 0.0010$ &$-1.95^{+0.29}_{-0.28}$
 & $1.000554^{+0.000057}_{-0.000058}$&$18.56 \pm 2.35$ \\
TESS UT 2023-10-10 (TESS) & 19 & $2460228.26460 \pm 0.00084$ &$-1.33^{+0.55}_{-0.51}$
 & $1.00073 \pm 0.00013$& $-7.69 \pm 2.76$ \\
\hline
\end{tabular}%
}
\tablefoot{$^{1}$ Units are in ppm. The added variance might be a negative value to account for the overestimated photometric errors \citep[][]{eastman2019}}
\label{tab:transitpars}
\end{table*}

\end{appendix}

\end{document}